\documentclass[aps,prd,10pt,twocolumn,showpacs,showkeys,preprintnumbers,floatfix,nofootinbib,superscriptaddress,longbibliography]{revtex4-2}

\usepackage{amsmath}
\usepackage{amssymb}
\usepackage{graphicx}
\usepackage{dcolumn}
\usepackage{bm}
\usepackage{bbm}
\usepackage{hyperref}
\usepackage{epsfig}
\usepackage{mathrsfs}
\usepackage[T1]{fontenc} 
\usepackage{xcolor}
\usepackage{booktabs} 
\usepackage{makecell} 
\usepackage{adjustbox}
\usepackage{slashed}

\newcommand{\tphy}{\mathop{t_0^{\rm Phy}}}
\newcommand{\wphy}{\mathop{w_0^{\rm Phy}}}
\newcommand{\fpifk}{\mathop{f_K / f_\pi}}
\newcommand{\fpiK}{\mathop{f_{\pi K }}}
\newcommand{\phiss}{\mathop{\phi_{s {\overline s} }}}
\newcommand{\Meff}{\mathop{M_{\rm eff}}}
\newcommand{\CPT}{\mathop{\chi}PT}
\newcommand{\MSbar}{\mathop{\overline{\rm MS}}}
\newcommand{\acorr}{\mathrm{corr.}}
\newcommand{\anaive}{\mathrm{naive}}
\newcommand{\tgf}{\mathop{t_{gf}}}

\newcommand{\meff}{M^{\rm eff}}

\begin{document}

\preprint{LA-UR-25-32114}

\title{The Spectrum and Scale Setting on 2+1-flavor NME Lattices}



\author{Jun-Sik Yoo}
\email[]{junsik.yoo@outlook.com}
\affiliation{Department of Physics, Korea University, Seoul 02841, Korea}
\author{June-Haak Ee}
\email[]{jhee@mit.edu}
\affiliation{Center for Theoretical Physics – a Leinweber Institute, Massachusetts Institute of Technology, Cambridge, MA 02139, USA}
\author{Sungwoo Park}
\email{sungwoo@sejong.ac.kr (Corresponding Author)}
\affiliation{Physical and Life Sciences Division, Lawrence Livermore National Laboratory, Livermore, CA 94550, USA}
\affiliation{Nuclear Science Division, Lawrence Berkeley National Laboratory, Berkeley, CA 94720, USA}
\author{Rajan Gupta}
\email[]{rg@lanl.gov}
\affiliation{Theoretical Division, Los Alamos National Laboratory, PO Box 1663, MS B283, Los Alamos, NM 87545, USA}
\author{Tanmoy Bhattacharya}
\email[]{tanmoy@lanl.gov}
\affiliation{Theoretical Division, Los Alamos National Laboratory, PO Box 1663, MS B283, Los Alamos, NM 87545, USA}
\author{Santanu~Mondal}
\email[]{santanu.sinp@gmail.com}
\affiliation{Ibsyn Scientific, 75C Park St, Kolkata, India 700016}
\author{B\'alint~Jo\'o}
\email[]{bjoo@nvidia.com}
\affiliation{NVIDIA Corporation, Santa Clara, CA 95051, USA.}

\author{Robert Edwards}
\email[]{edwards@jlab.org}
\affiliation{Jefferson Lab, 12000 Jefferson Avenue, Newport News, Virginia 23606, USA}

\author{Kostas Orginos}
\email[]{kostas@wm.edu}
\affiliation{Jefferson Lab, 12000 Jefferson Avenue, Newport News, Virginia 23606, USA}
\affiliation{Department of Physics, College of William and Mary, Williamsburg, Virginia 23187-8795, USA}

\author{Frank Winter}
\email[]{fwinter@jlab.org}
\affiliation{Jefferson Lab, 12000 Jefferson Avenue, Newport News, Virginia 23606, USA}

%




\date{\today}

\begin{abstract}
This paper describes the thirteen ensembles, named NME, generated with 2+1-flavor Wilson-clover 
fermions by the JLab/W\&M/LANL/MIT/Marseille collaborations, and presents 
an analysis of the meson and baryon spectrum, decay constants $f_\pi$ and $f_K$, flow 
scales $t_0$ and $w_0$, and time histories of the $\Theta$ and Weinberg operators 
under gradient flow. Using these quantities, the  
physical point values of the two flow scales, $\tphy$ and $\wphy$, and the ratio $\fpifk^{\rm Phy}$ 
are determined.  The masses of the octet and decuplet baryons are analyzed using both the 
next-to-leading order (NLO) and the next-next-to-leading order (NNLO) ansatz from heavy baryon chiral 
perturbation theory (HB$\chi$PT). The NNLO fit to the octet baryons, $M_N$, $M_\Sigma$, $M_\Lambda$ and  $M_\Xi$, 
is preferred  while the corresponding fits to the decuplet Omega mass, $M_\Omega$, 
are not distinguished. We also present a study of the autocorrelations 
in the data and show that there is no evidence,  even at large flow time, of the 
freezing of the topological charge or the Weinberg three-gluon operator. 

\end{abstract}


\maketitle




\section{Introduction}
\label{sec:intro}

The starting point of all lattice Quantum Chromodynamics (QCD) calculations is a 
set of ensembles of gauge configurations that cover a large range of values of 
the lattice spacing $a$, quark masses, and lattice volumes $L^3 \times T$  so that results can be 
extrapolated to the physical point defined by the physical quark masses, $\{m_u, m_d, m_s, m_c,m_b\}$,
infinite volume, $L \to \infty$, and the continuum limit, $a \to 0$. 
Over the last ten years, the 
JLab/W\&M/LANL/MIT/Marseille collaborations\footnote{The full list of contributors to lattice generation are Robert Edwards,  David Richards, and Frank Winter (JLAB); Bal\`int Jo\`o (NVIDIA); Kostas Orginos (W\&M); Tanmoy Bhattacharya, Rajan Gupta, Santanu Mondal, Sungwoo Park  and Boram Yoon (LANL); William Detmold, David Murphy, Phiala Shanahan (MIT); and Savvas Zafeiropoulos (Marseille). }
have generated thirteen ensembles of 2+1-flavors of Wilson-clover fermions 
using the Hybrid Monte Carlo (HMC) algorithm~\cite{Duane:1987de}. 
Simulations of 2+1-flavor lattice QCD involve four parameters: the lattice spacing $a$, two quark masses, light ($m_l = (m_u + m_d)/2$) and strange ($m_s$), and 
a large spatial lattice size $L$ with the understanding that for zero temperature 
simulations,  the temporal size $T \ge L$ in all cases. The lattice spacing is 
essentially fixed as a function of the gauge coupling $\beta \equiv 6/g^2$, with a small dependence on the quark masses for 
Wilson fermions because they break chiral symmetry explicitly~\cite{Bhattacharya:2005ss}. Our goal is to use these lattices to calculate the matrix elements of various quark bilinear operators within nucleon states and to also make them available to members of the USQCD collaboration for non-competitive calculations. First results from a subset of these ensembles have been presented in Refs.~\cite{Yoon:2016dij,Yoon:2016jzj,Park:2021ypf,Mondal:2020ela}.

The generated ensembles cover the  range $0.053$ fm $\leq a \leq 0.117$ fm 
in lattice spacing, $3.7 \leq M_\pi L \leq 6.2$ 
in lattice size and four values of the pion mass: 
$M_\pi \approx 290, 230, 175, 135$~MeV. The lattice parameters of these 13 ensembles 
are summarized in Tables~\ref{tab:ensembles} and~\ref{tab:Sparams}, and, in this work, we present an analysis of the meson and baryon spectrum, the decay constants $f_\pi$ and $f_K$, 
the setting of the lattice scale $a$, the behavior of the topological charge 
and Weinberg three-gluon operator under gradient flow, and results for the renormalization 
constants for all five quark bilinear local operators.  

The following sections provide a summary of the ensembles generated and analyzed, the lattice 
methodology for the quantities calculated, the quality of the signal obtained, 
the analysis strategy for the various observables, and the chiral-continuum 
extrapolation ansatz used to obtain data at the physical point, i.e., 
at $M_\pi=135$~MeV, $M_K=495$~MeV, $a = 0$ and $L = \infty$. 
The full calculation, including the generation of the lattices, the inversion of the Dirac operator, and the measurement of the two-point correlation functions was 
done using the CHROMA software package~\cite{Edwards:2004sx}.

\subsection{Methodology: Lattice Generation}
\label{sec:ensembles}

The generation of lattices using the 2+1-flavor Wilson-clover action was carried out 
using the rational hybrid Monte Carlo (RHMC) algorithm~\cite{Duane:1987de} as described in
Ref.~\cite{Yoon:2016jzj}. The action, set up by the JLab/W\&M collaborations~\cite{JLAB:2012}, 
is 
\begin{eqnarray}
    S & =& S_G + S_F \nonumber \\
    S_G (U) & =& \frac{a^4}{6 g^2}  \bigg[  \sum_{x,\mu, \nu} (1-8c_1)\ \Re Tr \big(1 - U_{\mu \nu} (x) \big) \nonumber \\
    &+& \sum_{x,\mu, \nu} \frac{c_1}{u_0^2} \ \Re Tr \big(1 - U^6_{\mu \nu} (x) \big) \bigg] \nonumber \\
    S_F^f (U, {\overline \psi}, \psi) & =& a^4 \sum_f \sum_x {\overline \psi}_f(x) ( D + m_f) \psi_f(x)  \nonumber \\
    &- & \frac{C_{SW}}{4} \sum_f \sum_{x,\mu,\nu}  {\overline \psi}_f(x) \sigma_{\mu \nu} G^{\mu \nu}\psi_f(x) \nonumber \\
    D & =& \frac{1}{2} \sum_{\mu=1}^4 \big[ (\bigtriangledown^+_\mu + \bigtriangledown^-_\mu) \gamma_\mu - 
     a \bigtriangledown^+_\mu  \bigtriangledown^-_\mu \big]
\end{eqnarray}
where $\beta \equiv 6/g^2$ is the gauge coupling, $U_{\mu \nu}$ is the plaquette, 
$U^6_{\mu \nu}$ is the $1 \times 2$ Wilson loop, 
$G_{\mu \nu}$ is the field strength tensor, $\sigma_{\mu \nu} = (1/2) [\gamma_\mu, \gamma_\nu]$,  $\bigtriangledown^\pm$ are the forward/backward covariant derivatives on the lattice,  and 
$m_f$ is the bare mass parameter for quark of flavor $f$. In this lattice formulation, 
\begin{itemize}
    \item the gauge part of the action, $S_G$, is the tree-level tadpole improved Symanzik with $c_1 = -1/12$.
    \item  The Sheikholeslami-Wohlert coefficient $C_{SW}$~\cite{Sheikholeslami:1985ij} in the fermion part of the action,  $S_F$, 
    is set to the tadpole improved tree-level value, 
    $c_{SW} = 1/u_0^3$~\cite{Bulava:2013cta}, 
    where $u_0$ is the fourth root of the plaquette
    expectation value. This choice is very close to the nonperturbative 
    value determined, a posteriori, using the Schr\"odinger functional
    method~\cite{Luscher:1996ug}. This closeness is attributed to the stout smearing of the links implemented in the action.
    \item The links entering the Dirac action are stout smeared using 
    one iteration with weight $\rho = 0.125$ for the sum of the staples~\cite{Morningstar:2003gk}.

    \item  In the molecular dynamics part of the HMC, we used a mass preconditioning scheme as  described in \cite{Hasenbusch:2001xh,Urbach:2005ji}. In this approach the fermion determinant for the light quarks is simulated as a chain of fermion determinant ratios: 
\begin{eqnarray}
&{}& \det (M^{\dagger}_0 M_0 ) = \nonumber \\
&{}& \left \{ \prod_{I=0}^{N-1} \frac{. \det (M^\dagger_i M_i ) }{ M^\dagger_{I+1}  M_{I+1} } \right \}  \det( M^\dagger_{N} M_{N} )  
\end{eqnarray}
where $M_0$ is the fermion matrix with the light quark mass and  the $M_i$ are auxiliary fermion matrices with  quark masses increasing with $I$.
The final term ends the chain by cancelling off the auxiliary determinant with the heaviest mass.
Up to 3-4 auxiliary determinant ratios were used  in the chain. 
    \item A nested multiple time-scale integration scheme \cite{Sexton:1992nu} with fourth order force-gradient integrators~\cite{Clark:2011ir} was implemented using the trick of \cite{Yin:2011np} at each time scale except potentially the innermost one. We integrated  the determinant ratios and clover actions on the slowest level, the single flavor term on the second level, and the gauge action and the heavy cancellation 2-flavor action on the innermost level.  While the cancellation term was computationally inexpensive due to its high mass, we further reduced its cost by utilizing a minimum residuum extrapolation “chronological” predictor \cite{Brower:1995vx}. 
    \item The Dirac matrix inversion was accelerated using a multi-grid solver \cite{Brannick:2007ue,Clark:2016rdz}  where the multi-grid subspace was developed for the lightest quark mass at the start of the trajectory and was updated if a solution took more than 72 iterations. Because both the chronological predictor and the reuse of the subspace through the trajectory break reversibility in principle, and to maintain stability of molecular dynamics,  a high accuracy (with a relative residuum of $O(10^{-10})$) was required for all solves.  
    \item The time step in the molecular dynamics evolution was adjusted to 
    obtain roughly 92\% acceptance rate for trajectory length of one unit.
\end{itemize}
The parameters of the 13 ensembles are given in Tables~\ref{tab:ensembles} and~\ref{tab:Sparams}.

\begin{table*}[h]
    \centering
    \setlength{\tabcolsep}{8pt} 
    \renewcommand{\arraystretch}{1.2} 
    \begin{tabular}{lccccc|ccccc}
     \hline \hline
    Ens. ID  & ${(\frac{L}{a}})^3 \times \frac{T}{a}$   & $\beta$ & $a m_l$   & $a m_s$   & $c_{SW}$   &  $a$   & $M_\pi$ & $M_\pi L$  & $L$    & Total \\
           &                  &         &           &           &            &   (fm)     &  (MeV)       &            &  (fm)  & Lattices\\
    \hline \hline
    $a117m310$   & $32^3\times 96$  & 6.1     & $-0.2850$ & $-0.2450$ & $1.249310$ &  0.117 & 310     & 5.85       & 3.7  & 4865    \\
     \hline
    $a087m290$     & $32^3\times 64$  & 6.3     & $-0.2390$ & $-0.2050$ & $1.205366$ &  0.087 & 289     & 4.1        & 2.8  & 3085    \\
     \hline
    $a087m290L$    & $48^3\times 128$ & 6.3     & $-0.2390$ & $-0.2050$ & $1.205366$ &  0.087 & 289     & 6.16       & 4.2  & 6223   \\
     \hline
    $a087m230$  & $48^3\times 128$ & 6.3     & $-0.2415$ & $-0.1974$ & $1.205366$ &  0.087 & 233     & 5.0        & 4.2  & 5200    \\
     \hline
    $a087m230X$   & $48^3\times 128$ & 6.3     & $-0.2406$ & $-0.2050$ & $1.205366$ &  0.087 & 228     & 5.0        & 4.2  & 3167    \\
     \hline
    $a086m180$     & $48^3\times 96$  & 6.3     & $-0.2416$ & $-0.2050$ & $1.205366$ &  0.086 & 179     & 3.7        & 4.1  & 3248    \\
     \hline
    $a086m180L$    & $64^3\times 128$ & 6.3     & $-0.2416$ & $-0.2050$ & $1.205366$ &  0.086 & 179     & 5.1        & 5.5  & 6622    \\
     \hline
    $a068m290$   & $48^3\times 128$ & 6.5     & $-0.2070$ & $-0.1750$ & $1.170082$ &  0.068 & 290     & 4.18       & 3.3  & 7620    \\
     \hline
    $a068m230$     & $64^3\times 192$ & 6.5     & $-0.2080$ & $-0.1788$ & $1.170082$ &  0.068 & 234     & 5.2        & 4.4  & 3604    \\
     \hline
    $a068m175$     & $72^3\times 192$ & 6.5     & $-0.2091$ & $-0.1778$ & $1.170082$ &  0.068 & 175     & 4.4        & 4.9  & 5556    \\
     \hline
    $a067m135$     & $96^3\times 192$ & 6.5     & $-0.2095$ & $-0.1793$ & $1.170082$ &  0.067 & 135     & 4.4        & 6.4  & 2904    \\
     \hline
    $a053m295$     & $64^3\times 192$ & 6.7     & $-0.1830$ & $-0.1650$ & $1.142727$ &  0.053 & 295     & 4.8        & 3.4  & 5301    \\
     \hline
    $a053m230$     & $72^3\times 192$ & 6.7     & $-0.1843$ & $-0.1640$ & $1.142727$ &  0.053 & 226     & 4.5        & 3.8  & 3331    \\
     \hline
    \end{tabular}
    \caption{Parameters of the thirteen isotropic 2+1-flavor Wilson-clover ensembles generated by the JLab/W\&M/LANL/MIT/Marseille collaborations using the highly tuned CHROMA code\cite{Clark:2016rdz}.    The acceptance rate in the HMC algorithm was tuned to roughly 92\%   on each ensemble.
    The bare light and strange quark masses are given by 
    $a m_{l,s}= 1/2\kappa_{l,s} - 4$ where $\kappa$ is the 
    hopping parameter in the SW action with the coefficient of the clover term given by $c_{SW}=1/u_0^3$. The last five columns give the approximate lattice spacing $a$ in fermi, the pion mass $M_\pi$ in MeV, the lattice size  $L$ in fermi, its dimensionless version $M_\pi L$,     
    and the total number of configurations  available for analysis.  The additional $L$ in $a087m290L$ and $a086m180L$, implies a larger volume. Ensembles $a087m230$  and $a087m230X$  differ mainly in the value of the strange quark mass. The ensemble IDs used here are updates from those 
    used in Refs.~\cite{Yoon:2016dij,Yoon:2016jzj,Park:2021ypf,Mondal:2020ela} to reflect the new values of the lattice spacing $a$, $M_\pi$  and $M_K$ obtained in this analysis. Updates of results 
    presented in Refs.~\cite{Park:2021ypf,Mondal:2020ela} are under progress.
   }
    \label{tab:ensembles}
\end{table*}

\begin{table*}[hbt!]
    \setlength{\tabcolsep}{5pt} 
    \renewcommand{\arraystretch}{1.2} 
\begin{tabular}{|l|c|c|c|c||c|c|c||c|c|c|c|}
\hline
ID           & $\beta$ & $a$   & $M_\pi$  & $M_K$ & $N_\text{KG}$ & $\sigma$ & $R_{RMS}$/a& \# Lattices & $N_{\rm HP}$ & $N_{\rm LP}$ & $\tau$                    \\
             &         & (fm)  & (MeV)  & (MeV)    &                                       &          &              &   Analyzed           &                &      &     \\
\hline
$a117m310$ & 6.1     & 0.117 & 310(6) & 517(10)   & 50            & 5        & 4.10(1)    & 3004     & 12,056       & 385,792      & \{8, 10, 12, 14\}         \\
\hline			                                   
$a087m290$ & 6.3     & 0.087 &        &          & 91            & 7        & 5.46(2)    & 2469     & 7,407        & 237,024      & \{10, 12, 14, 16\}        \\
\hline			                                   
$a087m290L$ & 6.3     & 0.087 & 289(5) & 550(10)   & 91            & 7        & 5.49(3)    & 5453     & 18,040       & 577,280      & \{10, 12, 14, 16, 18\}    \\
\hline			                                   
$a087m230$ & 6.3     & 0.087 & 233(4) & 585(11)   & 98            & 7        & 5.54(1)    & 2668     & 8,000        & 256,000      & \{10, 12, 14, 16, 18\}    \\
\hline			                                   
$a087m230X$ & 6.3     & 0.087 & 228(4) & 534(10)   & 98            & 7        & 5.56(1)    & 2003     & 8,024        & 256,768      & \{10, 12, 14, 16, 18\}    \\
\hline			                                   
$a086m180$ & 6.3     & 0.086 &        &          & 91            & 7        & 5.41(2)    & 4012     & 16,048       & 513,536      & \{8, 10, 12, 14, 16\}     \\
\hline			                                   
$a086m180L$ & 6.3     & 0.086 & 179(3) & 523(10)   & 91            & 7        & 5.49(3)    & 4014     & 20,000       & 640,000      & \{8, 10, 12, 14, 16\}     \\
\hline			                                   
$a068m290$ & 6.5     & 0.068 & 290(5) & 614(11)   & 150           & 9        & 6.96(1)    & 5295     & 18,880       & 604,160      & \{11, 13, 15, 17, 19\}    \\
\hline			                                   
$a067m230$ & 6.5     & 0.067 & 234(4) & 568(10)   & 200           & 10       & 7.53(1)    & 2350     & 14,100       & 225,600      & \{13, 15, 17, 19, 21\}    \\
\hline			                                   
$a067m175$ & 6.5     & 0.067 & 175(3) & 567(10)   & 185           & 10       & 7.57(2)    & 3771     & 22,632       & 362,112      & \{13, 15, 17, 19, 21\}    \\
\hline			                                   
$a067m135$ & 6.5     & 0.067 & 135(3) & 548(10)   & 200           & 10       & 7.57(1)    & 2500     & 15,000       & 240,000      & \{13, 15, 17, 19, 21\}    \\
\hline			                                   
$a053m295$ & 6.7     & 0.053 & 295(5) & 553(10)   & 365           & 14       & 10.09(1)   & 3250     & 16,200       & 259,200      & \{15, 18, 21, 24, 27\}    \\
\hline			                                   
$a053m230$ & 6.7     & 0.053 & 226(4) & 550(10)   & 392           & 14       & 10.18(1)   & 2485     & 15,300       & 244,800      & \{18, 21, 24, 27, 30\}    \\
\hline
\end{tabular}
\caption{The approximate value of the lattice spacing $a$ used to label the ensembles is taken from 
Table~\protect\ref{tab:lattice-spacing-decay-constant}. 
The large errors in $M_\pi$ and $M_K$ are mainly due to those in $a$.  
Columns 6 and 7 give the Wuppertal smearing parameters  
$\{N_\text{KG},\sigma\}$ used to construct the smeared sources 
for generating the quark propagators. The  
resulting root mean squared smearing radius in lattice units, $R_{RMS}/a$, 
is given in column 8. 
%
%
 Column 9 gives the number of configurations analyzed. The numbers $N_{\rm HP}$ ($N_{\rm LP}$) 
 give the total number of high (low) precision measurements made. The connection between these numbers to 
  the data given in Table~\protect\ref{tab:Confs} is---on  each configuration 
  $N_{\rm tsrc}$  maximally separated time slices were selected randomly, and on each  time slice 
  $N_{\rm src}^{LP}$ points were selected randomly as sources for low-precision quark 
  propagators and one for high-precision.  For completeness, the last column gives the values of source-sink separation $\tau$ used in the calculations of 3-point functions to be presented elsewhere.}\looseness-1 
\label{tab:Sparams}
\end{table*}

\begin{table*}[h]
    \centering
    \setlength{\tabcolsep}{4pt}
    \renewcommand{\arraystretch}{1.2}
\begin{tabular}{lllll|lllll}
 
\hline
Ens. ID & Streams: Traj & Configs & $t_0/a^2$ & $w_0/a$ & $N_l$ & $N_s$ & $N_\text{tsrc}$ & $N_\text{src}^\text{LP}$  & $N_{\rm bin}$   \\
\hline
$a117m310$& \makecell[l]{0: 1000--2108\\a: 1010--1870\\b: 1000--18340} & 4865 & 1.4953(44) & 1.3880(29) & 3004 & 1013 & 4 & 32  & 5 \\
\hline
$a087m290$ & 0: \makecell{2--13340} & 3335 & 2.6566(59) & 1.8709(45) &  2469 & 0 & 3 & 32 & 4 \\
\hline
$a087m290L$ & 0: 262--25154 & 6223 & 2.6517(13) & 1.86770(75) & 5453 & 944 & 4 & 32 & 5 \\
\hline
$a087m230$ & \makecell[l]{a: 0--12154\\b: 660--9244} & 5200 &  2.65879(73) & 1.86580(52) & 2668 & 1001 & 4 & 32 & 4 \\
\hline
$a087m230X$  & 0: 0 -- 12670 & 3167 & 2.7079(14) & 1.89853(90) & 2003 & 2003 & 4 & 32 & 4 \\
\hline
$a086m180$ & \makecell[l]{0: 10--5410\\a: 0--3560 \\b: 0--3720 \\c: 0--3276\\d: 40--2810\\e: 40--2180\\f: 40--3330\\g: 40-2420} & 6634 & 2.7442(30) & 1.9187(19) &  3941 & 0 & 4 & 32 &  5 \\
\hline
$a086m180L$ & \makecell[l]{0: 2--6896\\a: 2--5838\\b: 2--6970\\c: 2--6776} & 6622 & 2.74588(76) & 1.91977(44) & 4014 & 1014 & 5 & 32 & 4 \\
\hline
$a068m290$ & \makecell[l]{0: 800--7638\\a: 10--6612\\b: 10--17028} & 7620 & 4.3619(38) & 2.4115(19) & 5295 & 819 & 4 & 32 & 5 \\
\hline
$a068m230$ & \makecell[l]{0: 0--12150\\a: 802-3068} & 3604 & 4.4779(18) & 2.46372(95) & 2350 & 500 & 6 & 16 & 5 \\
\hline
$a068m175$ & \makecell[l]{0: 0--5640\\a: 2--4354\\b: 2--4072\\c: 2--5508} & 5556 & 4.5158(11) & 2.48351(68) & 3771 & 1273 & 6 & 16 & 4 \\
\hline
$a067m135$ & \makecell[l]{0: 0--6400\\a: 0--5210} & 2904 & 4.5620(10) & 2.50490(61) & 2500 & 1760 & 6 & 16 & 4 \\
\hline
$a053m295$ & \makecell[l]{0: 336--10536\\a: 100--5624\\b: 100--5376} & 5301 & 7.1233(41) & 3.1304(18) & 3250 & 550 & 6 & 16 & 5 \\
\hline
$a053m230$ & \makecell[l]{0: 2--11306\\a: 802--2816} & 3331 & 7.1904(55) & 3.1616(25) & 2485 & 2035 & 6 & 16 & 5 \\
\hline
\end{tabular}
\caption{The number of streams and thermalized trajectories (Traj) generated 
using the HMC algorithm and the number of configurations (Configs) stored at Oak Ridge. Neglecting lattices lost,  every second trajectory is stored and measurements were 
made on configurations mostly 
separated by four trajectories (6 for the rest). Columns four and five give 
the measured values of the flow parameters $t_0/a^2$ and $w_0/a$. $N_l$ 
gives the number of lattices analyzed for observables consisting of only light 
quarks ($M_\pi$, $f_\pi$, $M_\rho$, $M_N$ and $M_\Delta$), while  $N_s$ gives 
the number for observables 
with at least one strange quark ($M_K$, $f_K$, $M_{s \overline{s}}$, $M_\Sigma$,  $M_\Xi$, $M_{\Sigma^\ast}$,  $M_{\Xi^\ast}$, $M_\Omega$).  
On each configuration, $N_\text{tsrc}$ is the number of maximally separated  time slices on which sources were placed and $N_\text{tsrc}^\text{LP}$ is the number of 
    randomly selected sources points on each of these time slices from which low-precision (LP) measurements are made for a total of  
    $N_\text{src}^\text{LP} \times N_\text{tsrc}$ measurements. The 
    high-precision measurements are $N_\text{tsrc}$, i.e., only one point on each 
    of the $N_\text{tsrc}$ time slices was used. The last column gives the number of adjacent configurations, $N_{\rm bin}$, over which data were binned and averaged to construct the binned data used for the statistical analysis. 
    }
    \label{tab:Confs}
\end{table*}

\subsection{Methodology: Measurements}
\label{sec:measured}

\begin{figure}
      \includegraphics[trim=100 530 80 200, width=0.52\linewidth]{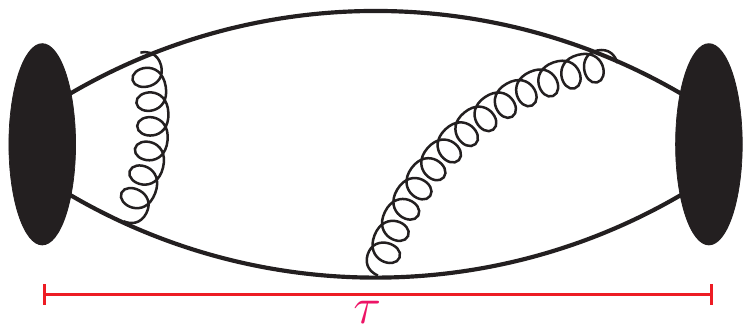}
      \includegraphics[trim=20 3 0 4, width=0.48\linewidth]{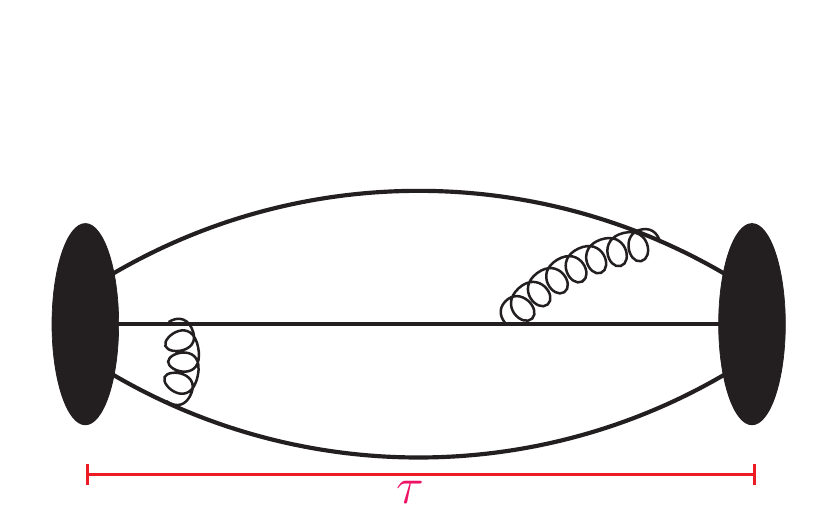}
    \caption{Quark line diagrams for meson (top) and baryon (bottom) two-point functions. The gluon 
    lines have been added only to remind the reader that all possible intermediate states 
    and gluon exchanges are included in  lattice calculations, i.e., they are fully non-perturbative.}
  \label{fig:quarkline}
\end{figure}

All spectral quantities presented here are extracted from fits to 
the spectral decomposition of the two-point correlation functions that 
describe the evolution of the meson and baryon states in 
Euclidean time $\tau$:
\begin{eqnarray}
      \Gamma_2(\tau) &\equiv &
  \langle  {\cal O}^\dagger (\tau) {\cal O} (0) \rangle = \sum_i \langle 0 | {\cal O}^\dagger | S_i \rangle \langle S_i | {\cal O} | 0 \rangle e^{-E_i \tau} \nonumber \\
  &=& \sum_i A_i^\ast A_i e^{-E_i \tau} \,,
    \label{eq:SD2pt}  
\end{eqnarray}
where ${\cal O} $ is the interpolating operator used to create/annihilate the 
states and $A_i \equiv \langle S_i | {\cal O} | 0 \rangle$ is the amplitude 
for $\cal O$ to create (or annihilate) the state $| S_i \rangle$ from the vacuum. (See Appendix~\ref{sec:Operators} for definitions of the operators used.)
The sum is over all states with the quantum numbers of ${\cal O}$ 
as one does not, apriori,  know which amplitudes are non-zero. 
The quark line diagrams representing the two-point functions 
for meson and baryon states are shown in  Fig.~\ref{fig:quarkline}.  
These are constructed by contracting the free spin, color and flavor 
indices in the non-perturbative Feynman quark propagator $P_F$ 
(the lines) with the quark fields in  $\cal O$ (the blobs), i.e., by taking 
appropriate spin and color traces. The gluon lines are shown only to indicate that all 
possible interactions (with momenta allowed by the lattice cutoff $1/a$) 
are included. Calculations of the correlation functions represented 
by these quark-line diagrams  are fully non-perturbative. 

The quark propagator $P_F \equiv {\cal D}^{-1}$ is the inverse of 
the lattice Dirac operator  $\cal D$. This inverse 
(used in both the construction of correlation functions, and in 
the generation of the lattices) is calculated by solving the equation 
$P_F|^{i,a}_{j,b} = {{\cal D}^{-1}}^{k,c}_{j,b} \xi_{k,c}^{i,a}$ using 
the efficient multigrid algorithm~\cite{Babich:2010qb,Clark:2009wm,Clark:2016rdz} built into the 
CHROMA software package~\cite{Edwards:2004sx}.  The Gaussian smeared source $\xi_{k,c}^{i,a}$ 
is produced by applying the Wuppertal 
procedure on a delta-function source~\cite{Gusken:1988yi,Yoon:2016dij,Gupta:2018qil,Mondal:2020cmt} 
with smearing  parameters $N_{\rm KG}$ and $\sigma$ given in 
Table~\ref{tab:Sparams}. 
To make these smeared sources smooth, the gauge links were first 
smoothed using twenty hits of the
stout algorithm with the sum of the spatial
staples added to the link with a weight factor $\rho=0.08$~\cite{Morningstar:2003gk}.   
The resulting root-mean-square size of the
smearing, $\sqrt{\int dr \, r^4 S^\dag S /\int dr \, r^2
S^\dag S} $ with $S(r)$ the value of the smeared source at radial
distance $r$, is between 0.5--0.57~fm (Table~\ref{tab:Sparams}). 

These quark propagators can be similarly smeared at the sink points, and using these we calculate two types of 2-point functions: the smeared-smeared (SS) using propagators 
with the same smearing at the source and sink points, and smeared-point (SP) 
using  unsmeared (point) propagators at the sink. Analyzing both using Eq.~\eqref{eq:SD2pt} provided checks on 
excited state contributions and increased the confidence in the 
extraction of the ground-state energy $M_0$ (see Figs.~\ref{fig:C13_nucleon}, ~\ref{fig:F6_nucleon}, ~\ref{fig:meff_baryons_1} and~\ref{fig:meff_baryons_2} in Appendix~\ref{sec:N3and4-state-fits} for the baryon states). By construction, all two-point 
correlation functions studied are gauge invariant, so no gauge 
fixing of the lattices is required in their calculation. 

For the spectral quantities studied in this work, the goal is to extract the ground-state 
mass $M_0$ and the amplitude $A_0$.  This is achieved in two ways: (i) analyzing $\Gamma_2(\tau)$ at  
large $\tau$ at which ground state domination is manifest, i.e., a long and stable plateau in $\meff(\tau)$ that, ideally, stretches to $\tau \to \infty$. 
Clear examples of this are the pseudoscalar states for which there is no degradation 
of the signal with $\tau$ as shown in Figs.~\ref{fig:C13} and~\ref{fig:F6}. (ii) For the remaining states, such as all vector mesons and baryons,  
for which the signal decays exponentially (as $e^{-(M_N - 1.5 M_\pi)}$ for nucleons as explained in~\cite{Parisi:1983ae,Hamber:1983vu,Lepage:1989hd}), 
the plateau with current statistics are not sufficiently long.  So we make fits to Eq.~\eqref{eq:SD2pt} truncated 
at $i = 2, 3, 4$ states depending on the statistical precision of the data. 
The  effects of this truncation are studied using  
the nucleon state for which, among baryons, the highest statistical precision data 
was collected (see Appendix~\ref{sec:N3and4-state-fits}).   

The quantities extracted from the analysis of the two-point functions include  
\begin{itemize}
    \item The pseudoscalar masses, $M_\pi$, $M_K$, and $M_{s\overline s}$, given in Table~\ref{tab:PStab}. 
    \item The pseudoscalar decay constants, $f_\pi$ and $f_K$, given in Table~\ref{tab:decayconst}.
    \item The vector meson masses, $M_\rho,\ M_{K^\ast},$ and $M_{\phi_{s\overline s}}$,  given in Table~\ref{tab:vectormasses}.
    \item The masses of octet baryons, $M_N$, $M_\Sigma$, $M_\Lambda$ and  $M_\Xi$, given in Table~\ref{tab:oBtab}. 
    \item The masses of decuplet baryons, $M_\Delta$, $M_{\Sigma^\ast}$,  $M_{\Xi^\ast}$ and $M_\Omega$, given in Table~\ref{tab:dBtab}. 
\end{itemize}

In addition, we present data for the flow time scales $w_0$ and $t_0$ 
in Table~\ref{tab:Confs}, and for 
the topological charge $Q$ and the Weinberg three-gluon operator $W$ 
as a function of the flow time in Appendix~\ref{sec:flowQandW}. 
The prescription for implementing the flow and the calculation of $w_0$, $t_0$ and the topological susceptibility $\chi_Q$ is given in Secs.~\ref{sec:t0w0def} and~\ref{sec:topology}.

Lastly, Table~\ref{tab:Zfactors} gives 
the renormalization factors in the $\MSbar$ scheme at 
scale $\mu = 2$~GeV for all five quark bilinear local 
operators for the 11 ensembles. Their calculation, using the 
regularization independent symmetric momentum subtraction
(RI-sMOM) intermediate scheme on the 
lattice~\cite{Martinelli:1994ty,Sturm:2009kb},  is described in Sec.~\ref{sec:Zfactors}. The matching between the RI-sMOM and $\MSbar$ schemes 
and the running in the continuum use 
results from continuum perturbation theory, whose evaluation requires knowing 
the lattice scale $a$ and $\alpha_s(1/a) $ on each ensemble. Results for these are given in Table~\ref{tab:lattice-spacing-decay-constant} and their 
extraction is discussed in Sec.~\ref{sec:lattice-spacing}.

\subsection{Neglecting Finite Volume Corrections}
\label{sec:noFV}

The two smaller volume ensembles, $a087m290$ and $a086m180$, were generated to evaluate finite 
volume corrections by comparing with results from their larger volume versions, $a087m290L$ and $a086m180L$. As shown in Tables~\ref{tab:PStab} and~\ref{tab:oBtab}, the errors on $M_\pi$ and $M_N$ 
from the smaller volume lattices are much larger, and within these errors, 
the two values for $M_\pi$ and $M_N$ overlap.
Noting this lack of resolution, we did not calculate any quantity involving strange quarks 
on these two ensembles. They are, therefore, not included in the analysis.  The remaining eleven ensembles have $M_\pi L \ge 4$ 
and $L > 3.4$~fm. On such large
lattices, the  finite volume corrections are expected to be  exponentially suppressed~\cite{Luscher:1986pf} and smaller than the statistical errors 
for both the pseudoscalar mesons and the baryons. A more detailed 
study of finite volume effects presented in Ref.~\cite{RQCD:2022xux}, 
with lattice parameters covering those simulated here, 
reached the same conclusion. We, therefore, 
neglect finite volume corrections in our analyses, i.e., the fits used to 
get the physical point values include only chiral-continuum (CC) corrections.  

\subsection{The Quality of the Signal}
\label{sec:quality}

The number of measurements made for all 
quantities with one or more strange quark are much fewer.  
Specifically, as shown in Table~\ref{tab:Confs}, note the 
difference between $N_l$ versus $N_s$ configurations used and 
$N_{\rm tsrc} \times N^{LP}_{\rm src}$ versus $N_{\rm tsrc}$ source 
points used for measuring correlation functions  composed of light quarks 
only versus those with one or more strange quarks. Nevertheless, 
the signal in two-point functions with even one strange quark is reasonable as shown in Figs.~\ref{fig:meff_baryons_1} and~\ref{fig:meff_baryons_2} in Appendix~\ref{sec:N3and4-state-fits}.

A necessary input in the calculation of the statistical errors is knowing the autocorrelation time for each observable. We  assess the integrated autocorrelation time (ICT), defined in Eq.~\eqref{eq:IAC}, using the data for $t_0/a^2$ and $w_0/a$ as these long-distance quantities show the largest correlations. Note that these data, presented in 
Table~\ref{tab:corr_time}  and shown in Appendix~\ref{sec:autocorr}, exhibit (i) significant variations in ICT 
between the various streams employed in lattice generation and (ii) an increase 
as $a \to 0$. To take into account these significant ICT we bin the data. Specifically, 
all two-point function data are first averaged over the $N_{\rm tsrc} \times N_{\rm src}^{LP}$ 
(or $N_{\rm tsrc} $ for strange) measurements made on a given configuration, and these 
averages are further binned over $N_{\rm bin}$ (given in  Table~\ref{tab:Confs})
configurations.  These binned data do not show any significant autocorrelations. Therefore, 
the errors coming out of the analysis of the binned data are not augmented by the ICT determined from 
$t_0/a^2$ or $w_0/a$. 

The  signal-to-noise (S2N) ratio is widely different for the pseudoscalar mesons versus vector mesons 
and the baryons. For both the $\rho$-meson and the nucleons, low lying multihadron 
states give rise to significant contributions and an exponentially decaying signal~\cite{Parisi:1983ae,Hamber:1983vu,Lepage:1989hd}, a well-known persisting challenge to their analysis. 

The main tool used to assess the S2N ratio for the various hadrons are the effective mass plots: $a M^{\rm  eff}(\tau) = \log (\Gamma_2(\tau)/\Gamma_2(\tau+1))$ versus $\tau$ 
where $\Gamma_2(\tau)$ is the two-point 
function at time separation $\tau$ as defined in Eq.~\eqref{eq:SD2pt}. 
To qualify for a robust signal, we require 
a long plateau in $a \meff(\tau)$ indicating that ground state domination has been achieved. 
A qualitative overview of the signal in the various channels is 

\begin{itemize}
    \item The signal in all two-point functions for pseudoscalar mesons is 
    excellent. There is no S2N degradation with $\tau$, i.e., the plateau 
    in the $M^{\rm  eff}(\tau)$ extends to all $\tau$. Final values for the 
    masses $M_\pi$, $M_K$ and $M_{{\overline s} s}$ are given in 
    Table~\ref{tab:PStab}, and for the decay constants, $f_\pi$ and $f_K$, 
    extracted from the amplitudes $A_0$ in Table~\ref{tab:decayconst}. The latter are     extracted from a simultaneous 
    fit to $P_{S}(0)P_{S}(\tau)$ and $P_{S}(0)A_{P}(\tau)$ 
    two-point correlation functions where $P= {\overline \psi} \gamma_5 \psi$ and 
    $A= {\overline \psi} \gamma_4 \gamma_5 \psi$ are the pseudoscalar and 
    axial vector interpolating operators, with the source placed at time 
    $t=0$ and sink at time $t=\tau$. The subscript $S$ denotes smeared and $P$ point sources. 
    The signal-to-noise (S2N) ratio in these quantities is approximately $O(10^3)$. 
    See Sec.~\ref{sec:fpi-fK} for their analysis and the plots for $a \Meff$ 
    for the pion in Appendix~\ref{sec:pi_decayplots}.
    \item Estimates of vector meson masses are given in Table~\ref{tab:vectormasses}, however, 
    the signal for the $\rho$ meson is poor (there is no credible plateau 
    in $M^{\rm  eff}(\tau)$ with single $\rho$ operator) while that for the $K^\ast$ and $\phi_{s {\overline s} }$ 
    states is reasonable. The multihadron $\pi \pi$  intermediate state afflicts 
    the signal in the $\rho$ correlator, with the S2N decaying as $e^{-(M_\rho - m_\pi)\tau}$, 
    whereas the situation for $K^\ast$ ($\phi_{s {\overline s} }$) is better because of 
    the much heavier $K \pi$ ($K K$) intermediate state. 
    To get a good signal for the rho meson, the method of choice is to use a variational ansatz for 
    the interpolating operator~\cite{Boyle:2024hvv,Boyle:2024grr,Dudek:2010wm,Thomas:2011rh}. This has not been implemented in this work. Given the lack of a good signal and because these states have strong decays and large widths, we do not present an analysis of their data.
    
    \item The signal in all four octet baryon states, \{$N, \Lambda, \Sigma, \Xi \}$, is similar and good. 
    The S2N for the nucleon degrades as $e^{-(M_N - 1.5 m_\pi)\tau}$ due to the variance being dominated 
    by the light three-pion state~\cite{Parisi:1983ae,Hamber:1983vu,Lepage:1989hd}. A second, related, 
    problem that impacts extracting the ground state properties of nucleons from 
    fits to the spectral decomposition given in Eq.~\eqref{eq:SD2pt} are the towers of 
    multi-hadron excited states beginning with the $N(\vec p) \pi(-{\vec p})$ states, with the 
    lowest having a  mass gap smaller than the radial state $N(1440)$. Again, the S2N is better for 
    the states with strange quarks, ($\Lambda,\ \Sigma,\ \Xi$), as the lowest multihadron 
    states are expected to be $K K \pi$ or $\phiss \pi \pi$ or $\phiss K K$.
    
    To further understand the signal, we have collected two-point data for a number of local interpolating
    operators defined in Table~\ref{tab:OP_list} and measured both smeared-smeared (SS) 
    and smeared-point (SP) correlation functions. Data for $\Meff$ are shown in 
    Figs.~\ref{fig:C13}--\ref{fig:F6} (pion); Figs.~\ref{fig:C13_nucleon}--\ref{fig:F6_nucleon} (nucleon); and Figs.~\ref{fig:meff_baryons_1} and~\ref{fig:meff_baryons_2} (the five baryon states). 
    
  The widths of $\{\Lambda,\ \Sigma,\ \Xi \}$, states are small as they decay 
  only via weak interaction, and 
  isospin breaking corrections are also small, i.e., a few MeV. We, therefore, 
  consider it appropriate to compare, after chiral-continuum extrapolation, 
   the lattice results given in Table~\ref{tab:oBtab}  with the experimental values 
   for the 2+1-flavor QCD given in Eq.~\eqref{eq:FLAG_inputs}. This is done  
   in Table~\ref{tab:spectrum-different-continuum}. 
     
    \item For the decuplet baryons,  a credible plateau in the effective mass plot at 
    large $\tau$ is found only for the $\Omega$ state. The S2N ratio for it is comparable 
    to that for the $\Lambda,\ \Sigma,$  and $\Xi$ octet states. 
    (These four states have the same number of smaller measurements, i.e.,  fewer configurations and  
    only HP source points,  compared to the nucleon.) In addition, the $\Delta$, ${\Sigma^\ast}$, 
    and  ${\Xi^\ast}$ have strong decays and their experimental masses have large widths. Thus, 
    to do a proper study of their masses, one needs to relate the resonance parameters 
    to the spectrum in a finite box.  This requires calculating correlation 
    functions with a much larger basis of operators including baryon-meson 
    ones. This is beyond the scope of this work. Our analysis of the decuplet baryons is, therefore, based on the data for only the $\Omega$ baryon 
    even though, for completeness, we give estimates for $aM_\Delta$ and $a M_{\Sigma^\ast}$ 
    in Table~\ref{tab:dBtab}. Data for $a M_{\Sigma^\ast}$ were, unfortunately, collected on only two ensembles due to an oversight. 

    \item The data for the flow time scales $w_0$ and $t_0$ is very precise with a S2N ratio of approximately $O(10^3)$.   
   \item The time history of $Q$ and $W$ presented in Appendix~\ref{sec:flowQandW} show no freezing of topology in any of the ensembles, and auto-correlations at even large flow times are small. The topological 
    susceptibility $\chi_Q$ on each ensemble is extracted with a S2N ratio of approximately $O(10^2)$.     
 \end{itemize}

The largest auto-correlations are observed in the flow variables $t_0$ and
$w_0$, which are listed in Table~\ref{tab:corr_time} and shown in Figs.~\ref{fig:C13}--\ref{fig:F6} in 
appendix~\ref{sec:autocorr}. Binned data, as described above, for meson and baryon two-point functions 
used in the analysis do not show any significant auto-correlations. Consequently, 
we do not augment the statistical errors coming out of the analysis based on the ICT observed in the $t_0$ and $w_0$ data.

For the quantities with a good S2N, the largest source of uncertainty in the final results is due to the 
chiral-continuum fits used to extrapolate the lattice data in the 
quark masses and to the continuum limit.  In all cases where data from multiple operators 
or correlation functions are available, the final results are obtained using 
an average weighted by their Akaike Information Criteria score
(AIC)~\cite{AIC}. 

\subsection{The chiral-continuum extrapolation}
\label{sec:CCstrategy}

To obtain the physical point values, i.e., at $M_\pi=135$~MeV, $M_K=495$~MeV, 
$a = 0$ and $L = \infty$, we need to specify the extrapolation 
ansatz used to address the chiral, continuum and finite volume corrections. 
While detailed descriptions will be provided in appropriate sections later, 
here we summarize the overall approach. 

\begin{figure}
    \centering
      \includegraphics[width=0.95\linewidth]{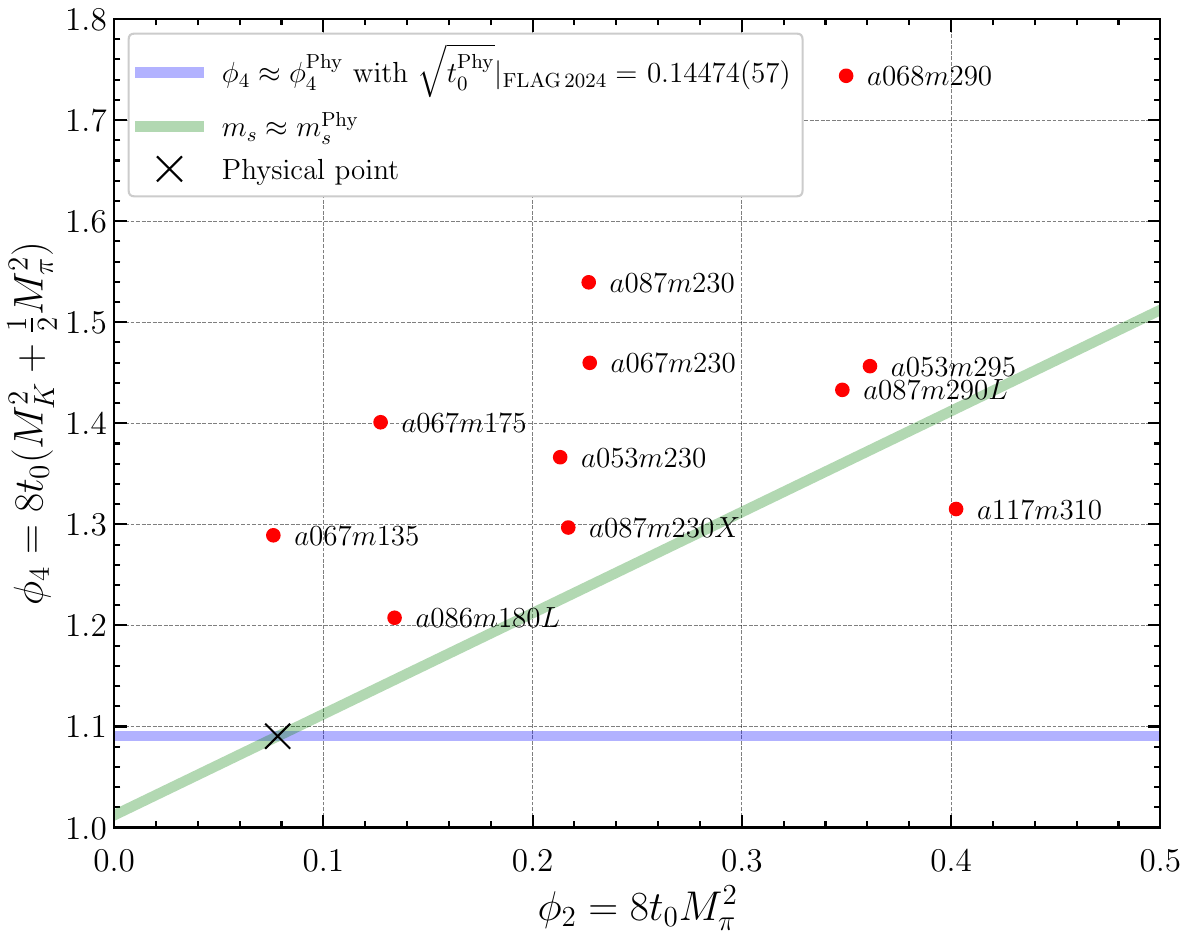}
    \caption{Overview of the 11 ensembles in the $\phi_2 - \phi_4$ plane. The blue line 
    corresponds to the choice $\phi_4 = \phi_4^\textrm{Phy}$, while the green line corresponds to 
    $\phi_4 = 4\tphy[2(M_K^\textrm{Phy})^2 - (M_\pi^\textrm{Phy})^2] + \phi_2$. 
    Since $m_s\propto 2M_K^2 - M_\pi^2$ in leading order chiral perturbation theory,     
    the strange-quark mass remains close to its physical value on the green line.
    The two lines intersect at the physical point (black cross), where the pion 
    and kaon have their physical masses.}
  \label{fig:ensemble_overview}
\end{figure}

The input parameters characterizing each of these 2+1-flavor 
isospin symmetric ensembles are four: the lattice 
spacing $a$ (or equivalently the gauge coupling $\beta = 6/g^2$), the light and strange 
quark masses, $m_l$ and $m_s$, and the spatial lattice size $L$. These 
are given in Table~\ref{tab:ensembles}. The intent during lattice generation 
was to tune the strange quark mass to roughly its physical value on 
each ensemble so as to reduce the number of free parameters 
to three. Unfortunately, the output kaon 
masses listed in Table~\ref{tab:Sparams} vary between $520 - 615 $~MeV. 
Therefore, the analysis of these $2+1$-flavor ensembles is carried out keeping $M_\pi$ and $M_K$ (equivalently $m_l$ and $m_s$) as independent parameters.  In the chiral fits, we use, instead of the quark masses, the pion and kaon masses expressed in units of the flow variable $t_0/a^2$  defined in Eq.~\eqref{eq:t0def}, i.e.,  
\begin{equation}
  \phi_2 \equiv 8 t_0 M_\pi^2 = 8(t_0/a^2)(aM_\pi)^2 \,,
  \label{eq:defPHI_2}
\end{equation} 
and the SU(3) symmetric mass parameter  
\begin{eqnarray}
\phi_4  &\equiv& 8 t_0(M_K^2 + M_\pi^2/2 ) \nonumber \\
&=& 8(t_0/a^2)\left[(aM_K)^2 + (aM_\pi)^2/2\right] \,.
  \label{eq:defPHI_4}
\end{eqnarray} 
The location of the 11 ensembles in the $\phi_2-\phi_4$ plane is shown in 
Fig.~\ref{fig:ensemble_overview}. The measured values of $t_0/a^2$ and  $w_0/a$ 
are given in Table~\ref{tab:Confs}.

We also carry out the full analysis using $w_0/a$,  defined in Eq.~\eqref{eq:w0def}, to construct dimensionless quantities as in Eqs.~\eqref{eq:defPHI_2} and~\eqref{eq:defPHI_4}.  Note, however, these two analyses (using  $t_0/a^2$ versus  $w_0/a$) are highly correlated. This is manifest from the results for the ratio 
$f_K/f_\pi$ in Table~\ref{tab:ratio-fit-results-BS}, and 
the chiral fit parameters for baryons in 
Table~\ref{tab:chiral-fit-parameters-t0-w0}.  

The chiral analysis of pseudoscalar mesons uses results from meson chiral perturbation 
theory ($\chi$PT)~\cite{Gasser:1984gg,Bar:2013ora,Durr:2016ulb}, while for  
baryons, results from heavy baryon $\chi$PT (HB$\chi$PT) are used \cite{Jenkins:1990jv, Ellis:1999jt}. On comparing outputs of next-to-leading order (NLO) and next-to-next leading order (NNLO) truncations, 
one of our main conclusions is that 
the NLO ansatz is not sufficient to fit the masses of the octet baryons.


The analysis of discretization errors uses the fact that 
the Wilson-clover fermion action we use is almost, but not fully, $O(a)$
improved. We, therefore, investigate discretization errors using three ansatz 
modeling the leading-order correction,  
$O(a)$ or $O(\alpha_s a)$ or $O(a^2)$. In most cases, the three 
sets of results overlap. Using any two terms together results in overparameterization. 

Lastly, as already stated in Sec.~\ref{sec:noFV}, our data are not extensive 
enough to resolve finite volume corrections. 
So we ignore them and restrict the analysis to CC fits to data from eleven ensembles that  
all have $M_\pi L > 4.0$. 
    
\subsection{Main Results}
\label{sec:main_results}
Using the data simulated and the various fit ansatz, we present results 
for the following quantities:
\begin{itemize}
    \item Time history for the topological charge $Q$ and Weinberg operator $W$. 
    Data presented in Appendix~\ref{sec:flowQandW} show no evidence 
    of freezing of $Q$ or $W$ on any of the 13 ensembles.
    \item The topological susceptibility $\chi_Q$ in the continuum limit calculated in Sec~\ref{sec:topology}. 
    \item The flow scales $\tphy$ and $\wphy$ extracted using 
    chiral-continuum fits to ${\sqrt{8 t_0}} f_{\pi K} \equiv {\sqrt {8 t_0}}(2f_K + f_\pi)/3$.
    Because the isovector renormalization constant $Z_A$ has no dependence on the lattice scale, 
    fits to $\fpiK$ involve only dimensionsless variables that have all been calculated on the lattice as discussed in Sec.~\ref{sec:fpiK}. 
        
    \item Using the value of $\tphy$ (or $\wphy$) extracted from either the fit to our $f_{\pi K}$ data 
    or using the FLAG 2024 values, we calculate the ratio, $f_K/f_\pi$, of pseudoscalar decay constants in Sec.~\ref{sec:fKbyfpi}. 
    
    \item A fit to the octet baryon data  versus $\phi_2$ and $\phi_4$ is used to extract second estimates of $t_0^{\rm Phy}$ and 
    $w_0^{\rm Phy}$ in Sec.~\ref{sec:octet_ChPT}. A comparison of results using the NLO versus NNLO ansatz 
    for the chiral behavior of the octet masses, shows that the NNLO ansatz is preferred.
    \item The NLO and the NNLO  fits to the $\Omega$ baryon mass versus $\phi_2$ and $\phi_4$, are comparable and NNLO is likely an overparameterization as presented in Sec.~\ref{sec:decuplet_ChPT}. These fits give a third estimate for $t_0^{\rm Phy}$ and $w_0^{\rm Phy}$.  
    \item
    Various estimates of $t_0^{\rm Phy}$ and $w_0^{\rm Phy}$ are summarized and compared in Table~\ref{tab:t0_w0_NLO_NNLO} and in Fig.~\ref{fig:t0-final}. 
\item The lattice spacing $a$, the coupling $\alpha_s(1/a)$, and the 
renormalization constants for quark bilinear local operators, $Z_\Gamma$, on each ensemble. 
The analysis of some of the spectral quantities presented in this 
paper do not require knowing the lattice scale $a$ on each ensemble. 
It is, however, needed when calculating (i) any  
quantity that has a scale dependence, for example matrix elements 
requiring scale-dependent renormalization constants, i.e., $Z_\Gamma$ specified in 
the continuum $\MSbar$ scheme 
at 2~GeV; or (ii) the analysis involves, explicitly, a mass scale in physical units 
such as we will encounter in the chiral logarithms in the analysis of $\fpifk$. 
To facilitate these scale dependent  calculations, we discuss scale setting in Sec.~\ref{sec:lattice-spacing}.

\end{itemize}

\subsection{Input Values of Physical Quantities}
\label{sec:Inputs}

Throughout this paper, when physical values for the spectrum or decay constants are needed, 
we take the numbers given in Ref.~\cite{RQCD:2022xux} that have been corrected 
for the isospin-breaking effects:
\begin{eqnarray}\label{eq:physical_inputs}
M_{\pi}^{\textrm{Phy}}&=&134.8(3)~\textrm{MeV},
\nonumber \\
M_{K}^{\textrm{Phy}}&=&494.2(3)~\textrm{MeV},
\nonumber \\
f_{\pi}^{\textrm{Phy}} &=& 130.56(13)~\textrm{MeV},
\nonumber \\
f_{K}^{\textrm{Phy}} &=& 157.2(5)~\textrm{MeV},
\nonumber \\
f_{\pi K}^\textrm{Phy} &\equiv& {(2f_{K}+f_\pi)}/{3} = 148.32(34)~\textrm{MeV},
\nonumber \\
M_{N}^{\textrm{Phy}}&=&937.54(6)~\textrm{MeV},
\nonumber \\
M_{\Lambda}^{\textrm{Phy}}&=&1115.68(1)~\textrm{MeV},
\nonumber \\
M_{\Sigma}^{\textrm{Phy}}&=&1190.66(12)~\textrm{MeV},
\nonumber \\
M_{\Xi}^{\textrm{Phy}} &=& 1316.9(3)~\textrm{MeV},
\nonumber \\
M_{\Omega}^{\textrm{Phy}} &=& 1669.5(3.0)~\textrm{MeV}.
\label{eq:physical_values}
\end{eqnarray}

In addition, we will use the following FLAG 2024 averages for the 2+1-flavor theory~\cite{FlavourLatticeAveragingGroupFLAG:2024oxs}:
\begin{eqnarray}\label{eq:FLAG_inputs}
\sqrt{t_{0}^\textrm{Phy}}|_{\rm FLAG} &= 0.14474(57)~{\rm fm} \,, \nonumber \\
w_{0}^\textrm{Phy}|_{\rm FLAG} &= 0.17355(92)~{\rm fm}\,.
\label{eq:FLAG_values}
\end{eqnarray}

\begin{table*}[]
    \setlength{\tabcolsep}{10pt} 
    \renewcommand{\arraystretch}{1.2} 
    \centering
\begin{tabular}{llll|ll}
\toprule
 Ens. ID &         $aM_\pi$ &      $aM_K$      &    $aM_{s\bar{s}}$ & $M_K/M_\pi$  & $M_{s\bar{s}}/M_\pi$ \\
\midrule
   $a117m310$&  0.18341(26)(1) &  0.30516(45)(10) &   0.38968(40)(5)  &  1.6621(47) & 2.1226(54)  \\
   $a087m290$  &   0.1279(4)     &                  &                   &             &             \\              							    
   $a087m290L$ &   0.12809(8)(1) &   0.24362(25)(3) &   0.32095(21)(2)  & 1.9004(41)  & 2.5030(51)  \\ 
  $a087m230$ &  0.10327(14)(0) &   0.25893(27)(7) &   0.35264(24)(1)  & 2.489(10)   & 3.391(14)   \\  
 $a087m230X$  &  0.10010(15)(3) &   0.23421(20)(4) &   0.31630(14)(7)  & 2.3402(55)  & 3.1612(71)  \\
    $a086m180$ &   0.0781(2)     &                  &                   &             &             \\					    
    $a086m180L$ &  0.07812(11)(1) &   0.22786(30)(1) &   0.31302(26)(0) 	&2.940(13)    & 4.040(17)   \\   
    $a068m290$ &   0.10013(6)(3) &   0.21204(29)(6) &   0.28461(21)(5) 	&2.1149(65)   & 2.8383(82)  \\ 
    $a068m230$ &   0.07966(6)(1) &   0.19385(31)(0) &  0.26384(26)(12) 	&2.4295(71)   & 3.3079(82)  \\
    $a068m175$ &   0.05938(8)(2) &   0.19240(19)(5) &   0.26717(14)(2) 	&3.246(11)    & 4.502(15)   \\   
    $a067m135$ &   0.04567(7)(4) &   0.18515(16)(1) &   0.25902(13)(2) 	&4.067(13)    & 5.691(18)   \\   
    $a053m295$ &   0.07962(6)(3) &  0.14963(33)(12) &   0.19765(29)(2) 	&1.8949(53)   & 2.5017(64)  \\ 
    $a053m230$ &   0.06088(6)(4) &  0.14799(13)(10) &  0.20148(10)(14) 	&2.4323(44)   & 3.3098(56)  \\ 
    Physical &           &                  &                   & 3.667       & 5.089       \\
\bottomrule
\end{tabular}
\caption{Results for pseudoscalar masses in lattice units.  These values
are obtained from weighted averages over two-state fits to SS and SP correlators. The first error is the overall analysis error and the second, a systematic error, is  estimated from the difference between SS and SP values. The strange spectrum, $M_K$ and $M_{s \overline s}$,  is computed on only a subset of the 
configurations (see Table~\protect\ref{tab:Confs} for the number of configurations 
analyzed) and using only the $N_{HP}$ measurements given in Table~\protect\ref{tab:Sparams}, 
however, the fractional error in them is comparable to that in $M_\pi$. Mass ratios are 
given in the two columns on the right. The "physical" values of the ratios are with  
$M_\pi = 135$ MeV, $M_K$ = $495$ MeV, and $M_{{\overline s}s} = 686$ MeV. The latter 
is estimated using $\chi$PT and the Gellmann-Oakes-Renner relation. The plots for the effective mass, $a \Meff^\pi$, are shown in Figs~\ref{fig:C13} and~\ref{fig:F6}.}
    \label{tab:PStab}
\end{table*}

\begin{table*}[]
    \setlength{\tabcolsep}{10pt} 
    \renewcommand{\arraystretch}{1.2} 
    \centering 
\begin{tabular}{llllll}
\toprule
Ens. ID & $i$  & $a M_\pi$ & $a M_K$  &  $a f_\pi$ &  $a f_K$ \\
\midrule
$a117m310$  & 1 &  0.18343(27) &   0.30518(40) &   0.10057(14)  &   0.11097(32) \\
      & 2 &  0.18351(26) &   0.30527(39) &   0.100656(96) &   0.11112(30) \\
\hline
$a087m230$ & 1 &  0.10315(13) &   0.25894(23) &   0.07139(25) &   0.08387(23) \\
      & 2 &  0.10315(15) &   0.25896(23) &   0.07136(16) &   0.08388(22) \\
\hline
$a087m230X$  & 1 &  0.10013(14) &   0.23430(16) &  0.069820(78) &   0.08096(17) \\
      & 2 &  0.10021(14) &   0.23427(18) &  0.069887(66) &   0.08096(18) \\
\hline
$a087m290L$   & 1 &  0.128067(80) &   0.24375(21) &  0.073575(44) &   0.08338(22) \\
      & 2 &  0.128176(75) &   0.24366(25) &  0.073668(31) &   0.08340(25) \\
\hline
$a086m180L$    & 1 &  0.07808(12) &   0.22794(24) &  0.067116(51) &   0.07894(18) \\
      & 2 &  0.07817(11) &   0.22787(28) &  0.067153(42) &   0.07888(21) \\
\hline
$a068m290$    & 1 &  0.100100(58) &   0.21206(24) &  0.056824(29) &   0.06601(26) \\
      & 2 &  0.100116(55) &   0.21199(30) &  0.056826(25) &   0.06592(32) \\
\hline
$a068m230$    & 1 &  0.079683(59) &   0.19396(26) &  0.054070(39) &   0.06305(27) \\
      & 2 &  0.079711(58) &   0.19383(34) &  0.054091(38) &   0.06309(35) \\
\hline
$a068m175$    & 1 &  0.059384(73) &   0.19243(14) &  0.051712(44) &   0.06196(13) \\
      & 2 &  0.059410(70) &   0.19244(15) &  0.051740(37) &   0.06195(15) \\
\hline
$a067m135$    & 1 &  0.045672(70) &   0.18526(12) &  0.050186(43) &   0.06082(11) \\
      & 2 &  0.045700(67) &   0.18522(12) &  0.050203(40) &   0.06076(11) \\
\hline
$a053m295$    & 1 &  0.079603(46) &   0.14969(21) &  0.043945(25) &   0.04970(22) \\
      & 2 &  0.079592(44) &   0.14962(30) &  0.043938(24) &   0.04966(29) \\
\hline
$a053m230$    & 1 &  0.060892(52) &  0.14809(9)  &  0.041897(28) &  0.04917(9) \\
      & 2 &  0.060850(52) &  0.14803(10) &  0.041877(27) &  0.04916(10) \\
\bottomrule
\end{tabular}    
\caption{Data for the pion and kaon masses and decay constants in lattice 
units obtained using simultaneous fits to the 
pseudoscalar-pseudoscalar (Smeared-Smeared) and the pseudoscalar-axial (Smeared-Point) correlators. See plots in Figs.~\ref{fig:C13} and~\ref{fig:F6}. 
The statistics for the pions and kaons are different as described in the text and given in Table~\ref{tab:Confs}. 
The number $i=1$ or $2$ in column two specifies that the results are based on an $i$-state truncation of Eq.~\protect\eqref{eq:SD2pt}.}
    \label{tab:decayconst}
\end{table*}

\begin{table*}[h]
    \setlength{\tabcolsep}{10pt} 
    \renewcommand{\arraystretch}{1.2} 
    \centering
    \begin{tabular}{lcccccc}
     \hline \hline
    Ens. ID & $aM_\rho$ & $aM_{K^*}$ & $aM_{\Phi}$ & $M_{K^*}/M_\rho$ & $M_{\Phi}/M_\rho$  \\
     \hline \hline
    $a117m310$&  0.4622(41) & 0.5202(20) & 0.5705(13) & 1.125(11) & 1.234(11) \\
    $a087m290L$ & 0.3548(32)  & 0.4081(17) &  0.45688(95) & 1.153(18) & 1.295(17) \\
    $a087m230$ & 0.3512(63) & 0.4176(34) & 0.4827(10) & 1.189(23) & 1.374(25) \\
    $a087m230X$  & 0.3330(25) & 0.3936(16) & 0.44893(56) & 1.182(10) & 1.348(10)\\
    $a086m180L$ & 0.3132(62) & 0.3901(15) & 0.44545(95) & 1.246(25) & 1.422(28)\\
    $a068m290$ &  0.2858(39) & 0.3388(17) & 0.3873(11) & 1.185(17) & 1.355(19)\\
    $a068m230$ & 0.2592(46) & 0.3193(40) & 0.3660(10) & 1.232(27) & 1.412(25)\\
    $a068m175$ & 0.2444(64)& 0.3169(13) & 0.36838(64) & 1.297(34) & 1.507(40) \\
    $a067m135$ & 0.229(11) & 0.31069(89) & 0.36145(40) & 1.357(65) & 1.578(76) \\
    $a053m295$ & 0.2252(39) & 0.2578(14) & 0.28624(94) & 1.145(21) & 1.271(22) \\
    $a053m230$ & 0.2041(48) & 0.25183(99) & 0.28707(50) & 1.234(29) & 1.407(33) \\
    \hline
    Physical &  &  &  &  1.15037(61) &  1.31525(58)  \\
    \hline
    \end{tabular}
    \caption{Data for vector meson masses using 2-state fits. The $\Phi$ meson assumed to be a pure ${\overline s} s$ state and the disconnected contributions are neglected. Note that (i) estimates of $M_\rho$ 
    are not robust as the effective mass plot, $M_\rho^{\rm eff}$, 
    does not exhibit a credible plateau, and (ii) the input value of the strange quark in the calculation of $M_{K^*}$ and $M_{\Phi}$ is heavier than the 
    physical. }
    \label{tab:vectormasses}
\end{table*}

\begin{table*}[]
    \setlength{\tabcolsep}{10pt} 
    \renewcommand{\arraystretch}{1.2} 
    \centering 
\begin{tabular}{l|llll|l}
\toprule
 Ens. ID &          $aM_N$ &  $aM_\Sigma$     &    $aM_\Lambda$  &   $aM_\Xi$       &  $M_N/M_\pi$ \\
\midrule
   $a117m310$&    0.6175(9)(1) &  0.7219(38)(10) &   0.6852(50)(5) &  0.7741(46)(15) & 3.414(23)  \\
    $a087m290$ &    0.468(5)     &                 &                 &                 &            \\
   $a087m290L$ &   0.4671(6)(10) &  0.5588(41)(10) &   0.5350(35)(1) &   0.6132(25)(1) & 3.654(39)  \\
  $a087m230$ &    0.4484(7)(0) &  0.5652(53)(12) &   0.5364(54)(9) &   0.6395(25)(4) & 4.448(51)  \\
 $a087m230X$  &   0.4368(14)(2) &   0.5411(33)(8) &   0.5107(37)(1) &   0.5986(18)(1) & 4.410(37)  \\
    $a086m180$ &    0.416(2)     &                 &                 &                 &            \\
    $a086m180L$ &    0.4166(7)(2) &  0.5291(60)(19) &   0.4918(92)(8) &  0.5895(31)(10) & 5.423(88)  \\
    $a068m290$ &    0.3726(7)(1) &  0.4593(50)(20) &  0.4376(39)(14) &  0.5144(25)(18) & 3.764(43)  \\
    $a068m230$ &    0.3482(8)(1) &  0.4306(73)(22) &  0.4095(42)(28) &   0.4856(25)(2) & 4.360(61)  \\
    $a068m175$ &  0.3262(29)(40) &   0.4268(42)(9) &   0.3983(42)(3) &   0.4854(16)(3) & 5.59(12)   \\
    $a067m135$ &   0.3161(15)(5) &   0.4227(30)(1) &  0.3884(46)(15) &  0.4730(23)(11) & 6.88(15)   \\
    $a053m295$ &    0.2935(4)(1) &  0.3494(27)(18) &   0.3338(26)(8) &   0.3820(18)(3) & 3.744(38)  \\
    $a053m230$ &    0.2749(6)(1) &   0.3424(14)(8) &   0.3221(13)(5) &    0.3800(9)(4) & 4.504(61)  \\
Physical&                &                 &                 &                 &  6.956     \\
\bottomrule                                                                    
\end{tabular}
\caption{Masses of octet baryons in lattice units from 3-state fits. The statistics of the nucleon data is the highest
  as discussed in the text. This allowed investigation of two-, three-, and four-state truncations of Eq.~\eqref{eq:SD2pt} with the three- and four-state fits compared in Figs.~\ref{fig:C13_nucleon} and~\ref{fig:F6_nucleon} in Appendix~\ref{sec:N3and4-state-fits}.}
    \label{tab:oBtab}
\end{table*}

\begin{table*}[]
    \setlength{\tabcolsep}{10pt} 
    \renewcommand{\arraystretch}{1.2} 
    \centering 
\begin{tabular}{l|llll|l}
\toprule
 Ens. ID &      $aM_\Delta$  & $aM_{\Sigma^*}$  & $aM_{\Xi^\ast}$ &  $aM_\Omega$  &  $M_\Omega/M_\pi$  \\
\midrule
   $a117m310$&    0.7859(24)(20) &   0.8553(72)(25) & 0.9009(63)(23) & 0.9473(59)(16) &   5.174(24)  \\
    $a087m290$ &                   &                  &                &                &              \\
   $a087m290L$ &      0.6077(9)(3) &  0.6528(125)(36) & 0.6988(92)(24) & 0.7454(66)(12) &   5.842(61)  \\
  $a087m230$ &    0.5931(28)(38) &  0.6499(135)(48) &                & 0.7883(38)(13) &   7.600(44)  \\
 $a087m230X$  &    0.5823(29)(18) &    0.6325(91)(5) &                &  0.7374(39)(7) &   7.398(31)  \\
    $a086m180$ &                   &                  &                &                &              \\
    $a086m180L$ &    0.5658(20)(57) &  0.6165(113)(31) &                & 0.7302(43)(47) &   9.425(63)  \\
    $a068m290$ &    0.4841(19)(38) &  0.5352(125)(12) &                & 0.6319(60)(66) &   6.292(56)  \\
    $a068m230$ &    0.4569(37)(62) &    0.5124(90)(4) &                & 0.6005(63)(3)  &   7.523(49)  \\
    $a068m175$ &    0.4501(23)(56) &  0.5046(99)(133) &                & 0.5992(35)(13) &   10.106(95) \\
    $a067m135$ &    0.4392(35)(20) &   0.4975(86)(68) &                & 0.5909(29)(22) &   13.058(55) \\
    $a053m295$ &    0.3810(15)(23) &  0.4076(115)(94) &                & 0.4689(44)(14) &   5.887(79)  \\
    $a053m230$ &    0.3664(19)(14) &   0.3917(51)(13) &                & 0.4704(15)(16) &   7.754(26)  \\
Physical&                  &                  &                &                &  12.39     \\

\bottomrule                      
\end{tabular}
\caption{Masses of decuplet baryons in lattice units. The signal in the 
effective mass plots for the $\Delta$, $\Sigma^\ast$ and $\Xi^\ast$ 
degrades into noise before a plateau is manifest, so these estimates, 
based on a 2-state fit, are not robust. For most ensembles, 
no data for ${\Xi^\ast}$ were  collected due to an oversight.}
    \label{tab:dBtab}
\end{table*}

\section{\label{sec:gradflow}Gradient Flow and the Scales $t_0$ and $w_0$}
\label{sec:t0w0def}

The gradient flow is a procedure for smoothing of gauge degrees of freedom (links matrices) that provides an  ultraviolet scale  characterizing lattice ensembles~\cite{Luscher:2010iy}.  The continuum limit 
of quantities calculated on the lattice can then be taken holding this scale fixed in physical units, say MeV. The smoothing procedure evolves the gauge fields in the flow time $t_{gf}$ according to a gauge-covariant diﬀusion equation towards static points of the lattice action. It involves updating each gauge link at site $i$ in direction $\mu$ in flow time $\tgf$ using 
\begin{equation}
    \frac{dV(\tgf)_{i, \mu}}{d\tgf} = -g_0^2\frac{\partial S(V)}{\partial V_{i, \mu}}V_{i, \mu}, \ \ V_{i, \mu}(0) = U_{i, \mu} \,.
\end{equation}
In this work, we employ the stout smearing process~\cite{Morningstar:2003gk}, 
\begin{equation}
U^{(n+1)}_\mu (x) = \exp( -i\rho Q_\mu(x) ) U^{(n)}_\mu (x) \,,
\end{equation}
that, for small $\rho$, has been shown to be equivalent to  the Wilson flow scheme in Refs.~\cite{Luscher:2010iy,Nagatsuka:2023jos}. Here  
$Q_\mu$ is the sum of the 6 staples in planes orthogonal to the direction $\mu$. 
The parameter $\rho$ serves as the small flow step size, $\delta t$, which we choose to be 
 $\delta t = 0.01$.  This procedure  ensures a gauge-covariant smoothing procedure with well-understood continuum behavior.

To calculate the scale $t_0$, the infinitesimal gauge-field smearing steps are performed 
up to a flow time $t_{gf}$ at which the product of the energy density and the square of the flow time reaches the predefined value 
\begin{equation}
    t_{gf}^2E(t_{gf}) |_{t_{gf} = t_0} = 0.3 \,.
    \label{eq:t0def}
\end{equation}
A second scale, $w_0$, is obtained using the condition, 
\begin{equation}
    t_{gf} \frac{d}{dt_{gf}} \{t_{gf}^2E(t_{gf})\}\Bigg|_{t_{gf} =w_0^2 } = 0.3 \,.
    \label{eq:w0def}
\end{equation}
The resulting values of  $t_0$ and $w_0$ for the 13 ensembles are given 
in Table~\ref{tab:Confs}. 
Either quantity,  $t_0$ or  $w_0$,  can be used to set the scale of the unflowed lattice configurations as discussed in Sec.~\ref{sec:lattice-spacing}.

\section{\label{sec:topo}The $\theta$ and Weinberg Operators}
\label{sec:topology}

Our long-term interest in calculating the CP symmetry violating topological 
charge $Q$ and the dimension-six Weinberg operator $W$ is to determine 
their contribution to the nucleon electric dipole moment (nEDM). To remove lattice artifacts, we calculate them as a function of the Wilson flow as it provides a gauge-covariant smoothing of the gauge field and suppresses ultraviolet
fluctuations while preserving the topological properties. 
Using the gluonic field strength tensor $G_{\mu \nu}(x, \tgf)$ at flow time $\tau$, the 
topological charge $Q$ is given by 
\begin{equation}
    Q(\tgf) = \int d^4x \frac{1}{32\pi^2} \epsilon^{\mu \nu \rho \sigma} \textrm{Tr}[G_{\mu \nu}(x, \tgf) G_{\rho \sigma}(x, \tgf)] \,,
\end{equation}
and the dimension-six Weinberg charge $W$ by
\begin{eqnarray}
&&\hspace{-0.2in} W(\tgf) \equiv \nonumber \\
&&\hspace{-0.2in} \int d^4x  \epsilon^{\rho \mu \eta \sigma} \textrm{Tr}
    [G_{\mu \nu}(x, \tgf) G_{\nu \rho}(x, \tgf) G_{\eta \sigma}(x, \tgf)] .
\end{eqnarray}
In the absence of lattice artifacts, the topological charge should be integer valued 
on each configuration. In our data, this convergence is not reached 
at even the largest flow time simulated. However, as shown in the 
 panels on the left in Figs.~\ref{fig:C13top}--\ref{fig:F6top}, 
the distribution of $Q$ stabilizes at relatively early flow time.  
Fortunately, as discussed in Ref.~\cite{Bhattacharya:2021lol},  
the property  relevant for reliably extracting the topological susceptibility 
and the correlation of $Q$ or $W$ with nucleon $n$-point functions needed 
to calculate their contribution to the nEDM is the stabilization of the distribution as shown in Figs.~\ref{fig:C13top} and~\ref{fig:F6top}. 

In this work we analyze only the renormalization group invariant topological susceptibility, 
\begin{equation}
    \chi_Q \equiv \langle \int d^4 x Q(x) Q(0)\rangle \equiv \frac{1}{V} \langle Q^2\rangle \quad {\rm with} \quad  \langle Q \rangle = 0. 
\end{equation}
since our calculation of the renormalization constant for $W$, which has a 
power divergent mixing with $Q$, is not yet complete. 

\begin{figure}
    \centering
    \includegraphics[trim=70 0 100 20, width=0.9\linewidth]{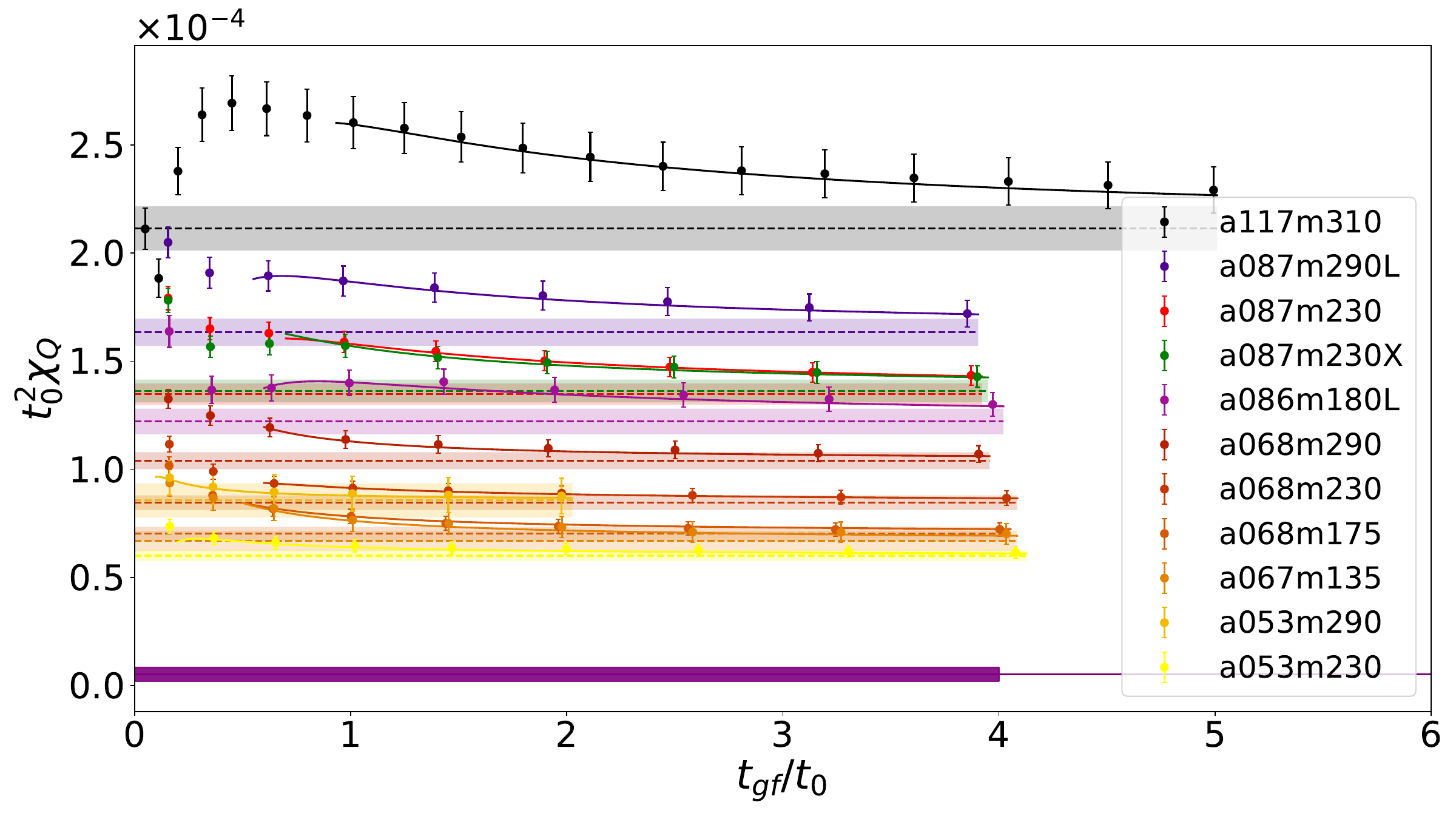}
    \vspace{-0.1in}
    \caption{The data for the dimensionless quantity $ t_0^2 \chi_Q$ is plotted as a function of the flow time $t_{gf} /t_{0}$ for the eleven ensembles. To fit all 
    the data on one plot, the x-axis for the $a117m310$ ensemble has been scaled and should be 
    read as $(t_{gf} /t_{0})/3$. To remove lattice artifacts, these data  
    are fit versus $1/t_{gf}$ using Eq.~\protect\eqref{eq:chifit_tf}. The $t_{gf}/t_0 \to \infty$ 
    values are shown by the horizontal bands that are then extrapolated to 
    the physical point using the chiral-continuum ansatz given in Eq.~\protect\eqref{eq:ChiQCC}. 
    This  physical point result is shown by the magenta horizontal band at the bottom.  
    }
    \label{fig:susc}
\end{figure}

In Fig.~\ref{fig:susc}, we plot the dimensionless quantity 
$ (t_0^2/a^4) (a^4 \chi_Q)$ as a function of the flow 
time $(t_0/a^2) (a^2 / \tgf)$ for the eleven ensembles.

Variation with the flow time $\tgf$ is a lattice artifact that, on each ensemble, is removed by fitting $\chi_Q$ versus $\tgf$ using the ansatz 
\begin{align}
  t_0^2 \chi_Q &= c_0 + c_1 a^2/\tgf +  c_2 (a^2/\tgf)^2 \,.
  \label{eq:chifit_tf}
\end{align}
This ansatz suffices (without overparameterization) for all eleven ensembles as shown in Fig.~\ref{fig:susc}.  The asymptotic values and the $\chi^2/{\rm dof}$ of the fits are given in Table~\ref{tab:chiQ_asymp}. Assuming finite volume corrections can be neglected, 
these numbers are then extrapolated to the 
chiral-continuum limit using a second order polynomial ansatz 
\begin{align}
    t_0^2 \chi_Q =   c_1 \frac{a^2}{t_0} &+ c_2 \phi_2 + c_3 \phi_2^2  +  c_4 \phi_2^2/ \phi_4 \nonumber \\
    &+ c_5 \phi_2 \frac{a^2}{t_0} + c_6 \phi_2 \phi_4 \,,
    \label{eq:ChiQCC}
\end{align}
that incorporates the requirement that the susceptibility $\chi_Q$ 
must vanish in the chiral-continuum limit. Possible discretization 
effects at finite lattice spacing $a$ are modeled by the term 
$c_1\frac{a^2}{t_0}$. Note that, throughout, this analysis is in terms of dimensionless quantities $t_0^2 \chi_Q$, $a^2/t_0$ (given in Table~\ref{tab:Confs}),  $\phi_2\equiv 8t_0 M_\pi^2$ and 
and $\phi_4\equiv 8t_0 (M_K^2 + M_\pi^2/2)$ that are calculated explicitly on the lattice  and do not require 
knowledge of the lattice scale $a$ of  each ensemble.

Once the $c_i$  are determined, 
Eq.~\eqref{eq:ChiQCC} is evaluated at $a = 0$, $M_\pi = 135$~MeV 
and with $t_0$ set to the  
already calculated $t_0^{\rm phy}$ to get $\chi_Q^{\rm phy}$ for the eleven 
models given by the various combinations of the $c_i$ specified in 
Table~\ref{tab:chiQ_models}.  Neglecting the two models, with 
non-zero $\{c_1,c_2\}$ and  $\{c_1,c_2,c_6\}$, that   
have large $\chi^2/{\rm dof}$, we perform an average weighted by $\sigma^2$ 
of the remaining nine to get the final value  $ \chi_Q = (65(10)\, \rm MeV)^4$. This is shown by the magenta horizontal band at the bottom of 
Fig.~\ref{fig:susc}, and is consistent with the result 
$ \chi_Q = (66(6)\,\text{MeV})^4$ obtained 
in Ref.~\protect\cite{Bhattacharya:2021lol} using the 
 same methodology but on ensembles with 2+1+1-flavors of HISQ fermions 
 provided by the MILC collaboration~\cite{Bazavov:2012xda}.

 
Phenomenologically, knowing $\chi_Q $ and especially its temperature dependence, is interesting 
for cosmology and whether the axion is a candidate for dark matter because 
of the relation:
\begin{equation}
    m_a^2 f_a^2 = \chi_Q \,,
\end{equation}
where $m_a$ and $f_a$ are the mass and the decay constant of the standard model 
axion that characterizes the axion coupling. To improve 
the estimate of $\chi_Q $, the main requirement is similarly precise data 
on many more ensembles to control   
the chiral-continuum extrapolation, and, hopefully, with increasingly 
more terms in Eq.~\eqref{eq:ChiQCC}. 

\begin{table}[hbt!]
    \setlength{\tabcolsep}{10pt} 
    \renewcommand{\arraystretch}{1.2} 
\begin{tabular}{|l|c|c|c|c||c|c|c||c|c|c|c|}
\hline
ID           & Configs  & $t_0^2 \chi_Q  \times 10^4 $   & $\chi^2/{\rm dof}$        \\
             & Analyzed & $(t_{gf} \to \infty)$  &                           \\
\hline
$a117m310$ & 998 &  2.12(10) & 0.56 \\
\hline			                                   
$a087m290L$ & 1794 & 1.634(63) & 0.79  \\
\hline			                                   
$a087m230$ & 2233 & 1.365(50) & 0.29  \\
\hline			                                   
$a087m230X$ & 2006 & 1.364(53) & 1.2   \\
\hline			                                   
$a086m180L$ & 1007 & 1.183(62) & 0.61   \\
\hline			                                   
$a068m290$ & 2112 & 1.040(39) & 1.7 \\
\hline			                                   
$a067m230$ & 2092 & 0.846(33) & 0.32 \\
\hline			                                   
$a067m175  $ & 1676 & 0.697(32) & 1.5  \\
\hline			                                   
$a067m135$ & 602 & 0.666(47) & 2.6 \\
\hline			                                   
$a053m295$ & 2441 & 0.856(79) & 0.3   \\
\hline			                                   
$a053m230$ & 1958 & 0.601(26) & 0.9  \\
\hline
\hline
\end{tabular}
\caption{The number of configurations used in the measurement of the 
    topological charge $Q$ and the Weinberg operator. The fit to $t_0^2\chi_Q$ versus $t_{gf/t_0}$, using Eq.~\protect\eqref{eq:chifit_tf} to obtain the  value in the limit $t_{gf} \to \infty$ 
for each ensemble, is shown in Fig.~\ref{fig:susc} and the results summarized here. \looseness-1 
}
\label{tab:chiQ_asymp}
\end{table}

Lastly, a goal of this paper is to characterize these 2+1-flavor 
clover ensembles and validate that they are a good representation of 
the QCD vacuum. One requirement is that topology is not frozen~\cite{Bernard:2017npd} during the lattice generation. 
This requirement is satisfied on all eleven ensembles as confirmed by the time histories of $Q$ and $W$  
at large flow time shown in Figs.~\ref{fig:C13top}--\ref{fig:F6top} 
(plots on the right) in Appendix~\ref{sec:flowQandW}. 

\begin{table*}[]
    \setlength{\tabcolsep}{7pt} 
    \renewcommand{\arraystretch}{1.2} 
    \centering
    \begin{tabular}{|c|c|c|c|c|c|c|c|}
    \hline
        $c_1$ & $c_2$ & $c_3$ & $c_4$ & $c_5$ & $c_6$ & $\chi_Q^{1/4}|^{Phy}\ (\textrm{MeV}) $ & $\chi^2/{\rm dof}$ \\
        \hline
         2.86 (12) &  1.10 (13) & & & & & 74.1(2.4) & 2.4\\
         \hline
         3.13(14) & 0.007(359) & 3.06(94)& & & & 51(15) & 1.2 \\
         \hline
         3.17(14) & -0.39(39) & & 6.6(1.7)& & & 40(31) & 0.50 \\
         \hline
         2.48(15) & 0.67(17) & & & 3.36(88) & & 65.3(4.3) & 0.69 \\
         \hline
         2.87(13) &0.93(66) & & & & 0.11(40) & 73.0(4.6) & 2.7 \\
         \hline
         3.15(14) &  -0.38(39) & -1.5(2.1) & 9.0(3.8) & & & 45(22)& 0.51\\
         \hline
         3.14(14) & 1.28(67) & 4.7(1.2) & & & -1.15(51) & 58(11) & 0.62 \\
         \hline
         2.35(18) & 1.54(68) & & & 3.86(96) & -0.58(44) & 70.8(5.0) & 0.51 \\
         \hline
         2.58(31) & 0.52(44) & 0.5(1.6) & & 2.9(1.5) & & 62.7(9.3) & 0.79 \\
         \hline
         2.89(31) & -0.05(50) & & 4.3(2.8) & 1.5(1.5) & & 51(17) & 0.41 \\
         \hline
         3.15(15) & -0.03(71) & & 6.9(1.7) & & -0.25(41) & 48(20) & 0.53 \\
         \hline
    \end{tabular}
    \caption{Results for the parameters  $c_i$, defined in Eq.~\protect\eqref{eq:ChiQCC}, for the eleven models used to perform the chiral-continuum extrapolation of the data for $\chi_Q$. The $\chi_Q^{1/4}|^{Phy}\ (\textrm{MeV})$ for each model is given in column seven and the $\chi^2/dof$ of the fit in column eight.
    } 
    \label{tab:chiQ_models}
\end{table*}


\section{Analysis of the Decay Constants $f_\pi $ and $f_K$}
\label{sec:fpi-fK}

The strength of the leptonic decays of pions ($\pi^+ \to \mu^+ \nu_\mu$) 
and kaons ($K^+ \to \mu^+ \nu_\mu$) via weak interactions is given by the decay 
constants  $f_\pi $ and $f_K$. 
These are obtained from the nuclear matrix elements, for example $\langle 0 | A_\mu | \pi (p) \rangle  = if_\pi p_\mu$, where $A_\mu$ is the 
renormalized axial part of the weak current and $\pi(p)$ is the pion state. 
These are, in turn, extracted efficiently and accurately from simultaneous fits to the lattice two-point functions (correlators) $\langle P_{S}(0)P_{S}(\tau)\rangle$ and $\langle P_{S}(0)A_{P}(\tau)\rangle$ with the interpolating 
operators $P = \overline{ \psi} \gamma_5 \psi$ and 
 $A_4 = \overline{ \psi} \gamma_4 \gamma_5 \psi$   placed at time $t=0$ (source) 
 and $\tau$ (sink). The subscript $S$ ($P$) denotes smeared (point) sources/sinks.
Since the S2N in these correlators does not degrade with $\tau$, we use the large $\tau$ 
limit of their spectral decomposition, 
\begin{equation}
  \langle  P_{S}(0)A_{P}(\tau) \rangle = \sum_i \langle 0 | P | \pi_i \rangle \langle \pi_i | A_4 | 0 \rangle e^{-E_i \tau} \,,
    \label{eq:SD2pt_dc}
\end{equation}
to isolate the ground state and obtain an accurate value of 
$\langle 0 | A_\mu | \pi (p) \rangle$. Two kinds of fits are done to 
check consistency: (i) a 1-state fit starting at large $\tau_{\rm min}$ 
where a single state domination is manifest, i.e., the plateau in $\meff$ is well established, and (ii) a 2-state fit starting earlier at a $\tau_{\rm min}$ where some excited state effects are present as shown in Figs.~\ref{fig:C13} and~\ref{fig:F6}. 
The results, summarized in Table~\ref{tab:decayconst}, 
show that the two fits give consistent, within $1 \sigma$, 
results for the masses, $M_\pi$ and $M_K$, and the decay constants, $f_\pi $ and $f_K$. 
For the final results, we take the estimates from the 1-state fits 
since the $M_0$ values from the individual fits to the 
two correlators are closer. 

The statistical analysis of the data was done by generating  two hundred bootstrap 
samples starting with the binned data on the eleven  
ensembles.  Noting that the number of binned data points $N_l/N_{\rm bin}$ 
(or $N_s/N_{\rm bin}$ for strange states) in each ensemble are different,   
$N_l/N_{\rm bin}$ (or $N_s/N_{\rm bin})$ values were drawn randomly from each. 
Estimates for $M_{\pi}$, $f_{\pi}$,  $M_{K}$, and $f_{K}$ on each of the 11 ensembles, 
were then calculated using Eq.~\eqref{eq:SD2pt}. 
These bare lattice values of $f_\pi $ and $f_K$ were renormalized using 
the  $Z_A$ calculated for the local isovector axial vector current on these ensembles. 
(The results for all 
five renormalization factors, $Z_\Gamma$, for all 11 ensembles,  calculated using the 
RI-sMOM intermediate scheme on the lattice, are summarized in Table~\ref{tab:Zfactors} 
and discussed in Sec.~\ref{sec:Zfactors}.) Lastly, results  at 
the physical point were obtained 
by doing a continuum-chiral fit. This process was repeated 200 times to get the 200 bootstrap samples used for estimating the errors.

The first application of these data is to extract the physical 
point value for the scales, $t_0^{\rm Phy}$ and $w_0^{\rm Phy}$. As  
described in Sec.~\ref{sec:fpiK}, this is done 
using the chiral-continuum (CC) fit ansatz given in Eq.~\eqref{eq:fit_fpiK} for 
the combination $f_{\pi K}$ defined as
\begin{equation}
af_{\pi K} = 
Z_A\left(\frac{2}{3}af_K + \frac{1}{3}af_\pi\right) \,.
\end{equation}
The second, described in Sec.~\ref{sec:fKbyfpi}, is to calculate the ratio $f_K / f_\pi|^{\rm Phy}$
using two ansatz for the CC fit given in 
Eqs.~\eqref{eq:log-chiral-ratio} and~\eqref{eq:poly-ratio-fit-formula} and the physical values given in Eq.~\eqref{eq:physical_values}.

Before presenting these analyses, we first discuss the calculation of the 
renormalization constants $Z_\Gamma$ for all local quark bilinear 
operators, of which  $Z_A$ is used to renormalize $A_\mu$ needed in the calculation of $f_\pi$ and $f_K$.

\subsection{The Renormalization Constants $Z_\Gamma$}
\label{sec:Zfactors}

The renormalization constants, $Z_\Gamma$, for quark bilinear local operators 
are calculated using the regularization independent
symmetric momentum subtraction (RI-sMOM) intermediate scheme 
on the lattice~\cite{Martinelli:1994ty,Sturm:2009kb}. We   
calculated the needed 2- and 3-point functions using external quark states on 
lattices fixed to the Landau gauge at a number of values of the  
four-momentum squared, $p^2_{\rm sym}$,  that are symmetric (the same) 
in all three legs of the 3-point functions. These $Z^{\rm RI-sMOM}$ for each 
$p^2_{\rm sym}$ are converted to the  $\MSbar$ scheme 
at scale $\mu$ using horizontal matching , i.e.,  with $\mu^2 = p^2_{\rm sym}$,  
and then 
run to $\mu = 2$~GeV. The last two steps use results from QCD perturbation theory. 
The end results at $\mu = 2$~GeV should be independent of $p^2_{\rm sym}$ for all the $Z_\Gamma$ , however, 
in practice, they exhibit a small dependence 
on $p^2_{\rm sym}$ that is a lattice artifact arising due to discretization effects, 
in particular due to the Lorentz group being broken to $90^o$ rotations. 
This artifact is removed by making a fit versus $p^2_{\rm sym}$ over a 
range specified in physical units, i.e., in GeV. 
Such a fit introduces the need to know,  even for $Z_V$ and $Z_A$,  $a$ in GeV and $\alpha_s(1/a)$ for each ensemble. These are given in Table~\ref{tab:lattice-spacing-decay-constant} 
and their extraction is discussed in Sec.~\ref{sec:lattice-spacing}.

Further details of this procedure for calculating $Z_\Gamma$
for the 2+1-flavor theory  are given in the Appendix B in Ref.~\cite{Park:2025rxi} and we direct readers to it.  
While we give results for $Z_\Gamma$ 
calculated using both $a^\anaive$ and  $a^\acorr$ in Table~\ref{tab:Zfactors}, 
all our final results are obtained using $a^{\rm naive}$. 
Using these $Z_A$ and $a^\anaive$, the estimates for the 
decay constants on the 11 ensembles are given in Table~\ref{tab:decayconst_ren}. 

\begin{table*}[]
    \centering \footnotesize
        \setlength{\tabcolsep}{6pt} 
    \renewcommand{\arraystretch}{1.2} 
\begin{tabular}{l c ll ll ll ll l}

Ens. ID &   & $Z_A$ &  $Z_S$ &  $Z_P$ &  $Z_T$ &  $Z_V$ & $Z_A/Z_V$ & $Z_S/Z_V$ & $Z_P/Z_V$ & $Z_T/Z_V$ \\

\hline
$a117m310$ & n &   0.876(9) &  0.820(11) &  0.739(34) &  0.881(12) &  0.796(17) &  1.097(13) &  1.021(23) &  0.925(49) &  1.106(10) \\
      & c &  0.879(10) &  0.841(11) &  0.763(31) &  0.871(16) &  0.799(21) &  1.096(15) &  1.043(27) &  0.951(49) &  1.091(10) \\


\hline
$a087m290L$ & n &  0.884(11) &   0.790(8) &  0.717(16) &  0.923(16) &  0.823(16) &   1.075(9) &  0.957(22) &  0.871(30) &   1.122(8) \\
      & c &  0.885(12) &  0.807(10) &  0.736(13) &  0.913(15) &  0.824(17) &   1.074(9) &  0.977(23) &  0.892(29) &   1.109(8) \\

\hline
$a087m230$ & n &   0.883(7) &   0.791(9) &  0.720(21) &  0.924(16) &  0.825(15) &  1.071(11) &  0.958(26) &  0.873(40) &   1.122(8) \\
      & c &   0.884(8) &  0.804(11) &  0.733(20) &  0.917(16) &  0.826(16) &  1.071(12) &  0.973(27) &  0.888(41) &   1.112(8) \\

\hline
$a087m230X$ & n &  0.874(14) &   0.776(7) &  0.710(13) &  0.919(19) &  0.820(19) &   1.065(8) &  0.944(22) &  0.865(30) &   1.121(7) \\
      & c &  0.875(16) &   0.786(7) &  0.721(12) &  0.913(20) &  0.821(22) &   1.065(9) &  0.954(24) &  0.877(30) &   1.113(7) \\


\hline
$a086m180L$ & n &  0.872(11) &  0.782(10) &  0.712(15) &  0.918(13) &  0.823(15) &   1.059(9) &  0.946(19) &  0.863(24) &   1.116(8) \\
      & c &  0.873(11) &  0.787(10) &  0.717(14) &  0.915(13) &  0.823(16) &   1.060(9) &  0.952(19) &  0.869(24) &   1.113(8) \\

\hline
$a068m290$ & n &   0.896(7) &  0.770(11) &  0.721(10) &  0.957(16) &  0.853(12) &   1.052(8) &  0.902(26) &  0.846(25) &   1.123(6) \\
      & c &   0.897(7) &  0.787(10) &   0.738(9) &  0.946(16) &  0.853(13) &   1.052(9) &  0.923(26) &  0.865(24) &   1.110(6) \\

\hline
$a067m230$ & n &   0.891(5) &  0.757(13) &  0.701(11) &   0.959(9) &   0.852(9) &   1.046(7) &  0.886(22) &  0.821(18) &   1.126(7) \\
      & c &   0.892(5) &  0.769(13) &  0.712(10) &   0.951(9) &   0.852(8) &   1.046(7) &  0.900(22) &  0.834(17) &   1.117(7) \\

\hline
$a066m175$ & n &   0.895(6) &  0.761(11) &  0.713(12) &  0.961(13) &  0.852(12) &   1.050(8) &  0.891(24) &  0.835(24) &   1.129(6) \\
      & c &   0.895(6) &  0.767(11) &  0.719(12) &  0.958(13) &  0.852(12) &   1.050(8) &  0.898(23) &  0.842(24) &   1.124(6) \\

\hline
$a066m135$ & n &  0.894(14) &  0.768(39) &  0.714(29) &   0.970(9) &  0.853(16) &  1.047(20) &  0.891(44) &  0.830(34) &  1.138(20) \\
      & c &  0.897(12) &  0.771(37) &  0.715(30) &   0.969(8) &  0.857(13) &  1.045(19) &  0.892(47) &  0.828(35) &  1.132(17) \\

\hline
$a053m295$ & n &   0.900(6) &  0.737(10) &  0.692(11) &  0.989(12) &   0.865(9) &   1.040(5) &  0.850(19) &  0.798(19) &   1.143(6) \\
      & c &   0.900(7) &   0.752(9) &   0.706(9) &  0.979(11) &  0.866(10) &   1.040(5) &  0.866(19) &  0.815(19) &   1.132(6) \\

\hline
$a053m230$ & n &   0.902(5) &  0.745(10) &  0.705(10) &  0.984(14) &   0.870(9) &   1.038(6) &  0.856(19) &  0.810(20) &   1.132(6) \\
      & c &   0.903(5) &  0.754(10) &  0.715(10) &  0.978(14) &   0.870(9) &   1.038(6) &  0.866(19) &  0.821(19) &   1.125(6) \\

\end{tabular}
    
    \caption{The renormalization constants $Z_\Gamma^{\MSbar}(2\text{GeV})$ for the isovector quark bilinear operators. The steps in their calculation are  (i)  calculate $Z_\Gamma^{\rm RI-sMOM}$ nonperturbatively on the lattice  using the RI-sMOM scheme at various values of symmetric momentum $p^2_{\rm sym}$; (ii) match to the continuum $\MSbar$ scheme at $\mu^2=p^2_{\rm sym}$;  (iii) run these results to 2~GeV; and (iv) remove a lattice artifact, the dependence on $p^2_{\rm sym}$, by making fits as explained in the text. The labels "n" and "c" in the second column specify whether the results were obtained using $a^{\rm naive}$ or $a^\acorr$, respectively. 
    }
    \label{tab:Zfactors}
\end{table*}

\subsection{Extracting the scales $t_0^{\rm Phy}$ and $w_0^{\rm Phy}$ using  the data for $f_{\pi K}$}
\label{sec:fpiK}

To extract $t_0^{\rm Phy}$ and $\wphy$ from the data for $f_{\pi K}$, a fit is made using 
the following ansatz: 
\begin{widetext}
\begin{equation}
\sqrt{8t_0}f_{\pi K}(M_\pi, M_K, a)
= \frac{A}{4\pi}
\left[
1-\frac{7}{6} L\left(\frac{\phi_2}{A^2}\right) - \frac{4}{3}
L\left(\frac{\phi_4 - \frac{1}{2}\phi_2}{A^2}\right)
-
\frac{1}{2}
L\left(\frac{\frac{4}{3}\phi_4 - \phi_2}{A^2}\right)
+ B\phi_4
\right]
\left(
1 + C f(\frac{a}{\sqrt{8t_0}})
\right),
\label{eq:fit_fpiK}
\end{equation}
\end{widetext}
with the first part on the right hand side modeling,  
motivated by $\chi$PT~\cite{Gasser:1984gg,Bar:2013ora,Strassberger:2021tsu}, 
 the chiral behavior and the second part the discretization  corrections. 
 
Here $L(x)=x\ln x$ and  $A,\ B,\ C$ are fit parameters that, in general, 
depend on $\phi_2$, $\phi_4$ and other low-energy constants of $\chi$PT. 
The ideal situation for the adequacy of such a low-order ansatz would be 
if the ensembles had been generated along the green 
line in Fig.~\ref{fig:ensemble_overview} implying a roughly constant $m_s$. 
Our ensembles, unfortunately, have a large variation 
with $m_s > m_s^{\rm Phy}$. Future lattice 
generation should target the green line. 

For parameterizing the discretization errors, we investigate three possibilities: 
\begin{equation}
f(a) = 
\frac{a^2}{8t_0},
\quad
\frac{\alpha_s a}{\sqrt{8t_0}},
\quad
\frac{a}{\sqrt{8t_0}}\,,
\label{eq:fa3cases}
\end{equation}
with ${\alpha_s a}/{\sqrt{8t_0}}$ expected to be the closest  for our almost $O(a)$ improved action.

As shown in Eq.~\eqref{eq:fit_fpiK}, the analysis is done making all terms 
dimensionless using $t_0/a^2$ or $w_0/a$ that are also calculated on the lattice. 
When using $t_0/a^2$,  $A = 4 \pi \sqrt{8t_0} \fpiK^{\rm chiral\ limit}$. 
Doing the full analysis using  $w_0^2$ instead of $t_0/a^2$
in Eq.~\eqref{eq:fit_fpiK} gives $\wphy$, however, the results 
for $\tphy$ and $\wphy$ are highly correlated.

\begin{figure}
    \begin{center}
    \includegraphics[width=0.49\linewidth]{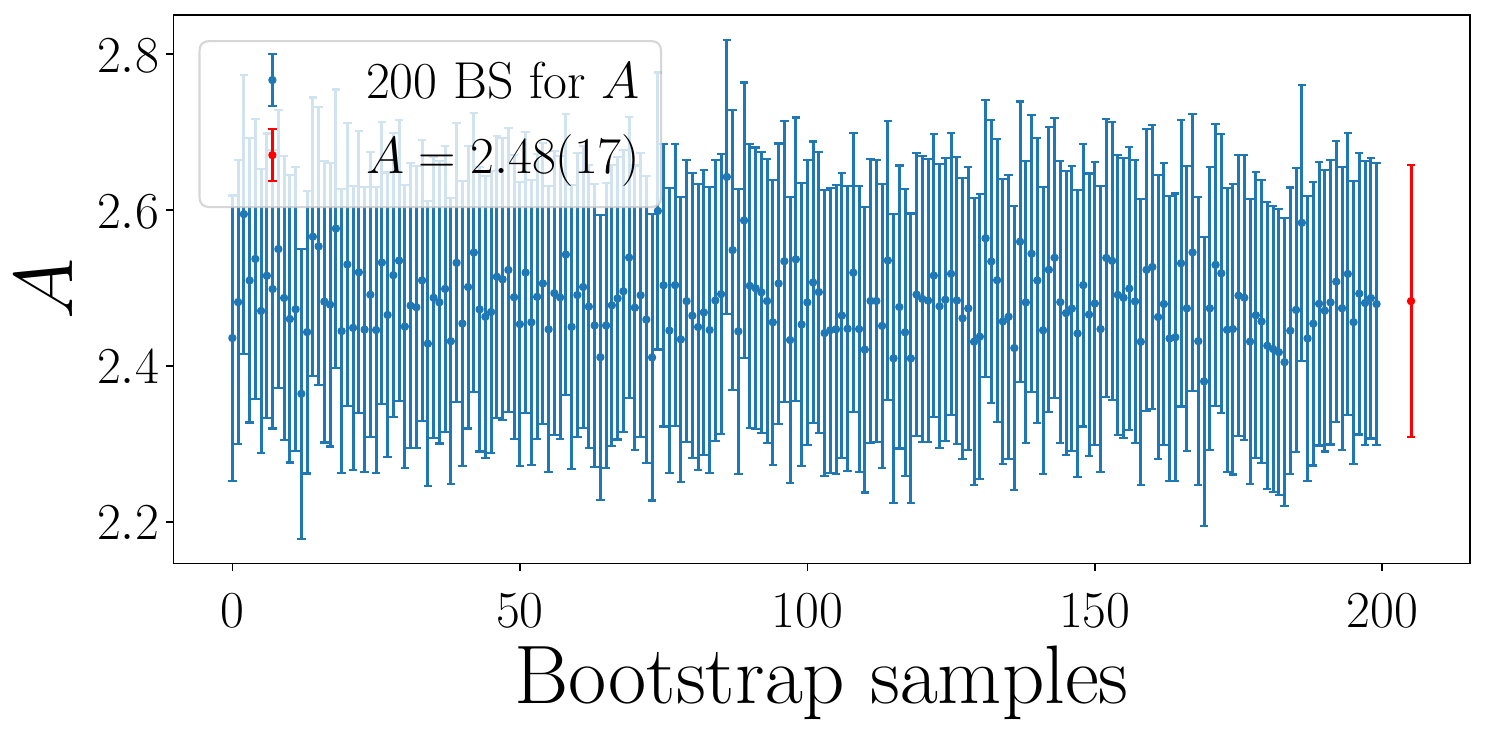}
    \includegraphics[width=0.49\linewidth]{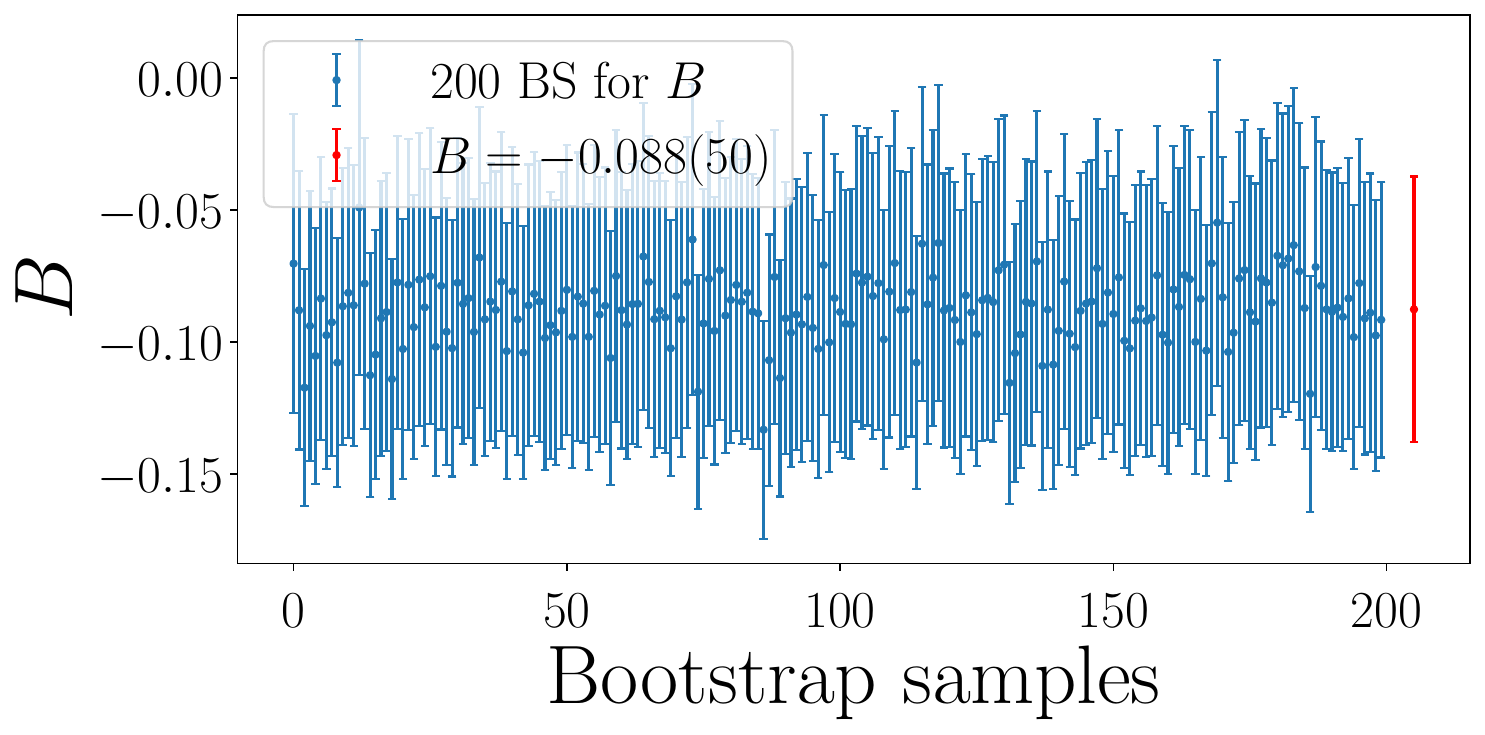} \\
    \includegraphics[width=0.49\linewidth]{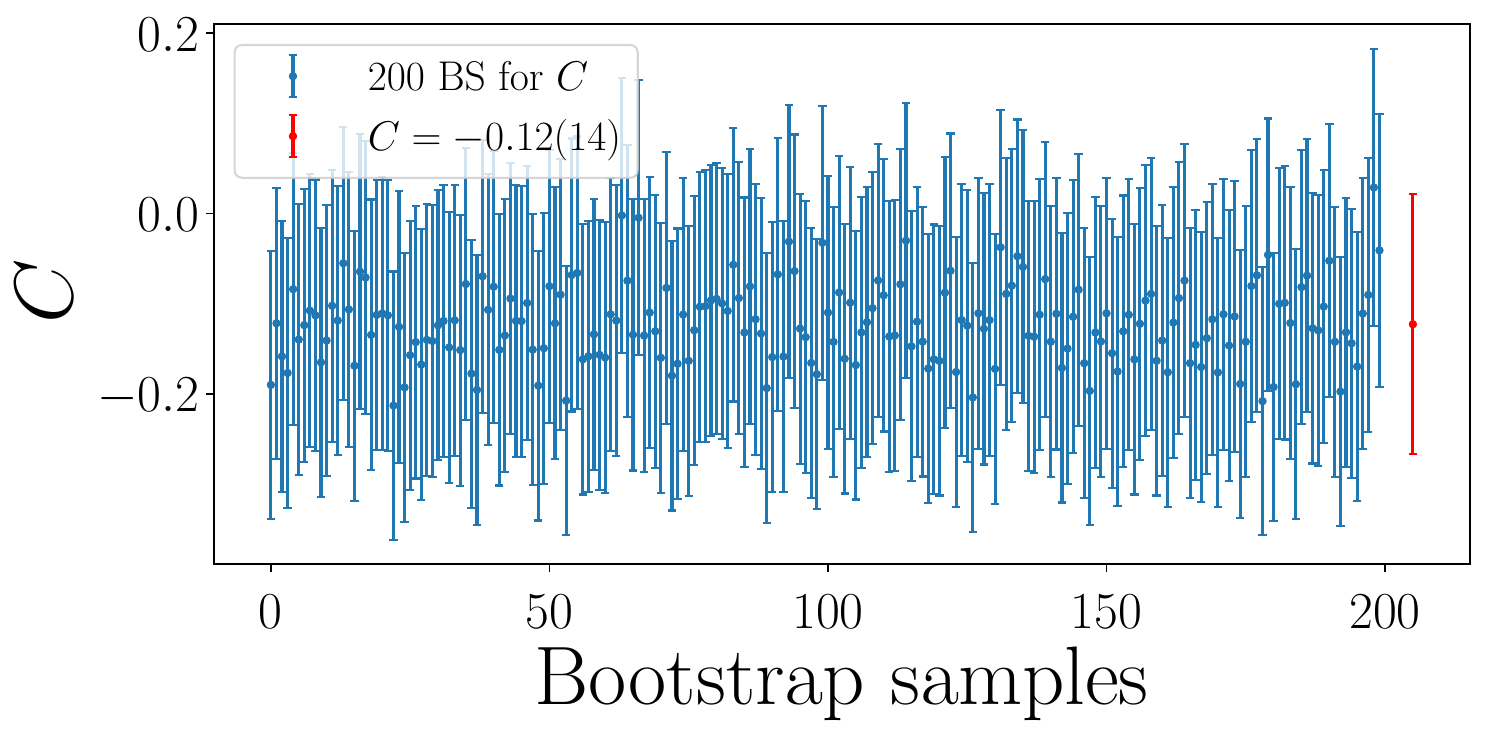}
    \includegraphics[width=0.49\linewidth]{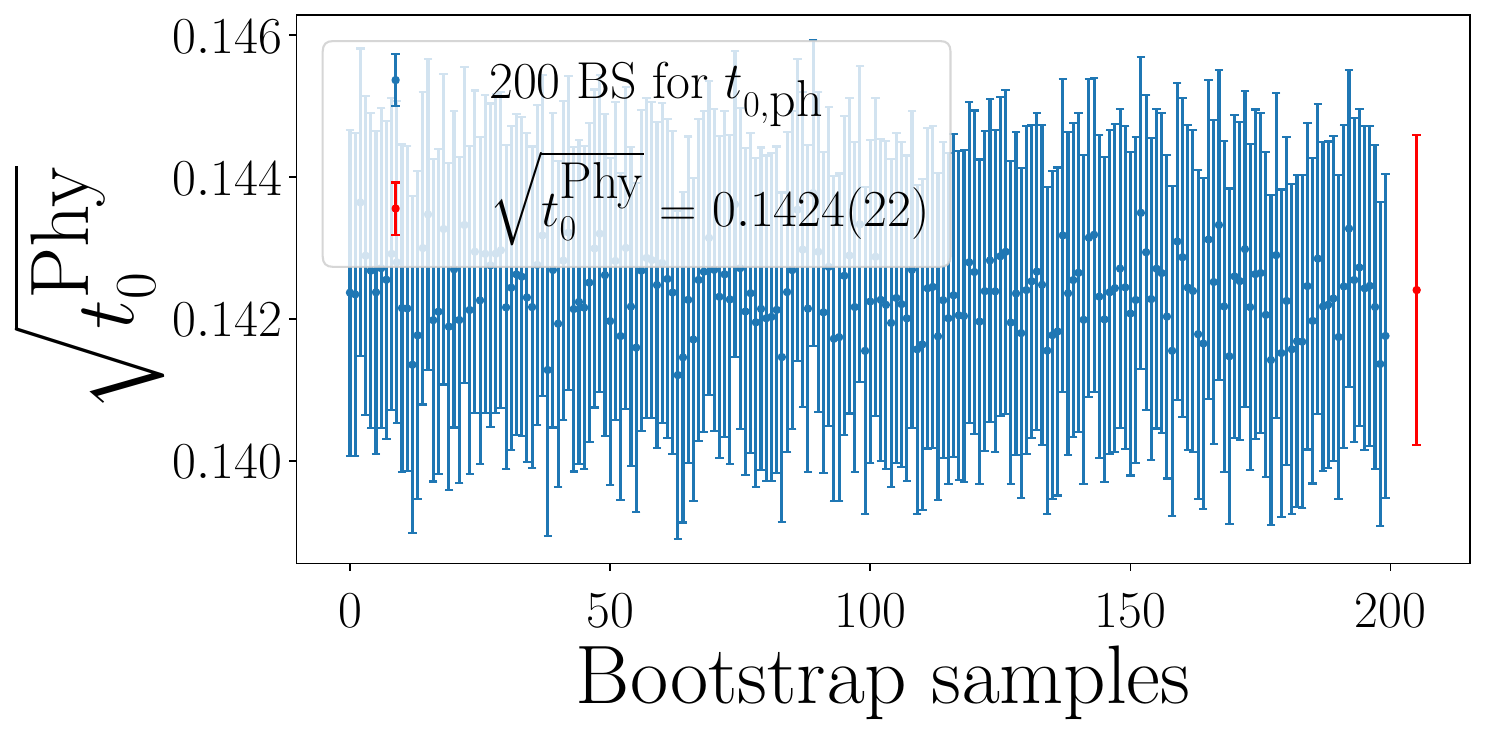}
    \end{center}
    \caption{Results for the fit parameters $\{A, \ B, \ C \}$ defined in Eq.~\eqref{eq:fit_fpiK} for 
    the 200 bootstrap samples (BS) used for the statistical analysis. The red point on the 
    right in each panel shows the bootstrap average. In these fits, $t_0$ 
    was used to make quantities dimensionless. 
    }
    \label{fig:fpiK-ABC-BS-fit}
\end{figure}

Our central analysis is done assuming $\{A,\ B,\ C\}$ are constants.  
The statistical analysis is done by generating 200 bootstrap samples. 
The values of $\{A,\ B,\ C\}$, using $t_0$ to make quantities dimensionless, 
are shown in Fig.~\ref{fig:fpiK-ABC-BS-fit}.

\begin{table*}
    \centering \footnotesize
    \setlength{\tabcolsep}{6pt} 
    \renewcommand{\arraystretch}{1.2} 
    \begin{tabular}{l|c|c|c|c|c||c|c|c|c|c}
        \hline
        \textbf{$O(a)$ } & $A$ & $B$ & $C$ & \textbf{$\sqrt{\tphy}$} &\textbf{$\chi^2_{\mathrm{red}}$}
        & $A$ & $B$ & $C$ & \textbf{$\wphy$} & \textbf{$\chi^2_{\mathrm{red}}$}\\
        \hline
        $a^2$  
        & $2.48(17)$ & $-0.088(51)$ & $-0.12(14)$ & $0.1421(20)$ & $0.96$ 
        & $3.17(21)$ & $-0.100(29)$ & $-0.72(18)$ & $0.1702(23)$ & $0.80$
        \\
        \hline
        $\alpha_S a$  
        & $2.48(17)$ & $-0.088(50)$ & $-0.12(14)$  & $0.1424(22)$ & $0.96$ 
        & $3.20(22)$ & $-0.100(29)$ & $-0.64(16)$ & $0.1718(25)$ & $0.75$
        \\
        \hline
        $a$ 
        & $2.49(18)$ & $-0.087(50)$ & $-0.050(55)$ & $0.1430(26)$ & $0.95$ 
        & $3.25(22)$ & $-0.097(28)$ & $-0.243(59)$ & $0.1744(29)$ & $0.69$
        \\
        \hline
    \end{tabular}
    \caption{
        Estimates for the fit parameters $A$, $B$, and $C$ defined in Eq.~\eqref{eq:fit_fpiK}, and the
        extracted values of $\sqrt{\tphy}|_{f_{\pi K}}$ and $\wphy|_{f_{\pi K}}$. Results are presented for the  three 
        ansatz used for modeling the discretization error.
    }
    \label{tab:t0_C_dependence-new}
\end{table*}

\begin{figure}    
    \centering
    \includegraphics[width=0.98\linewidth]{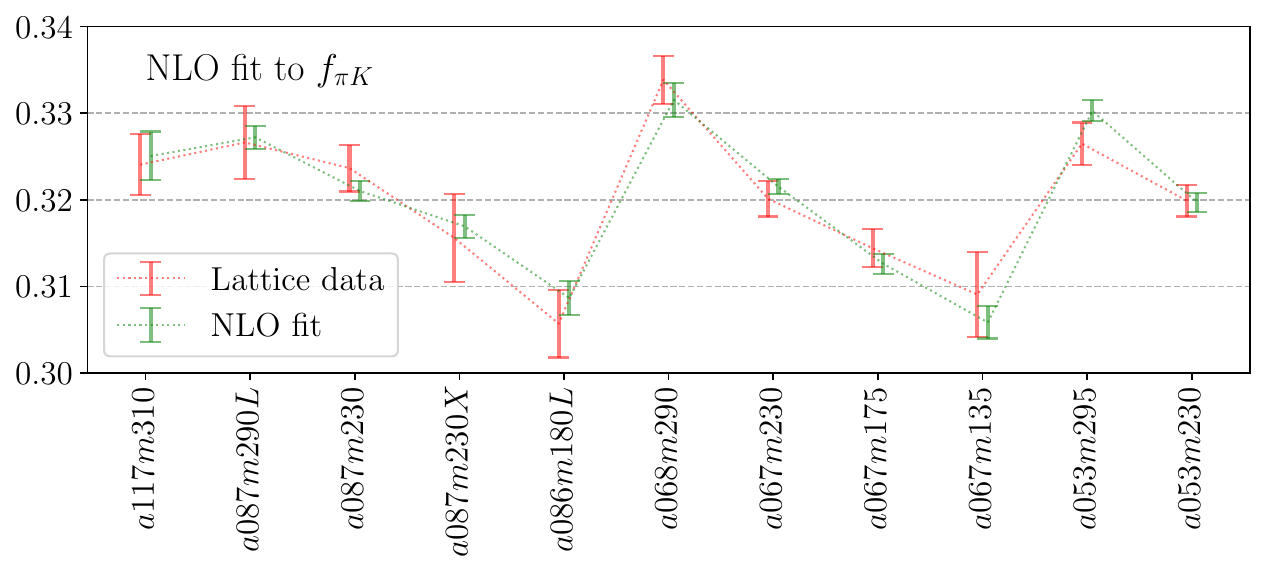}
    \caption{This figure illustrates how well data for $\sqrt{8t_0} f_{\pi K}$ are fit by 
    the ansatz in Eq.~\eqref{eq:fit_fpiK}. The fit shown 
    is for one bootstrap sample with $\tphy$ used to set the scale and the $\mathcal{O}(\alpha_s a)$ continuum correction factor. It has $\chi^2/\textrm{dof}=1.20$.
    }
    \label{fig:fpiK-t0-fit-one-BS}
\end{figure}

With $A,\ B,\ C$ determined, the relation in Eq.~\eqref{eq:fit_fpiK} 
is solved for $t_0^{\rm Phy}$ (or $\wphy$)
using physical point values for $f_{\pi K}$, $\phi_2$ and $\phi_4$. 
The results for the three ansatz used for modeling discretization errors overlap as shown in 
Table~\ref{tab:t0_C_dependence-new}, however, there is an  
insignificant increase in the value of $\tphy$ 
(and similarly of $\wphy$) under $a^2 \to \alpha_s a \to a$. 
An example,  one of the 200 bootstrap samples, of how well the ansatz in Eq.~\eqref{eq:fit_fpiK} fits the data is 
shown in Fig.~\ref{fig:fpiK-t0-fit-one-BS}.

Considering that 
the Dirac action used in this study is close to $O(a)$ improved, we choose 
the  $f(a/\sqrt{8t_0}) = (\alpha_s a)/\sqrt{8t_0}$ ansatz for the central value 
and the larger of the three errors to get 
\begin{eqnarray}
\sqrt{t_{0}^\textrm{Phy}}|_{f_{\pi K}} &=& 0.1424(26)~{\rm fm} \,, \nonumber \\
w_{0}^\textrm{Phy}|_{f_{\pi K}} &=& 0.1718(25)~{\rm fm} \,.
\label{eq:scales_fpiK}
\end{eqnarray}
These results agree, within $1\sigma$, with the FLAG 2024 averages  
$\sqrt{t_{0}^\textrm{Phy}}|_{\rm FLAG} = 0.14474(57)~{\rm fm} $ 
and $w_{0}^\textrm{Phy}|_{\rm FLAG} = 0.17355(92)~{\rm fm}$ for the 2+1-flavor theory~\cite{FlavourLatticeAveragingGroupFLAG:2024oxs}, however, 
the errors in our results are much larger. 

With eleven data points, we have investigated a larger parameter space by 
allowing the $A$ and $ B$ to depend on $\phi_4$. The ansatz 
$ A(\phi_4) = A_0 + A_1 \phi_4$ changed the results by $\sim 10\%$, however the 
error in $\tphy$ increased by a factor of about 5. The ansatz 
$ B(\phi_4) = B_0 + B_1 \phi_4$ led to unphysical results. We, therefore, conclude 
that these additional  terms 
lead to overparameterization and our best estimates from the 11 ensembles 
are with keeping $A$ and $B$ constants.

\begin{table}
    \centering
    \begin{tabular}{lccccc}
     \hline \hline
Ens. ID & $f_\pi^{\textrm{bare}}$ & $f_K^{\textrm{bare}}$ &  $f_\pi^{\textrm{ren}}$ & $f_K^{\textrm{ren}}$  \\
& MeV & MeV  & MeV & MeV \\
 \hline \hline
$a117m310$  & 170.3(3.1) & 188.0(3.4) & 149.2(3.1) & 164.7(3.4) \\
$a087m290L$ & 161.2(3.0) & 189.4(3.5) & 142.5(3.2) & 167.4(3.7) \\
$a087m230$  & 157.8(2.9) & 183.0(3.4) & 139.4(2.8) & 161.6(3.2) \\
$a087m230X$ & 167.8(3.1) & 190.2(3.6) & 146.7(3.6) & 166.2(4.1) \\
$a086m180L$ & 154.2(2.9) & 181.3(3.4) & 134.4(3.0) & 158.1(3.6) \\
$a068m290$  & 164.4(2.9) & 191.0(3.4) & 147.3(2.8) & 171.1(3.4) \\
$a068m230$  & 158.5(2.8) & 184.9(3.4) & 141.3(2.6) & 164.7(3.2) \\
$a068m175$  & 152.3(2.7) & 182.5(3.3) & 136.3(2.6) & 163.3(3.1) \\
$a067m135$  & 148.5(2.7) & 179.9(3.3) & 132.7(3.2) & 160.9(3.8) \\
$a053m295$  & 162.5(3.0) & 183.8(3.4) & 146.3(2.8) & 165.4(3.3) \\
$a053m230$  & 155.7(2.8) & 182.7(3.4) & 140.6(2.7) & 165.0(3.2) \\
\hline
    \end{tabular}
    \caption{Bare and renormalized values for the pion and kaon decay constant using 
    the renormalization constant $Z_A$ specified in Table~\ref{tab:Zfactors}  
    calculated using the lattice 
    spacing $a^{\rm naive}$ given in Table~\ref{tab:lattice-spacing-decay-constant}. For comparison, the values at the physical point given in FLAG 2024~\cite{FlavourLatticeAveragingGroupFLAG:2024oxs} are $f_\pi = 130.2(0.8)$ MeV and $f_K = 155.7(0.7)$ MeV for the $N_f = 2+1$ theory. }
    \label{tab:decayconst_ren}
\end{table}

\subsection{Setting the lattice scale $a$ using $\sqrt{t_{0}^\textrm{Phy}}$ and $w_{0}^\textrm{Phy}$}
\label{sec:lattice-spacing}

In the determination of the scales $\tphy$ and $\wphy$ using  $\fpiK$ 
(see Sec.~\ref{sec:fpiK}) or the octet or decuplet baryon masses, 
the value of the lattice spacing $a$ in physical units 
on each ensemble is not needed as the chiral part of the CC ansatz 
can be written in terms of dimensionless quantities and 
the calculation of $Z_A$ does not depend on 
knowing $\alpha_s(1/a)$ and, thus, on $a$. 
A dependence on $a$ is, however, introduced if the continuum part of the CC ansatz 
includes terms of $O(\alpha_s a)$. In this work we will neglect this (circular) 
effect as it is tiny as evident from the agreement of results obtained by considering the two 
values of $\alpha_s$ from two significantly different estimates 
of $a$ given in Table~\ref{tab:lattice-spacing-decay-constant}. 
Knowing the value of the scale $a$ on each ensemble is, however, necessary 
in other analyses, for example in the calculation of the renormalization constants 
presented in Sec.~\ref{sec:Zfactors} or if  scales in physical units 
are used to regulate chiral logarithms as in Eq.~\eqref{eq:log-chiral-ratio}. 

In this section, we therefore determine the lattice spacing $a$ on each ensemble 
in physical units based on the $f_{\pi K}$ fit. 
The first is the "naive" choice of $a$ using the extracted value of $\tphy$ given in Eq.~\eqref{eq:scales_fpiK} and the measured $t_0/a^2$:
\begin{equation}
  a_{f_{\pi K}}^\mathrm{naive} = (t_0/a^2)^{-1/2}\sqrt{\tphy}|_{f_{\pi K}} \,.
  \label{eq:a_naive}
\end{equation}
This definition implies  that the value of $a$ is adjusted so that $t_0=\tphy$ on each ensemble. The second, $a_{f_{\pi K}}^\mathrm{corr.}$, includes the corrections factors given by the fit for $\fpiK$:
\begin{widetext}
\begin{eqnarray}
a_{f_{\pi K}}^\mathrm{corr.} &=& a_{f_{\pi K}}^\mathrm{naive} \times C_1 \times C_2 \nonumber \\
C_1 &\equiv& 1 + C \frac{\alpha_s a}{\sqrt{8t_0}},
\nonumber \\
C_2 &\equiv&
\frac{
1-\frac{7}{6} L\left(\frac{\phi_2}{A^2}\right) - \frac{4}{3}
L\left(\frac{\phi_4 - \frac{1}{2}\phi_2}{A^2}\right)
-
\frac{1}{2}
L\left(\frac{\frac{4}{3}\phi_4 - \phi_2}{A^2}\right)
+ B\phi_4}
{1-\frac{7}{6} L\left(\frac{\phi_2^\mathrm{Phy}}{A^2}\right) - \frac{4}{3}
L\left(\frac{\phi_4^\mathrm{Phy} - \frac{1}{2}\phi_2^\mathrm{Phy}}{A^2}\right)
-
\frac{1}{2}
L\left(\frac{\frac{4}{3}\phi_4^\mathrm{Phy} - \phi_2^\mathrm{Phy}}{A^2}\right)
+ B\phi_4^\mathrm{Phy}},
\label{eq:correction-factors}
\end{eqnarray}
\end{widetext}
where $\phi_2^\mathrm{Phy}$ and $\phi_4^\mathrm{Phy}$ are computed using the physical pion and kaon masses. 
These two estimates of the lattice spacing are summarized in Table~\ref{tab:lattice-spacing-decay-constant}. 

\begin{table*}[h!]
    \centering 
        \setlength{\tabcolsep}{8pt} 
    \renewcommand{\arraystretch}{1.2} 
\begin{tabular}{l|ll|llll}
\toprule
Ens. ID 
& $a_{f_{\pi K}}^\mathrm{naive}$ [fm] & $\alpha_s(1/a)$
& $C_1$
& $C_2$
& $a_{f_{\pi K}}^\mathrm{corr.}$ [fm] & $\alpha_s(1/a)$ 
\\
\midrule
$a117m310$  & 0.1165(21) & 0.324(4)   & 0.998(13)  & 1.090(11)  & 0.1255(28)   & 0.338(6) \\
$a087m290L$   & 0.0874(16) & 0.278(3)   & 0.9926(87)  & 1.090(12)  & 0.0946(18)   & 0.289(3) \\
$a087m230$ & 0.0873(16) & 0.278(3)   & 0.9926(87)  & 1.069(12)  & 0.0927(18)   & 0.286(4) \\
$a087m230X$  & 0.0865(16) & 0.277(3)   & 0.9927(86)  & 1.0566(87) & 0.0908(16)   & 0.283(3) \\
$a086m180L$    & 0.0859(16) & 0.276(3)   & 0.9928(85)  & 1.0301(62) & 0.0879(15)   & 0.279(3) \\
$a068m290$    & 0.0682(12) & 0.249(2)   & 0.9948(61) & 1.099(16)  & 0.0745(15)   & 0.258(3) \\
$a068m230$    & 0.0673(12) & 0.247(2)   & 0.9949(59) & 1.067(11)  & 0.0714(13)   & 0.253(2) \\
$a068m175$    & 0.0670(12) & 0.247(2)   & 0.9950(59) & 1.0384(84) & 0.0692(11)   & 0.250(2) \\
$a067m135$    & 0.0667(12) & 0.246(2)   & 0.9950(59) & 1.0167(61) & 0.0674(10)   & 0.248(2) \\
$a053m295$    & 0.0534(10) & 0.226(2)   & 0.9963(43) & 1.093(13)  & 0.0581(11)   & 0.233(2) \\
$a053m230$    & 0.0531(10) & 0.225(2)   & 0.9964(43) & 1.0597(95)  & 0.05607(92)   & 0.230(2) \\
\bottomrule
\end{tabular}    
\caption{
The lattice spacing $a_{f_{\pi K}}^\mathrm{naive}$ is obtained using 
 Eq.~\eqref{eq:a_naive}, which corresponds to requiring the value 
of $t_0$ on each ensemble to equal $\tphy$ obtained from the $f_{\pi K}$ fit.   The second estimate, 
$a_{f_{\pi K}}^\mathrm{corr.}$, is extracted 
from $a_{f_{\pi K}}^\mathrm{naive}$ by including 
the continuum and chiral 
correction factors $C_1$ and $C_2$ as defined in Eq.~\eqref{eq:correction-factors}. This 
scale corresponds to requiring $f_{\pi K}$ on each ensemble equals $f_{\pi K}^{\rm Phy}$.}
    \label{tab:lattice-spacing-decay-constant}
\end{table*}

Two notable points regarding the second definition, $a^\acorr$, 
given in Eq~\eqref{eq:correction-factors}.  
First, most of the correction comes from the chiral behavior, i.e., from the term $C_2$.
Second, the net effect of including the correction factors $C_i$ in 
extracting the lattice spacing $a^\acorr$ 
is to set the quantity $f_{\pi K} = f_{\pi K}^\mathrm{Phy}$ on each ensemble 
to within the accuracy of the fit ansatz. Thus, using $a^\acorr$ to take the continuum limit for other quantities implies $f_{\pi K}$ is being kept constant in those  fits. 

In other words, these two definitions of the lattice scale specify which mass dimension quantity is held fixed in physical units and used to construct the dimensionless variables used in the CC fits to take the continuum limit, i.e., $a/\sqrt{8t_0}$ or $a \fpiK$. To any given order of the CC ansatz both  are valid. 
Differences reflect the fact that (i) truncated ansatz are used to make the CC fits and (ii) a limited number 
of data points are available.

The last issue is--which value of $a$ should be used in the calculation of $\alpha_s(1/a)$ that enters in modeling discretization errors if terms 
proportional to $\alpha_s a$ are used and in the scale dependent renormalization constants $Z_{S,T}$?  For example, how to handle the difference in the  results for the renormalized scalar charge of the nucleon, $Z_S g_S$,  on each ensemble and in the final result $g_{S}^{\rm Phy}$ after the continuum-chiral extrapolation using a given ansatz? 
Again our contention is, at a given order, both are valid. One should, however, use 
$\alpha_s(1/a^{\rm naive})$ [$\alpha_s(1/a^\acorr)$] with $a^{\rm naive}$ [$a^\acorr$]. 
Note that in the two fits with the same CC ansatz and different $a$ values, the fit parameters will  
absorb part of the differences. Only an exact fit ansatz applied to much more extensive data  
is expected to give the same $g_{S}^{\rm Phy}$. Based on this understanding, in this work we 
use $a^\mathrm{naive}$ for the calculation of 
$\alpha_s$, and the renormalization factors discussed in Sec.~\ref{sec:Zfactors}. 
We neglect the differences in results from the full analysis with  $a^\anaive$ versus $a^\mathrm{corr.}$ as 
they are much smaller than the other errors we quote.

\subsection{Calculating the ratio $f_K/f_\pi$}
\label{sec:fKbyfpi}

Having used $\fpiK$ to set the lattice scale, we can determine the second independent 
quantity $\fpifk$ using the data for $f_\pi$ and $f_K$. For this, we investigate two 
ansatz to carry out the chiral-continuum extrapolation. The first is the 
next-to-leading order chiral fit given in Ref.~\cite{Durr:2016ulb} that is  
based on the SU(3) $\CPT$ in Ref.~\cite{Gasser:1984gg}:
\begin{eqnarray}
\label{eq:log-chiral-ratio}
\frac{f_K}{f_\pi}
=
1
&+&
\frac{c_0}{2}
\bigg[
\frac{5}{4}\left(8t_0 M_\pi^2\right)\log\frac{(aM_\pi)^2}{(a\mu)^2}
 \nonumber \\
&-&  \frac{1}{2}\left(8t_0 M_K^2\right)\log\frac{(aM_K)^2}{(a\mu)^2}
\nonumber \\
&-& \frac{3}{4}\left(8t_0 M_\eta^2\right)\log\frac{(aM_\eta)^2}{(a\mu)^2}
 \nonumber \\
&+&
c_1
8t_0
\left(
M_K^2 - M_\pi^2
\right)
\bigg],
\end{eqnarray}
where $M_\eta^2 = (4M_K^2 - M_\pi^2)/3$ and 
$\mu$ is a fixed renormalization scale of the chiral effective theory in physical 
units, which we choose to be $\mu=770~\mathrm{MeV}$. Then, to keep all terms 
in Eq.~\eqref{eq:log-chiral-ratio} dimensionless  requires  knowing the lattice spacing $a$ on each ensemble to 
converting $\mu$ to $a \mu$. 
We use $a^{\rm naive}$ determined in Sec.~\ref{sec:lattice-spacing}. 

\begin{figure}      
    \centering
   \includegraphics[width=0.98\linewidth]{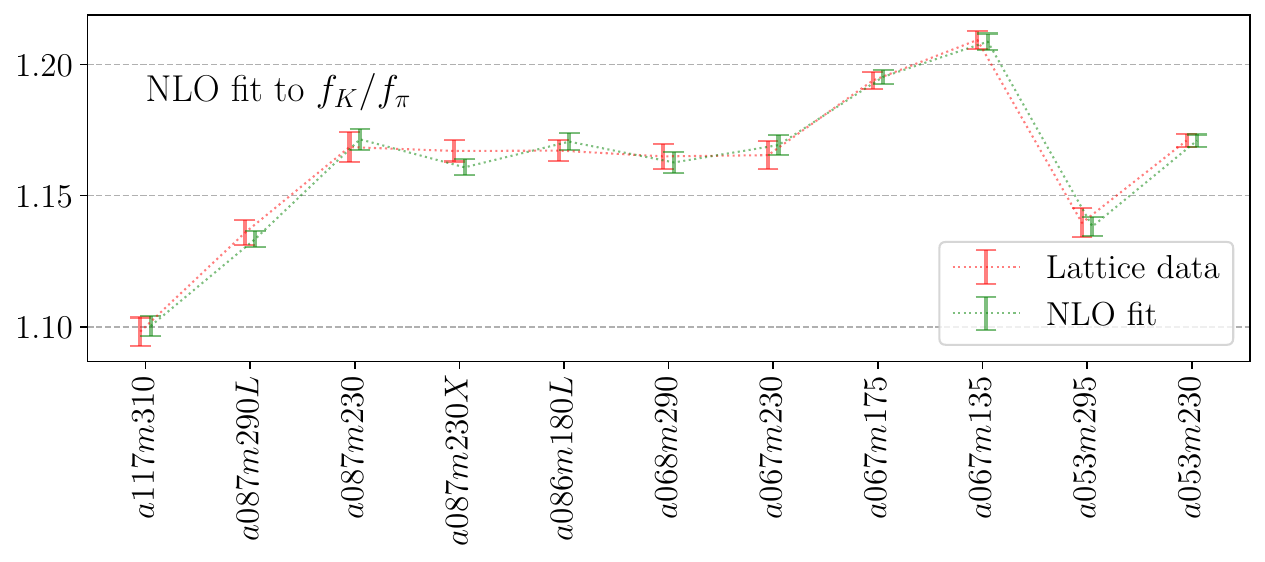}
    \caption{Result of the fit to $f_{K}/f_{\pi}$ using Eq.~\eqref{eq:log-chiral-ratio} with $\mathcal{O}(\alpha_s a)$ discretization ansatz and scale set by $\tphy$. A single bootstrap sample was used to make this plot. The fit has $\chi^2/\textrm{dof}=1.09$. Plots using $\mathcal{O}(a)$ and $\mathcal{O}(a^2)$ are 
    almost identical\looseness-1.
    }
    \label{fig:fpiK-t0-fit}
\end{figure}

We also consider the 4-parameter polynomial ansatz used in Ref.~\cite{Durr:2016ulb}:
\begin{eqnarray}
\label{eq:poly-ratio-fit-formula}
\frac{f_K}{f_\pi}
=
1
&+&
8t_0
\left(
M_K^2 - M_\pi^2
\right) 
\bigg[
c_0
+
c_1
8t_0
\left(
M_K^2 - M_\pi^2
\right) \nonumber \\
&+&
c_2
8t_0
M_\pi^2
+
c_3
8t_0
M_\pi^4
\bigg].
\end{eqnarray}

\begin{figure}    
    \centering
    \includegraphics[width=0.9\linewidth]{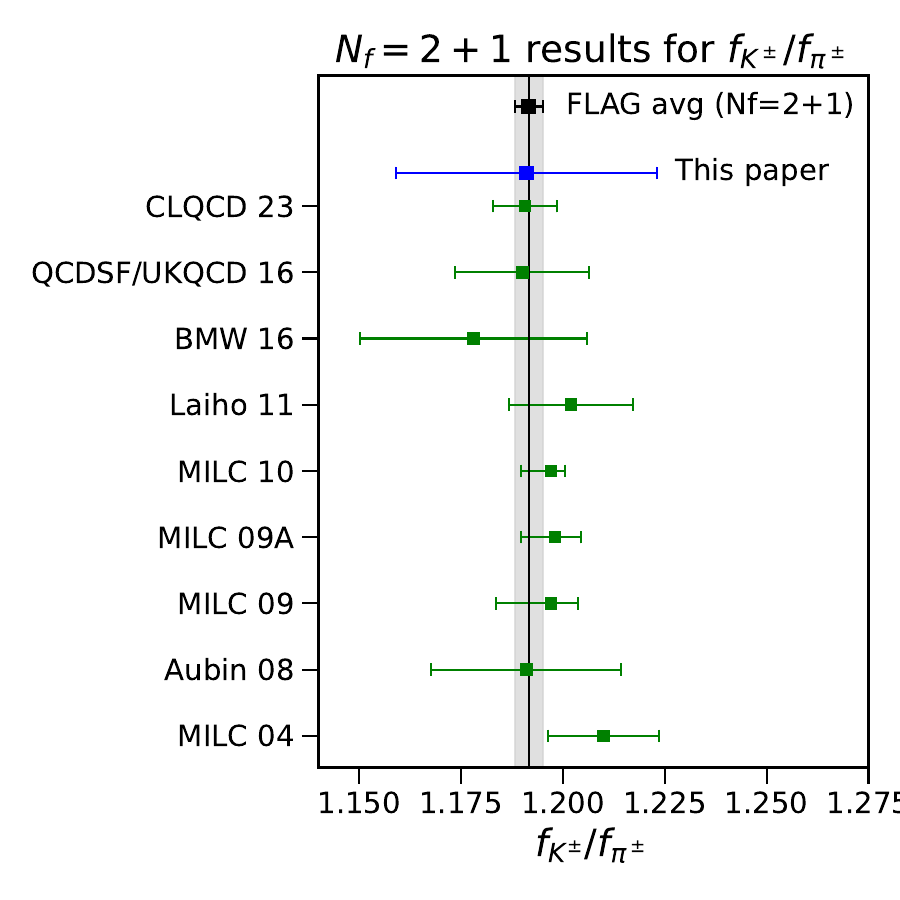}
\caption{
Comparison of \( \frac{f_{K^\pm}}{f_{\pi^\pm}} \) (blue square) obtained in 
this work with results from previous lattice studies with 2+1-flavor QCD: 
CLQCD 23~\cite{CLQCD:2023sdb}, 
QCDSF/UKQCD 16~\cite{QCDSF-UKQCD:2016rau}, 
BMW 16~\cite{Durr:2016ulb},
Laiho 11~\cite{Laiho:2011np}, 
MILC~10~\cite{MILC:2010hzw}, 
MILC~09A~\cite{MILC:2009ltw}, 
MILC~09~\cite{MILC:2009mpl}, 
Aubin~08~\cite{Aubin:2008ie}, 
and MILC~04~\cite{MILC:2004qnl}.
The grey band shows the FLAG 2024 average given in  Ref.~\cite{FlavourLatticeAveragingGroupFLAG:2024oxs}. 
}    
\label{fig:ratio-decay}
\end{figure}

Table~\ref{tab:ratio-fit-results-BS} gives the 
results for $f_K/f_\pi |^{\rm Phy}$ obtained using different CC fit models: the 
three ansatz for discretization errors, the two ansatz for the chiral correction, 
and the two values of the lattice spacing, $a_{f_{\pi K}}^\mathrm{naive}$ and $a_{f_{\pi K}}^\mathrm{corr.}$.
 We choose $\fpifk=1.193(33)$ given in Table~\ref{tab:ratio-fit-results-BS} for 
the $\mathcal{O}(\alpha_s a)$ continuum correction and $a^{\rm naive}$. To compare this isosymmetric 2+1-flavor theory result 
 with the phenomenological extraction,  ${f_{K}^\pm}/{f_{\pi}^\pm}$, requires applying the 
 isospin breaking correction $\delta_{\textrm{SU}(2)}$~\cite{Cirigliano:2011tm,FlavourLatticeAveragingGroupFLAG:2024oxs} using the relation:
\begin{equation}
    \frac{f_{K^\pm}}{f_{\pi^\pm}} = \frac{f_{K}}{f_{\pi}} \sqrt{1 + \delta_{\textrm{SU}(2)}} \,.
\end{equation}
We find $\delta_{\textrm{SU}(2)}= -0.0039(11) \Longrightarrow \frac{f_{K^\pm}}{f_{\pi^\pm}} = 1.191(32)$. As shown in Fig.~\ref{fig:ratio-decay}, while our central value is consistent, the error is large compared with other lattice determinations that enter the FLAG average and the FLAG 2024 average  $\frac{f_{K^\pm}}{f_{\pi^\pm}}=1.1917(37)$.
 
\begin{table*}[h!]
    \centering
    \setlength{\tabcolsep}{10pt} 
    \renewcommand{\arraystretch}{1.2} 
    \begin{tabular}{c|l|c|c}
        \hline
        \textbf{$f(a/\sqrt{8t_0})$ } 
        & Fit formula 
        & $f_K/f_\pi |^{\rm Phy}$  
        & $f_K/f_\pi |^{\rm Phy}$ 
        \\
                & 
        &  using $\sqrt{\tphy}|_{f_{\pi K}}$ 
        &  using $\wphy|_{f_{\pi K}}$
        \\
        \hline
        $a^2$ & Eq.~\eqref{eq:log-chiral-ratio} with $a=a_{f_{\pi K}}^\mathrm{naive}$
        & $1.192(33)$ & $1.195(35)$
        \\
        $a^2$ & Eq.~\eqref{eq:log-chiral-ratio} with $a=a_{f_{\pi K}}^\mathrm{corr.}$
        & $1.200(44)$ & $1.202(46)$
        \\
        $a^2$ & Eq.~\eqref{eq:poly-ratio-fit-formula}
        & $1.186(22)$ & $1.190(27)$
        \\
        \hline
        $\alpha_S a$ & Eq.~\eqref{eq:log-chiral-ratio} with $a=a_{f_{\pi K}}^\mathrm{naive}$
        & $1.193(33)$ & $1.193(35)$
        \\
        $\alpha_S a$ & Eq.~\eqref{eq:log-chiral-ratio} with $a=a_{f_{\pi K}}^\mathrm{corr.}$
        & $1.202(45)$ & $1.201(47)$
        \\
        $\alpha_S a$ & Eq.~\eqref{eq:poly-ratio-fit-formula}
        & $1.188(23)$ & $1.191(29)$
        \\
        \hline
        $a$ & Eq.~\eqref{eq:log-chiral-ratio} with $a=a_{f_{\pi K}}^\mathrm{naive}$
        & $1.196(35)$ & $1.189(35)$
        \\
        $a$ & Eq.~\eqref{eq:log-chiral-ratio} with $a=a_{f_{\pi K}}^\mathrm{corr.}$
        & $1.206(48)$ & $1.199(47)$
        \\
        $a$ & Eq.~\eqref{eq:poly-ratio-fit-formula}
        & $1.190(25)$ & $1.191(30)$
        \\
        \hline
    \end{tabular}
    \caption{
        Results for  the ratio $f_{K}/f_{\pi}|^{\rm Phy}$ for the three ansatz used for 
        removing discretization errors, the two ansatz for the chiral correction, and two values of 
        the lattice spacing, $a_{f_{\pi K}}^\mathrm{naive}$ and $a_{f_{\pi K}}^\mathrm{corr.}$.
        The statistical errors shown within the parentheses were calculated using the 
        200 bootstrap samples described in Sec.~\protect\ref{sec:fpi-fK}. 
        Results are shown for the two cases: scale set by $t_0^{\mathrm{Phy}}$ (the $\chi^2/{\rm dof} \approx 0.63(31)$ for all the fits) and $w_0^\mathrm{Phy}$ (the $\chi^2/{\rm dof} \approx 0.70(34)$ for all the fits).
    }
    \label{tab:ratio-fit-results-BS}
\end{table*}

\section{Analysis of the octet and decuplet baryon spectrum}

Our analysis covers the 4 octet baryon states, $O \in \{N,\, \Sigma,\,\Lambda,\,\Xi\}$, 
and only one decuplet state, $D = \{\Omega\}$ 
that have a good S2N ratio as discussed in Sec.~\ref{sec:quality}.
The physical point values are obtained using a simultaneous CC fit
as discussed in sections~\ref{sec:noFV} and~\ref{sec:CCstrategy} with the full analysis done 
using $t_0$ and $w_0^2$ independently to make quantities dinmensionless.
The chiral parts of the fits, i.e., behavior versus pion and kaon masses, 
use predictions from heavy baryon chiral perturbation theory (HB$\chi$PT) 
at NLO (accurate to $\mathcal{O}(m_q)$) and at NNLO (accurate to $\mathcal{O}(m_q^{3/2})$) 
order~\cite{Jenkins:1990jv, Ellis:1999jt} as discussed in Sec.~\ref{sec:CCstrategy}. For discretization effects, three possible ansatz for the leading order correction term are explored.
The inputs for this CC analysis are 
\begin{itemize}
\item Lattice measurements of meson and baryon masses, $\{aM_\pi, aM_K, aM_O, aM_D\}$, 
the decay constants $f_\pi$ and $f_K$, and the Wilson flow scales $\{t_0/a^2, w_0/a\}$ for each ensemble. 
\item HB$\chi$PT formulae, made dimensionless using $t_0/a^2$ (or $w_0/a$) \cite{Bar:2013ora}, describing the mass dependence of the octet ($O$) and the decuplet ($D$) 
baryons~\cite{Jenkins:1990jv, Ellis:1999jt}, in terms of $M_{\pi}$ and $M_{K}$.  
\item The physical values, $M_{\pi}^{\textrm{Phy}}$,  $M_{K}^{\textrm{Phy}}$, and one other quantity given in 
Eq.~\eqref{eq:physical_inputs} to set the quark masses and the lattice scale. One can choose it to be a baryon mass, or $\tphy$ or $\wphy$. 
\end{itemize}

Predictions for the octet and decuplet baryon spectrum are then extracted from these CC fits in the following two ways:
\begin{itemize}
    \item Set the scale using $\tphy$ (or $\wphy$) extracted in Sec.~\ref{sec:fpiK}.  
    \item Use one baryon mass to set the scale and then predict  $\tphy$ (or $\wphy$), $\fpiK$ and $\fpifk$ and all but  
    that one baryon mass. 
\end{itemize}

\subsection{Analysis of the octet baryon masses}
\label{sec:octet_ChPT}

The CC extrapolation formula for the octet baryon masses, with all variables made dimensionless using $t_0/a^2$, is:
\begin{eqnarray}\label{eq:continuum-form-for-octet}
&&
\sqrt{8t_0} M_O(M_\pi, M_K, a)
\nonumber \\
&=&
\sqrt{8t_0} M_O(M_\pi, M_K, a=0)
\nonumber \\
&&
\times
\left\{
1+f(a)
\left[
c_o + \bar{c}_o(8t_0 \overline{M}^2)
+\delta c_O(8t_0 \delta{M}^2)
\right]
\right\},
\phantom{XX}
\end{eqnarray}
where $O \in \{N,\Lambda,\Sigma,\Xi\}$. For the factor $f(a)$ modeling discretization corrections, 
we investigate the same three cases given in Eq.~\eqref{eq:fa3cases}. 
The continuum part of the ansatz has 
six parameters: the $c_o$ and $\bar{c}_o$ are common while $\delta c_O$ are different for 
each octet baryon.
The two mass parameters $\overline{M}^2$ and $\delta M^2 $ are 
\begin{equation}\label{eq:mass-simp}
\overline{M}^2
\equiv
\frac{1}{3}(2M_K^2 + M_\pi^2)
,
\quad
\delta M^2 \equiv 2(M_K^2 - M_\pi^2).
\end{equation}

The chiral part of the ansatz, at NNLO $[\mathcal{O}(m_q^{3/2})]$ and again in units of $t_0$, is 
\begin{eqnarray}
\sqrt{8t_0}
{M}_O(M_\pi, M_K)
&=&
\bm{M_0} + 
\bm{\bar{b}} \left(8t_0\overline{M}^2\right) + \bm{\delta b_O} \left(8t_0\delta {M}^2\right)
\nonumber \\
&&
+
\sum_{P = \pi, K, \eta_8}
\bm{g_{O,P}} f_O\left(\frac{\sqrt{8t_0}{M}_P}{\bm{M_0}}\right),
\nonumber \\
\label{eq:a-NNLO-ChPT-fit-octet}
\end{eqnarray}
where the first three terms on the right define the NLO $[\mathcal{O}(m_q)]$ ansatz. 
The various dimensionless parameters (bold letters), and their relation to 
commonly used parameters, are 
\begin{eqnarray}
\bm{M_0} &\equiv& \sqrt{8t_0}M_0\,,
\nonumber \\
\displaystyle
\bm{\bar{b}} &\equiv& \frac{\bar{b}}{\sqrt{8t_0}}\,,
\qquad
\bm{\delta b_O} \equiv \frac{\delta b_O}{\sqrt{8t_0}}\,, \nonumber \\
\bm{g_{O,P}}
&\equiv&
\sqrt{8t_0}
\frac{M_0^3}{(4\pi F_0)^2}{g}_{O,P}\,,
\end{eqnarray}
with $F_0$ the pion decay constant in the limit of $N_f=3$ massless quarks and $P \in \{\pi,\ K, \ \eta_8 \}$.
Note that in Ref.~\cite{RQCD:2022xux}, the same factor ${a}/{\sqrt{8t_0^*}}$ 
was used for all ensembles with the same inverse coupling $\beta$. We  use the $a/\sqrt{8t_0}$ calculated on each ensemble.
The parameter $\bm{M_0}$ is the common octet baryon mass in the chiral-continuum  limit, $\bm{\bar{b}}$ is the common fit parameter which controls the $\overline{M}^2$ dependence, and $\bm{\delta {b_O}}$ defines  
the SU(3) breaking part that is different for each state. The EOMS loop-function $f_O(x)$ \cite{Gegelia:1999gf, Geng:2009hh, MartinCamalich:2010fp}
is given by\looseness-1
\begin{equation}
f_O(x) = -2x^3
\left[
\sqrt{1-\frac{x^2}{4}}
\arccos\left(\frac{x}{2}\right)
+\frac{x}{2}\ln x
\right].
\end{equation}
The $\eta_8$ mass can be written in terms of the pion and kaon masses as
\begin{equation}
M_{\eta_8}^2\approx
\frac{4M_K^2 - M_\pi^2}{3}
=
\overline{M}^2 + \frac{1}{3}\delta M^2,
\end{equation}
where we used Eq.~\eqref{eq:mass-simp} to get the second relation.

The NLO ansatz is specified by the four parameters, $M_0, b_0, b_D, b_F$ (or equally, $M_0, \bar{b}, b_D, b_F$) defined as 
\begin{eqnarray}
\label{eq:ChPT-octet-NLO-param}
\bar{b} &=& -6 b_0 -4 b_D,
\quad
\delta b_N = \frac{2}{3}(3b_F-b_D),
\nonumber \\
\delta b_\Lambda &=& -\frac{4}{3}b_D,
\qquad\quad\ 
\delta b_\Sigma = \frac{4}{3}b_D,
\nonumber \\
\delta b_\Xi &=& -\frac{2}{3}(3b_F + b_D) \,.
\end{eqnarray}
The LECs at NNLO introduce 
only two additional parameters $D$ and $F$, usually expressed as 
$D/(4\pi F_0)$ and $F/(4\pi F_0)$, in terms of which 
\begin{widetext}
\begin{align}
g_{N,\pi}&=\frac{3}{2}(D+F)^2,
&
g_{N,K}&=\frac{5}{3}D^2 - 2DF +3F^2,
&
g_{N,\eta_8}&=\frac{1}{6}(D-3F)^2,
\nonumber \\
g_{\Lambda,\pi}&=2D^2,
&
g_{\Lambda,K}&=\frac{2}{3}D^2+6F^2,
&
g_{\Lambda,\eta_8}&=\frac{2}{3}D^2,
\nonumber \\
g_{\Sigma,\pi}&=\frac{2}{3}D^2 + 4F^2,
&
g_{\Sigma,K}&=2D^2 + 2F^2,
&
g_{\Sigma,\eta_8}&=\frac{2}{3}D^2,
\nonumber \\
g_{\Xi,\pi}&=\frac{3}{2}(D-F)^2,
&
g_{\Xi,K}&=\frac{5}{3}D^2+2DF+3F^2,
&
g_{\Xi,\eta_8}&=\frac{1}{6}(D+3F)^2.
\end{align}
\end{widetext}
Their dimensionless versions, $\bm{D}$ and $\bm{F}$, are 
\begin{eqnarray}
\bm{D}^2 &=& \sqrt{8t_0}
\frac{M_0^3}{(4\pi F_0)^2} D^2,
\nonumber \\
\bm{F}^2 &=& \sqrt{8t_0}
\frac{M_0^3}{(4\pi F_0)^2} F^2.
\end{eqnarray}

\begin{figure}
    \centering
    \includegraphics[width=0.98\linewidth]{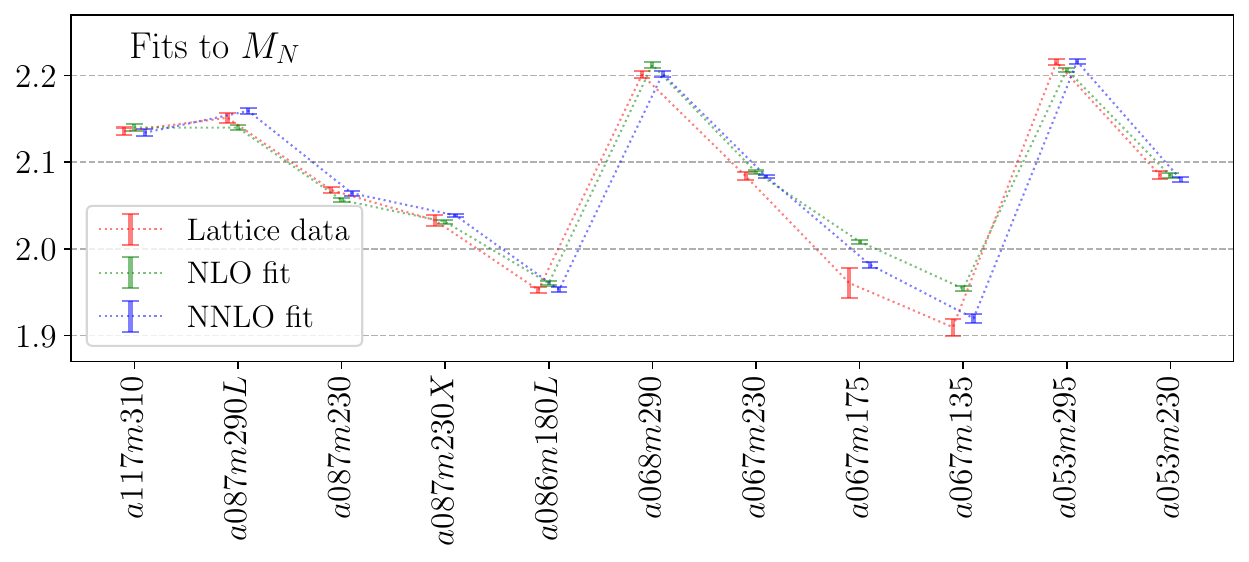}
    \includegraphics[width=0.98\linewidth]{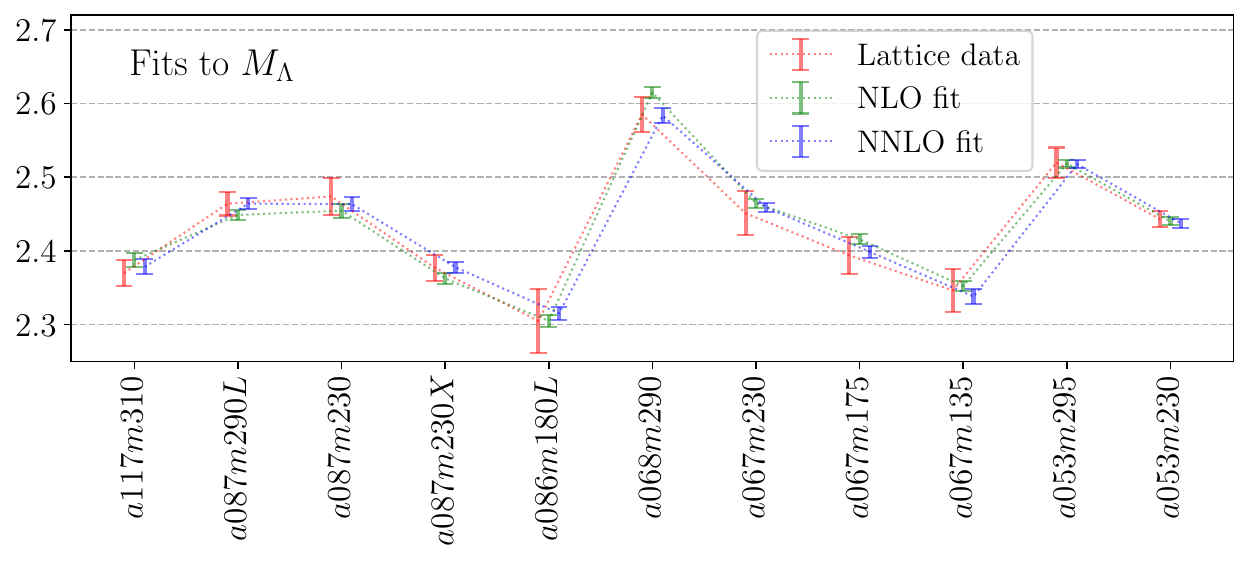}
    \includegraphics[width=0.98\linewidth]{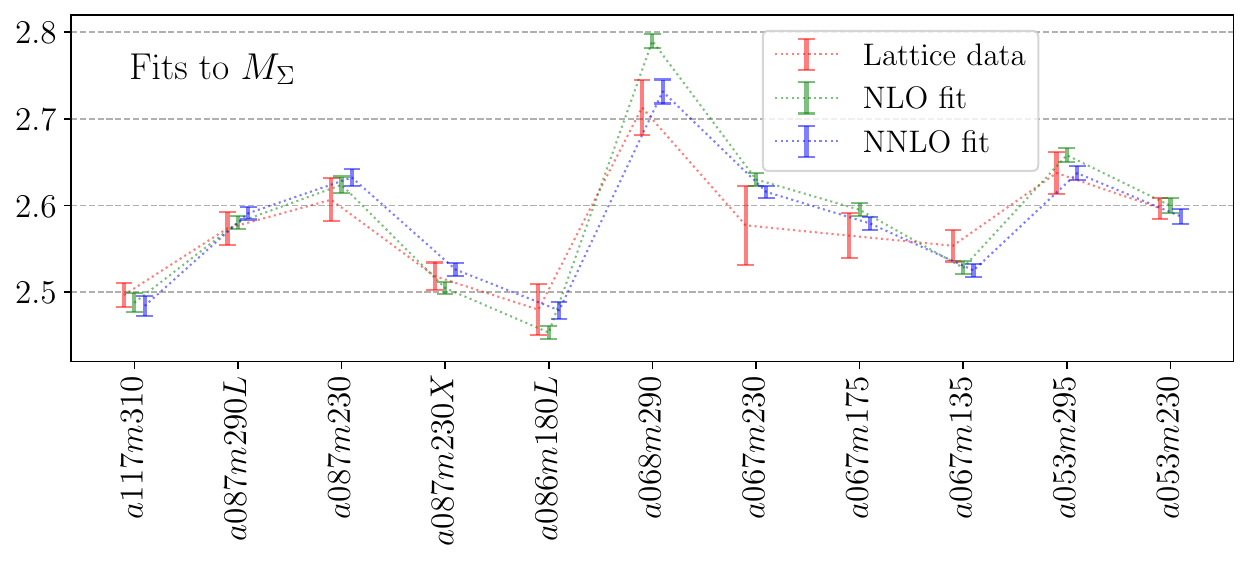}
    \includegraphics[width=0.98\linewidth]{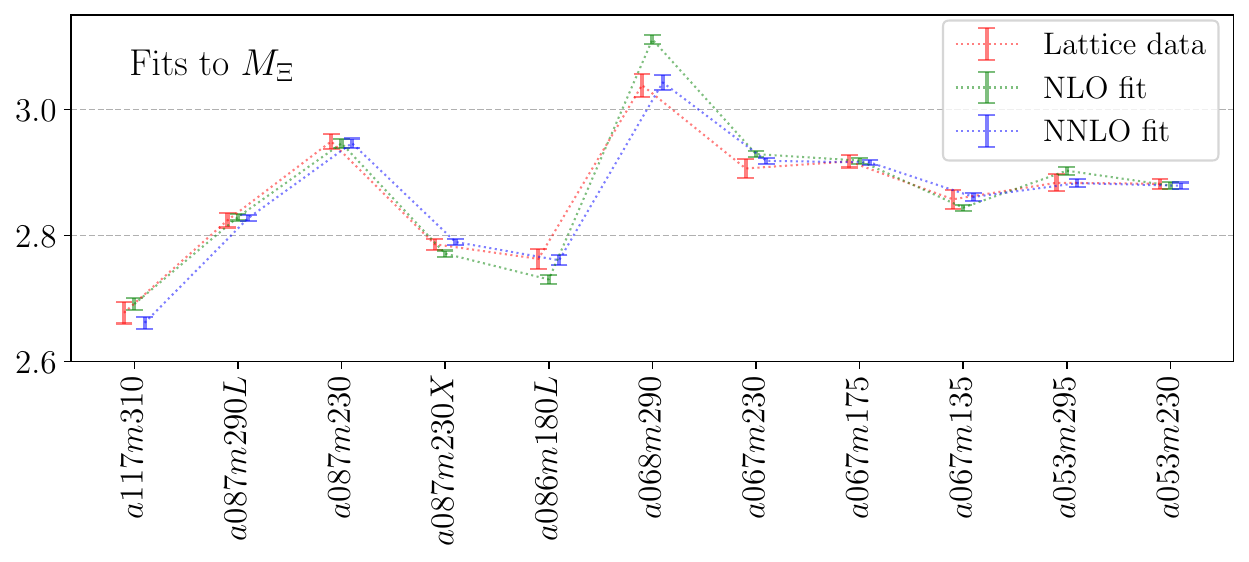}
    \caption{Comparison of NLO versus  NNLO chiral fits to the four octet baryon masses in units of $t_0$, i.e., the y-axis labels are $\{ \sqrt{8t_0}M_N, \sqrt{8t_0}M_\Lambda,\ \sqrt{8t_0}M_\Sigma, \ \sqrt{8t_0}M_\Xi\}$. The 
    $\mathcal{O}(\alpha_s a)$ ansatz is used to remove discretization errors and $\tphy$ is used to set the scale. 
    }
    \label{fig:octet-t0-fit}
\end{figure}

Thus, the NLO (NNLO) ansatz has a total of 10 (12) free parameters.
We obtain these by fitting the 44 octet masses measured on the eleven ensembles, 
and the results are given in Table~\ref{tab:chiral-fit-parameters-t0-w0} for both $t_0$ 
and $w_0$ to make quantities dimensionless. How well the fits agree with the 
data for the masses on the 11 ensembles is shown in Figs.~\ref{fig:octet-t0-fit} using $\tphy$. 
From these plots, we infer that the NNLO fit is preferred.
With these fit parameters in hand, we 
\begin{itemize}
    \item extract the scales $\tphy$ or $\wphy$ using  any one 
    baryon mass, say $M_N$, to set the lattice scale. Thereafter, the analysis is similar to that presented 
    in  Sec.~\ref{sec:fpiK} with $\fpiK$ replaced by $M_N$. The resulting values of 
    $\tphy$ and $\wphy$ are given in Table~\ref{tab:t0_w0_NLO_NNLO} and shown in Figs.~\ref{fig:t0-final}.
    \item Use the scales $\tphy$ or $\wphy$ extracted in Sec.~\ref{sec:fpiK} to predict the 
    octet spectrum. The results are given in Table~\ref{tab:spectrum-different-continuum} and shown in Figs.~\ref{fig:t0-final}.
\end{itemize}

\begin{figure}
    \centering
    \includegraphics[width=0.98\linewidth]{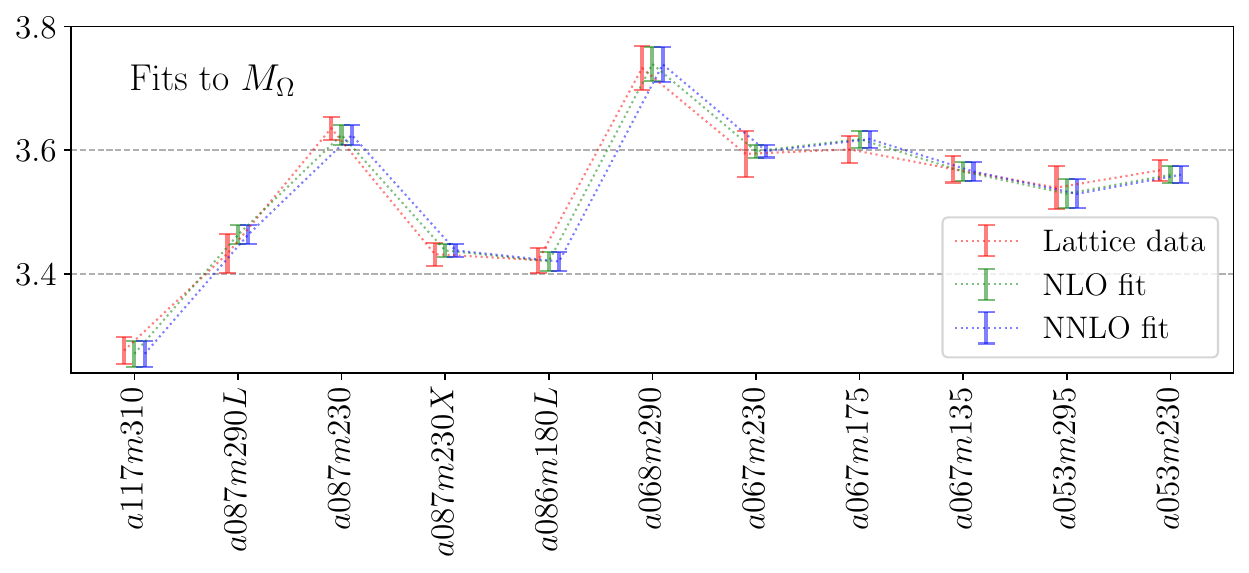}
    \caption{Comparison of the NLO versus  NNLO chiral fits to the Omega  baryon mass 
    in units of $t_0$, i.e., $\sqrt{8t_0}M_\Omega$. The  ansatz used to remove 
    discretization errors is the $\mathcal{O}(\alpha_s a)$ and $\tphy$ is used to set the scale. 
    }
    \label{fig:omega-t0-fit}
\end{figure}

\subsection{Analysis of the decuplet baryon masses}
\label{sec:decuplet_ChPT}

The chiral-continuum fit ansatz used for the decuplet masses is 
\begin{eqnarray}
&&\sqrt{8t_0}
{M}_D(M_\pi, M_K, a)
\nonumber \\
&=&
\sqrt{8t_0}{M}_D(M_\pi, M_K)
\nonumber \\
&&
\times
\left\{1+f(a)
\left[
c_d + \bar{c}_d \left(8t_0\overline{M}^2\right)
+\delta c_D\left(8t_0\delta{M}^2\right)
\right]
\right\}.
\nonumber \\
\label{eq:decuplet_fit}
\end{eqnarray}
The $[\mathcal{O}(m_q^{3/2})]$ NNLO ChPT formula for the decuplet $D=\{\Delta,\, \Sigma^*,\,\Xi^*,\,\Omega\}$ baryons, using bold letters for dimensionless quantities, 
is~\cite{Jenkins:1990jv, Ellis:1999jt}.
\begin{eqnarray}
&&
\sqrt{8t_0}{M}_D(M_\pi, M_K)
\nonumber \\
&=&
\bm{M_{D0}} + \bm{\bar{t}} \left(8t_0\overline{M}^2\right)  
+ \bm{\delta t_D} \left(8t_0\delta {M}^2\right) \nonumber \\
&&
+
\sum_{P = \pi, K, \eta_8}
\bm{g_{D,P}} f_D\left(\frac{\sqrt{8t_0}{M}_P}{\bm{M_{D0}}},\frac{\sqrt{8t_0}{M}_P}{\bm{M_0}}\right).
\label{eq:a-NNLO-ChPT-fit-decuplet}
\end{eqnarray}
Again, the first three terms define the $[\mathcal{O}(m_q)]$ NLO ansatz. 
The parameter $\bm{M_{D0}}$ is the decuplet baryon mass in the chiral limit; $\bm{\bar{t}}$ is a common fit parameter that controls the $\overline{M}^2$ dependence; $\bm{\delta t_D}$ is the SU(3) 
breaking term that is different for each 
decuplet state; and the EOMS loop-function is \cite{Gegelia:1999gf, Geng:2009hh, MartinCamalich:2010fp}
\begin{eqnarray}
f_D(x,y) &=& -2x^3
\bigg\{
\left(1-\frac{x^2}{4}\right)^{5/2}
\arccos\left(\frac{x}{2}\right) \nonumber \\
 &+&
\frac{x}{64}
\left[
17-2x^2+2(30-10x^2+x^4)\ln y
\right]
\bigg\}.
\nonumber \\
\end{eqnarray}
Note the $M_0$ dependence in the second arguments of the loop function. 
For fits using only the $\Omega$ mass, $\bm{M_0}$ in the functions $f_D$ 
in Eq.~\eqref{eq:a-NNLO-ChPT-fit-octet} is not constrained. Therefore, 
we replace $\bm{M_0}$ by $\bm{M_{D0}}$. This replacement effects  
higher-order corrections, i.e., beyond the order we are working 
at (see the comment below Eq.~(5.23) in Ref.~\cite{RQCD:2022xux}).

The relation between the dimensionless parameters and the commonly used parameters are 
\begin{eqnarray}
\bm{M_{D0}} &\equiv& \sqrt{8t_0} M_{D0},
\quad
\displaystyle
\bm{\bar{t}} \equiv \frac{\bar{t}}{\sqrt{8t_0}},
\bm{\delta t_D} \equiv  \frac{\delta{t}_D}{\sqrt{8t_0}}, \nonumber \\
\bm{g_{D,P}}
&\equiv&
\sqrt{8t_0}
\frac{M_0^3}{(4\pi F_0)^2}{g}_{D,P}.
\end{eqnarray}
These NLO parameters can be written in terms of the decuplet LECs $t_{D0}$ and $t_D$ as 
\begin{eqnarray}
\bar{t} = 3t_{D0} + 3t_D,
\quad
\delta{t}_\Delta &=& -t_D,
\quad\, 
\delta{t}_{\Sigma^*} = 0,
\nonumber \\
\delta{t}_{\Xi^*} &=& t_D,
\qquad\ 
\delta{t}_{\Omega} = 2t_D \,,
\end{eqnarray}
and the NNLO parameters in terms of the LEC $\mathcal{H}$  
\begin{align}
g_{\Delta,\pi}&=\frac{25}{54}\mathcal{H}^2,
&
g_{\Delta,K}&=\frac{5}{27}\mathcal{H}^2,
&
g_{\Delta,\eta_8}&=\frac{5}{54}\mathcal{H}^2,
\nonumber \\
g_{\Sigma^*,\pi}&=\frac{20}{81}\mathcal{H}^2,
&
g_{\Sigma^*,K}&=\frac{40}{81}\mathcal{H}^2,
&
g_{\Sigma^*,\eta_8}&=0,
\nonumber \\
g_{\Xi^*,\pi}&=\frac{5}{54}\mathcal{H}^2,
&
g_{\Xi^*,K}&=\frac{5}{9}\mathcal{H}^2,
&
g_{\Xi^*,\eta_8}&=\frac{5}{54}\mathcal{H}^2,
\nonumber \\
g_{\Omega,\pi}&=0,
&
g_{\Omega,K}&=\frac{10}{27}\mathcal{H}^2,
&
g_{\Omega,\eta_8}&=\frac{10}{54}\mathcal{H}^2.
\end{align}
Its dimensionless version $\bm{H}$ is 
\begin{equation}
\bm{H}^2 = \sqrt{8t_0}
\frac{M_{D0}^3}{(4\pi F_0)^2} \mathcal{H}^2.
\end{equation}

\begin{table*}
\centering
\footnotesize
\begin{tabular}{c|c|c|c|c|c|c|c}
\hline
\hline
\rule{0pt}{2.8ex}
fit & $\chi^2/N_\textrm{DF}$ & $c_o$ & $\bar{c}_o$ & $\bm{\delta}c_N$ & $\bm{\delta}c_\Lambda$ & $\bm{\delta}c_\Sigma$ & $\bm{\delta}c_\Xi$ \\[0.6ex]
\hline
\hline
\rule{0pt}{3.4ex}
octet NLO with $t_0$  & 3.75 & 1.75(66) & -1.84(78) &  -0.413(55) & -0.370(66) & -0.399(64) &  -0.332(56) 
\\[0.6ex]
\rule{0pt}{2.8ex}
octet NNLO with $t_0$ & 0.57 &  -1.07(65) & 0.13(78) & 0.157(93) & 0.22(11) & 0.27(11) & 0.236(86) 
\\[0.6ex]
\rule{0pt}{3.4ex}
octet NLO with $w_0$  & 3.02 & 0.28(11) & -0.259(99) &  -0.0498(78) & -0.0388(87) & -0.0396(82) & -0.0289(76)
\\[0.6ex]
\rule{0pt}{2.8ex}
octet NNLO with $w_0$ & 0.79 &  -0.03(11) &-0.10(10) & -0.0009(116) & 0.009(13) & 0.015(12) & 0.015(11) \\[0.6ex]
\hline
\hline
\rule{0pt}{2.8ex}
fit & $\chi^2/N_\textrm{DF}$ & $\bm{M_0}$ & $\bm{\bar{b}}$ & $\bm{b_D}$ & $\bm{b_F}$ & $\bm{D}$ & $\bm{F}$ 
\\[0.6ex]
\hline
\rule{0pt}{3.4ex}
octet NLO with $t_0$ & 3.75 & 1.520(84) & 1.160(89) &  0.0301(40) & -0.0954(16) & - & - 
\\[0.6ex]
\rule{0pt}{2.8ex}
octet NNLO with $t_0$  & 0.57 & 1.654(88) & 2.14(19) & 0.073(26) & -0.195(13) & 0.752(63) & 0.276(63)
\\[0.6ex]
\rule{0pt}{3.4ex}
octet NLO with $w_0$ & 3.02 & 1.77(12) & 1.022(92) &  0.0253(35) & -0.0802(14) & - & - 
\\[0.6ex]
\rule{0pt}{2.8ex}
octet NNLO with $w_0$  & 0.79 & 1.90(12) & 1.75(18) & 0.053(26) & -0.157(12) & 0.725(79) & 0.284(81)
\\[0.6ex]
\hline
\hline
\rule{0pt}{2.8ex}
fit & $\chi^2/N_\textrm{DF}$ & $c_d$ & $\bar{c}_d$ & $\bm{\delta}c_\Delta$ & $\bm{\delta}c_{\Sigma^*}$ & $\bm{\delta}c_{\Xi^*}$ & $\bm{\delta}c_\Omega$
\\[0.6ex]
\hline
\hline
\rule{0pt}{3.4ex}
$\Omega$ NLO with $t_0$  & 0.57 & -1.3(2.1) & 1.0(2.4) & - & - & - &  -0.10(39) 
\\[0.6ex]
\rule{0pt}{2.8ex}
$\Omega$ NNLO with $t_0$ & 0.63 &  -1.4(2.1) & 1.1(2.4) & - & - & - & -0.10(39) 
\\[0.6ex]
\rule{0pt}{3.4ex}
$\Omega$ NLO with $w_0$  & 0.88 & -0.7(2.6) & 0.2(2.5) & - & - & - &  -0.30(38)
\\[0.6ex]
\rule{0pt}{2.8ex}
$\Omega$ NNLO with $w_0$ & 1.02 &  -0.7(2.6) & 0.3(2.5) & - & - & - & -0.30(38)
\\[0.6ex]
\hline
\rule{0pt}{2.8ex}
fit & $\chi^2/N_\textrm{DF}$ & $\bm{M_{D0}}$ & $\bm{\bar{t}}$ & \multicolumn{2}{c|}{$\bm{t}_D$} & \multicolumn{2}{c}{$\bm{H}$}
\\[0.6ex]
\hline
\rule{0pt}{3.4ex}
$\Omega$ NLO with $t_0$  & 0.57 & 2.79(43) & 0.37(45) & \multicolumn{2}{c|}{0.119(47)} & \multicolumn{2}{c}{-}
\\[0.6ex]
\rule{0pt}{2.8ex}
$\Omega$ NNLO with $t_0$ & 0.63 &  2.80(43) & 0.36(45) & \multicolumn{2}{c|}{0.118(47)} & \multicolumn{2}{c}{Not constrained}
\\[0.6ex]
\rule{0pt}{3.4ex}
$\Omega$ NLO with $w_0$  & 0.88 & 3.07(38) & 0.41(30) & \multicolumn{2}{c|}{0.115(30)} & \multicolumn{2}{c}{-}
\\[0.6ex]
\rule{0pt}{2.8ex}
$\Omega$ NNLO with $w_0$ & 1.02 &  3.08(38) & 0.40(31) & \multicolumn{2}{c|}{0.115(30)} & \multicolumn{2}{c}{Not constrained}
\\[0.6ex]
\hline
\hline
\end{tabular}
\caption{Results for the dimensionless fit parameters, converted using $\sqrt{8t_0}$ (upper table) or using $\sqrt{8}w_0$ (lower table), 
 in the CC extrapolation ansatz used 
for the octet and $\Omega$ baryon masses as defined in 
Eqs.~\protect\eqref{eq:a-NNLO-ChPT-fit-octet}, 
\protect\eqref{eq:decuplet_fit} 
and~\protect\eqref{eq:a-NNLO-ChPT-fit-decuplet}. 
Results are given for both the NLO and NNLO chiral ansatz, and organized 
into separate blocks for the  
continuum and the chiral parameters. Note that $\bm{H}$ is not constrained by the fit.
The discretization errors were modeled using the $\mathcal{O}(\alpha_s a)$ ansatz.}
\label{tab:chiral-fit-parameters-t0-w0}
\end{table*}

\begin{table*}
    \setlength{\tabcolsep}{10pt} 
    \renewcommand{\arraystretch}{1.2} 
\centering
\begin{tabular}{c|c|c|c|c|c|}
\hline
\hline
\rule{0pt}{2.8ex}
 & \multicolumn{5}{c|}{Physical baryon mass used as input} \\
Flow scales [fm] & $N$ & $\Lambda$ & $\Sigma$ & $\Xi$ & $\Omega$
\\[0.6ex]
\hline
\hline
\rule{0pt}{3.4ex}
$\sqrt{\tphy}$ (NLO) & 0.1457(26) & 0.1394(44) & 0.1384(57) & - & 0.1489(58)
\\[0.6ex]
\rule{0pt}{2.8ex}
$\sqrt{\tphy}$ (NNLO) & 0.1461(17) & 0.1453(26) & 0.1456(21) & 0.1501(22) & 0.1491(56)
\\[0.6ex]
\rule{0pt}{2.8ex}
$\wphy$ (NLO) & 0.1756(31) & 0.1681(58) & 0.1672(77) & 0.161(11) & 0.1778(98)
\\[0.6ex]
\rule{0pt}{2.8ex}
$\wphy$ (NNLO) & 0.1738(22) & 0.1730(38) & 0.1731(28) & 0.1791(30) & 0.1783(92)
\\[0.6ex]
\hline
\end{tabular}
\caption{Results for the flow scales $\sqrt{\tphy}$ and $\wphy$ determined 
using fits with the NLO and NNLO ansatz from HB$\chi$PT and the physical 
mass  of one of the octet or the $\Omega$ baryon given in Eq.~\eqref{eq:physical_inputs}. 
The analysis of the decuplet spectrum  is done only for the $M_\Omega$ 
data as explained in the text. The ansatz $\mathcal{O}(\alpha_s a)$ is 
used to model the discretization correction. We did not find a solution 
for $\tphy$  using the NLO ansatz with $M_\Xi$ as input.
}
\label{tab:t0_w0_NLO_NNLO}
\end{table*}

\begin{table*}
\centering
\setlength{\tabcolsep}{4.5pt} 
\renewcommand{\arraystretch}{1.2} 
\begin{tabular}{l|c|c|c|c|c|c}
\hline
\hline
\rule{0pt}{2.8ex}
 & $m_N$  & $m_\Lambda$ & $m_\Sigma$ & $m_\Xi$ & $m_\Omega$ & $f_{\pi K}$
\\[0.6ex]
\hline
\hline
\rule{0pt}{3.4ex}
\textrm{Pheno: Eq.~\protect\eqref{eq:physical_values}} & 937.54(6) & 1115.68(1) & 1190.66(12) & 1316.9(3) & 1669.5(3.0) & 148.32(34) 
\\[0.6ex]
\rule{0pt}{2.8ex}
\textrm{Fit with $a^2$} & 939.3(7.7) & 1114(11) & 1188.7(8.3) & 1333.7(8.8) & 1682(34) & 146.3(1.5)
\\[0.6ex]
\rule{0pt}{2.8ex}
\textrm{Fit with $\alpha_s a$} & 945(10) & 1118(14) & 1195(11) & 1340(12) & 1698(46) & 146.5(1.6)
\\[0.6ex]
\rule{0pt}{2.8ex}
\textrm{Fit with $a$} & 955(14) & 1126(19) & 1206(16) & 1350(17) & 1737(72) & 147.0(1.9)
\\[0.6ex]
\hline
\end{tabular}
\caption{Comparing our results for the mass spectrum and the decay constant 
$\fpiK$ in units of MeV obtained using the NNLO chiral ansatz and for 
the three possible  ansatz for discretization corrections given in Eq.~\protect\ref{eq:fa3cases}. The 
phenomenological values in row one  for the 2+1-flavor theory used for 
comparison are reproduced from Eq.~\protect\eqref{eq:physical_values}.   
The scale is set using 
$\sqrt{t_{0}^\textrm{Phy}}|_{\rm FLAG}=0.14474(57)$ fm taken from the FLAG 2024 report~\protect\cite{FlavourLatticeAveragingGroupFLAG:2024oxs}. 
}
\label{tab:spectrum-different-continuum}
\end{table*}

\begin{figure*}      
    \begin{center}
      \includegraphics[width=0.49\linewidth]{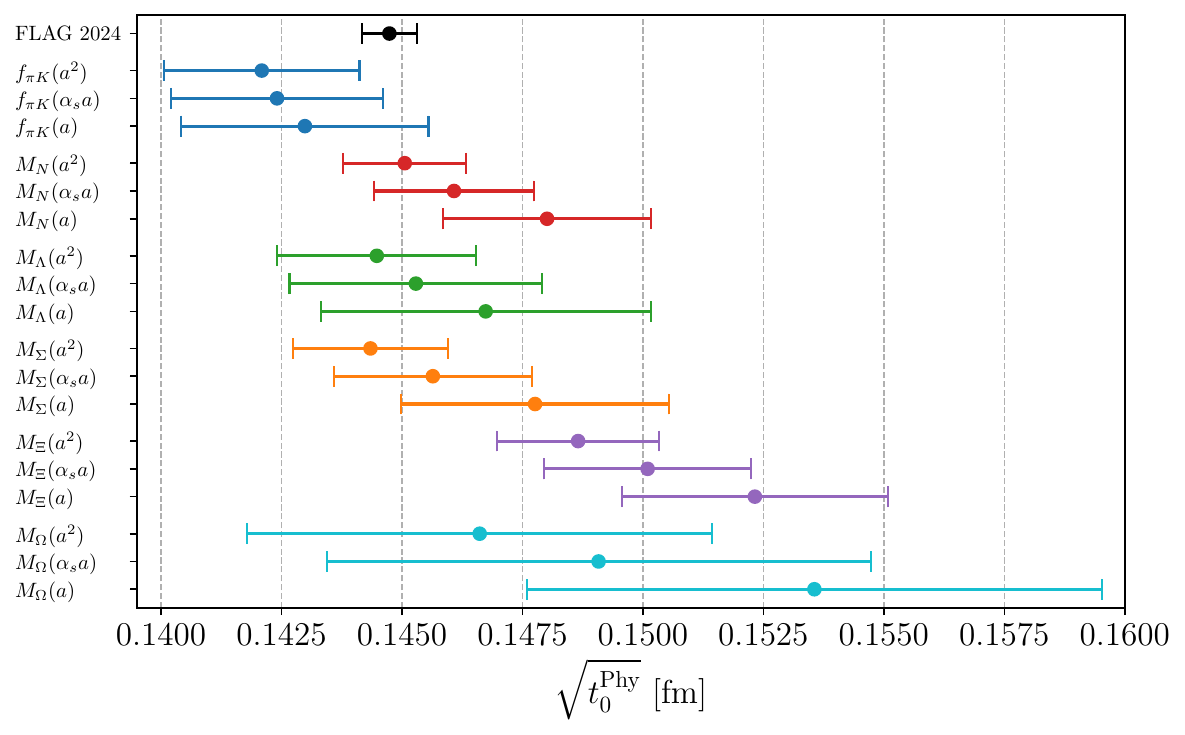}
      \includegraphics[width=0.49\linewidth]{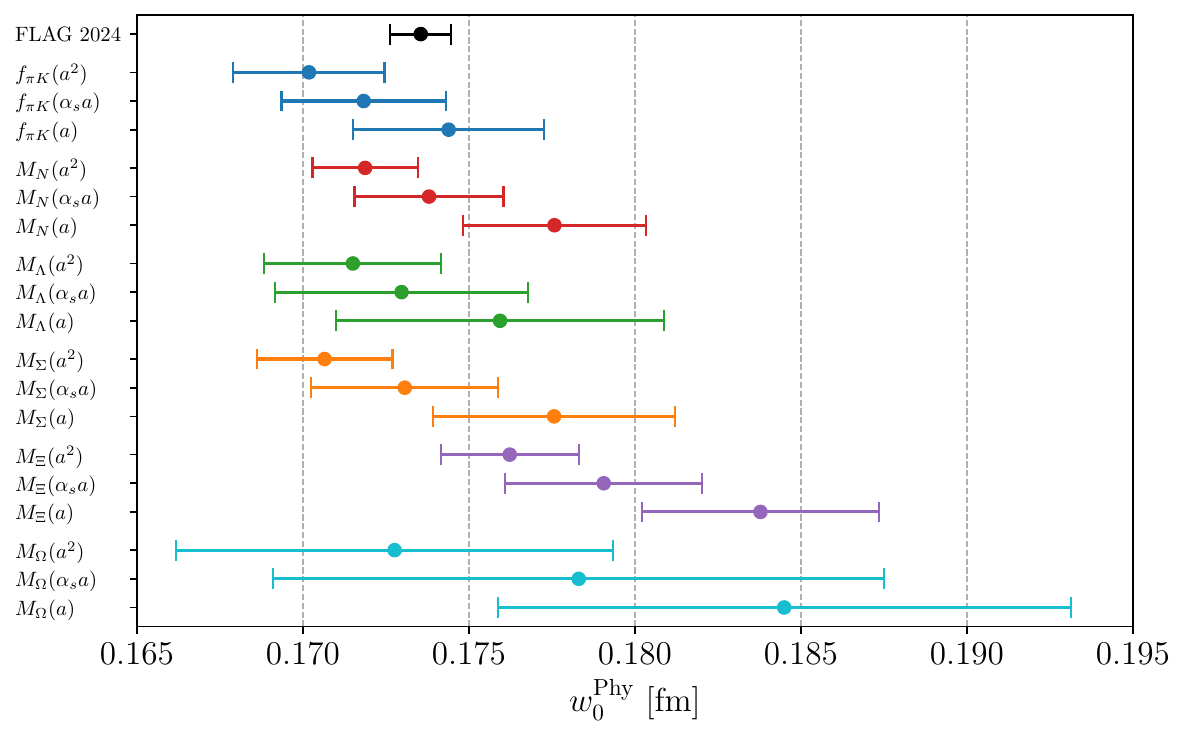}
    \end{center}
    \caption{Our results for $\tphy$ (left) and $\wphy$ (right) 
    using various measured quantities specified in the y-axis labels and for the 
    three ansatz used to remove discretization errors. 
    For comparison, the FLAG 2024 average value~\cite{FlavourLatticeAveragingGroupFLAG:2024oxs} is shown at the top.
    }      
    \label{fig:t0-final}
\end{figure*}

\begin{figure*}     
    \centering
    \includegraphics[width=1.0\linewidth]{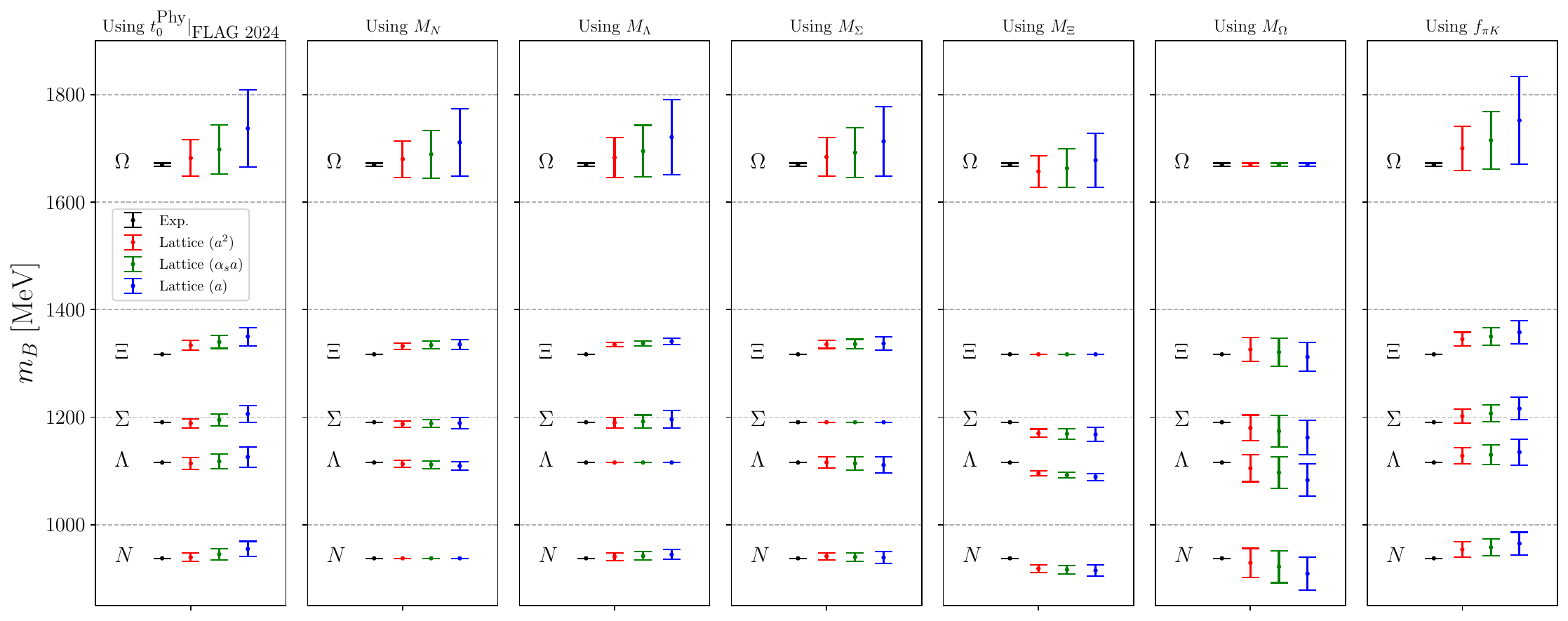}
    \includegraphics[width=1.0\linewidth]{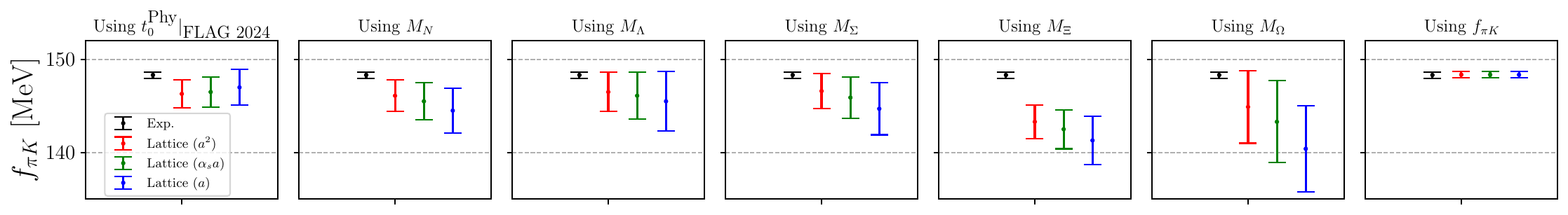}

    \includegraphics[width=1.0\linewidth]{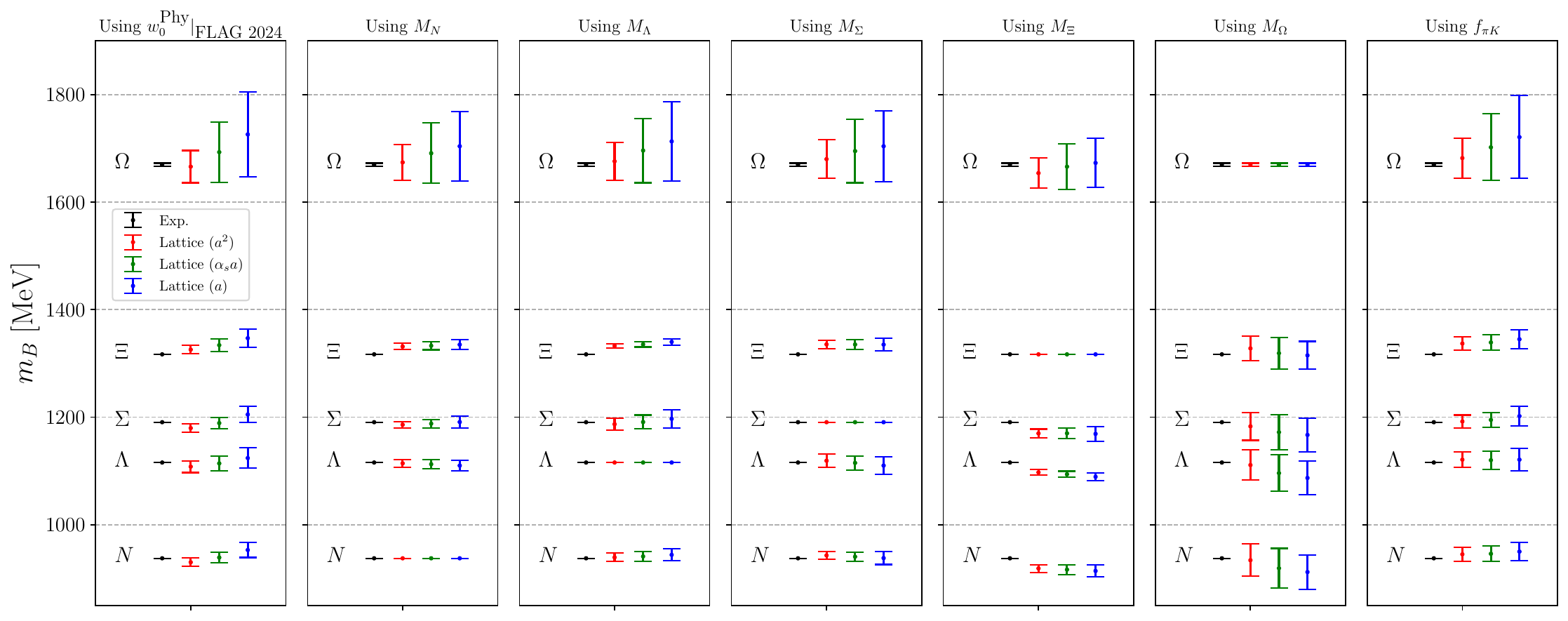}
    \includegraphics[width=1.0\linewidth]{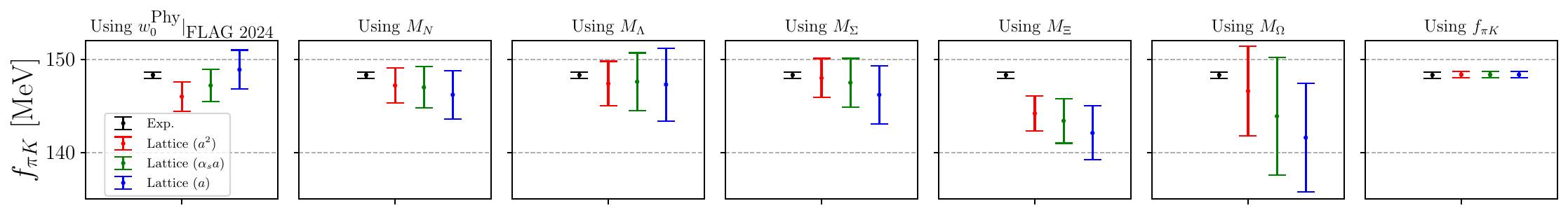}
    \caption{Predictions of the baryon spectrum and $f_{\pi K}$. The 
    scale is set using $\tphy$ (top 2 rows) and $\wphy$ (bottom two rows) taken from the 
    FLAG 2024 report~\protect\cite{FlavourLatticeAveragingGroupFLAG:2024oxs} or our calculation 
    of the quantity specified on top of the figures. Also all 
    quantities are made dimensionless using $t_0$ (top) $w_0$ (bottom).
    Results are given for the three ansatz used to model discretization errors. 
    While the diagonal panels, e.g., predicting $M_N$ using $M_N$, are shown, the 
    data are trivial. 
    The experimental and FLAG values are taken from Eqs.~\protect\eqref{eq:physical_values} 
    and~\protect\eqref{eq:FLAG_inputs}.
    }
    \label{fig:spectrum_t0w0phys}
\end{figure*}

Thus, there are 6 (7) dimensionless fit parameters in Eq.~\eqref{eq:physical_inputs} 
for the $\Omega$ NLO (NNLO) fits, respectively. These are given in Table~\ref{tab:chiral-fit-parameters-t0-w0} 
for both $t_0$ and $w_0$ to make quantities dimensionless. How well the NLO and NNLO 
fits agree with the $M_\Omega$ data on the 11 ensembles is shown in Fig.~\ref{fig:omega-t0-fit} using $\tphy$. We find no significant difference and prefer the NLO ansatz since the $\chi^2/\textrm{dof}$ increases with the NNLO.
With these dimensionless fit parameters in hand, we determine 
\begin{itemize}
\item the flow scales $\tphy$ (or $\wphy$)  using, as input, $M_\Omega^{\rm Phy}$. 
Results are given in Table~\ref{tab:t0_w0_NLO_NNLO} and 
compared with other determinations in Fig.~\ref{fig:t0-final}. The errors in our determinations are 
large with that from $M_\Omega$ being the largest. This is because while the fit for the octet 
uses 44 data points, only 11 $M_\Omega$ are available for the decuplet.
\item 
$M_\Omega^{\rm Phy}$ using $\tphy$. The results are given in 
Table~\ref{tab:spectrum-different-continuum} 
and shown in  
Fig.~\ref{fig:spectrum_t0w0phys}.
\end{itemize}

\section{Comparison with other efforts to generate and characterize 2+1-flavor lattice ensembles. }
\label{sec:comparison}

A short overview of various efforts worldwide to generate lattice ensembles is being 
provided at the annual international conferences on lattice QCD~\cite{Bali:2022mlg,Aoki:2025etz}.
A number of collaborations are generating and analyzing 2+1-flavor Wilson-clover lattices. 
Using our set of 11 NME ensembles (one with physical pion mass), we present a 
 study similar to that by the RQCD collaboration in Ref.~\cite{RQCD:2022xux} based on 
45 ensembles generated by the Coordinated
Lattice Simulations (CLS) effort~\cite{Bulava:2013cta,Bruno:2014jqa,Bruno:2016plf} and some additional ones 
by the RQCD collaboration. Others efforts are 
(i) the HALQCD collaboration~\cite{Aoyama:2024cko} (one physical pion mass ensemble)
(ii) the QCDSF collaboration~\cite{QCDSFUKQCDCSSM:2023qlx} (21 ensembles with $M_\pi^{\rm lightest}=220$~MeV; 
(iii) the PACS collaboration~\cite{Aoki:2025taf} (three physical pion mass ensembles); 
and (iv) the CLQCD collaboration~\cite{CLQCD:2023sdb} (16 ensembles with three  at physical pion mass). 
One of our goals is to reach high statistics to determine whether one can demonstrate a 
data driven method for removing excited-state contributions in 3-point functions 
using multi-state fits~\cite{Park:2025rxi}. Towards that goal,  
the analysis of 3-point functions for extracting nucleon charges and form 
factors from these ensembles is under progress and results will be reported 
separately. The analyses here 
highlight the need for also having a much larger set of ensembles to control the chiral-continuum 
fits, especially with higher order ansatz in each. Lastly, we plan to add 
companion ensembles to the existing 
eleven with the strange quark mass tuned to its physical value. 

\section{Conclusions}
\label{sec:conclusions}

This work describes the eleven NME ensembles with 2+1-flavors of Wilson-clover fermions
generated mostly by us as part of a larger effort by the JLab/W\&M/LANL/MIT/Marseille 
collaborations in Sec.~\ref{sec:intro}, and presents  
data on the flow scales $ t_0$ and $w_0$ (Secs.~\ref{sec:t0w0def} and~\ref{sec:fpiK}, 
and~Appendix~\ref{sec:flowQandW}),  scale setting on the lattices (Sec.~\ref{sec:lattice-spacing}), the decay 
constants $f_\pi$ and $f_K$ (Sec.~\ref{sec:fpi-fK}), the behavior of the topological charge and 
the Weinberg three-gluon operator under Wilson flow (Sec.~\ref{sec:topology}, and Appendix~\ref{sec:flowQandW}), 
the meson and baryon spectrum (Secs.~\ref{sec:octet_ChPT} and~\ref{sec:decuplet_ChPT}), a study of 
autocorrelations in the data, which is needed for the statistical analysis of the 
various quantities (Appendix~\ref{sec:autocorr}), and results for the renormalization 
constants for all five quark bilinear local operators (Sec.~\ref{sec:Zfactors}). 

In Sec.~\ref{sec:fpi-fK}, we use the data for $\fpiK \equiv (2f_K + f_\pi)/3$ to 
calculate the physical point values, $\tphy$ and $\wphy$. 
In Sec.~\ref{sec:fKbyfpi}, we calculate $\fpifk^{\rm Phy}$, the second independent 
quantity using the data for $f_\pi$ and $f_K$. The result  
$\frac{f_{K^\pm}}{f_{\pi^\pm}} = 1.191(32)$
agrees with the FLAG 2024 average, however, the errors in our estimate are larger as shown in 
Fig.~\ref{fig:ratio-decay}. 

The scales $\tphy$ and $\wphy$ are also extracted using the data for each of the masses of the four 
octet states, $\{N,\ \Lambda,\ \Sigma,\ \Xi\}$, and the decuplet state $\Omega$. 
The resulting estimates are summarized in Fig.~\ref{fig:t0-final}. 
The errors in our estimates are significantly larger than those in the FLAG 2024 averages, 
mostly because the chiral-continuum extrapolation based on 11 ensembles does not provide enough constraints on the parameters. Our best 
estimates of $\tphy$ and $\wphy$ are from the analysis of $\fpiK$. 

The lattice scale $a$ on each ensemble is determined using the values of 
$\tphy$ and $\wphy$ and the fit to $\fpiK$.  We give two values, $a^{\rm naive}$ and $a^{\rm corr.}$ 
that correspond to taking the continuum limit holding $t_0 = \tphy$ 
and $\fpiK = \fpiK^{\rm Phy}$, respectively, on each ensemble. 
Both are equally well motivated and the differences reflect residual discretization errors and having most  
ensembles at 
non-physical values of the light and strange quark masses. The impact of the 
differences in the two sets of lattice scale $a$ in the CC fits used to obtain physical point values for the 
other observables is small---the parameters of 
the CC fits adjust to absorb most of the differences. 

We choose to present all final results using $a^{\rm naive}$ but do give results for 
both  so that the reader can assess the effect. 
An overview of the status and issues with various methods used to set the lattice  
scale $a$ and determine $\tphy$ and $\wphy$ have been reviewed in the FLAG 2024 report~\cite{FlavourLatticeAveragingGroupFLAG:2024oxs}.

Fits to get the values of the octet spectrum, $\{M_N,\ M_\Lambda,\ M_\Sigma,\ M_\Xi\}$, at 
the physical point were carried out using both NLO and NNLO expressions from HB$\chi$PT.  
As shown in Fig.~\ref{fig:octet-t0-fit}, the NNLO ansatz fit does a better job. 
For discretization errors, we explored three ansatz $\{\mathcal{O}( a),\ \mathcal{O}(\alpha_s a), \ \mathcal{O}( a^2) \}$. They give overlapping results, and 
the final result is presented with the $\mathcal{O}(\alpha_s a)$ ansatz as our action is 
almost $O(a)$ improved.  

Fits to decuplet data are less robust as we can only use $M_\Omega$. The NLO and NNLO fits 
are equally good and the output values consistent as shown in Fig.~\ref{fig:omega-t0-fit}, however, 
since no reduction in $\chi^2/{\rm dof}$ is observed due to the 
extra parameter in the NNLO ansatz, we pick the NLO result.

The time history of the $\Theta$ and Weinberg operator at large flow time
is shown in Appendix~\ref{sec:flowQandW} and the data show no evidence for frozen 
topology in the generation of any of the 11 ensembles.

In Appendix~\ref{sec:autocorr}, we present a study of  autocorrelations in the long-distance variables 
$ t_0$ and $w_0$, and find them to be large. To offset these, all analyses of spectral quantities 
and decay constants, is done using data binned over a sufficiently large number of 
measurements such that the binned data do not show significant autocorrelations.

Overall, the results from these eleven ensembles are consistent with previous lattice 
calculations  by other collaborations and experimental values.  
The main limitation in obtaining high precision results are the chiral-continuum fits, i.e., the need for a larger set of ensembles. 
This work validates the efficacy of using these ensembles for 
further studies of other observables and our goal 
of improving precision by 
generating and analyzing more ensembles, especially ones with $M_K = M_K^{\rm Phy}$.

\begin{acknowledgments}
We thank Will Detmold and Stefan Meinel for discussions over the years 
on lattice generation and with Stefan Schaefer on scale setting. 
The calculations used the Chroma
software suite~\cite{Edwards:2004sx}.  This research used resources at
(i) the National Energy Research Scientific Computing Center, a DOE
Office of Science User Facility supported by the Office of Science of
the U.S.\ Department of Energy under Contract No.\ DE-AC02-05CH11231;
(ii) the Oak Ridge Leadership Computing Facility, which is a DOE
Office of Science User Facility supported under Contract
DE-AC05-00OR22725, through the ALCC program project LGT107 and HEP145, and the
INCITE program project HEP133; (iii) the USQCD collaboration, which is
funded by the Office of Science of the U.S.\ Department of Energy; and
(iv) Institutional Computing at Los Alamos National Laboratory.
This work was prepared in part by LLNL under Contract DE-AC52-07NA27344. 
S.~Park acknowledges the support from the ASC COSMON project.
T.~Bhattacharya and R.~Gupta were partly
supported by the U.S.\ Department of Energy, Office of Science, Office
of High Energy Physics under Contract No.\
DE-AC52-06NA25396. T.~Bhattacharya, R.~Gupta, S.~Mondal, S.~Park,
and J.~Yoo were partly supported by the LANL LDRD program, and
S.~Park by the Center for Nonlinear Studies. 
J.~Yoo was partly supported by Brain Pool program funded by the Ministry of Science and ICT through 
the National Research Foundation of Korea (R2528521).
\end{acknowledgments}


\clearpage

\appendix

\begin{widetext}

\section{Effective mass plots for  the pion and 2-state fits}
\label{sec:pi_decayplots}

Figures~\ref{fig:C13}--\ref{fig:F6} show the $\Meff$ for the pion 
and the results of 2-state fits to the 
$\langle P_S (0) P_S(\tau) \rangle$, $\langle P_S (0) P_P(\tau) \rangle$, 
and $\langle P_S (0) A_P(\tau) \rangle$  two-point correlators on the eleven ensembles.  
The final results for $M_\pi^{(2)}$ and $f_\pi^{(2)}$,  
presented in Table~\ref{tab:decayconst}, are obtained from simultaneous 
fits to $\langle P_S (0) P_S(\tau) \rangle$ and 
$\langle P_S (0) A_P(\tau) \rangle$.

\begin{figure*}[h!]   
    \centering
    \includegraphics[width=.32\textwidth]{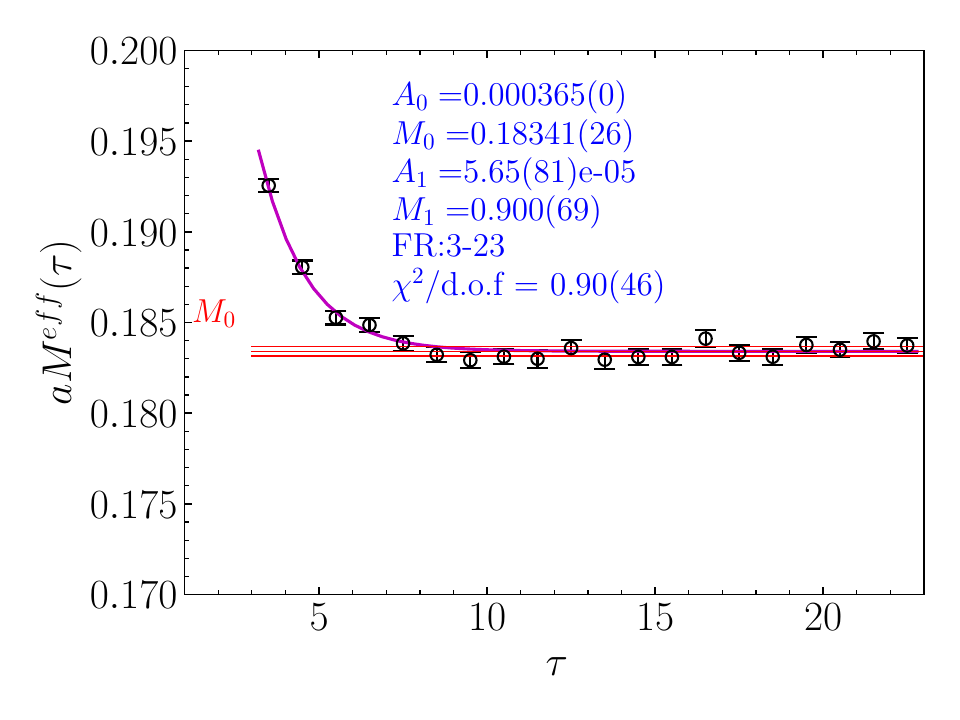}%
    \includegraphics[width=.32\textwidth]{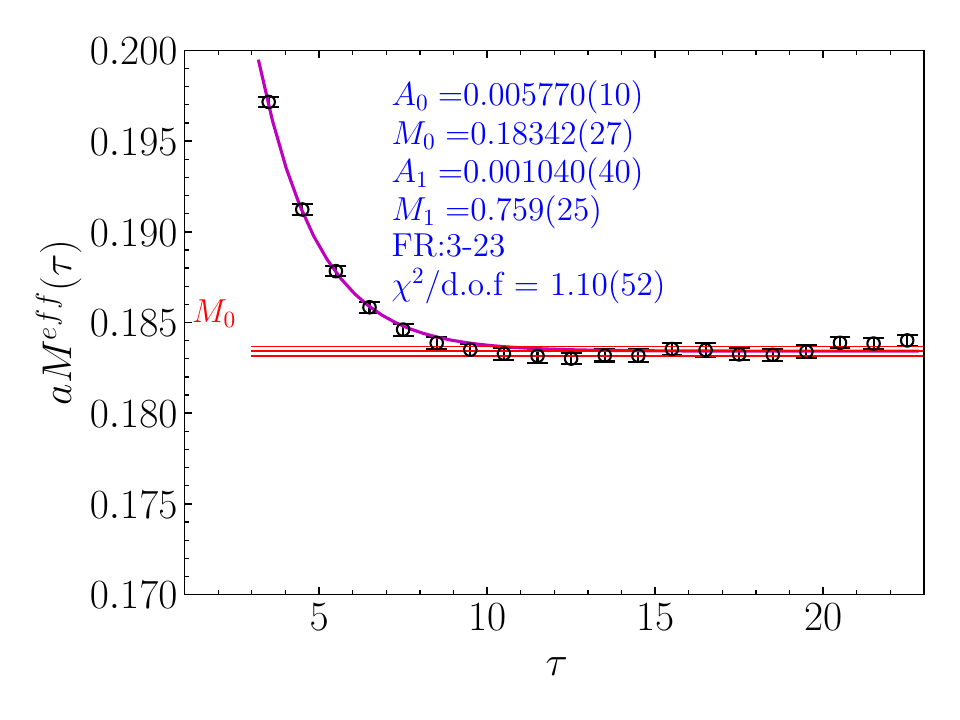}%
    \includegraphics[width=.32\textwidth]{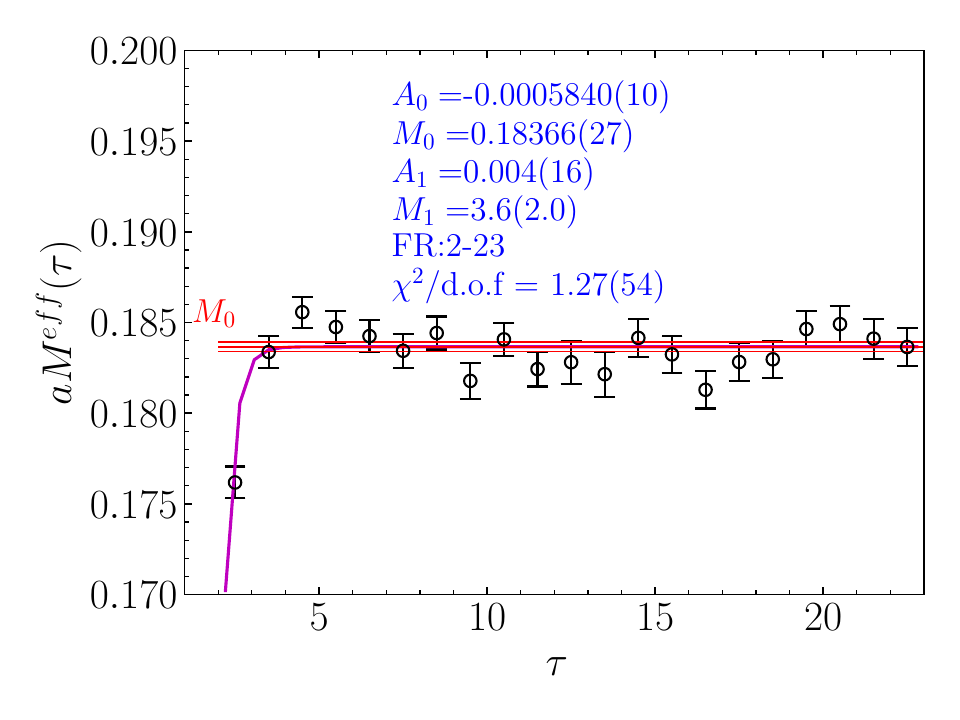}\\
    \vspace{-0.5cm}
    \includegraphics[width=.32\textwidth]{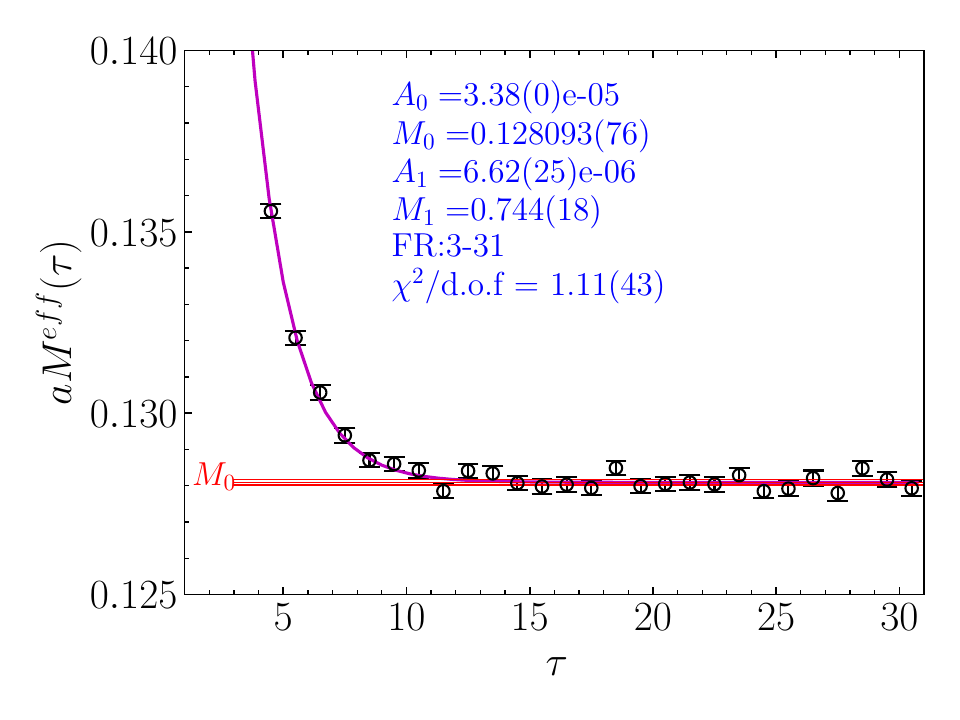}%
    \includegraphics[width=.32\textwidth]{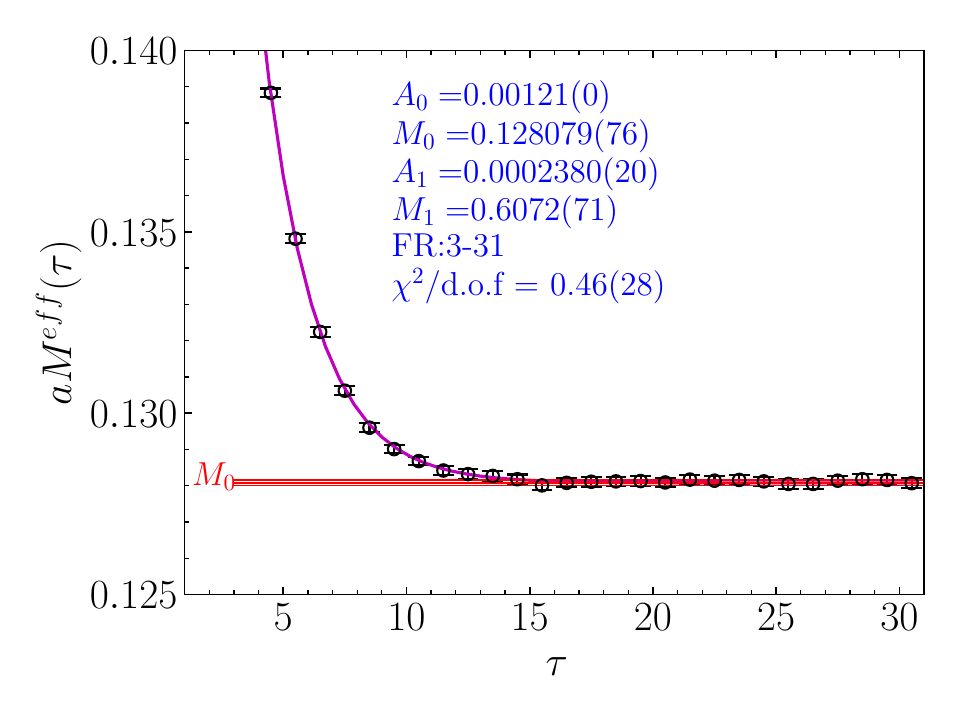}%
    \includegraphics[width=.32\textwidth]{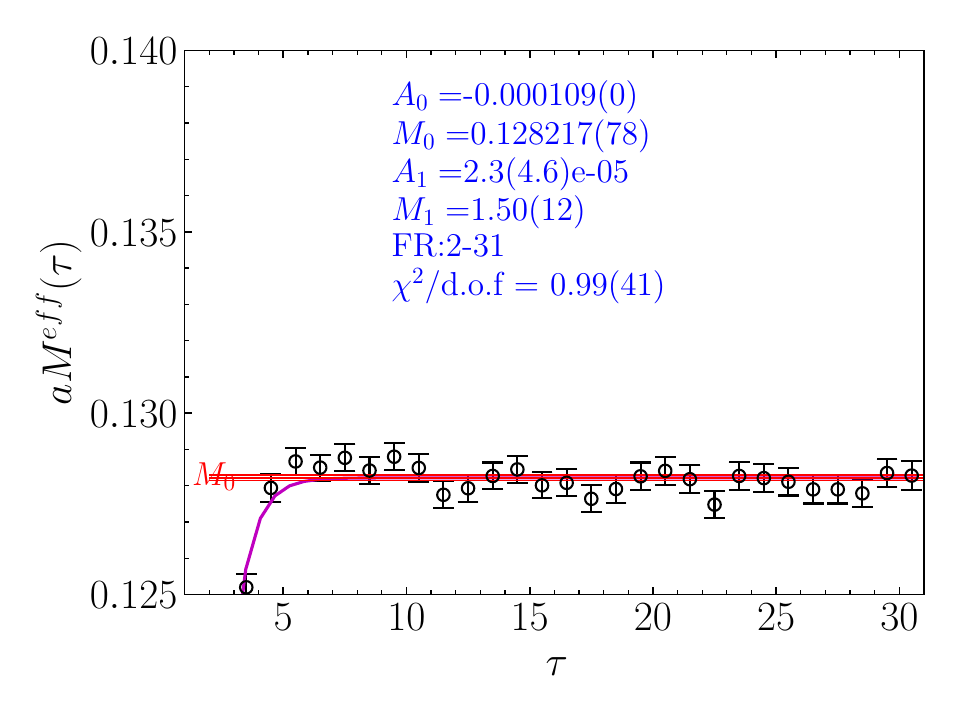}\\
    \vspace{-0.5cm}
    \includegraphics[width=.32\textwidth]{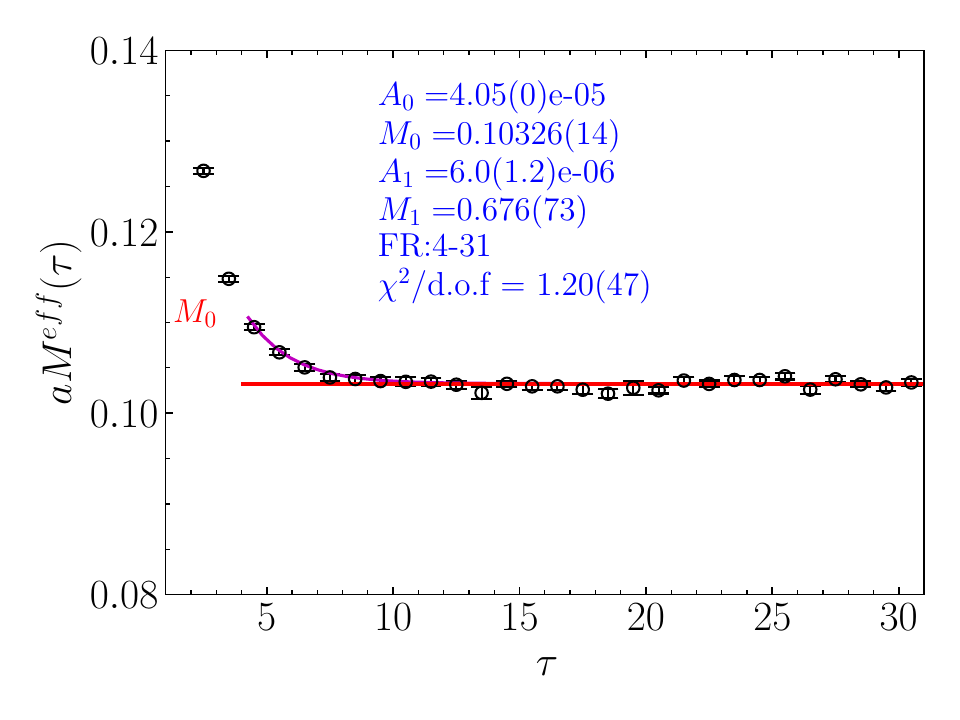}%
    \includegraphics[width=.32\textwidth]{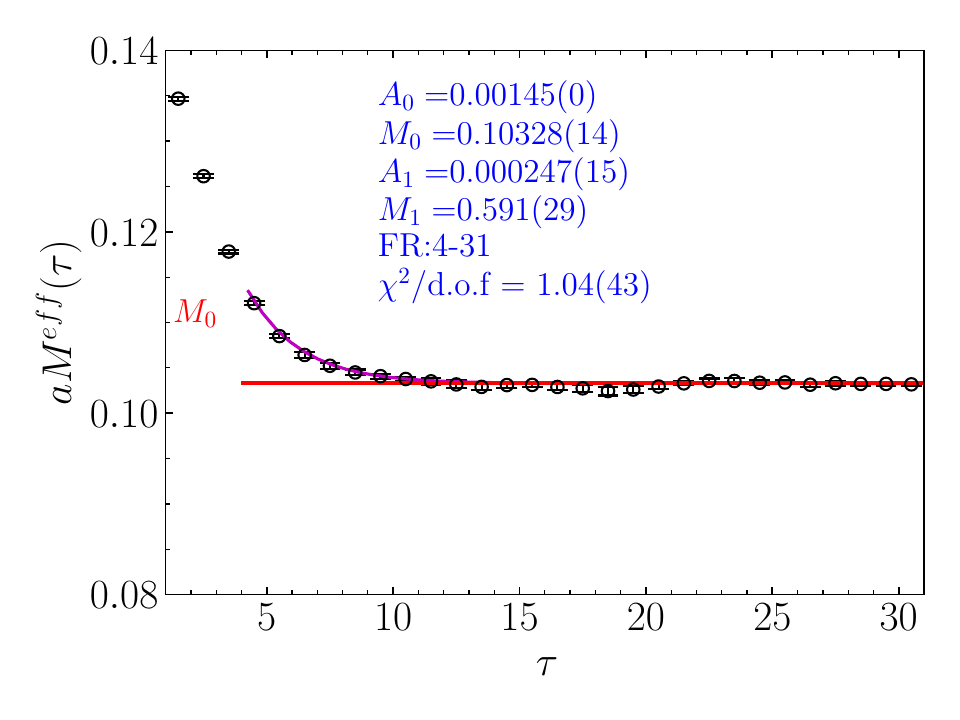}%
    \includegraphics[width=.32\textwidth]{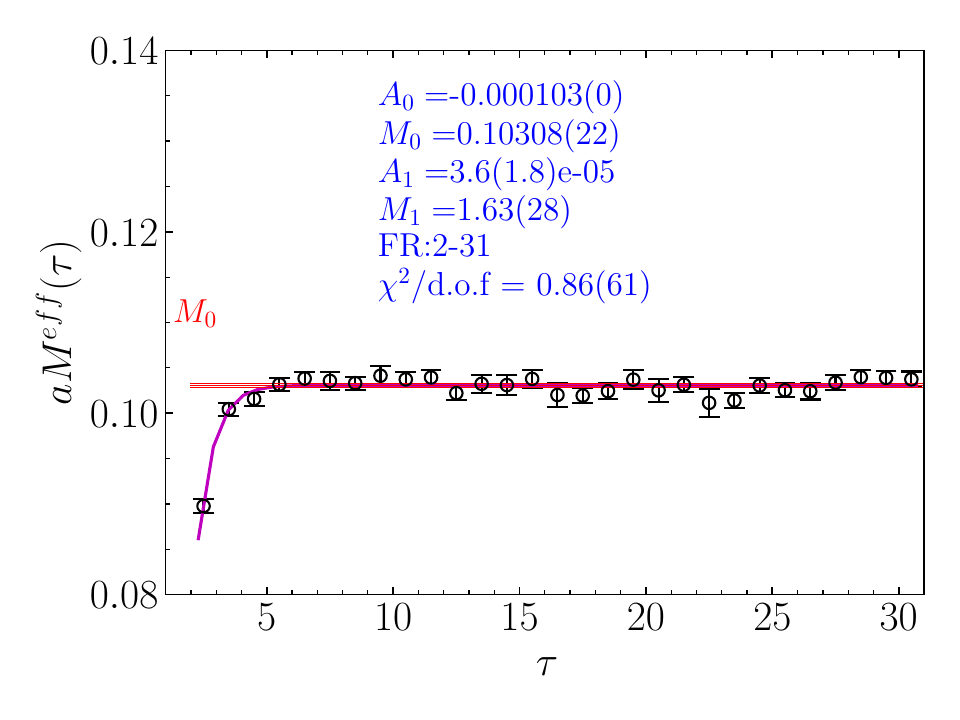}\\
    \vspace{-0.5cm}
    \includegraphics[width=.32\textwidth]{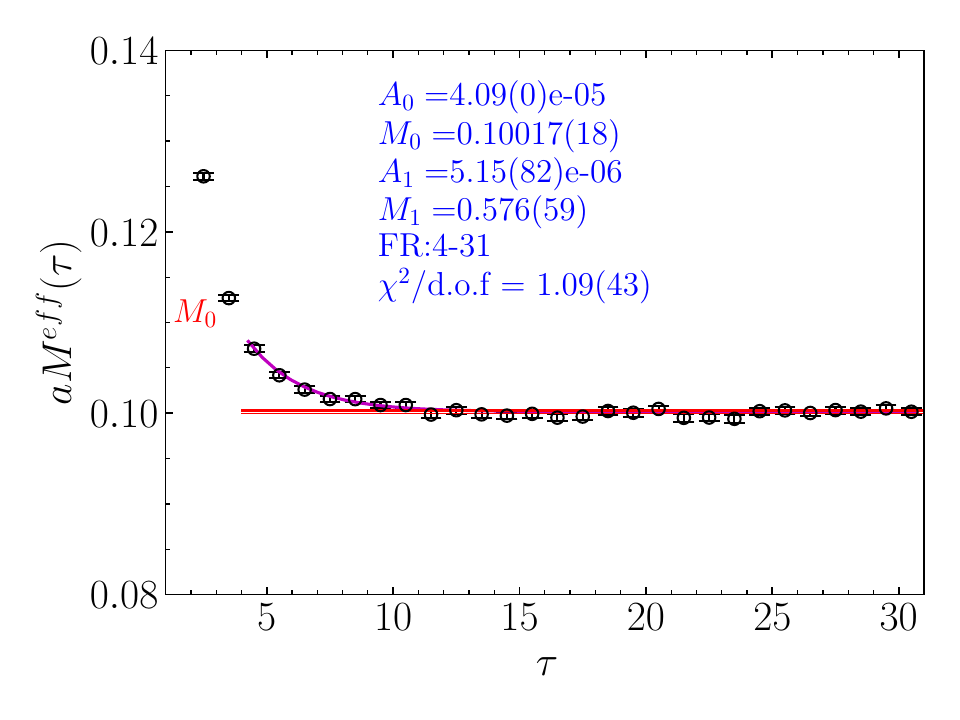}%
    \includegraphics[width=.32\textwidth]{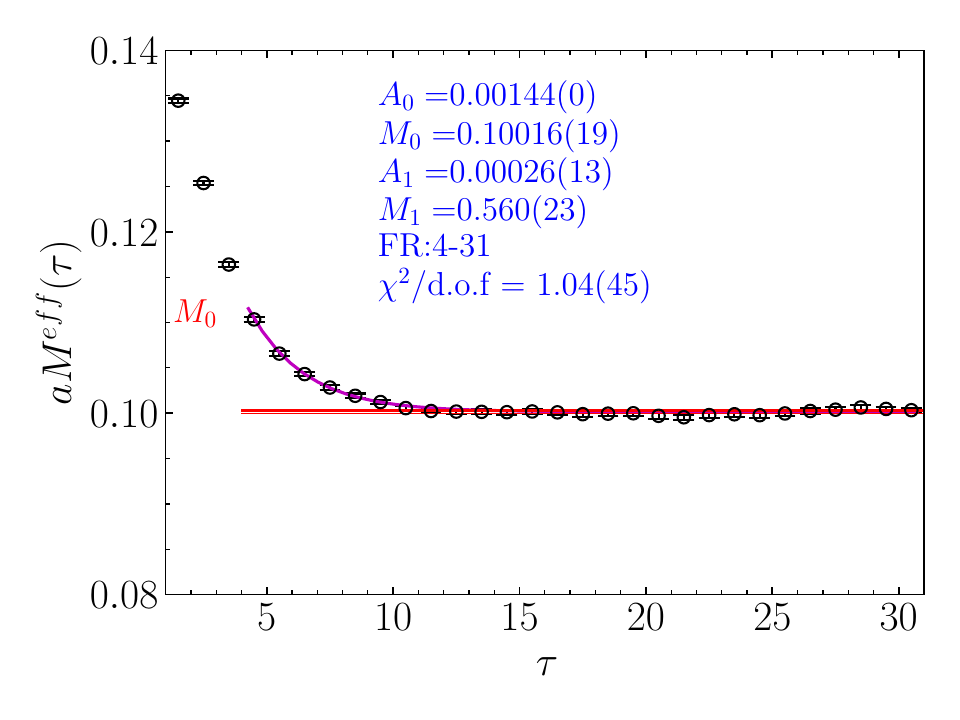}%
    \includegraphics[width=.32\textwidth]{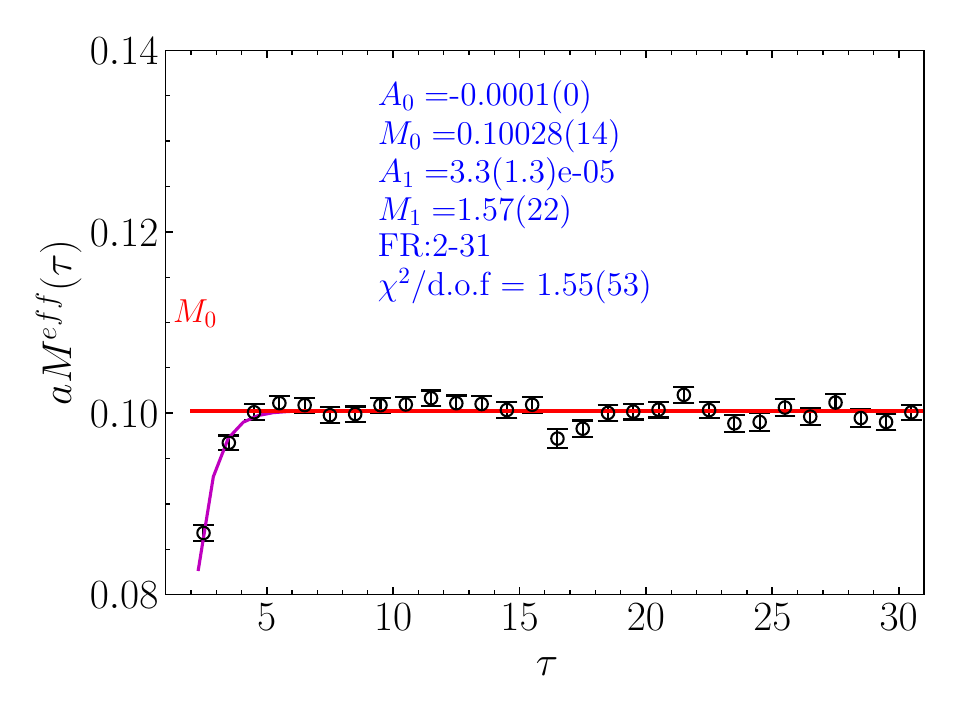}\\
    \vspace{-0.5cm}
    \includegraphics[width=.32\textwidth]{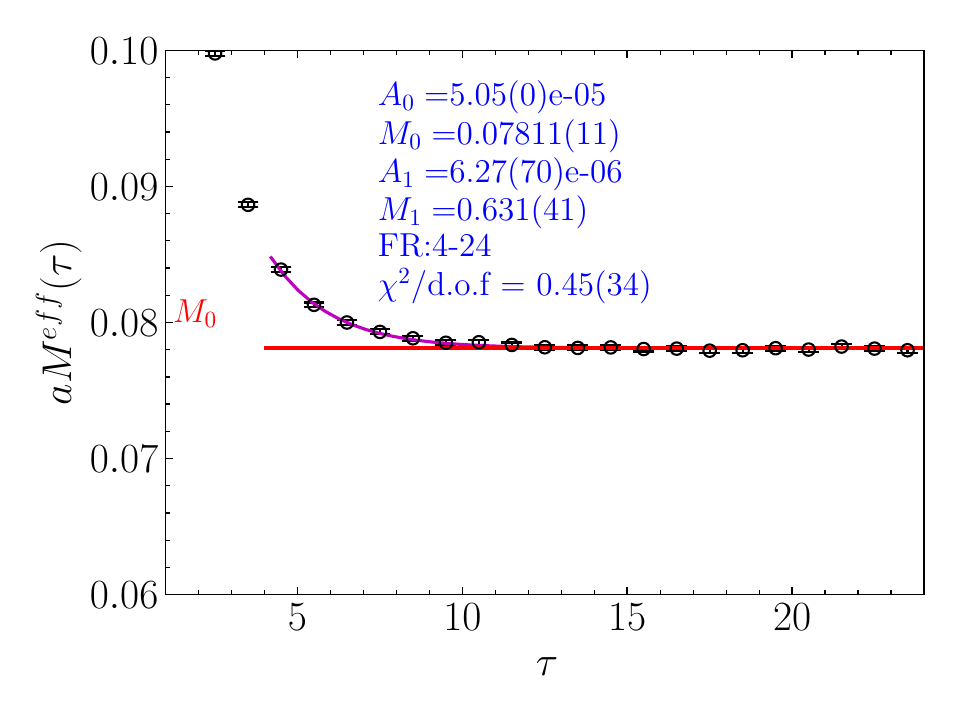}%
    \includegraphics[width=.32\textwidth]{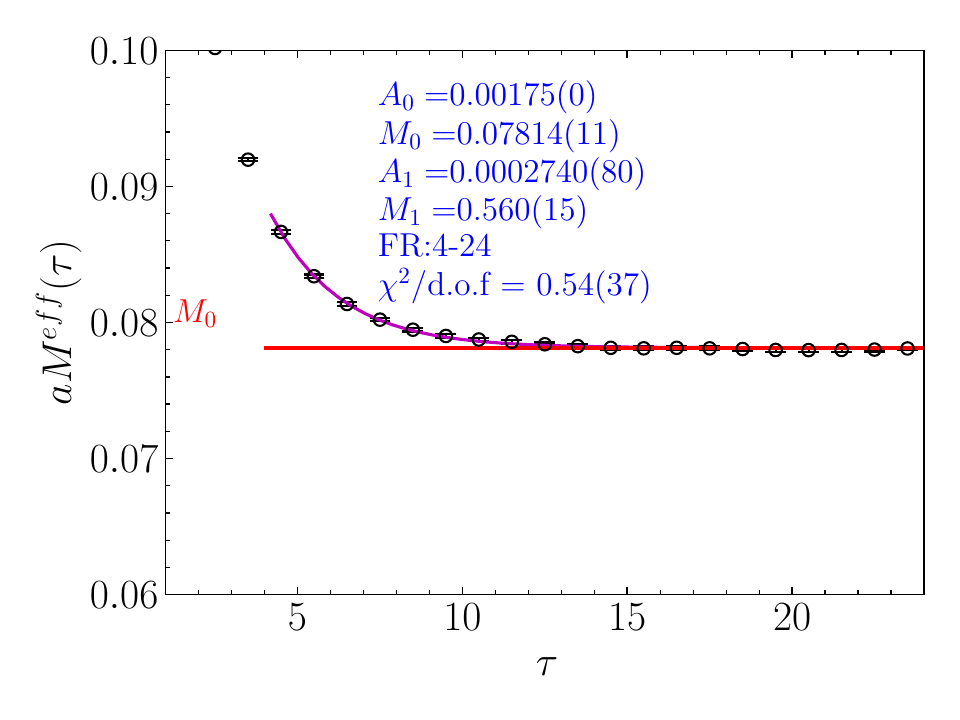}%
    \includegraphics[width=.32\textwidth]{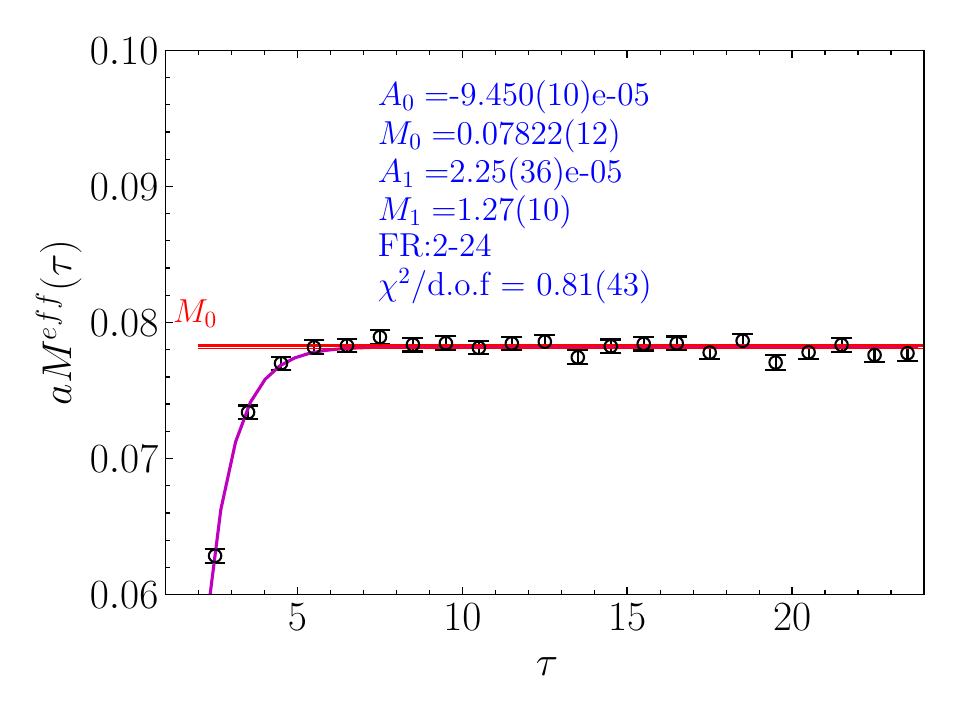}
    \vspace{-0.2cm}
    \caption{The pion effective mass in lattice units, $a \meff$, and two-state fits  to the  $\langle P_S (0) P_S(\tau) \rangle$ (left), $\langle P_S (0) P_P(\tau) \rangle$ (middle), and $\langle P_S (0) A_P(\tau) \rangle$ (right)  correlators on the $a117m310$(top), $a087m290L$ (second), $a087m230$ (third) and $a087m230X$ (fourth), and $a086m180L$ (bottom) ensembles. The legend gives the two-state fit parameters, the fit range (FR), and the $\chi^2/{\rm d.o.f.}$ of the fit. 
    The horizontal red lines give $M_0$, also in lattice units.}
    \label{fig:C13}
\end{figure*}

\begin{figure*}[h]    
    \centering
    \includegraphics[width=.32\textwidth]{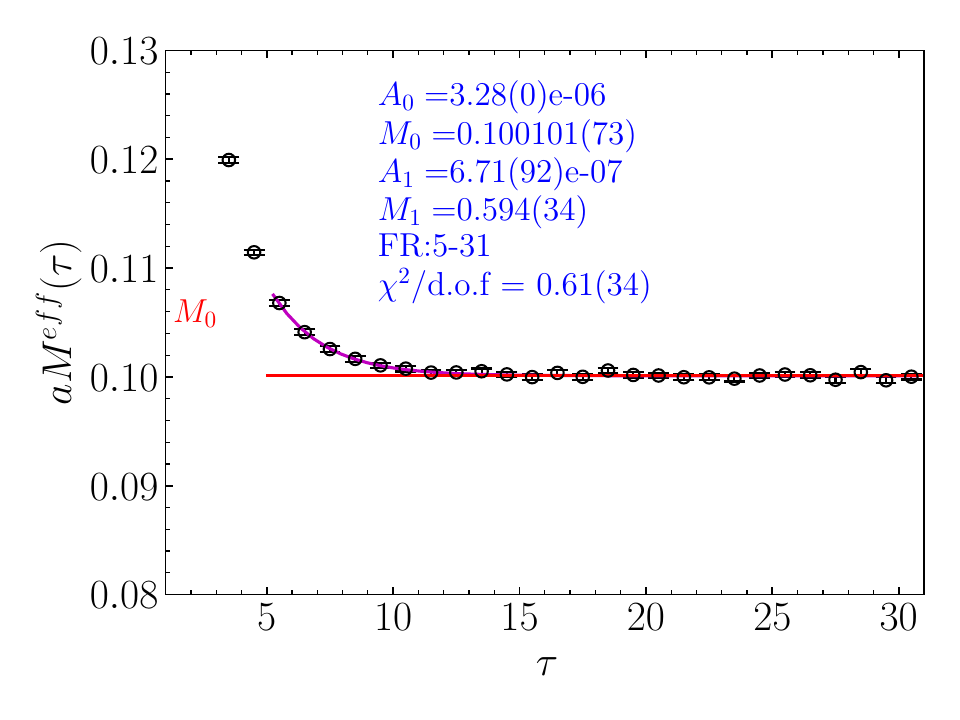}%
    \includegraphics[width=.32\textwidth]{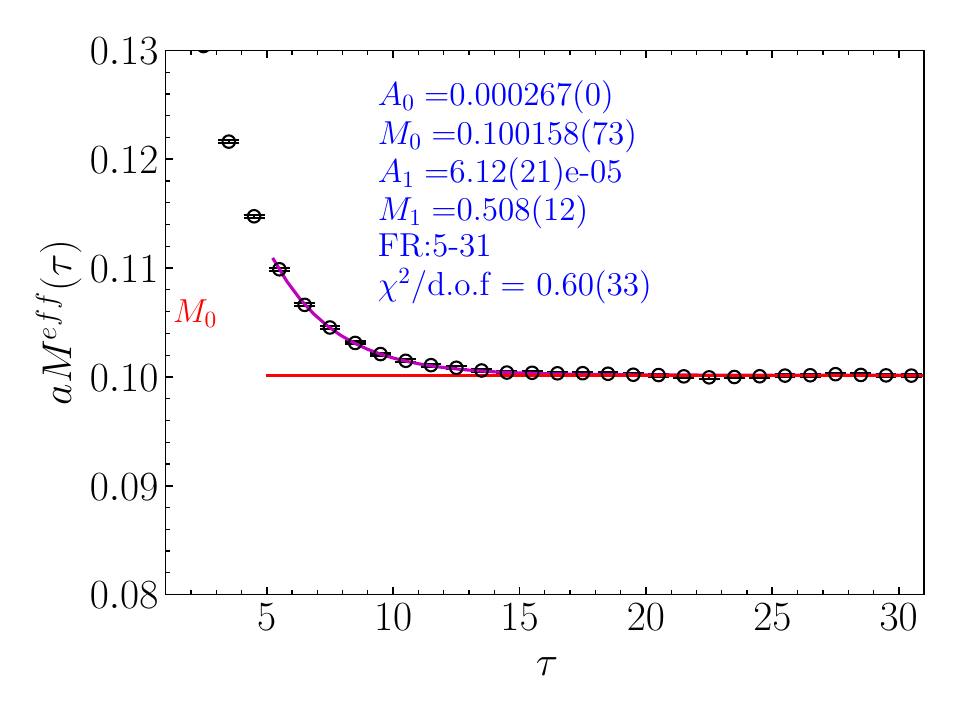}%
    \includegraphics[width=.32\textwidth]{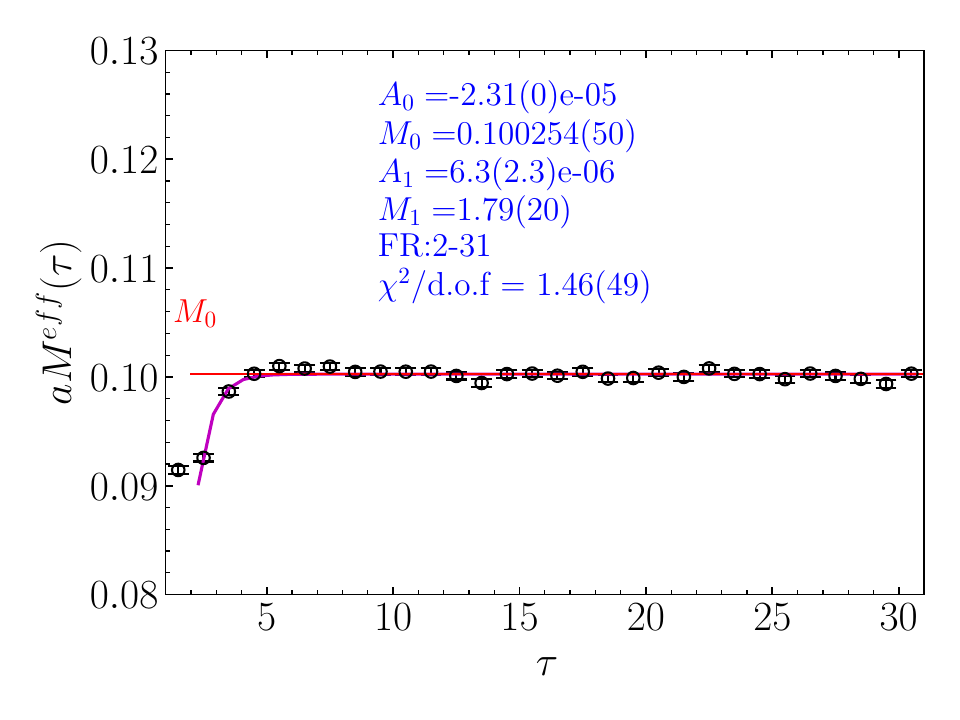} \\
        \vspace{-0.5cm}
    \includegraphics[width=.32\textwidth]{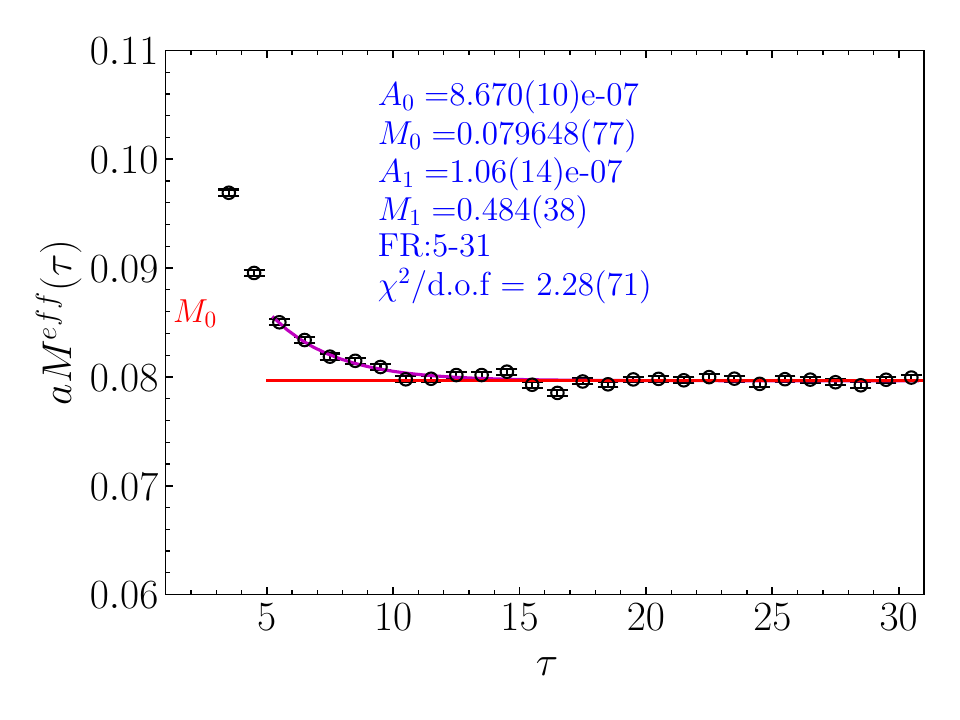}%
    \includegraphics[width=.32\textwidth]{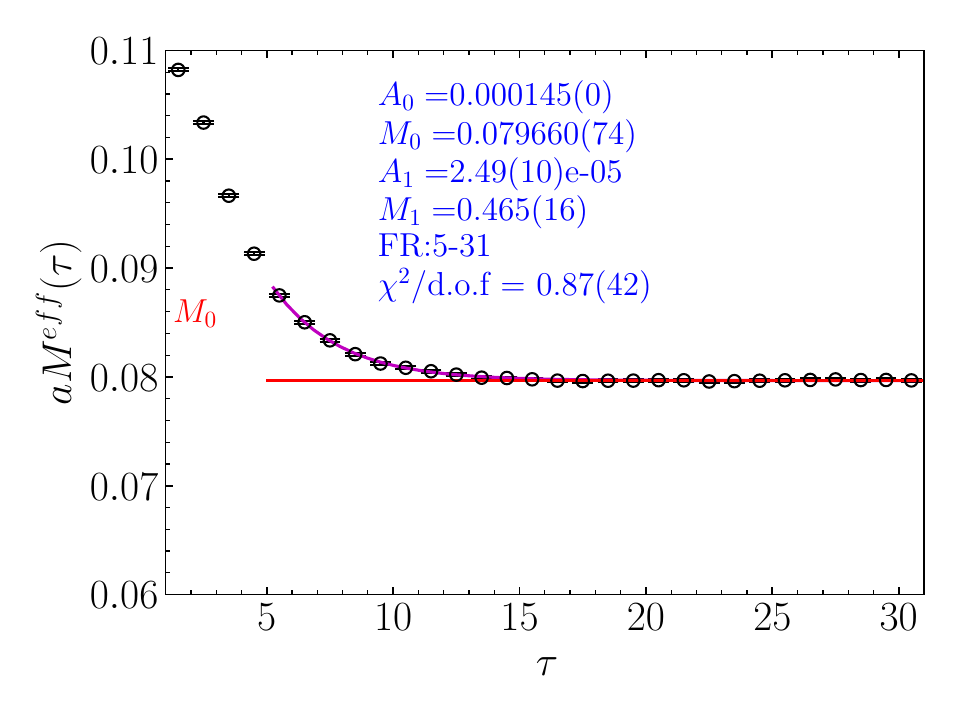}%
    \includegraphics[width=.32\textwidth]{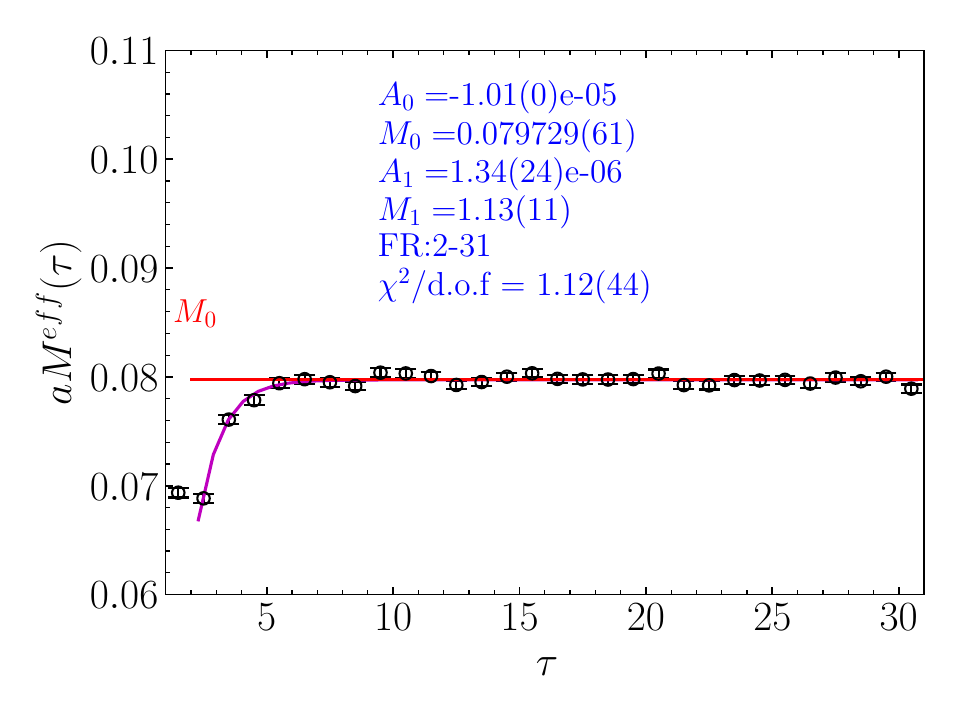} \\
        \vspace{-0.5cm}
    \centering
    \includegraphics[width=.32\textwidth]{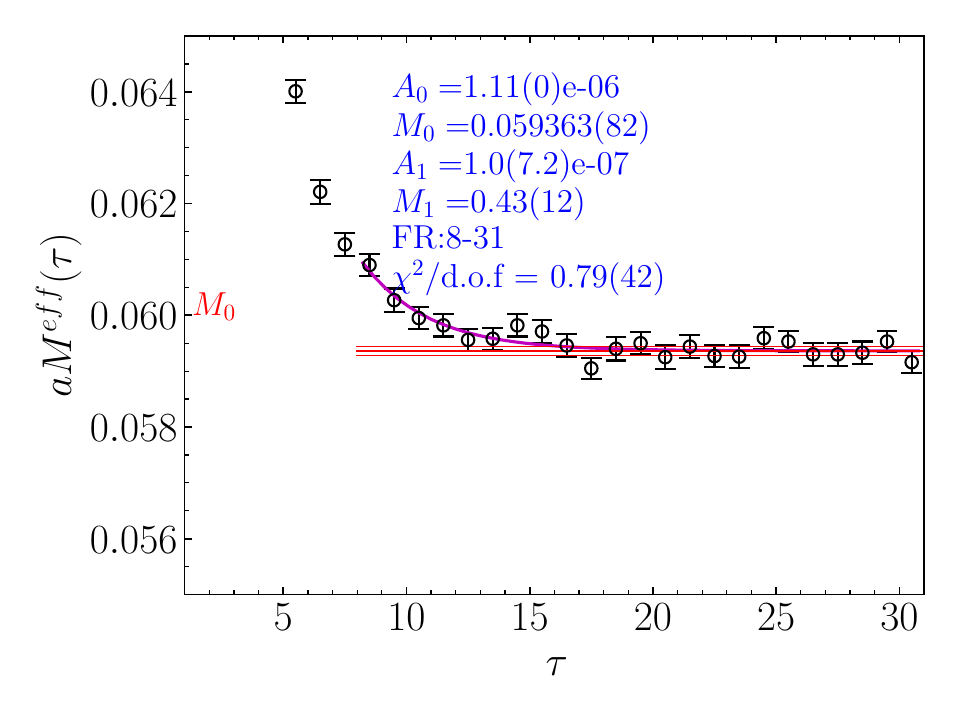}%
    \includegraphics[width=.32\textwidth]{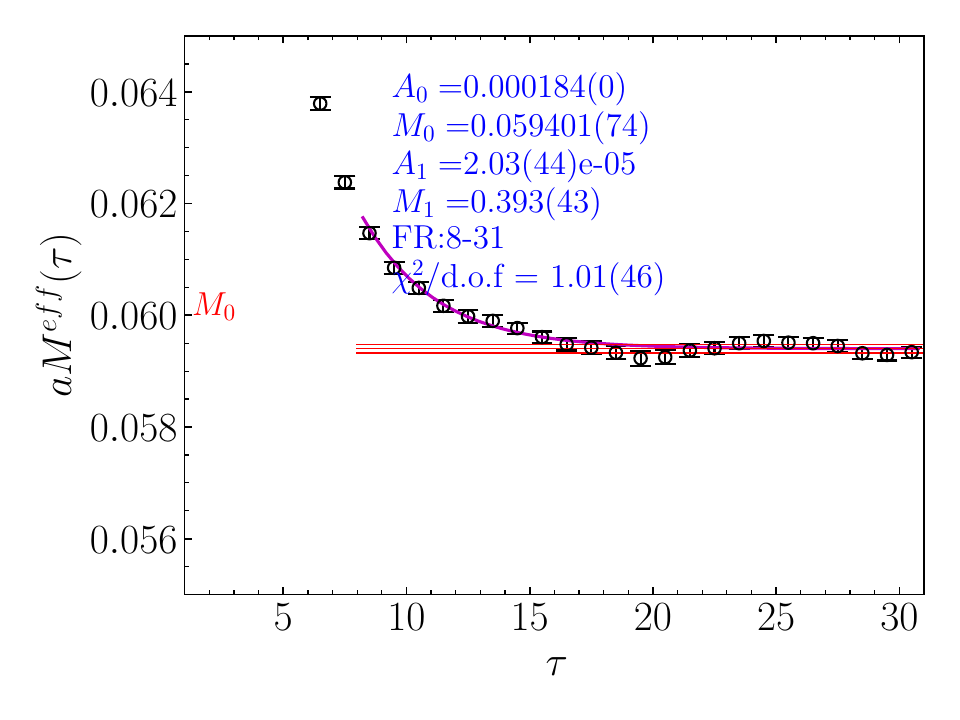}%
    \includegraphics[width=.32\textwidth]{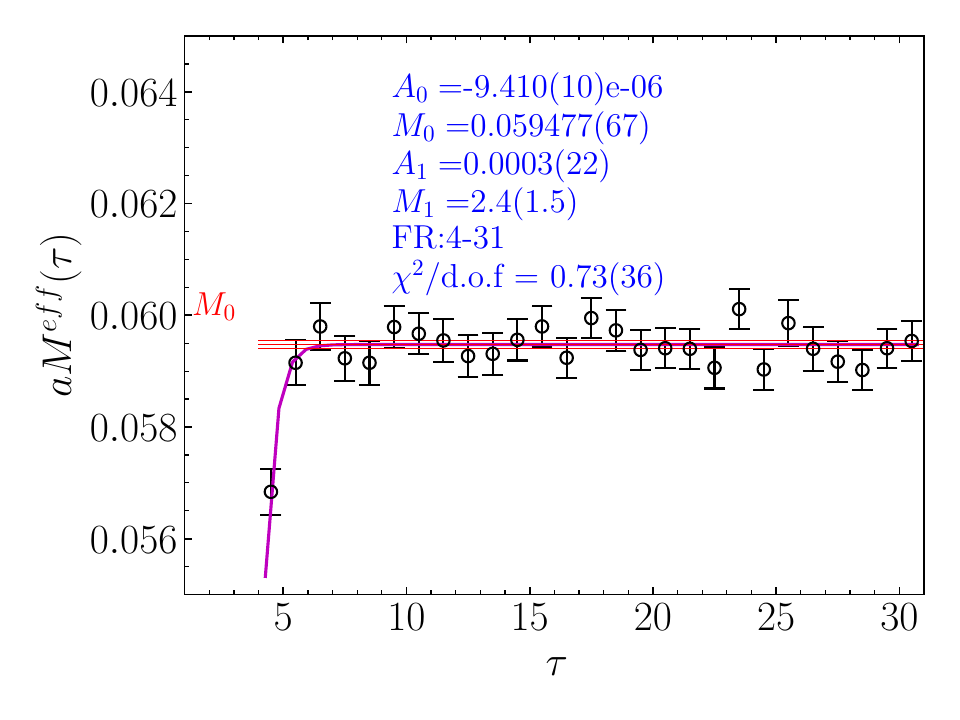} \\
        \vspace{-0.5cm}
    \includegraphics[width=.32\textwidth]{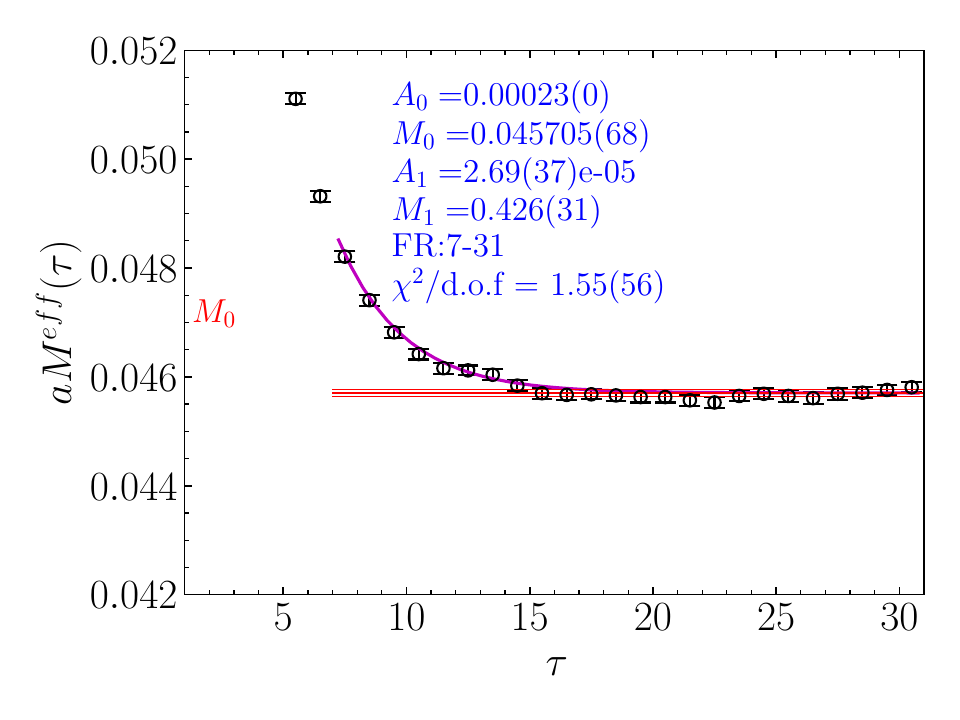}%
    \includegraphics[width=.32\textwidth]{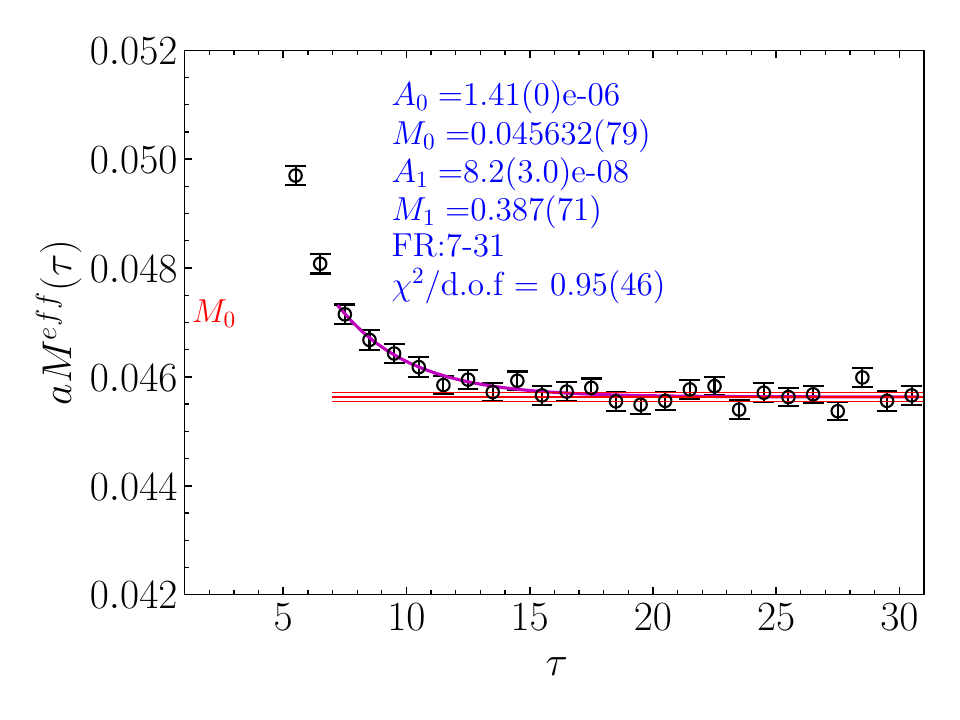}%
    \includegraphics[width=.32\textwidth]{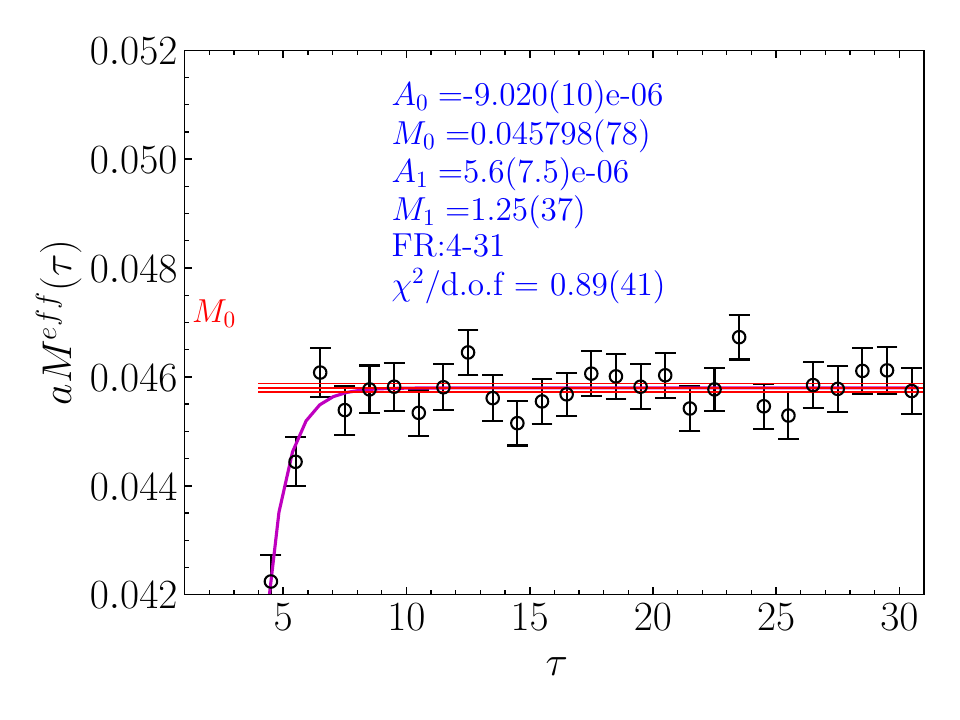} \\
        \vspace{-0.5cm}
    \includegraphics[width=.32\textwidth]{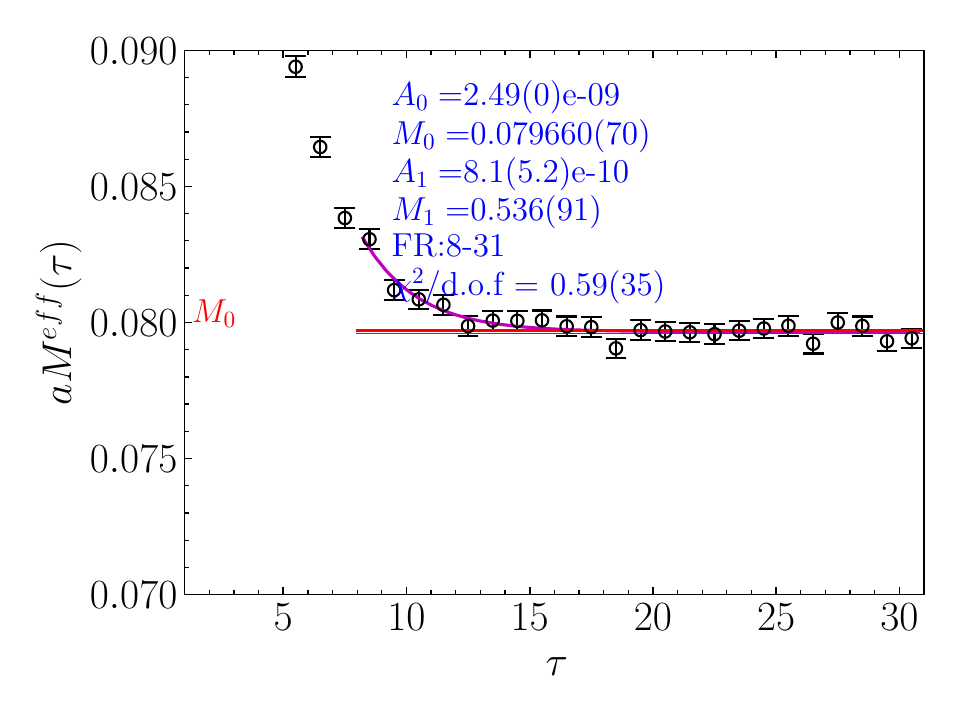}%
    \includegraphics[width=.32\textwidth]{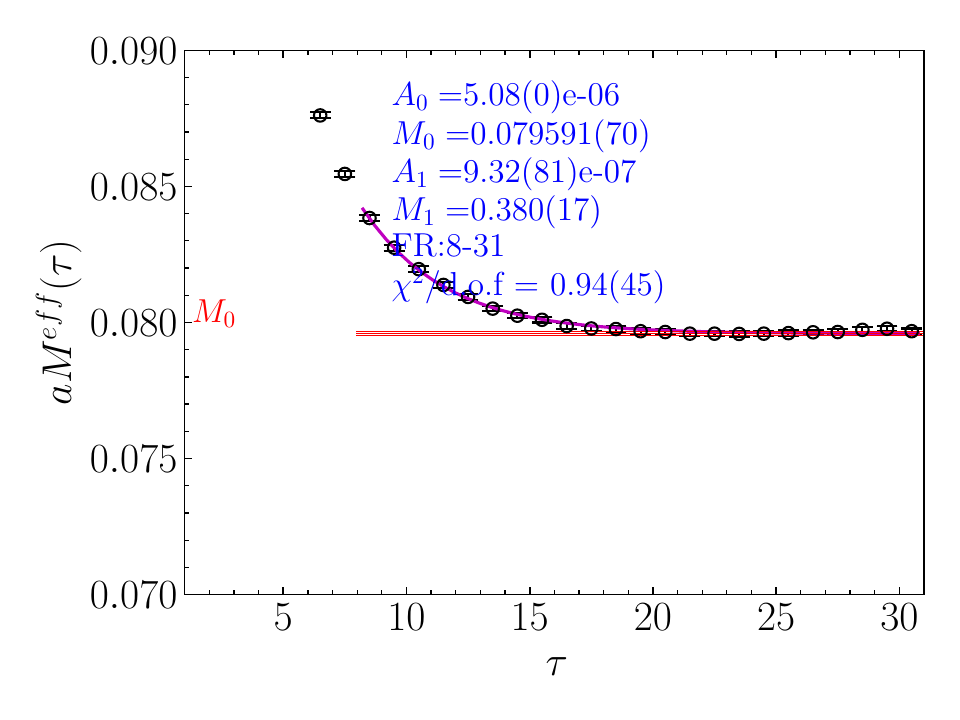}%
    \includegraphics[width=.32\textwidth]{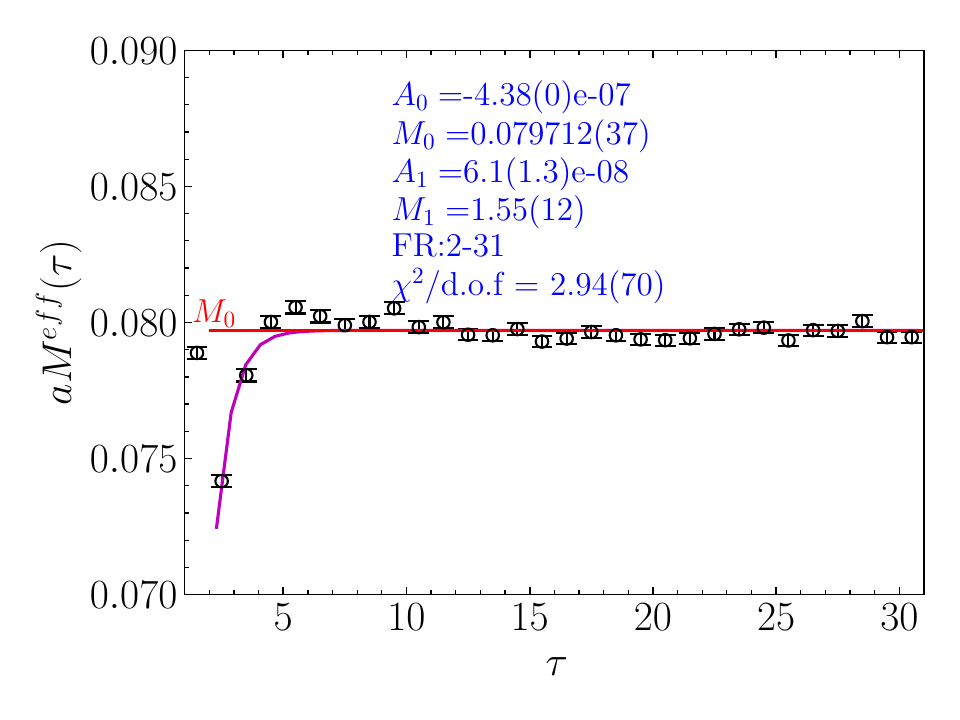} \\
        \vspace{-0.5cm}
    \includegraphics[width=.32\textwidth]{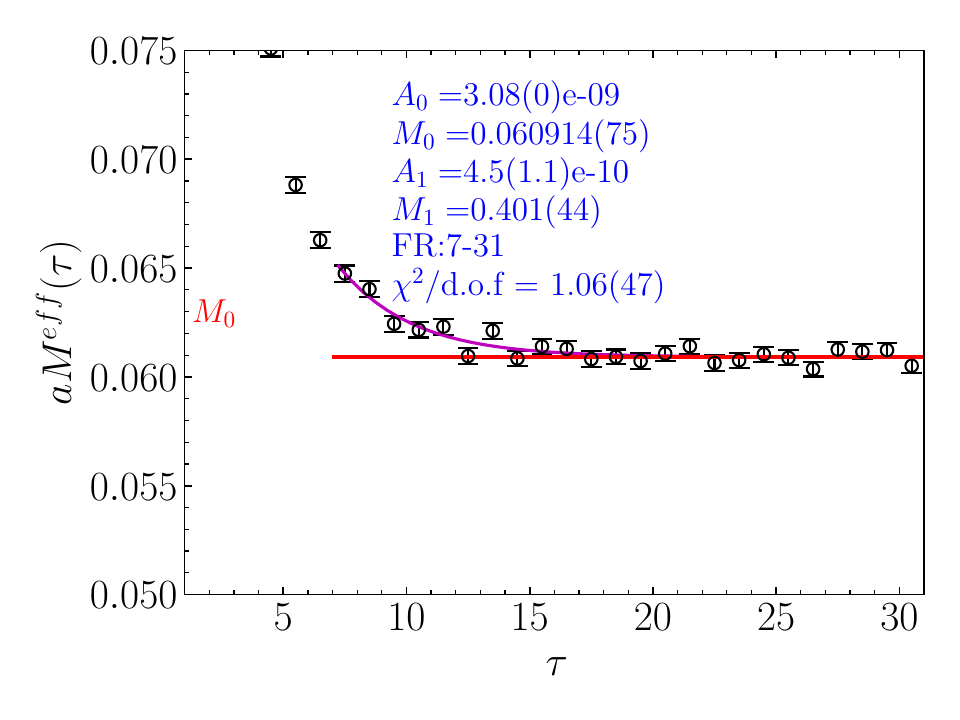}%
    \includegraphics[width=.32\textwidth]{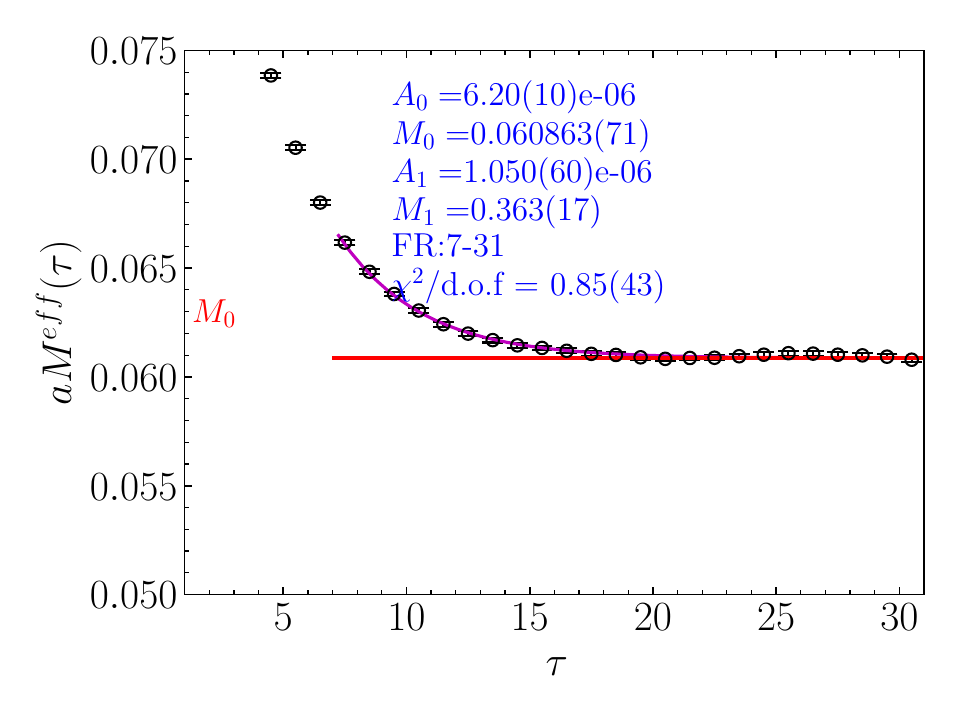}%
    \includegraphics[width=.32\textwidth]{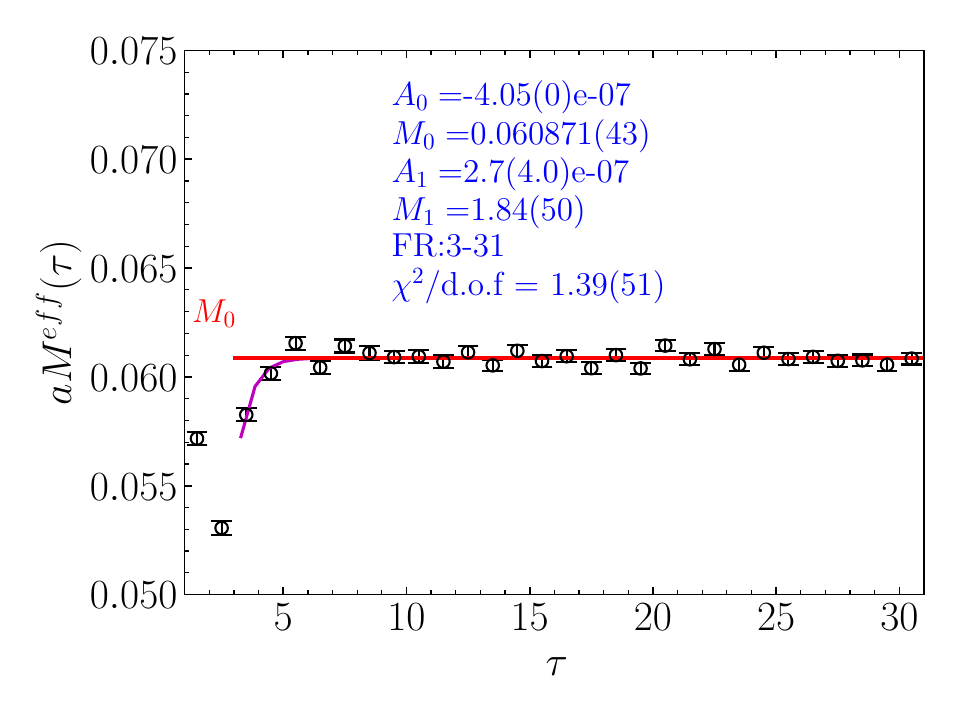}%
        \vspace{-0.5cm}
    \caption{The pion effective mass data and the two-state fit on the $a068m290$ (top), $a068m230$ (second), 
    $a068m175$ (third), $a067m135$ (fourth), $a053m295$ (fifth), and $a053m230$ (bottom) ensembles. The rest 
    is the same as in Fig.~\protect\ref{fig:C13}.}
    \label{fig:F6}
\end{figure*}
    
\end{widetext}

\clearpage    
\begin {widetext}
\section{Theta and Weinberg operators under gradient flow}
\label{sec:flowQandW}

This Appendix presents, in Figs.~\ref{fig:C13top}--\ref{fig:F6top}, the evolution 
of the distribution of $Q$ as a function of the flow time $t_{gf}$, and 
the time histories of the topological charge $Q$  
and the Weinberg operator $W$ at large flow times for the eleven  ensembles. 
From these figures, we infer the following features:
\begin{itemize}
    \item No freezing of the $Q$ or $W$ is observed as $M_\pi \to 135$~MeV or/and as $a \to 0$ even at large flow times. 
    \item The distribution of $Q$ stabilizes much before it converges to an integer. This, 
    as described in Ref.~\cite{Bhattacharya:2021lol}, is sufficient for extracting a 
    robust value for the topological susceptibility and for the measurement of the  correlation of $Q$ 
    with nucleon 3-point functions needed, for example, for calculating the contribution of the 
    $\Theta$ and Weinberg operators to the neutron electric dipole moment. 
    \item 
    There is little change in the distribution profile of $Q$ in the three 
    ensembles $a087m290L$, $a087m230$ and $a087m230X$ 
that have the same lattice volume ($48^3 \times 128$) (see Fig.~\ref{fig:D7top}), whereas it broadens with an increase in the volume as inferred from Fig.~\ref{fig:E9top} by moving from  $a068m290 \to a067m230 \to a067m175 \to a067m135$ data.
\end{itemize}

\begin{figure*}[h]  
  \begin{minipage}[t]{0.45\linewidth}
      \hrule width0pt
    \includegraphics[trim=0 0 10 10, width=\linewidth]{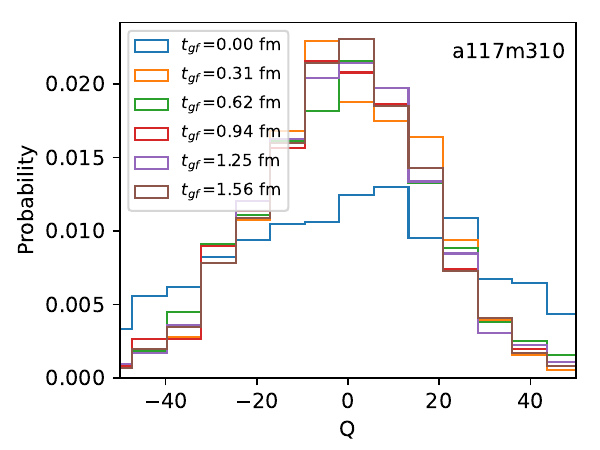}
  \end{minipage}\hfill
  \begin{minipage}[t]{0.50\linewidth}
      \hrule width0pt
      \hspace{-0.1in} \includegraphics[trim=0 20 0 10, width=\linewidth]{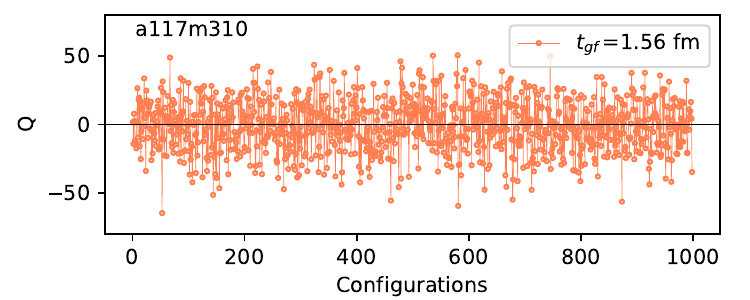}
      \label{subfig:C13Q}

    \vspace{-0.08in} \hspace{-0.13in}
      \includegraphics[trim=0 10 0 10,width=1.0\linewidth]{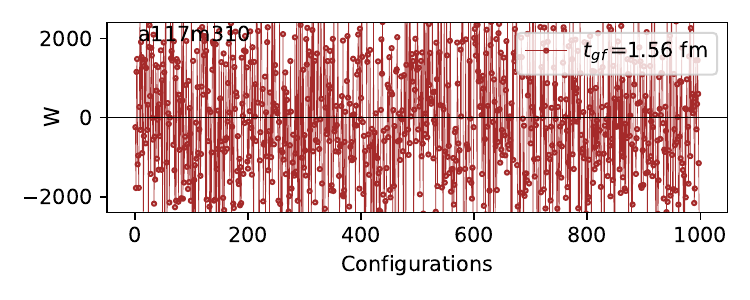}
      \label{subfig:C13W}
   \end{minipage}
    \vspace{-0.1in}
\caption{(Left) Histogram of the distribution of the topological charge $Q$ 
at different flow times $\tau_{gf}$ specified in the legend for the $a117m310$ ensemble.
(Right) The Monte Carlo time history of the Q (top) and the Weinberg operator $W$ (bottom). The flow times, $t_{gf}$ in fermi, at which data are presented are specified in the labels.
 }
    \label{fig:C13top}
\end{figure*}

\begin{figure*}      
   
    \begin{minipage}[t]{0.45\linewidth}
    \hrule width0pt
    \includegraphics[trim=0 0 10 10, width=\linewidth]{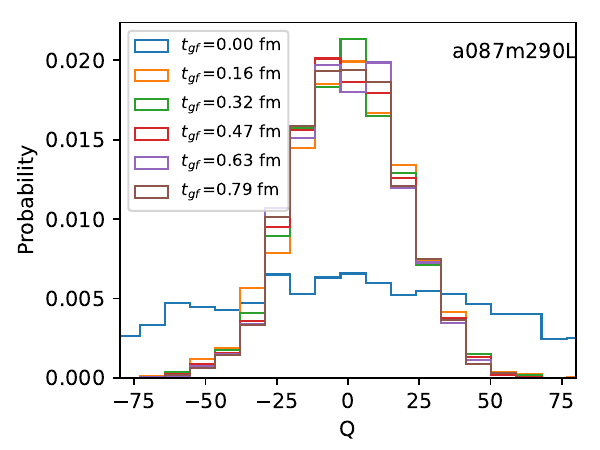}
  \end{minipage}\hfill
  \begin{minipage}[t]{0.50\linewidth}
    \hrule width0pt
      \includegraphics[trim=0 50 0 10, width=\linewidth]{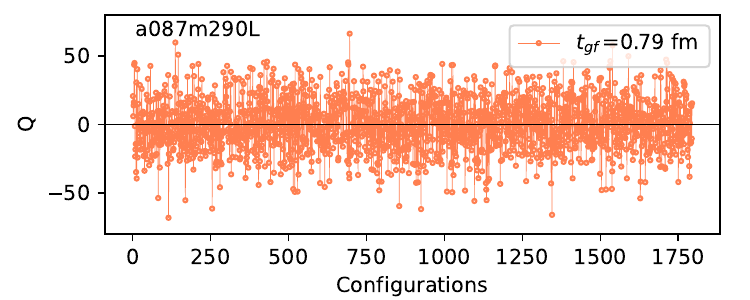}
      \label{subfig:D5lQ}

      \vspace{-0.08in} \hspace{-0.03in}
      \includegraphics[trim=0 10 0 10, width=\linewidth]{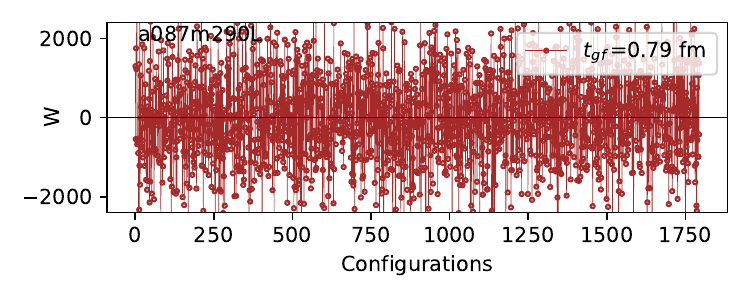}
      \label{subfig:D5lW}
  \end{minipage}
   \vspace{-0.02in}

  \begin{minipage}[t]{0.45\linewidth}
    \hrule width0pt
    \includegraphics[trim=0 0 10 10, width=\linewidth]{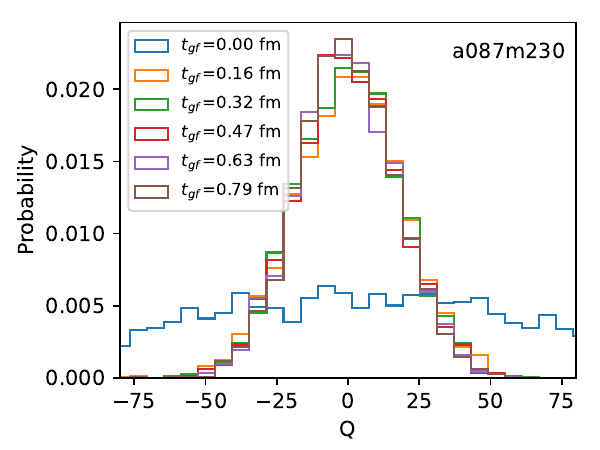}
  \end{minipage}\hfill
  \begin{minipage}[t]{0.50\linewidth}
    \hrule width0pt
      \includegraphics[trim=0 50 0 10, width=\linewidth]{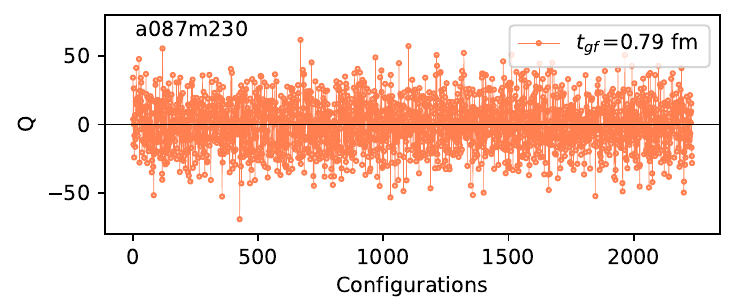}
      \label{subfig:D220Q}

      \vspace{-0.08in} \hspace{-0.02in}
      \includegraphics[trim=0 10 0 10, width=\linewidth]{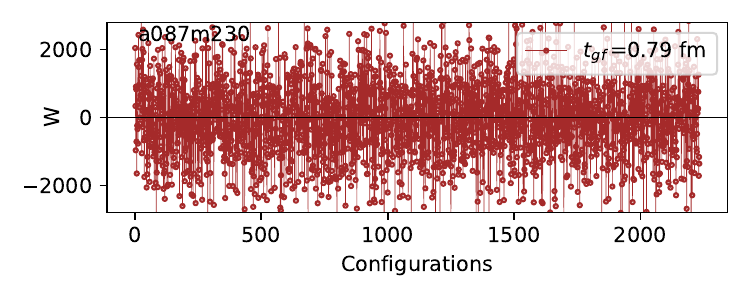}
      \label{subfig:D220W}
  \end{minipage}
  \label{fig:D220top}
 \vspace{-0.02in}
 
  \begin{minipage}[t]{0.45\linewidth}
    \hrule width0pt
    \includegraphics[trim=0 0 10 10, width=\linewidth]{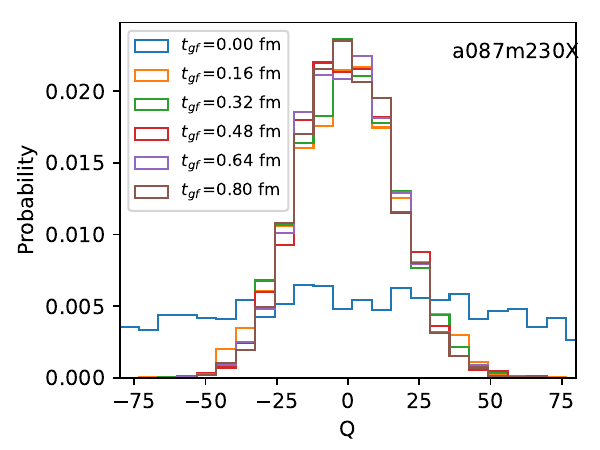}
  \end{minipage}\hfill
  \begin{minipage}[t]{0.50\linewidth}
    \hrule width0pt
      \includegraphics[trim=0 50 0 10, width=\linewidth]{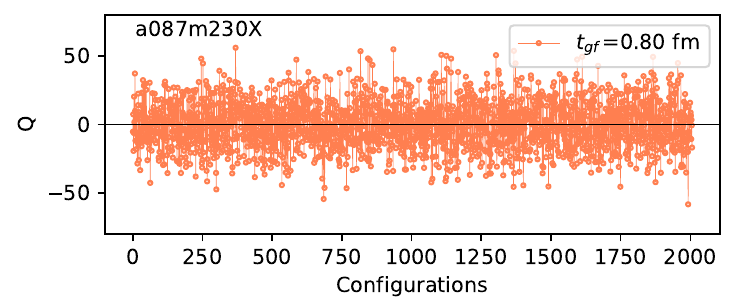}
      \label{subfig:D220XQ}

      \vspace{-0.08in} \hspace{-0.02in}
      \includegraphics[trim=0 10 0 10, width=\linewidth]{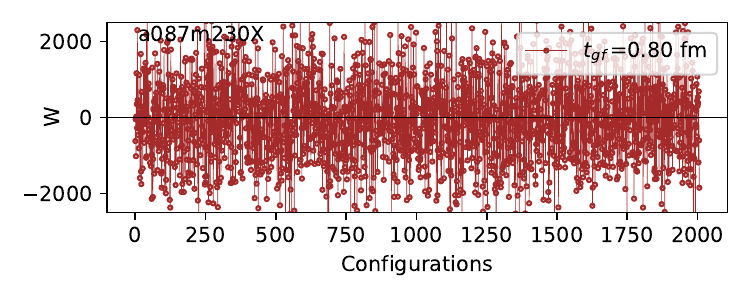}
      \label{subfig:D220XW}
  \end{minipage}
  \label{fig:D220Xtop}
 \vspace{-0.02in}

  \begin{minipage}[t]{0.45\linewidth}
    \hrule width0pt
    \includegraphics[trim=0 0 10 10, width=\linewidth]{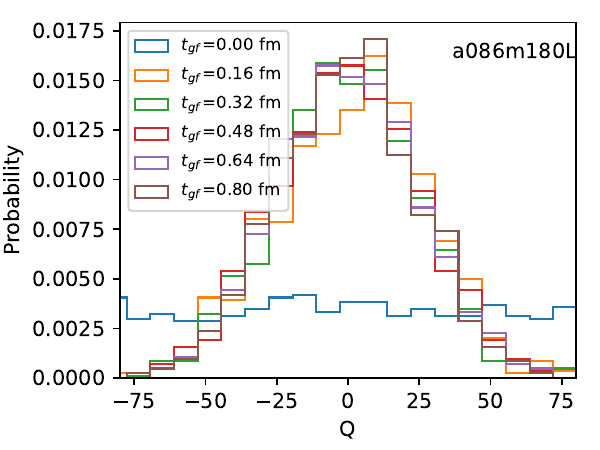}
  \end{minipage}\hfill
  \begin{minipage}[t]{0.50\linewidth}
    \hrule width0pt
      \includegraphics[trim=0 50 0 10, width=\linewidth]{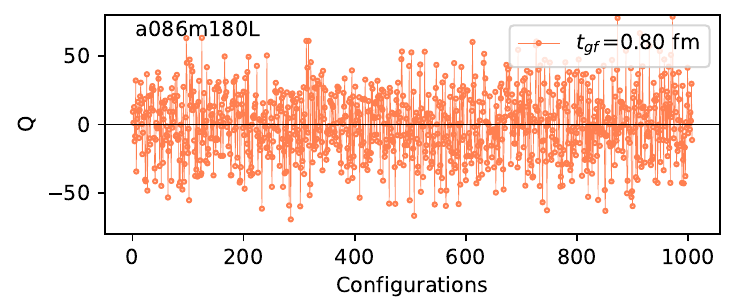}
      \label{subfig:D7Q}

      \vspace{-0.08in} \hspace{-0.02in}
      \includegraphics[trim=0 10 0 10, width=\linewidth]{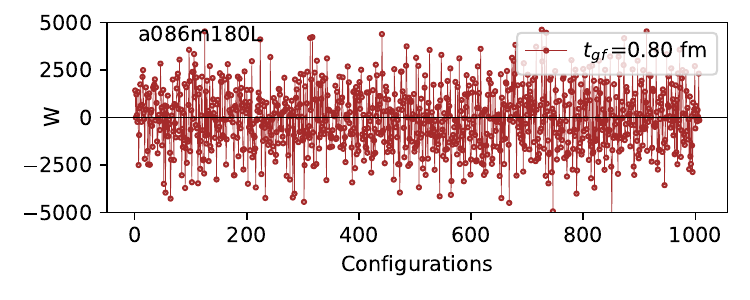}
      \label{subfig:D7W}
  \end{minipage}
     \vspace{-0.2in}
    \caption{Same as in Fig.~\protect\ref{fig:C13top} except data are for ensembles $a087m290L$ (top);  $a087m230$ (second); $a087m230X$ (third); and $a086m180L$ (bottom). The flow times, $t_{gf}$ in fermi, at which data are presented are specified in the labels. }
  \label{fig:D7top}
\end{figure*}

\begin{figure*}   
  \begin{minipage}[t]{0.45\linewidth}
    \hrule width0pt
    \includegraphics[trim=0 0 10 10, width=\linewidth]{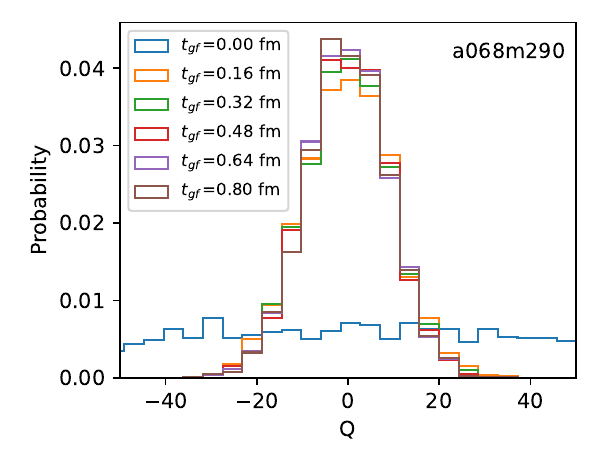}
  \end{minipage}\hfill
  \begin{minipage}[t]{0.50\linewidth}
    \hrule width0pt
      \includegraphics[trim=0 50 0 10, width=\linewidth]{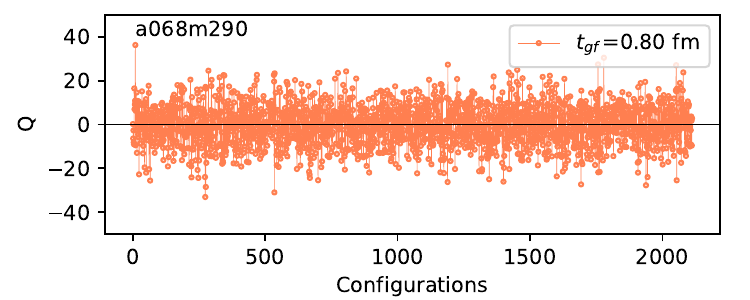}
      \label{subfig:E5Q}

      \vspace{-0.08in} \hspace{-0.02in}
      \includegraphics[trim=0 10 0 10, width=\linewidth]{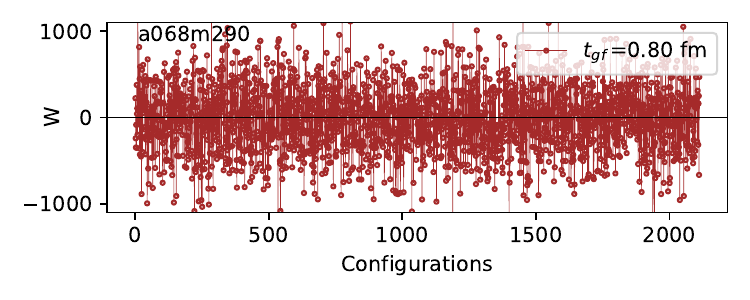}
      \label{subfig:E5W}
  \end{minipage}
  \label{fig:E5top}
 \vspace{-0.02in}
 
  \begin{minipage}[t]{0.45\linewidth}
    \hrule width0pt
    \includegraphics[trim=0 0 10 10, width=\linewidth]{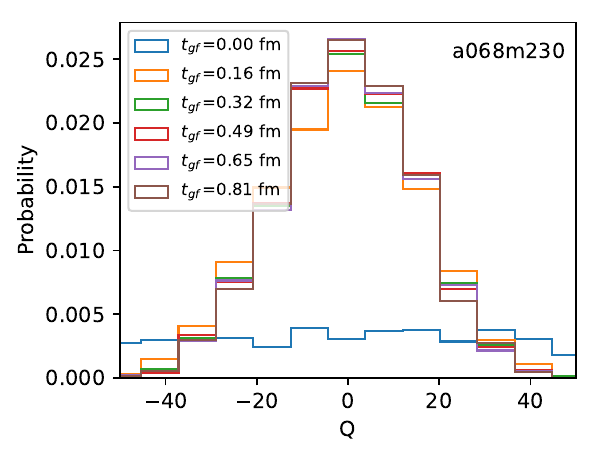}
  \end{minipage}\hfill
  \begin{minipage}[t]{0.50\linewidth}
    \hrule width0pt
      \includegraphics[trim=0 50 0 10, width=\linewidth]{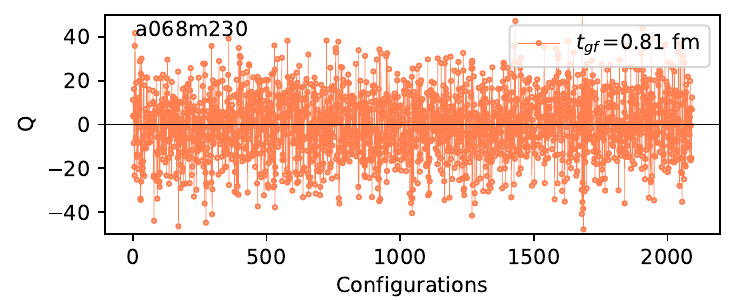}
      \label{subfig:E6Q}

      \vspace{-0.08in} \hspace{-0.02in}
      \includegraphics[trim=0 10 0 10, width=\linewidth]{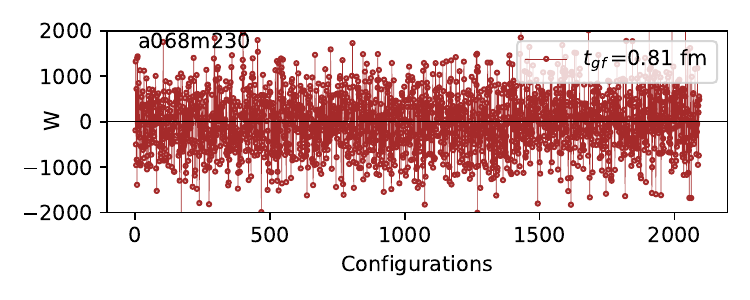}
      \label{subfig:E6W}
  \end{minipage}
  \label{fig:E6top}
 \vspace{-0.02in}
  \begin{minipage}[t]{0.45\linewidth}
    \hrule width0pt
    \includegraphics[trim=0 0 10 10, width=\linewidth]{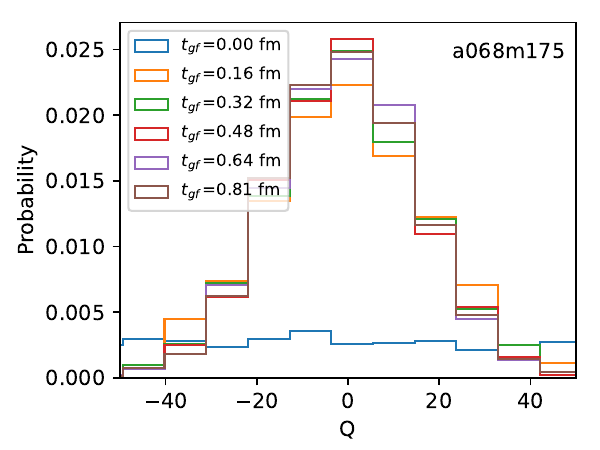}
  \end{minipage}\hfill
  \begin{minipage}[t]{0.50\linewidth}
    \hrule width0pt
      \includegraphics[trim=0 50 0 10, width=\linewidth]{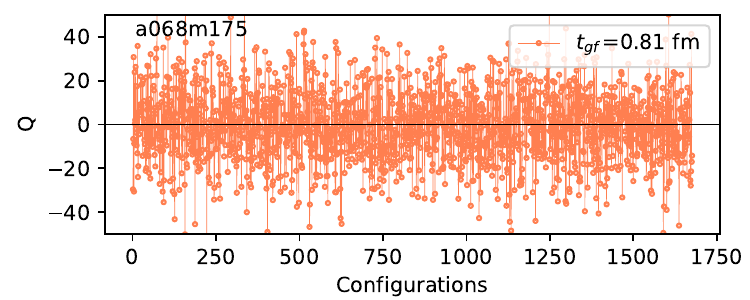}
      \label{subfig:E7Q}

      \vspace{-0.08in} \hspace{-0.02in}
      \includegraphics[trim=0 10 0 10, width=\linewidth]{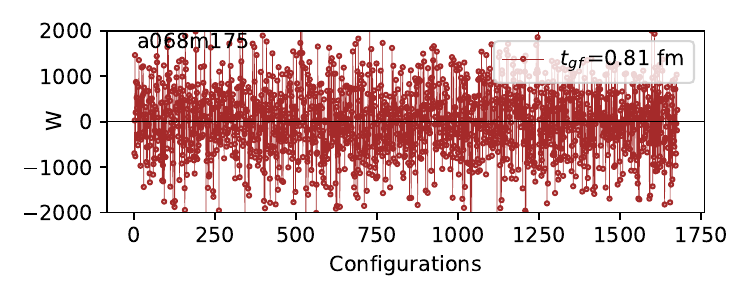}
      \label{subfig:E7W}
  \end{minipage}
  \label{fig:E7top}
 \vspace{-0.02in}
 
  \begin{minipage}[t]{0.45\linewidth}
    \hrule width0pt
    \includegraphics[trim=0 0 10 10, width=\linewidth]{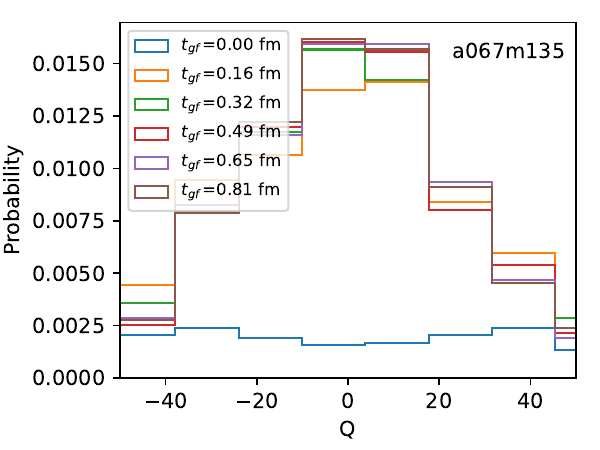}
  \end{minipage}\hfill
  \begin{minipage}[t]{0.50\linewidth}
    \hrule width0pt
      \includegraphics[trim=0 50 0 10, width=\linewidth]{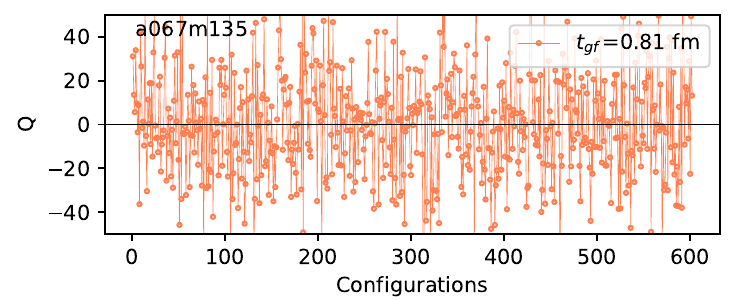}
      \label{subfig:E9Q}

      \vspace{-0.08in} \hspace{-0.02in}
      \includegraphics[trim=0 10 0 10, width=\linewidth]{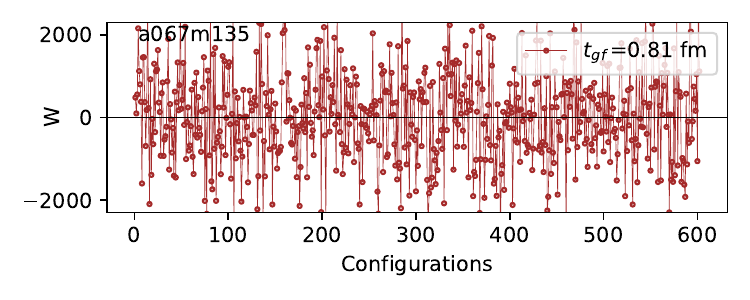}
      \label{subfig:E9W}
  \end{minipage}
   \vspace{-0.2in}
      \caption{Same as in Fig.~\protect\ref{fig:C13top} but for ensembles $a068m290$ (top);  $a067m230$ (second); $a067m175$ (third); and $a067m135$ (bottom). The flow times, $t_{gf}$ in fermi, at which data are presented are specified in the labels.}
  \label{fig:E9top}
\end{figure*}

\begin{figure*}      
  \begin{minipage}[t]{0.45\linewidth}
    \hrule width0pt
    \includegraphics[trim=0 0 10 10, width=\linewidth]{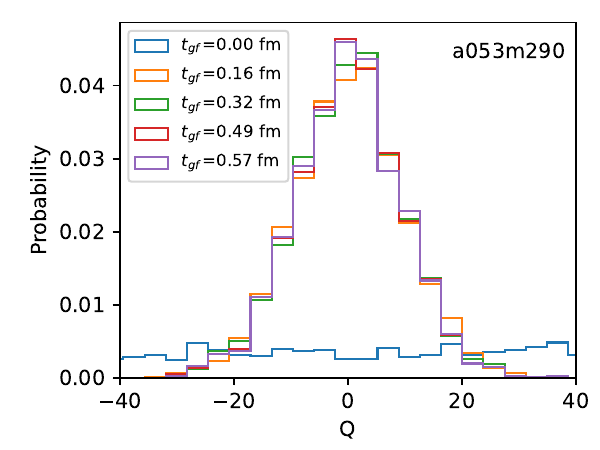}
  \end{minipage}\hfill
  \begin{minipage}[t]{0.50\linewidth}
    \hrule width0pt
      \includegraphics[trim=0 50 0 10, width=\linewidth]{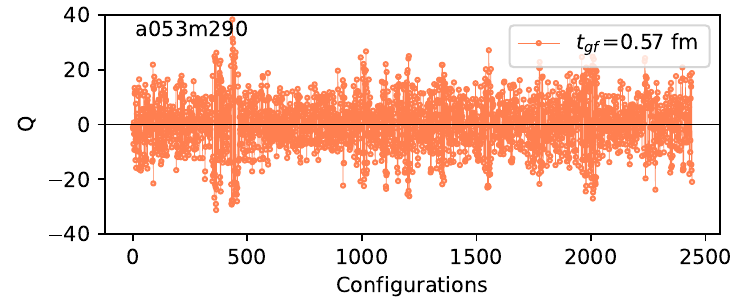}
      \label{subfig:F5Q}

      \vspace{-0.08in} \hspace{-0.02in}
      \includegraphics[trim=0 10 0 10, width=\linewidth]{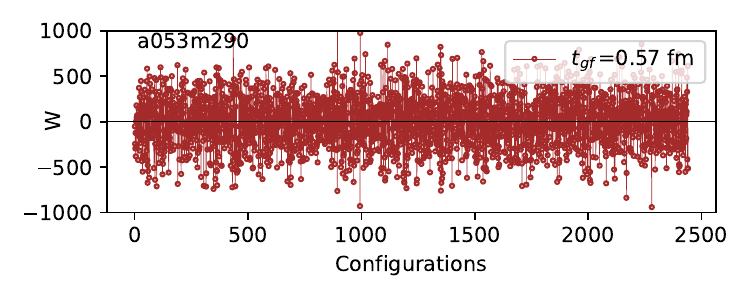}
      \label{subfig:F5W}
  \end{minipage}
  \label{fig:F5top}
  \vspace{-0.02in}
  
  \begin{minipage}[t]{0.45\linewidth}
    \hrule width0pt
    \includegraphics[trim=0 0 10 10, width=\linewidth]{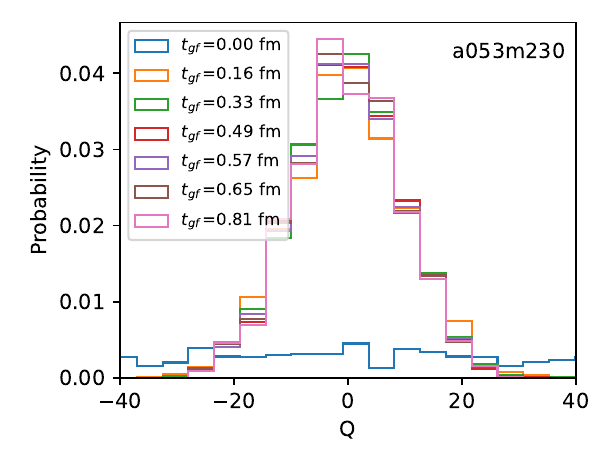}
  \end{minipage}\hfill
  \begin{minipage}[t]{0.50\linewidth}
    \hrule width0pt
      \includegraphics[trim=0 50 0 10, width=\linewidth]{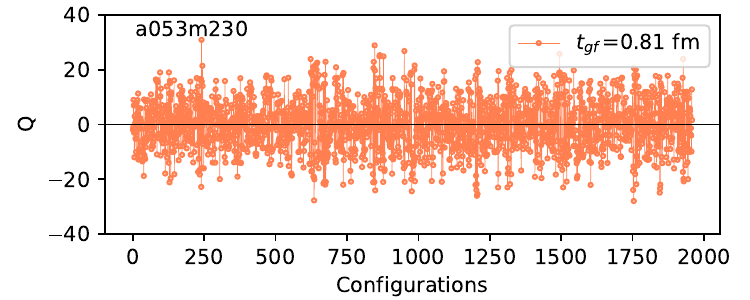}
      \label{subfig:F6Q}

      \vspace{-0.08in} \hspace{-0.02in}
      \includegraphics[trim=0 10 0 10, width=\linewidth]{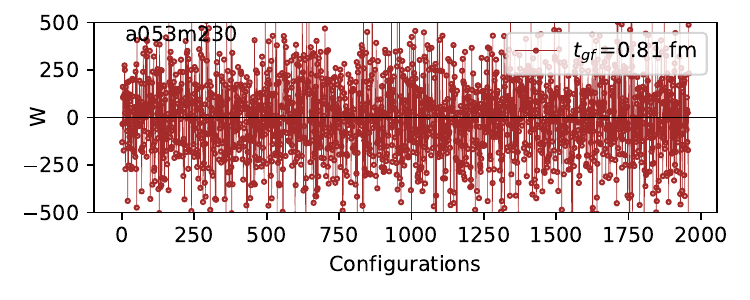}
      \label{subfig:F6W}
  \end{minipage}
    \vspace{-0.1in}
        \caption{Same as Fig.~\protect\ref{fig:C13top} but for ensembles $a053m295$ (top)  and $a053m230$ (bottom).  The flow times, $t_{gf}$ in fermi, at which data are presented are specified in the labels. }
  \label{fig:F6top}
\end{figure*}    
\end{widetext}

\clearpage

\section{Handling Autocorrelations in the Data}
\label{sec:autocorr}

Measurements made on configurations generated sequentially in Monte Carlo time are very often 
correlated. To determine the size of these correlations from $N$ sequential measurements, 
the autocorrelation function of an observables A is calculated as
\begin{equation}
    \Gamma_{A} (t) = \frac{1}{N- t} \sum_{t_0= 1}^{N-t} 
    (A_{t_0+t} -\langle A \rangle)(A_{t_0} -\langle A \rangle) \,,
\end{equation}
and modeled as 
\begin{equation}
    \rho_{A} (t) = \frac{\Gamma_{A}(t)}{\Gamma_{A}(0)} 
    = \sum_{k } c_{A,k}^2 e^{-|t|/\tau_k}  
\end{equation}
where $\tau_k$ is the autocorrelation time for mode k. Using this, the integrated autocorrelation time (ICT),  $\tau_{A, int}$, is taken to be 
\begin{align}
    \tau_{A, ICT} &\equiv \frac{1}{2} \int^{\infty}_{-\infty} dt \rho_{A}(t) \nonumber \\
    &= \sum_{k} c_{A, k}^2 \tau_k \,.
    \label{eq:IAC}
\end{align}

\begin{figure*}[t]    
    \centering
    \includegraphics[width=.34\textwidth]{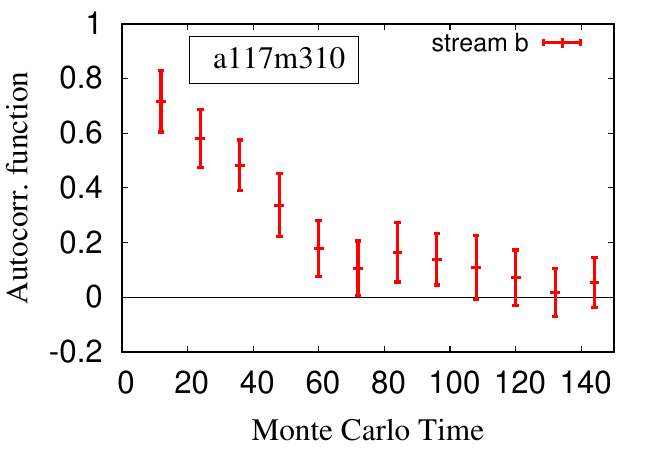} \hspace{0.5in} 
    \includegraphics[trim=0 0 0 0, clip, width=.34\textwidth]{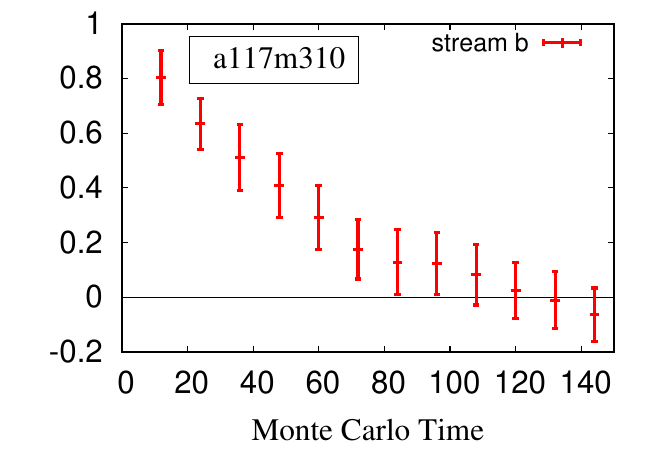}%
    \\
    
        \includegraphics[width=.34\textwidth]{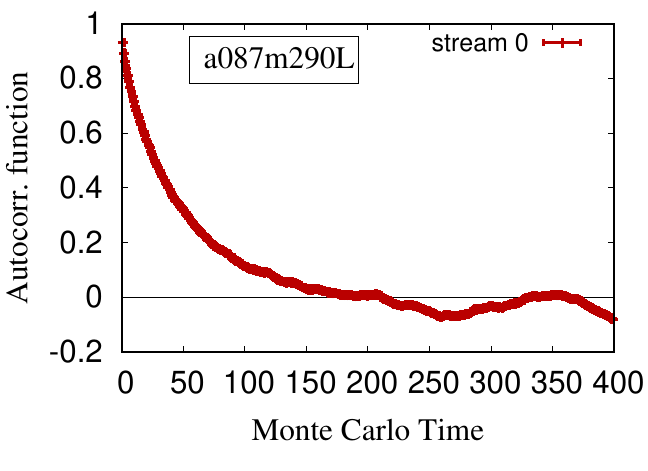} \hspace{0.5in} 
    \includegraphics[trim=0 0 0 0, clip,width=.34\textwidth]{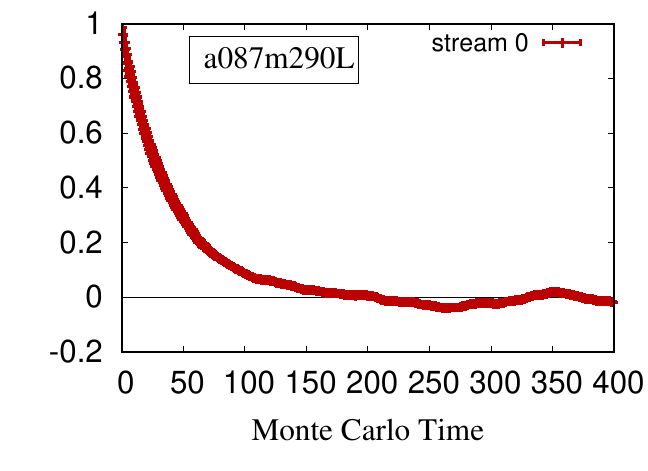}%
    \\
    \includegraphics[width=.34\textwidth]{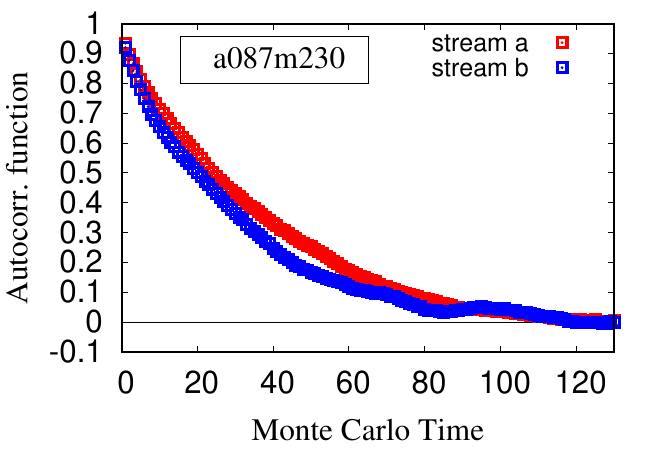} \hspace{0.5in} 
    \includegraphics[width=.34\textwidth]{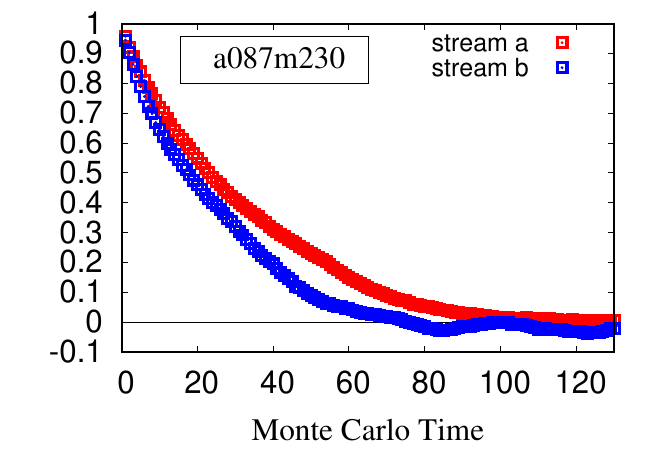}%
    \\
    \includegraphics[width=.34\textwidth]{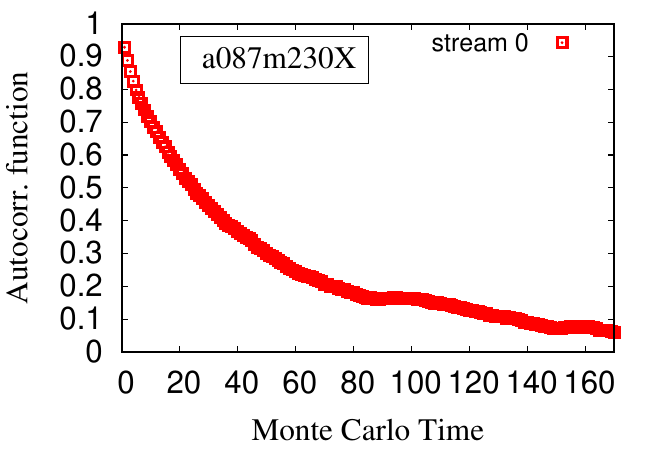} \hspace{0.5in} 
    \includegraphics[width=.34\textwidth]{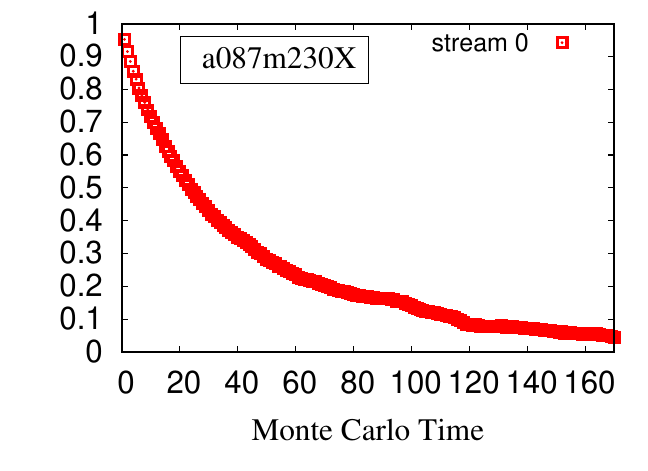}%
    \\
    
    \includegraphics[width=.34\textwidth]{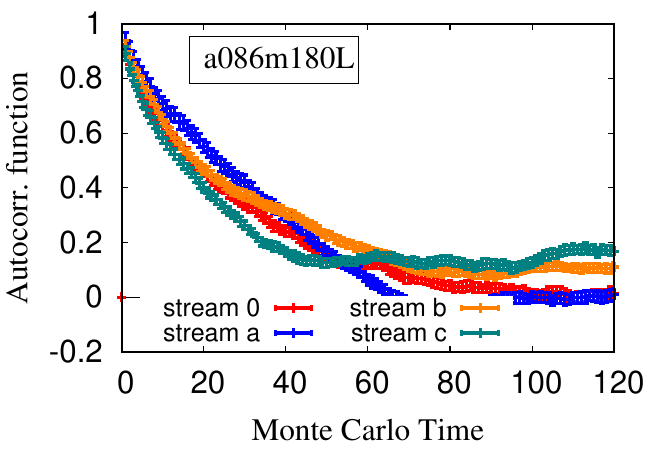} \hspace{0.5in} 
    \includegraphics[width=.34\textwidth]{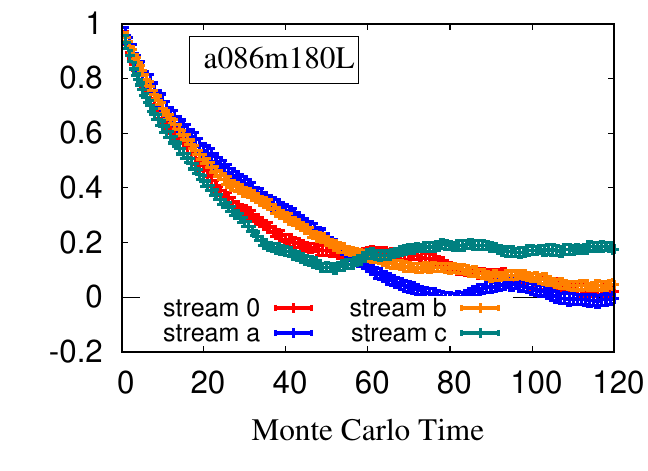}%
    \\
    \caption{The autocorrelation function for $w_0$ (left) and $t_0$ (right) plotted versus the Monte Carlo (trajectory) time for ensembles $a117m310$ (top), $a086m290$ (second), $a087m230$ (third), $a087m230X$ (fourth) and 
    $a086m180L$ (bottom). The integrated autocorrelation time is  calculated from it using Eq.~\protect\eqref{eq:IAC} and given in Table~\ref{tab:corr_time}. 
    }
    \label{fig:AC_C13}
\end{figure*}

\begin{figure*}[t]     
    \centering
    \includegraphics[width=.34\textwidth]{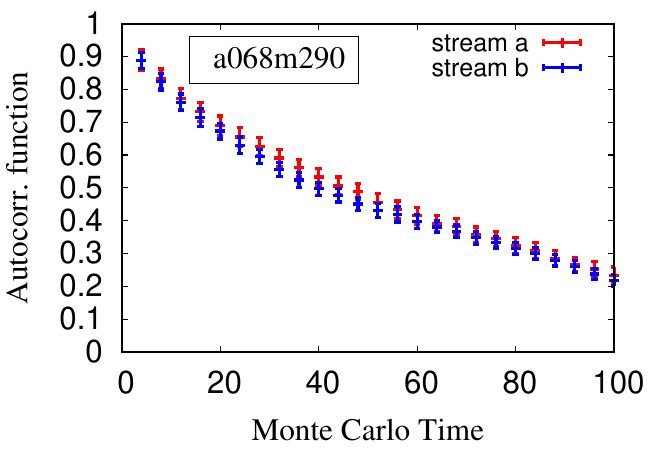} \hspace{0.5in} 
    \includegraphics[width=.34\textwidth]{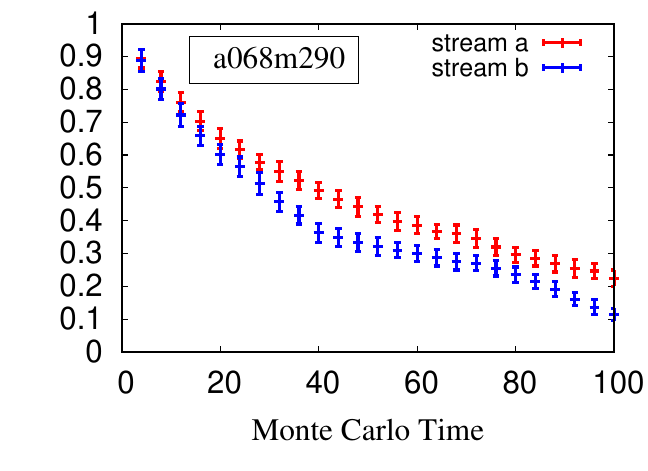}%
    \\
    \vspace{-0.1in}
    \includegraphics[width=.34\textwidth]{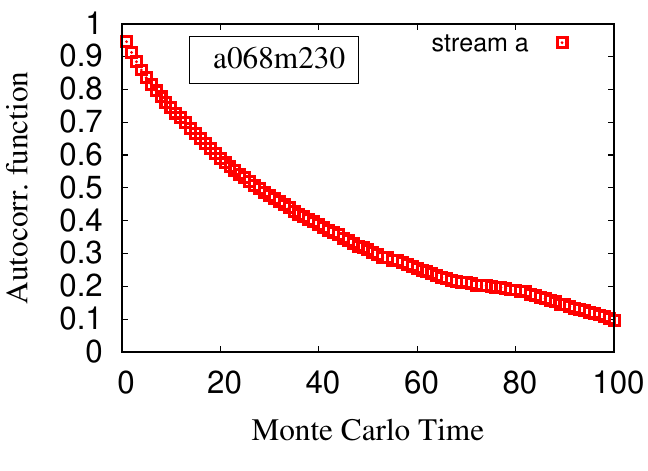} \hspace{0.5in} 
    \includegraphics[width=.34\textwidth]{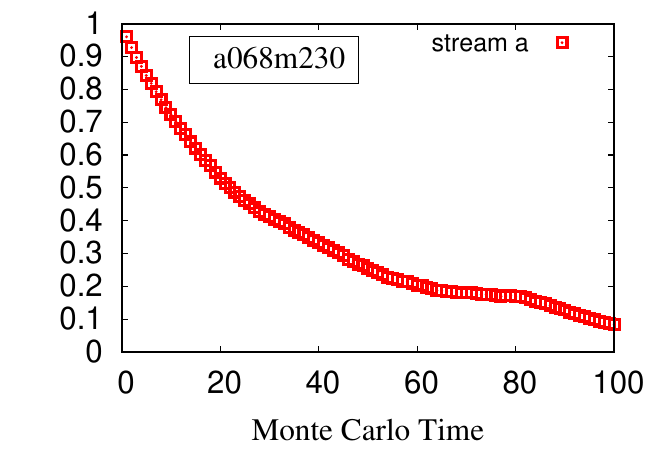}%
    \\
    \vspace{-0.1in}
    \includegraphics[width=.34\textwidth]{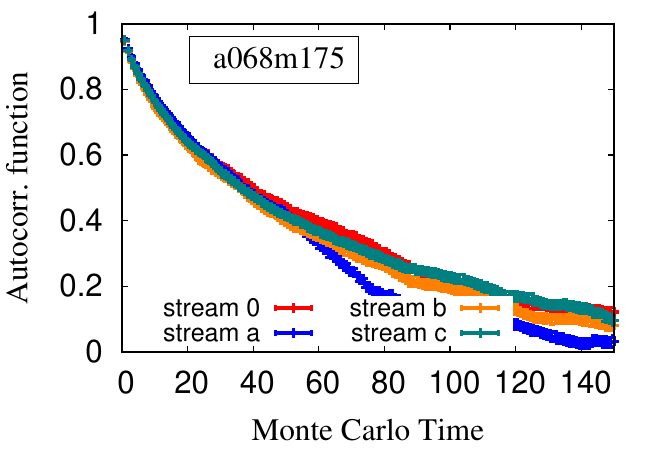} \hspace{0.5in} 
    \includegraphics[width=.34\textwidth]{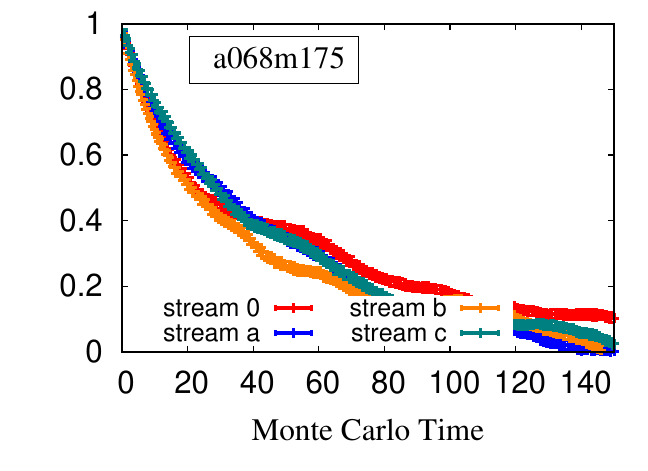}%
    \\
    \vspace{-0.1in}
    \includegraphics[width=.34\textwidth]{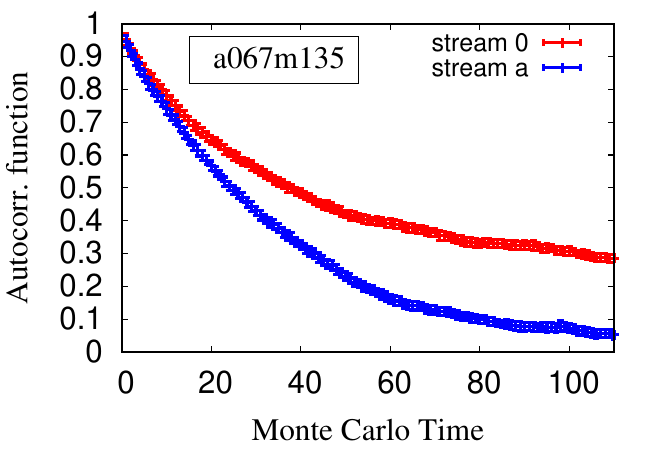} \hspace{0.5in} 
    \includegraphics[width=.34\textwidth]{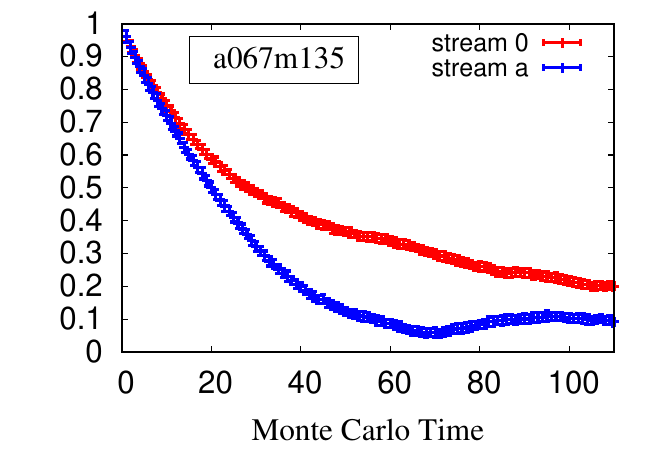}%
    \\
    \vspace{-0.1in}
    \includegraphics[width=.34\textwidth]{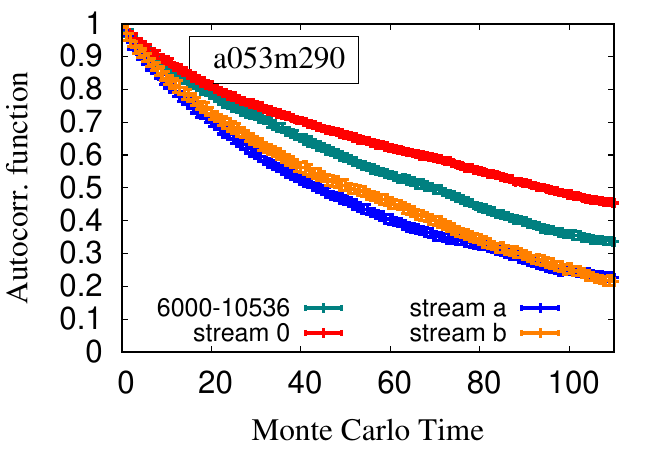} \hspace{0.5in} 
    \includegraphics[width=.34\textwidth]{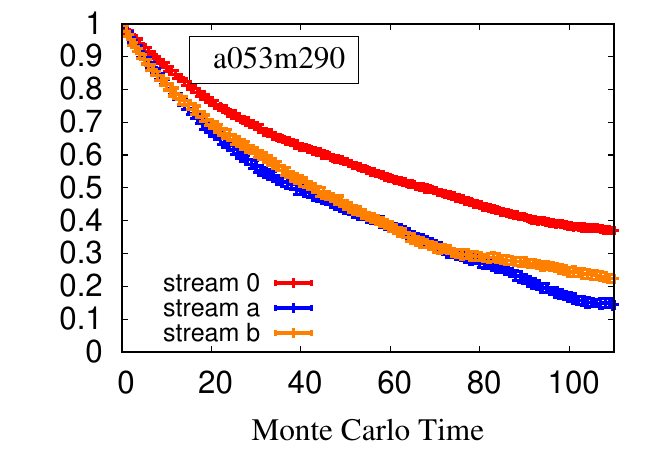}%
    \\
    \vspace{-0.1in}
    \includegraphics[width=.34\textwidth]{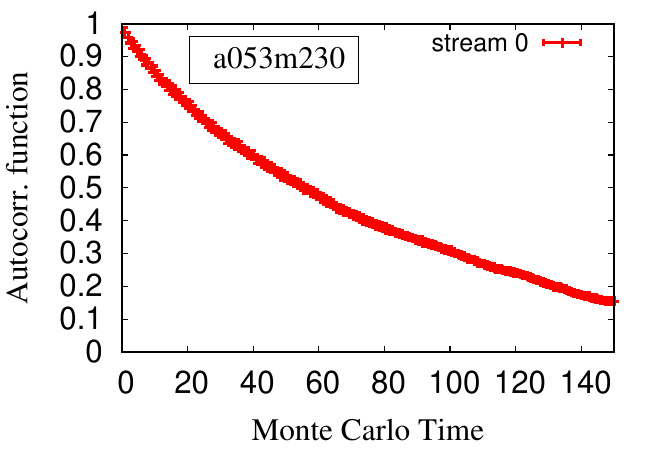} \hspace{0.5in} 
    \includegraphics[width=.34\textwidth]{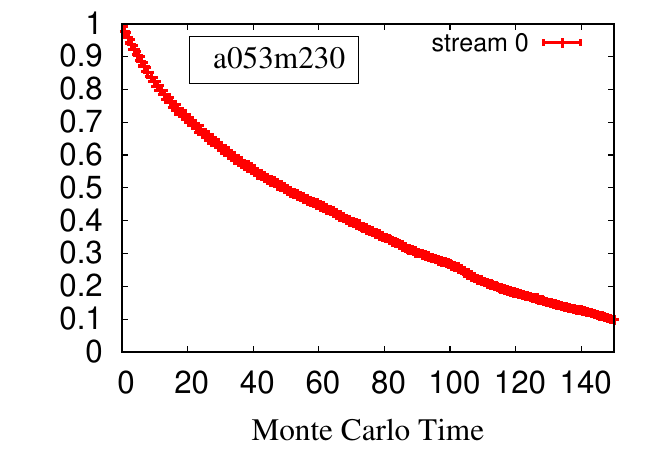}%
    \vspace{-0.1in}
     \caption{The autocorrelation function for $w_0$ (left) and $t_0$ (right) plotted versus the Monte Carlo (trajectory) time for ensembles $a068m290$ (top), $a068m230$ (second), $a068m175$ (third), $a067m135$ (fourth),  $a053m295$ (fifth) and 
    $a053m230$ (bottom). The integrated autocorrelation time is  calculated from it using Eq.~\protect\eqref{eq:IAC} and given in Table~\ref{tab:corr_time}. 
    }
    \label{fig:AC_F6}
\end{figure*}

We show the autocorrelation function using the precise data for the  flow variables $w_0$ and $t_0$ stored after each trajectory, except for five streams specified in Table~\ref{tab:corr_time}, for each stream of the 11 ensembles in Figs.~\ref{fig:AC_C13}--\ref{fig:AC_F6}. The calculated ICT for $w_0$ and $t_0$,  given in Table~\ref{tab:corr_time},  are large. Also note the still un-understood  variations in the autocorrelation function for the different streams generated on the same computer and over roughly the same interval of time. These are 
large in the $a067m135$ and $a053m295$ data. 

For the analysis of the data for mesons and baryons, we start with configuration averages, 
i.e., we first average over the $ N_\text{src}^\text{LP} \times N_\text{tsrc}$ measurements. For  $\Gamma_{\pi}(\tau=1.4)$~fm data, the ICT was found to be smaller 
than $10$ trajectories, and for the nucleon using 
$\Gamma_{N}(\tau=0.9)$~fm data it was less than $5$ trajectories. To account for these correlations, we further bin the configuration average data 
using the bin size $N_{\rm bin}$ specified in Table~\ref{tab:Confs}. On performing all analyses 
using these binned data, no augmentation of the resulting statistical errors is considered necessary.

\begin{table*}[]
    \centering
    \begin{tabular}{|l|c|c|c|c|c|}
    \hline
 Ens. ID  &  streams  &  trajectory interval     &  Files with $t_0$ and $w_0$ &  \multicolumn{2}{c|}{IAC time} \\
          &           &  used for         &  data stored every  &  $\tau_{w_0}$ &  $\tau_{t_0}$ \\
          &           &  calculating $\tau_{int}$  & \# trajectories     &  &   \\
 \hline
 $a117m310\dag$  &  0, a  &    &    &    & \\
 $a117m310^*$  &  b  &   2374 - 3310  &  12  &  43 & 47 \\
 $a087m290$  &  0   &  2010 -6020   &  4,6   &   45 & 49 \\
 $a087m290L$  &  0   &  1 - 14340  &  1   &   44 & 42 \\
 $a087m230$ &  a  &  1-12154  &  1  &  34 & 33 \\
 $a087m230$ &  b  &  659-9245  &  1  &  28 & 26 \\
 $a087m230X$  &  s  &  1-12670$\dag \dag$  &  1  &  46 & 45 \\
 $a086m180$ $\dag$  &  0, a, b    &     &     &   &  \\
 $a086m180$  &  c   &  4-2848   &  4   &  23 & 22 \\
 $a086m180$ $\dag$  &  d,e,f,g      &    &     &    & \\
 $a086m180L$  &  0  &  4047 - 6896   &  1  &  27 & 26\\
 $a086m180L$  &  a  &  3323 - 5838   &  1  &  31 & 33\\
 $a086m180L$  &  b  &  2947 - 6970   &  1  &  34 & 32\\
 $a086m180L$  &  c  &  3287 - 6776   &  1  &  22 & 23\\
 $a068m290$ $\dag$  &  0  &    &    &    & \\
 $a068m290$  &  a  &  200-8224  &  4  &    55 & 63\\
 $a068m290$  &  b  &  4-13796  &  4  &   68 & 45 \\
 $a068m230$  &  0  &  1 - 10896  &  1  &  40 & 40 \\
 $a068m230$ $\dag$  &  a  & 802-3068   &   &    & \\
 $a068m175$  &  0  &  1 - 348, 541-5640  &  1  &  62 & 56 \\
 $a068m175$  &  a  &  1 - 5516  &  1  &  50 & 45 \\
 $a068m175$  &  b  &  1 - 5544  &  1  &  54 & 40 \\
 $a068m175$  &  c  &  1 - 5508  &  1  &  58 & 46 \\
 $a067m135$  &  0  &  767 - 6400  &  1  &  85 & 70 \\
 $a067m135$  &  a  &  1 - 5210  &  1  &  34 & 26 \\
 $a053m295$  &  0  &  337-10536  &  1  &  181 & 108 \\
 $a053m295$  &  a  &  1-5624  &  1  &  63 & 63 \\
 $a053m295$  &  b  &  1-5376  &  1  &  74 & 62 \\
 $a053m230$  &  0  &  1-11306  &  1  &  79 & 76 \\
 $a053m230$ $\dag$  &  a  &  802-2816    &     &    & \\
 \hline
 \end{tabular}
    \caption{ The integrated autocorrelation (IAC) time $\tau$, in units of the molecular dynamics (MD) trajectories of roughly unit length, calculated using $w_0$ and $t_0$. A symbol $\dag$ in the ensemble ID implies log files with flow data were lost. The $\tau$ on C13(b) is evaluated using 75 configurations separated by 12 MD units.  The $\tau$ for stream F5(0) is much larger, mostly arising from correlations on the first 6000 trajectories. The spectrum and topology data on these trajectories do not, however, show any significant deviation-from-mean behavior. We also note that estimates from $w_0$ and $t_0$ are similar in almost all cases. The calculations of 2- and 3-point functions is  done on configurations separated by  four trajectories except for ensemble $a068m290$ where measurements were made every 6 trajectories. 
    }
    \label{tab:corr_time}
\end{table*}

\clearpage
\begin{widetext}
\section{Baryon Effective Mass Plots and Extracting Masses}
\label{sec:N3and4-state-fits}

In Figs.~\ref{fig:C13_nucleon}--~\ref{fig:F6_nucleon}, we present the data for $M_{eff}$ 
for the nucleon along with 3- and 4-state fits using Eq.~\eqref{eq:SD2pt}. We expect 
good consistency between estimates for the ground state results, $M_0$ and $A_0$, and 
rough agreement for the first excited-state values, $M_1$ and $A_1$. A 
reasonable length plateau is observed in $M_{eff}$ for all the ensembles and the 
estimates of the ground-state mass, $M_0$, and the amplitude, $A_0$, agree 
between the 3-state and 4-state fits. The $\chi^2/{\rm dof}$ increases slightly 
for the 4-state fits while the estimate for $M_0$ is $(1-2) \sigma$ larger in the two-state fits. 
Using both the Akaike criteria (requiring $\chi^2$ to decrease by two units for each extra parameter) 
and the observed convergence in $M_0$ from above, 
we take results of the 3-state fits for our best estimates. 

The data also show that the results for the first excited state, $\Delta M_1 \equiv M_1 - M_0$ and $R_1 \equiv A_1 /A_0$, are also consistent except for the $a087m290L$ and $a067m175$ ensembles. We are 
investigating these cases, however, note that in this paper we only use the ground-state parameters. 
The estimates for the second excited state differ.  This is not unexpected---for any $n$-state fit, 
the parameters of the $n^{th}$ state try to accommodate all the higher states. In short, only the 
ground and first excited state parameters in 3-state fits are expected to be useful.


A comparison of the effective mass plots for the five baryon states, $\{\Omega_3,\ \Xi_1, \ \Sigma_6,\ \Lambda_3, \ N_6 \}$, is shown in Figs.~\ref{fig:meff_baryons_1} and~\ref{fig:meff_baryons_2}. The labels specify the ensemble and the subscript on the state, e.g., in $\Omega_3$, specifies the interpolation operator used as defined in Table~\ref{tab:OP_list}. 
The horizontal dashed lines give the ground-state mass, $M_0$, obtained from 2-state fits.
Comparing the data from the SS (left) and SP (right) correlators, we note that the errors in $\Meff(\tau)$ at a given $\tau$ are smaller in the SP data whereas the plateau sets in slightly 
earlier in $\tau$ in the SS data. Estimates of $M_0$ overlap.


\begin{figure*}[t]     
    \centering

    \includegraphics[width=.32\textwidth]{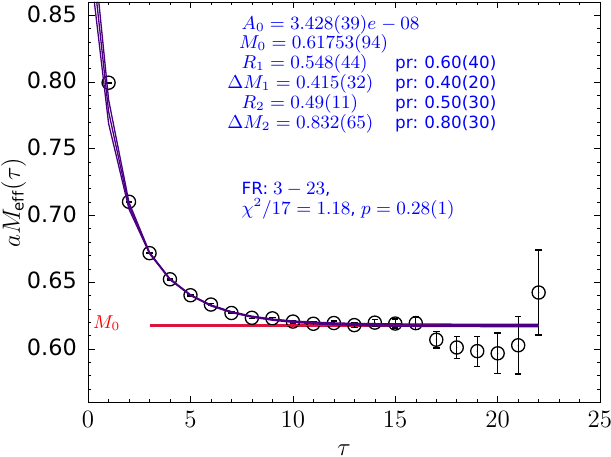} \hspace{0.5in} %
    \includegraphics[width=.32\textwidth]{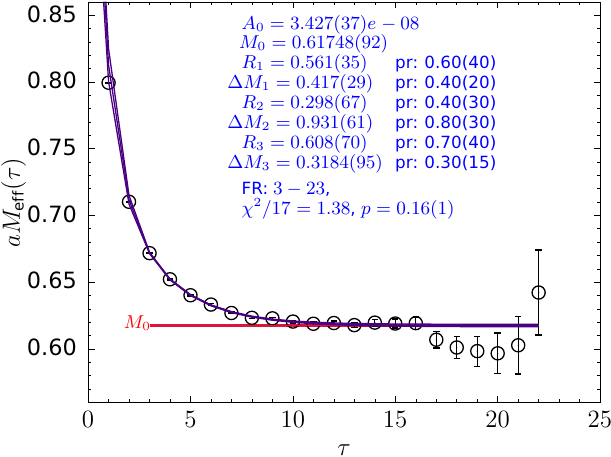}
\\
\vspace{-0.1in}
    \includegraphics[width=.32\textwidth]{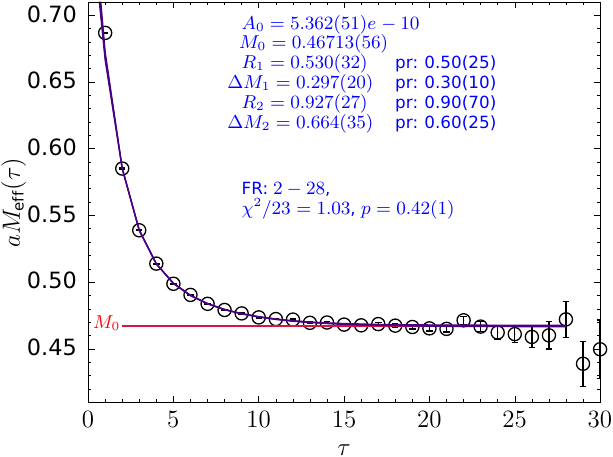}  \hspace{0.5in}%
    \includegraphics[width=.32\textwidth]{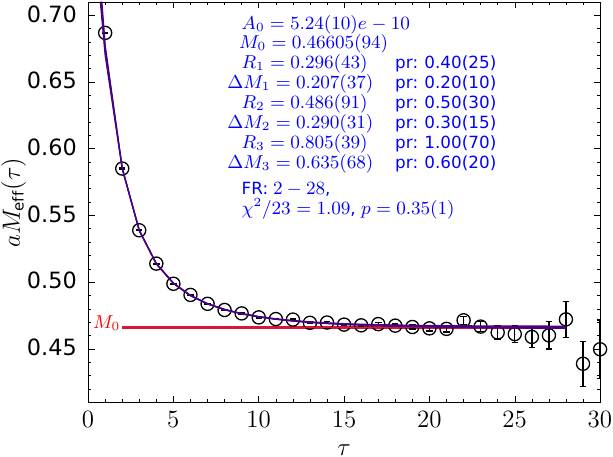}%
\\
\vspace{-0.1in}
    \includegraphics[width=.32\textwidth]{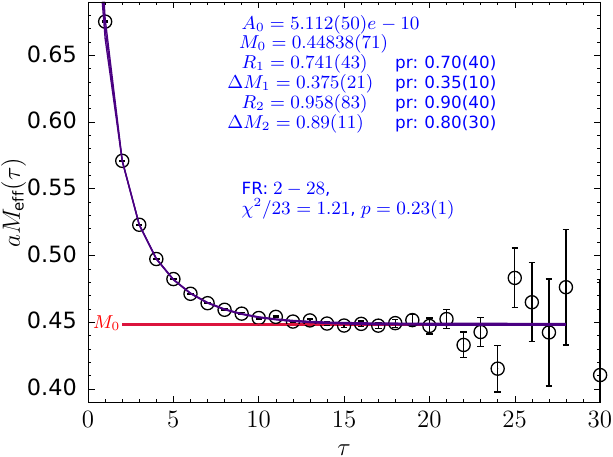}  \hspace{0.5in} %
    \includegraphics[width=.32\textwidth]{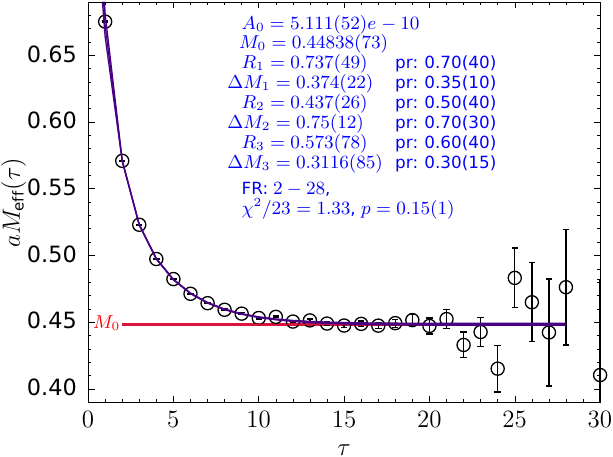}%
\\
\vspace{-0.1in}
    \includegraphics[width=.32\textwidth]{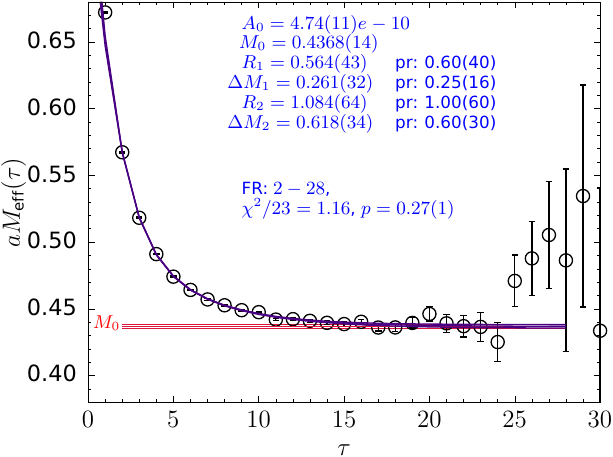}  \hspace{0.5in} %
    \includegraphics[width=.32\textwidth]{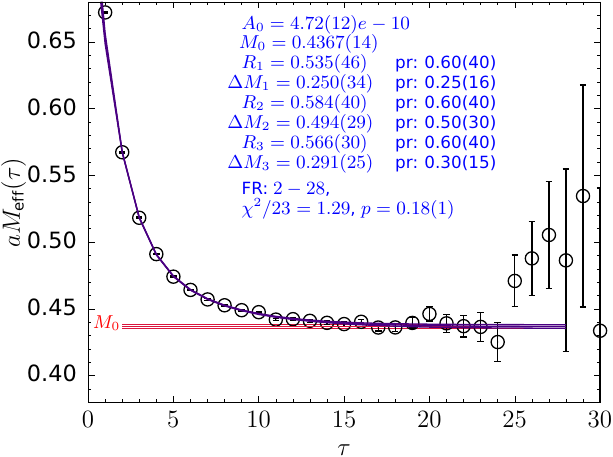}
\\
\vspace{-0.1in}
    \includegraphics[width=.32\textwidth]{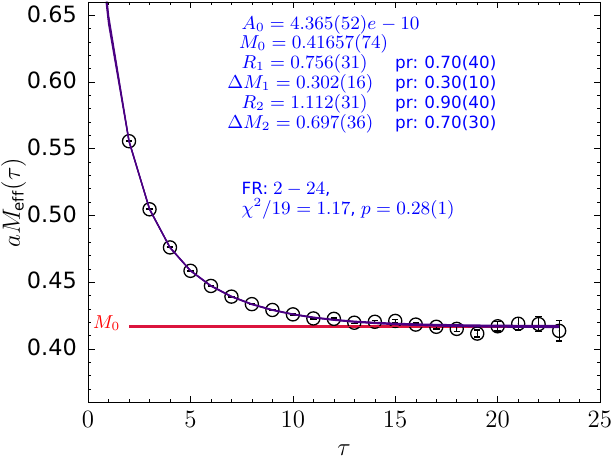} \hspace{0.5in} %
    \includegraphics[width=.32\textwidth]{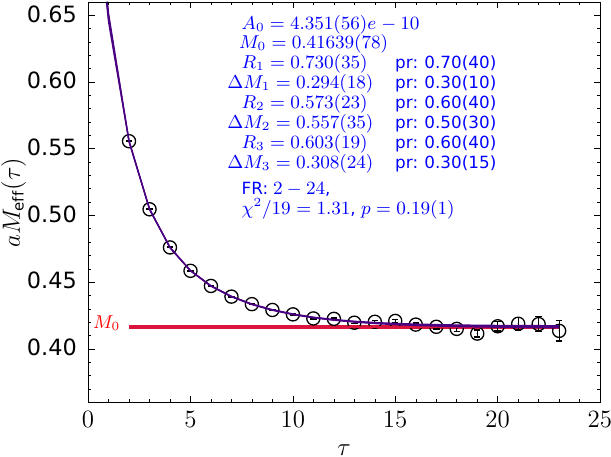}%
    \caption{Data for $a M_{eff}(\tau)$ for the nucleon using the $N_3$ operator and SS correlators, along with 3-state (left) and 4-state  (right) fits on  
    a117m310 (top), a086m290 (second), a087m230 (third),
a087m230X (fourth) and a086m180L (bottom)  ensembles.  The fit parameters and the input priors (pr) are given in the labels with $\Delta M_i \equiv M_i - M_{i-1}$, $R_i \equiv A_i / A_0$. FR is short for the fit range and $M_0$ is the ground-state mass.}
    \label{fig:C13_nucleon}
\end{figure*}

\begin{figure*}[h]    
    \centering
    \includegraphics[width=.32\textwidth]{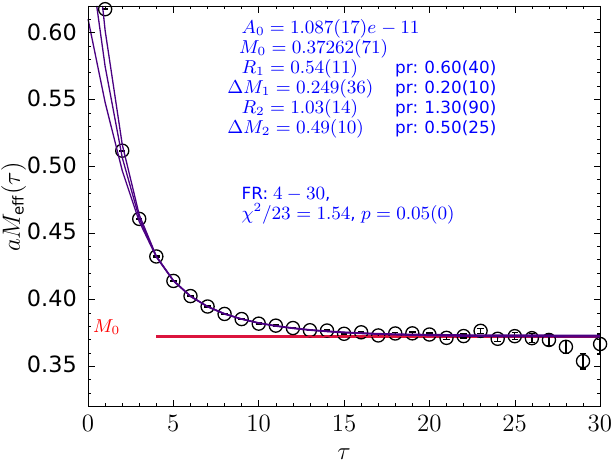}  \hspace{0.5in}%
    \includegraphics[width=.32\textwidth]{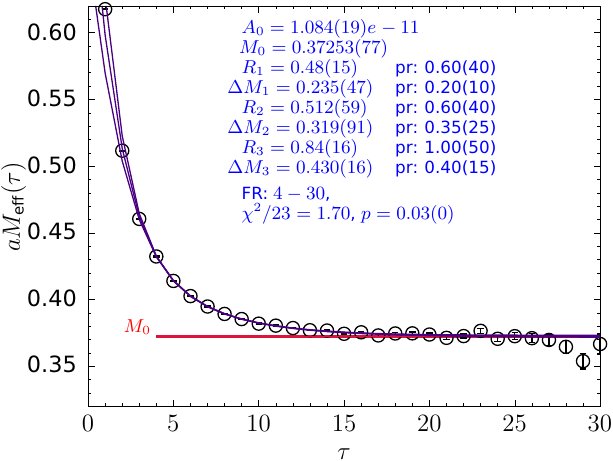}%
\\
\vspace{-0.1in}
    \includegraphics[width=.32\textwidth]{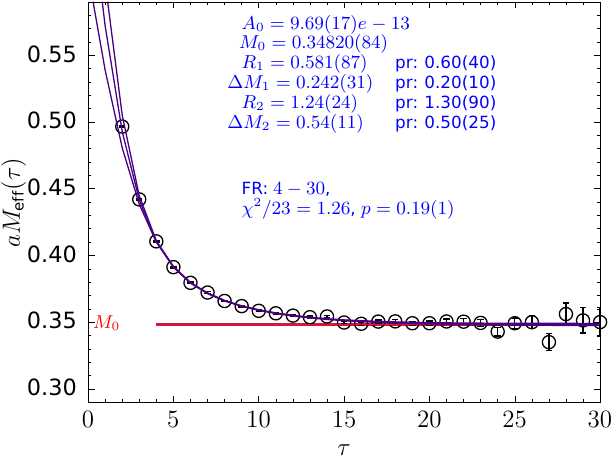}  \hspace{0.5in} %
    \includegraphics[width=.32\textwidth]{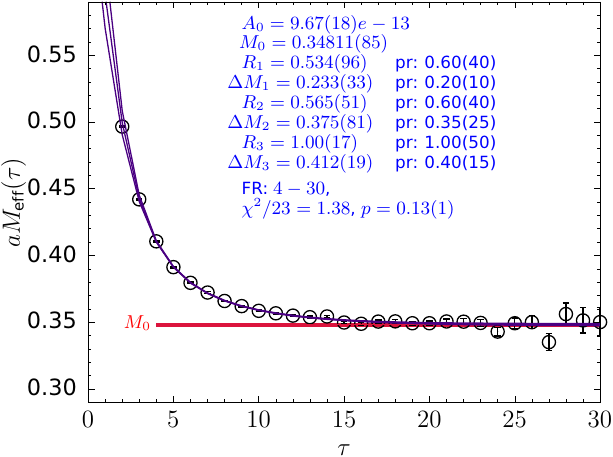}%
\\
\vspace{-0.1in}
    \includegraphics[width=.32\textwidth]{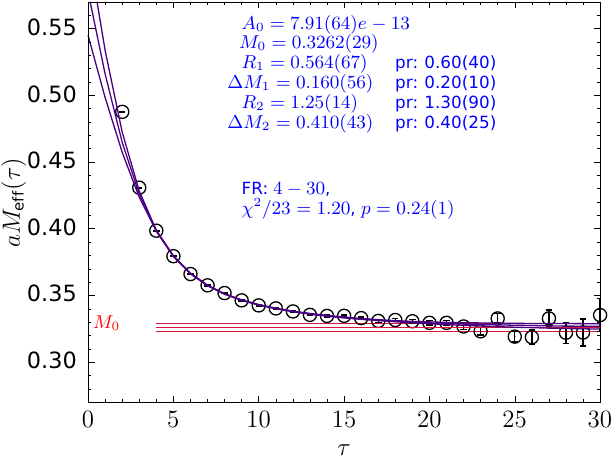}  \hspace{0.5in} %
    \includegraphics[width=.32\textwidth]{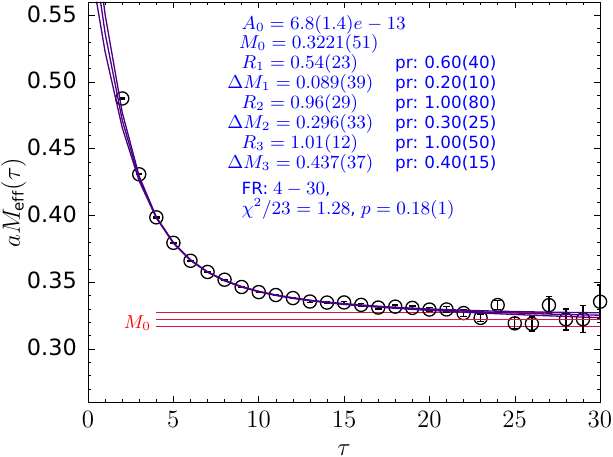}%
\\
\vspace{-0.1in}
    \includegraphics[width=.32\textwidth]{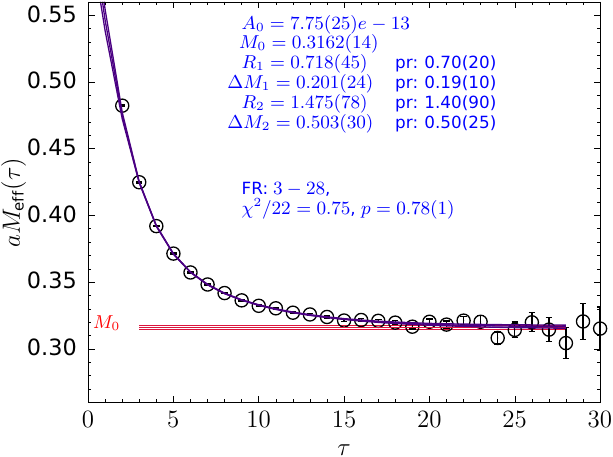}  \hspace{0.5in}%
    \includegraphics[width=.32\textwidth]{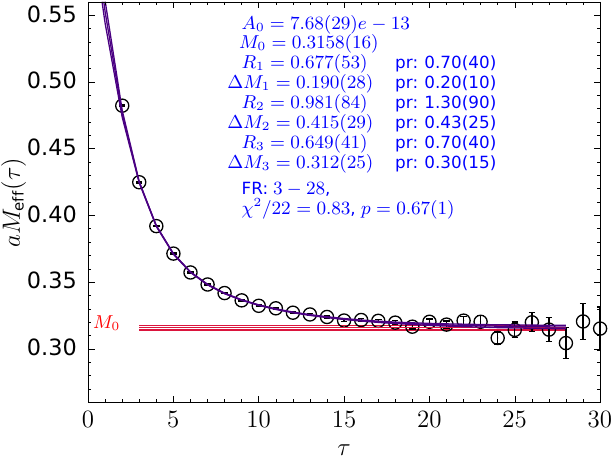}%
\\
\vspace{-0.1in}
    \includegraphics[width=.32\textwidth]{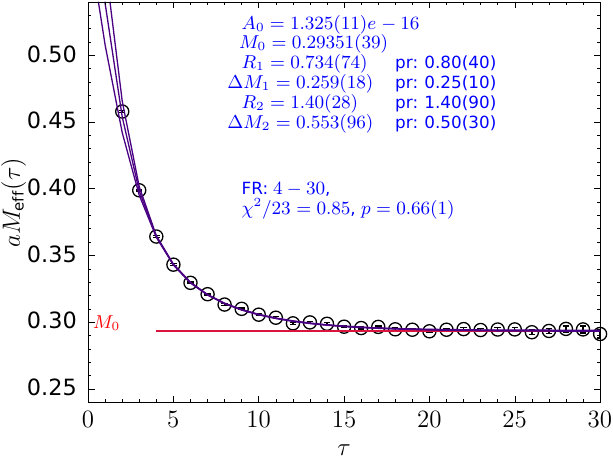}  \hspace{0.5in} %
    \includegraphics[width=.32\textwidth]{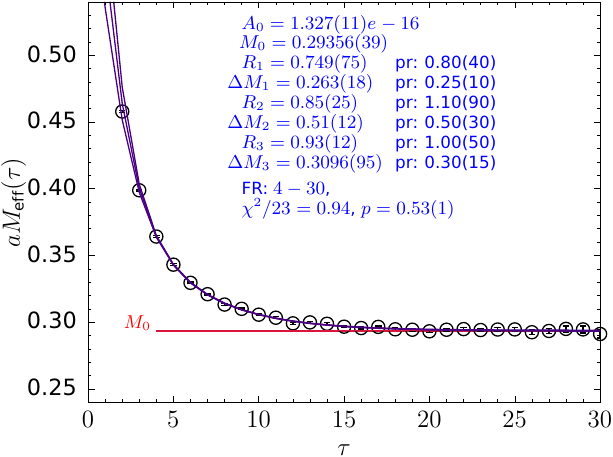}%
\\
\vspace{-0.1in}
    \includegraphics[width=.32\textwidth]{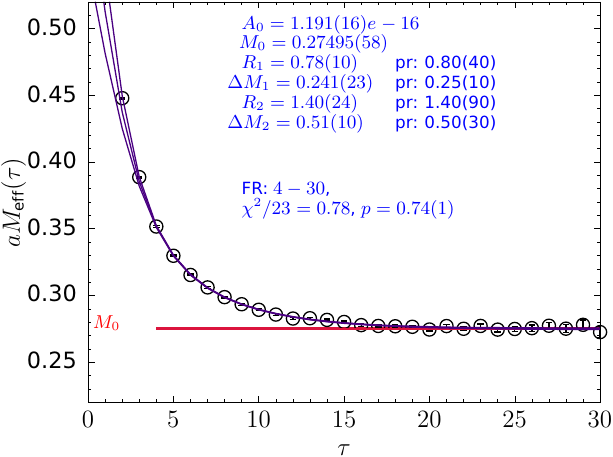} \hspace{0.5in} %
    \includegraphics[width=.32\textwidth]{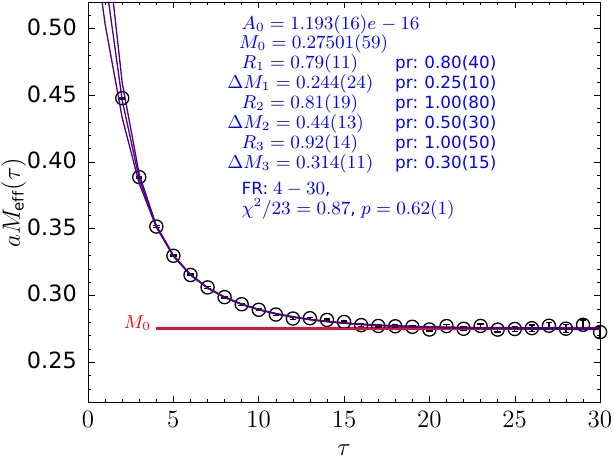}%
    \vspace{-0.1in}
    \caption{Data for $a M_{eff}^{\rm nucleon}(\tau)$ with 3-state (left) and 4-state  (right) fits on  
    $a068m290$ (top), $a068m230$ (second), $a068m175$ (third), $a067m135$ (fourth),  $a053m295$ (fifth) and 
    $a053m230$ (bottom)   ensembles.  The rest is same as in Fig.~\protect\ref{fig:C13_nucleon}.}
    \label{fig:F6_nucleon}
\end{figure*}


\begin{figure*}[h!]     
    \centering
    \includegraphics[width=0.32\linewidth]{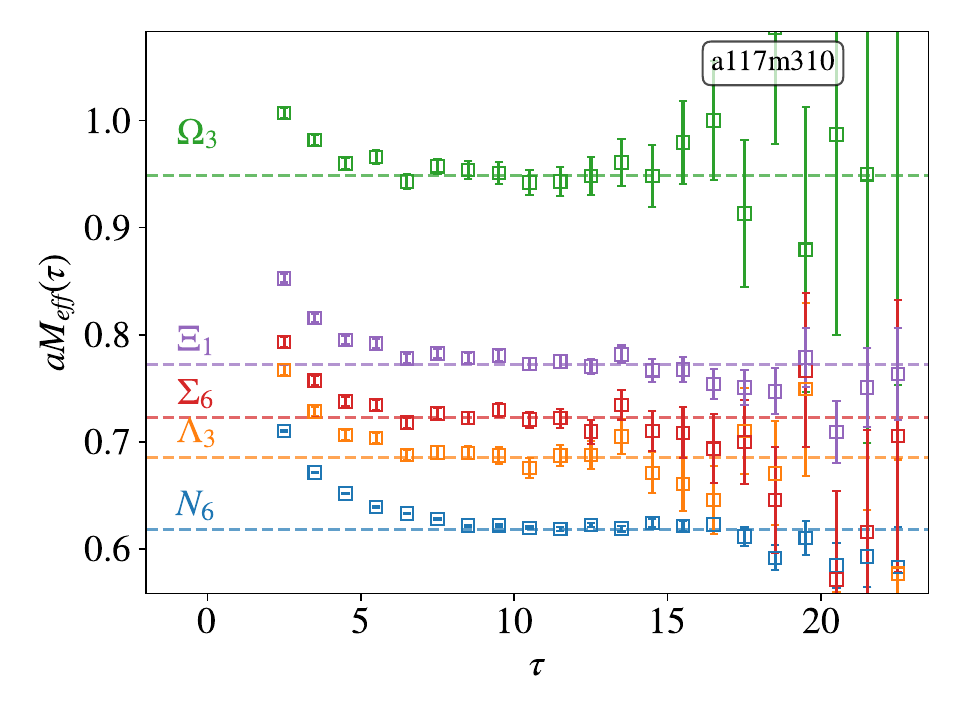} \hspace{0.5in}%
    \includegraphics[width=0.32\linewidth]{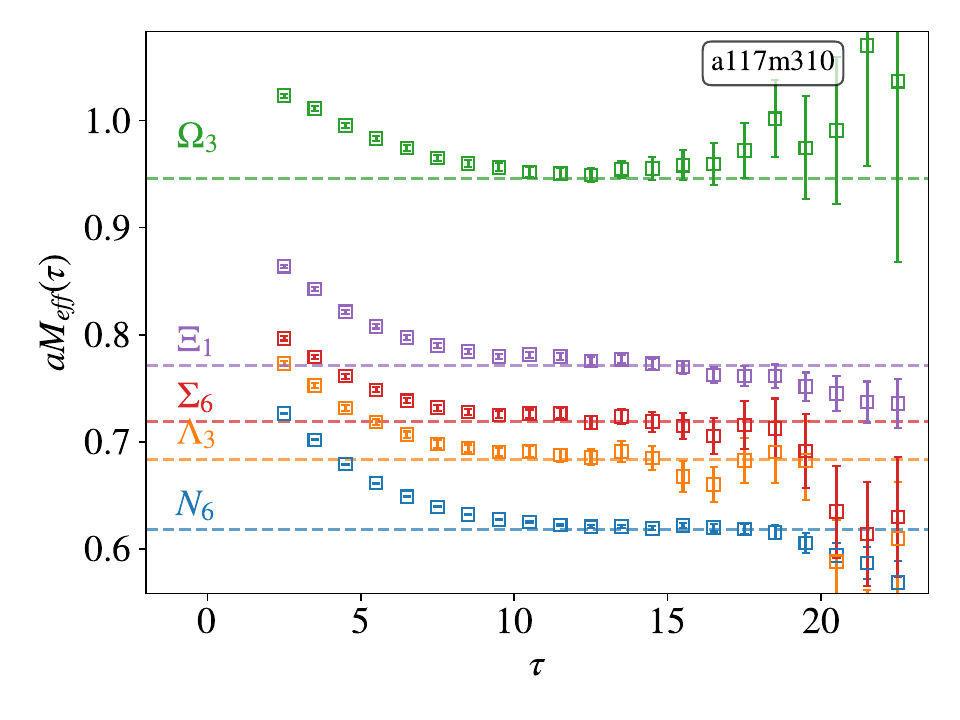}%
    \\
 \vspace{-0.1in}   
    \includegraphics[width=0.32\linewidth]{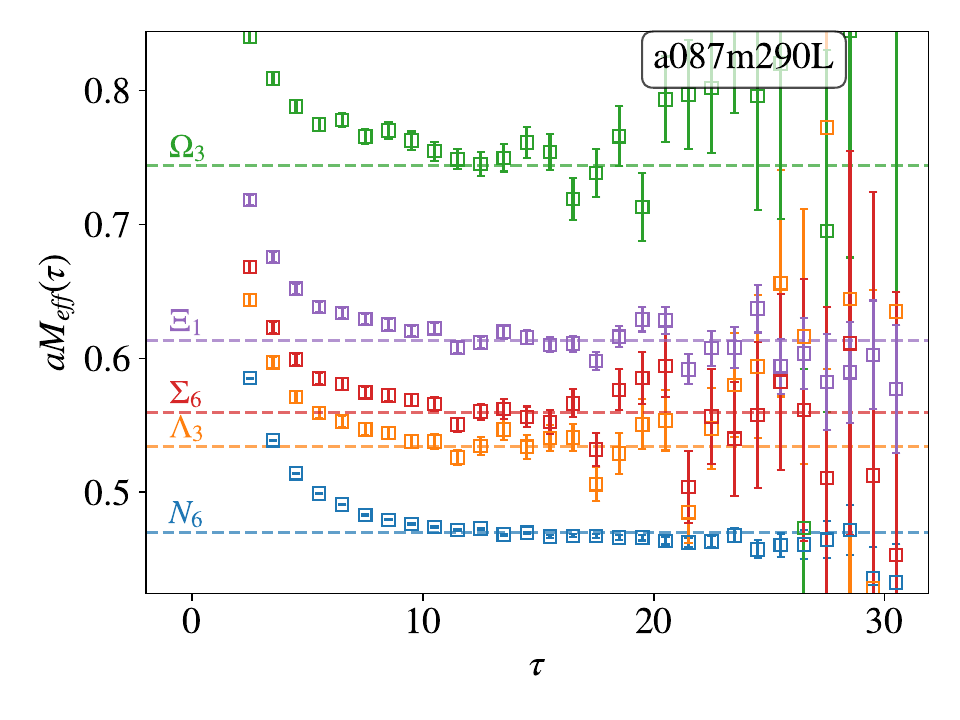} \hspace{0.5in}%
    \includegraphics[width=0.32\linewidth]{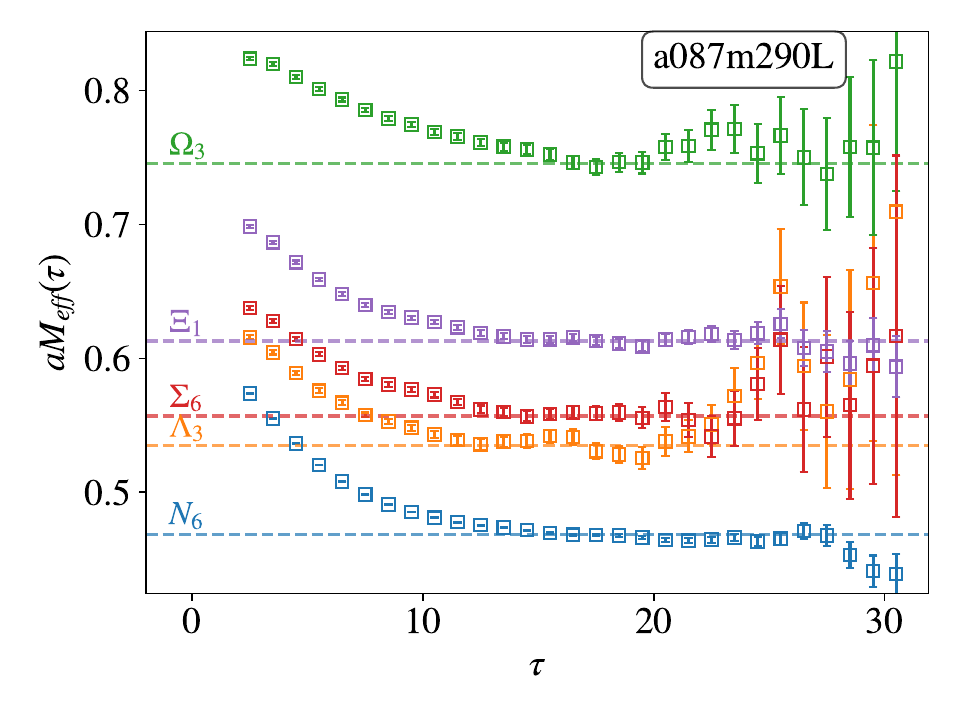}%
    \\
 \vspace{-0.1in}   
    \includegraphics[width=0.32\linewidth]{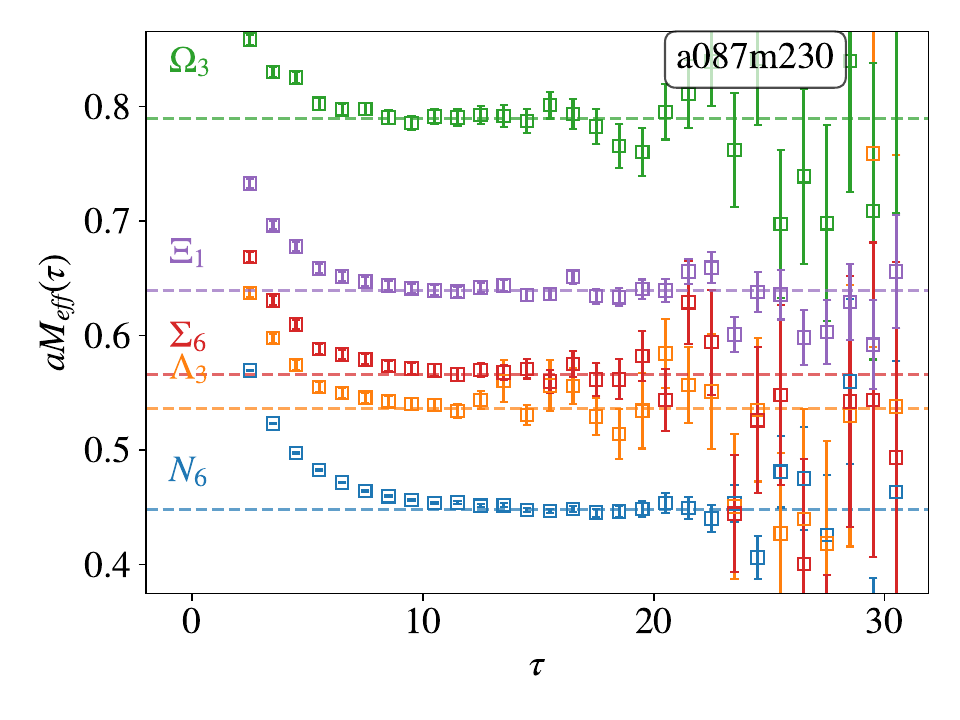} \hspace{0.5in}%
    \includegraphics[width=0.32\linewidth]{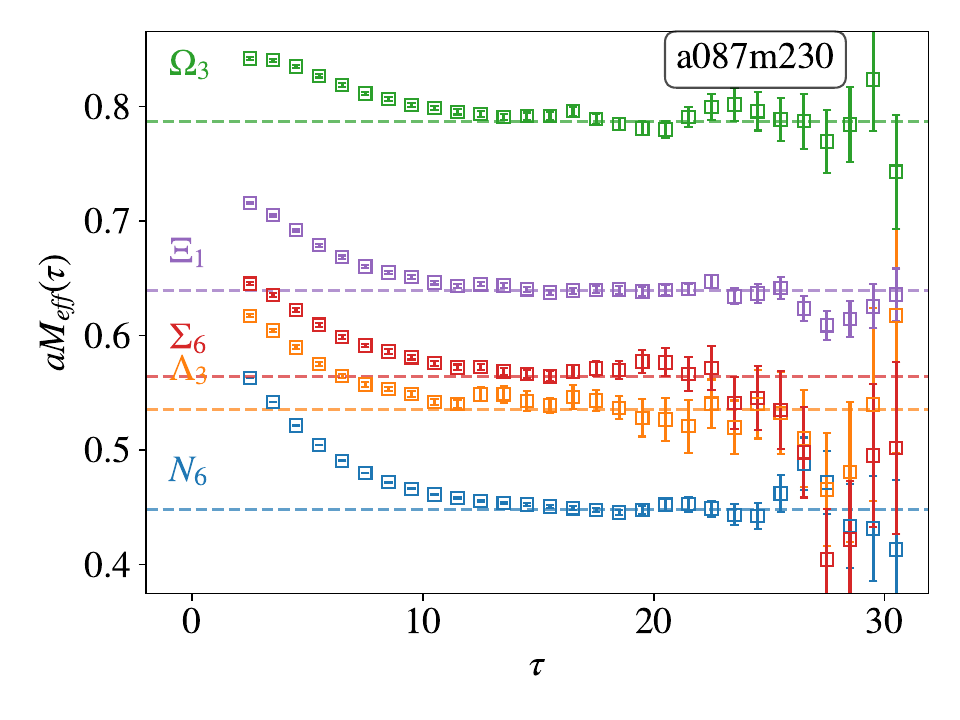}%
    \\
     \vspace{-0.1in}
    
    \includegraphics[width=0.32\linewidth]{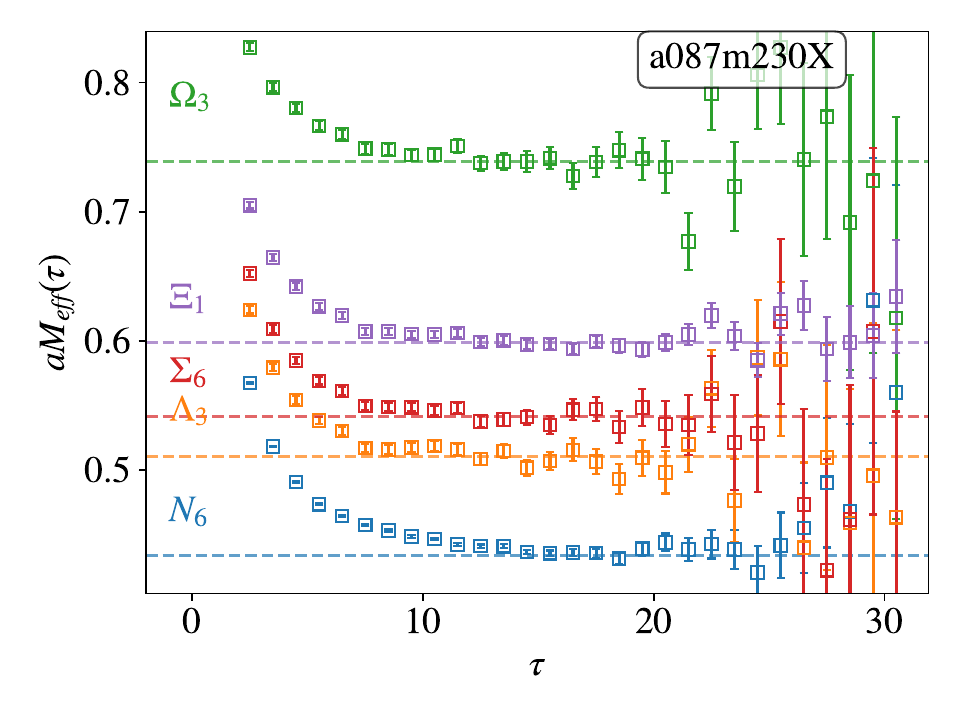} \hspace{0.5in}%
    \includegraphics[width=0.32\linewidth]{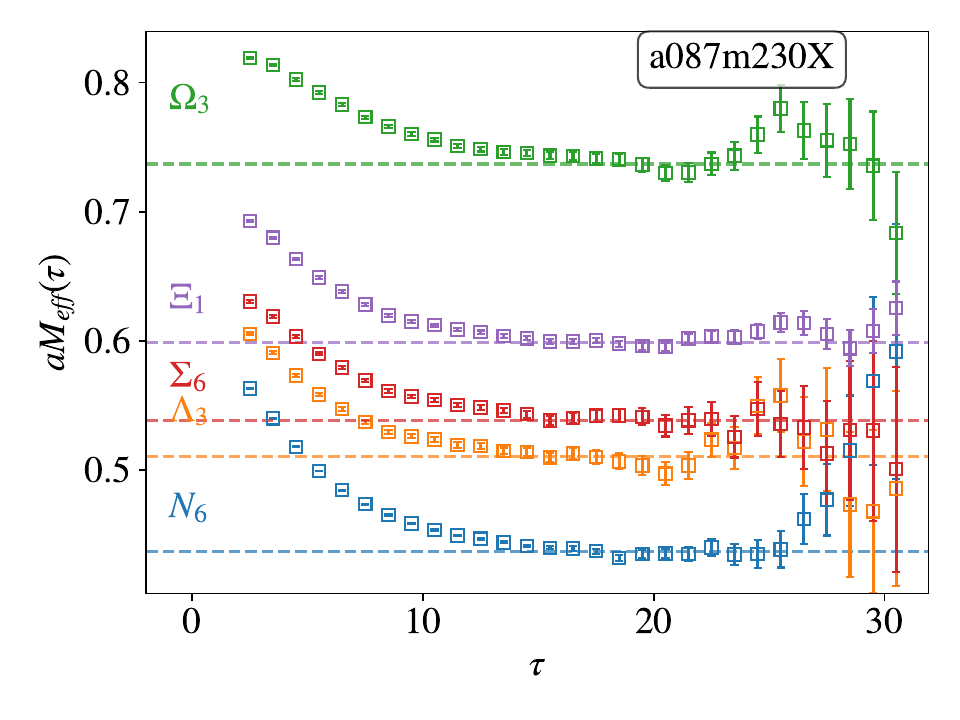}%
    \\
 \vspace{-0.1in}    
    \includegraphics[width=0.32\linewidth]{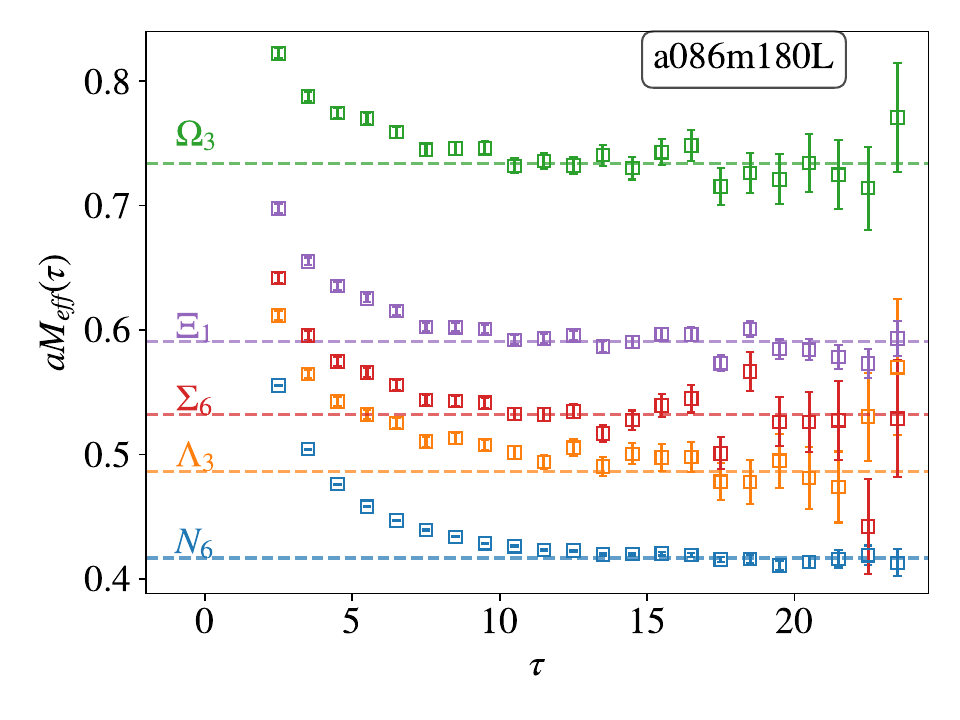} \hspace{0.5in}%
    \includegraphics[width=0.32\linewidth]{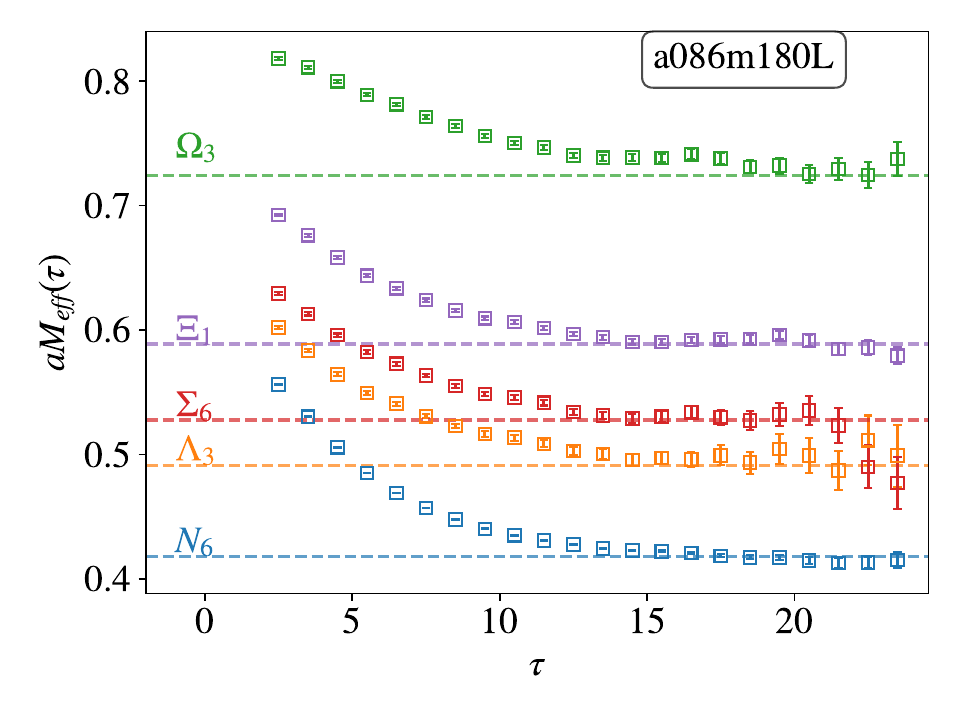}%
\vspace{-0.1in}
    \caption{Effective mass plots for the five baryon states, $\{\Omega_3,\ \Xi_1, \ \Sigma_6,\ \Lambda_3, \ N_6 \}$, using the SS (left) and SP (right) correlators.  The labels specify the ensemble and the baryon state, e.g., $\Omega_3$, specifies the interpolation operator used as defined in Table~\ref{tab:OP_list}. The horizontal dashed lines give the ground-state mass obtained 
    from 2-state fits.}
    \label{fig:meff_baryons_1}
\end{figure*}

\begin{figure*}[h!]
    \centering
    \includegraphics[width=0.32\linewidth]{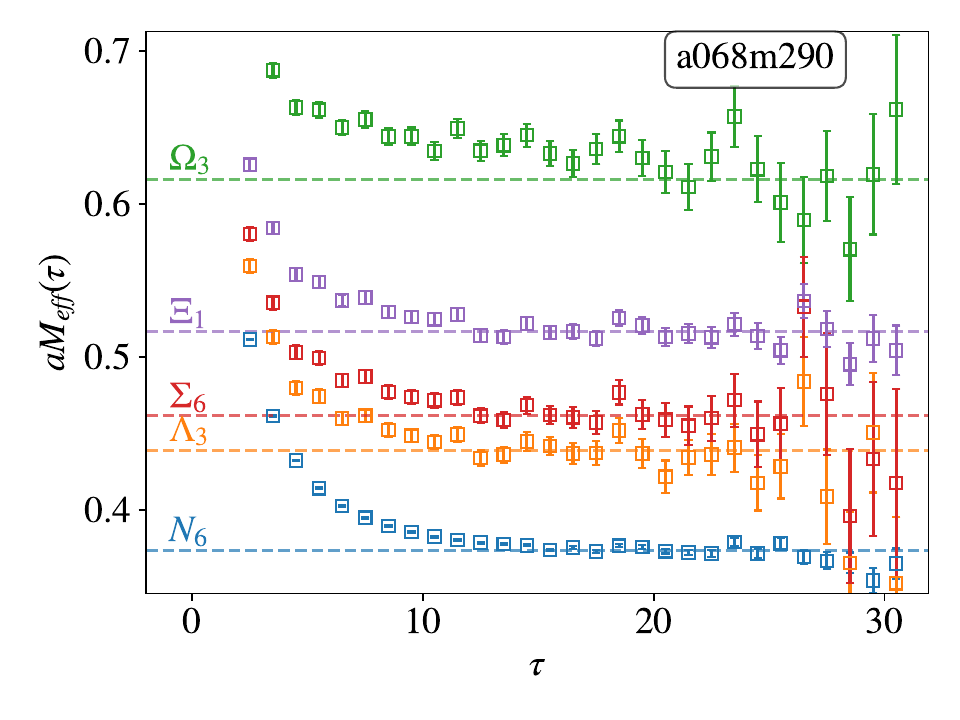} \hspace{0.5in} %
    \includegraphics[width=0.32\linewidth]{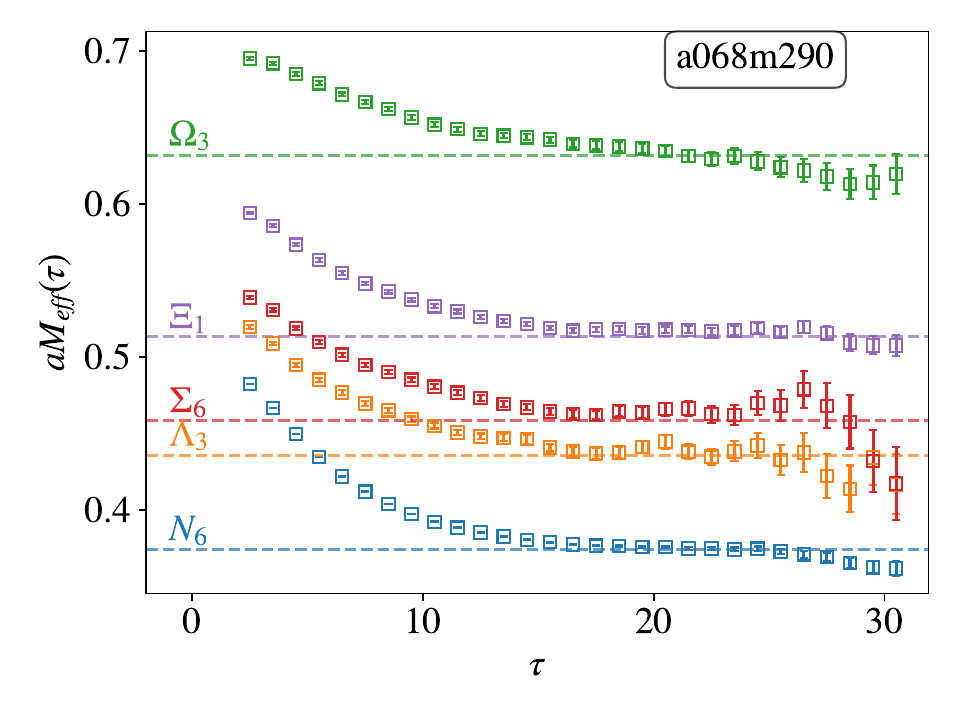} \\
\vspace{-0.15in}
    \includegraphics[width=0.32\linewidth]{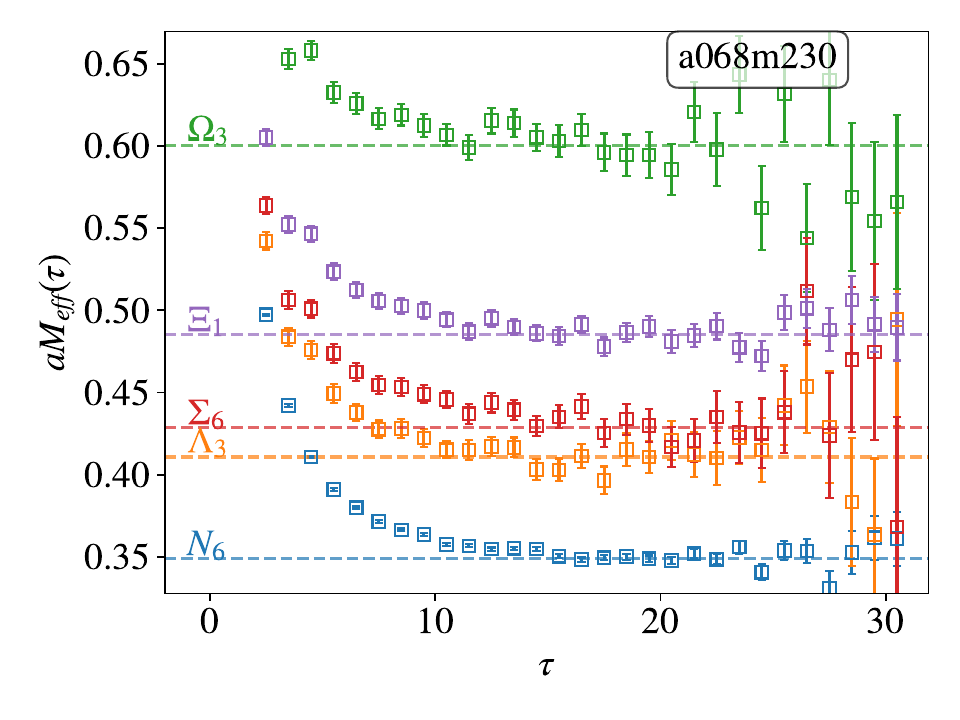} \hspace{0.5in} %
    \includegraphics[width=0.32\linewidth]{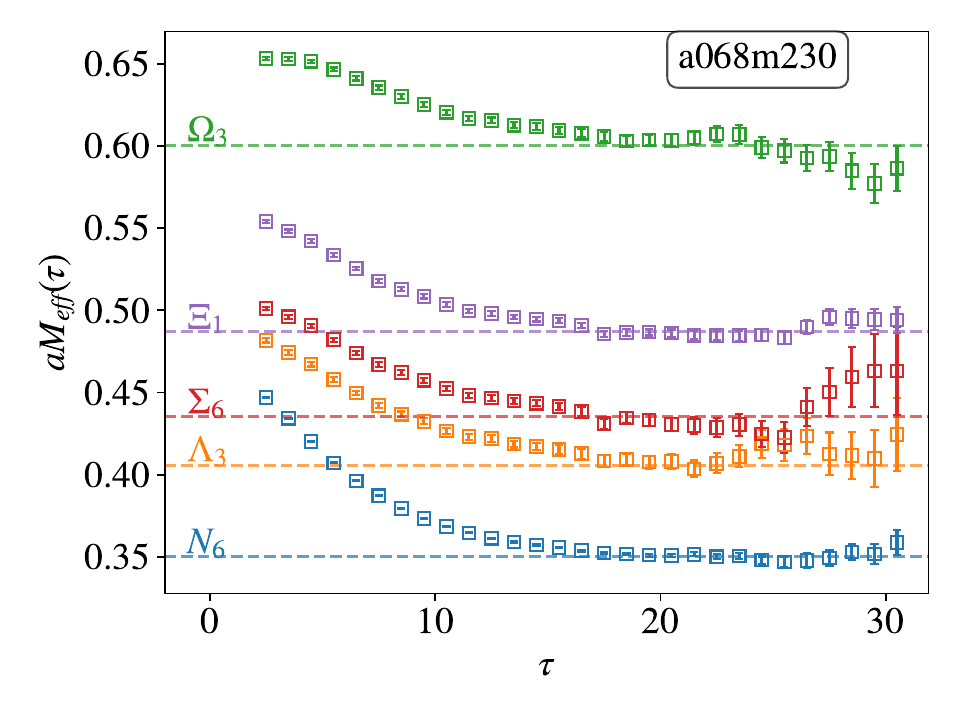}%
    \\
  \vspace{-0.15in}  
    \includegraphics[width=0.32\linewidth]{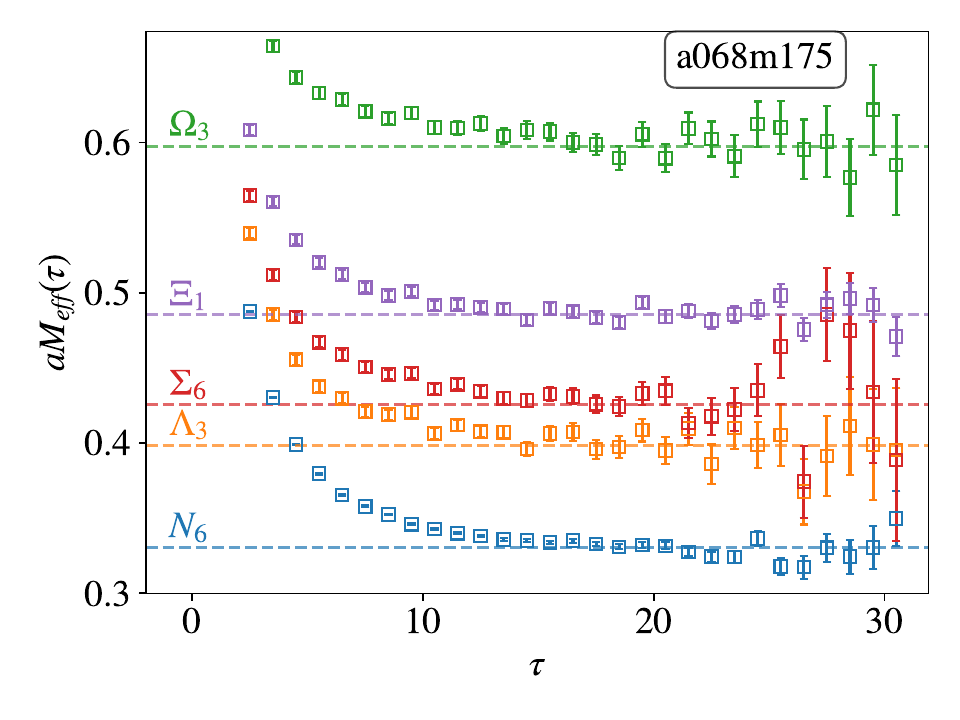}\hspace{0.5in} %
    \includegraphics[width=0.32\linewidth]{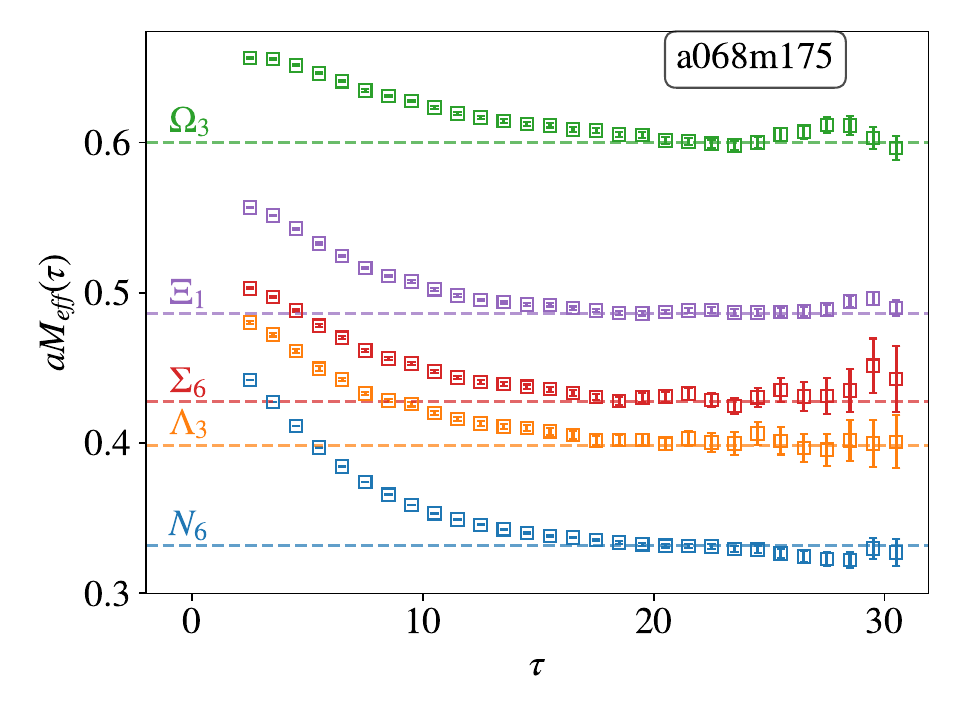}%
    \\
    \vspace{-0.15in}
    \includegraphics[width=0.32\linewidth]{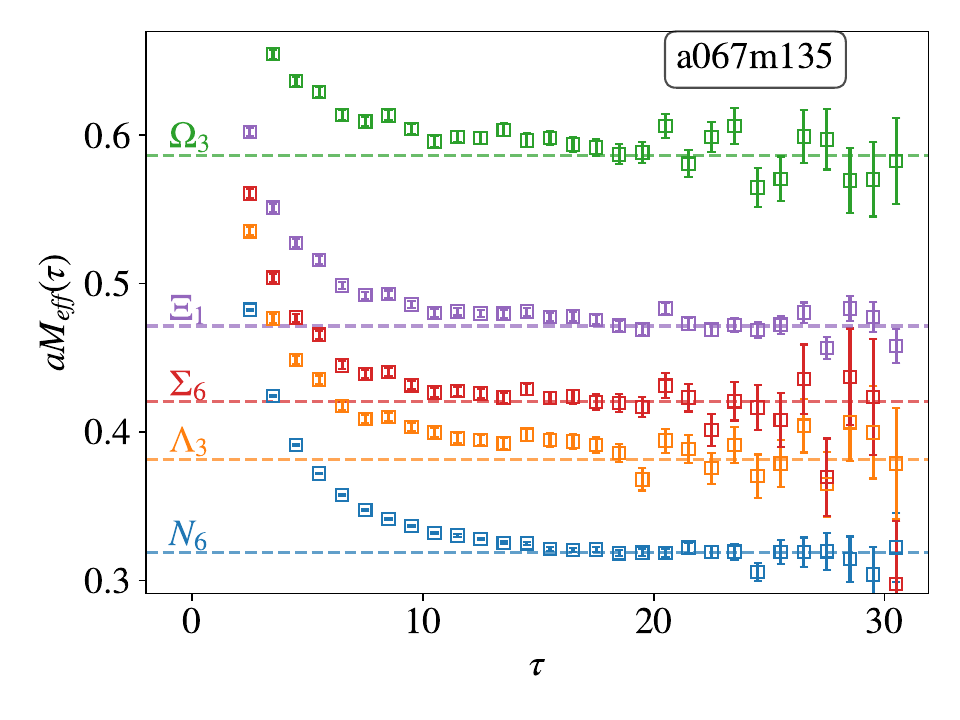} \hspace{0.5in} %
    \includegraphics[width=0.32\linewidth]{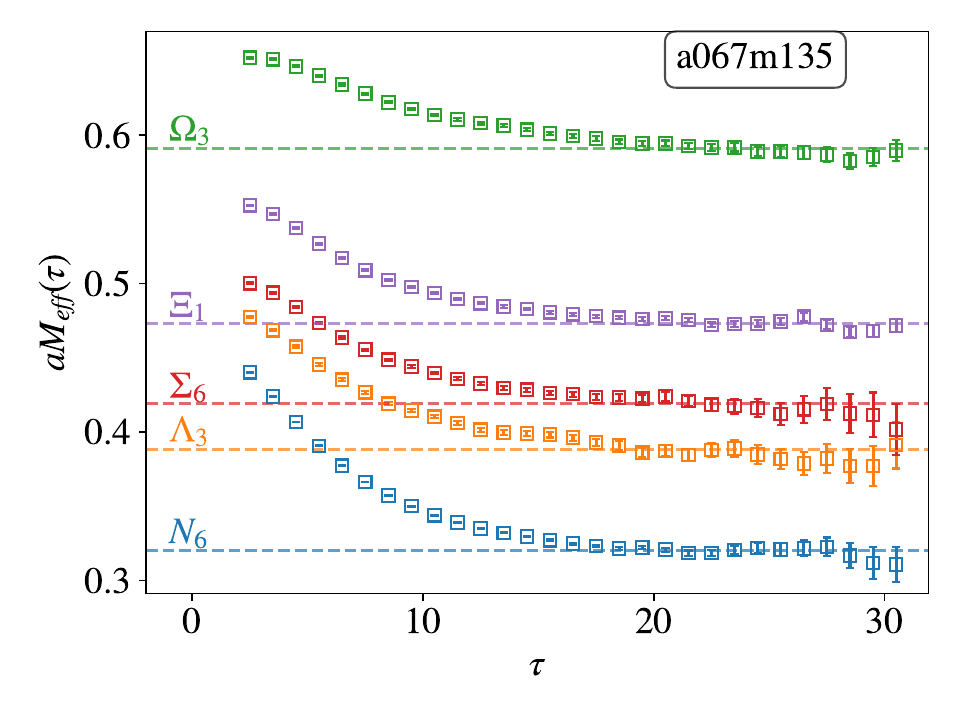} %
    \\
\vspace{-0.15in}
    \includegraphics[width=0.32\linewidth]{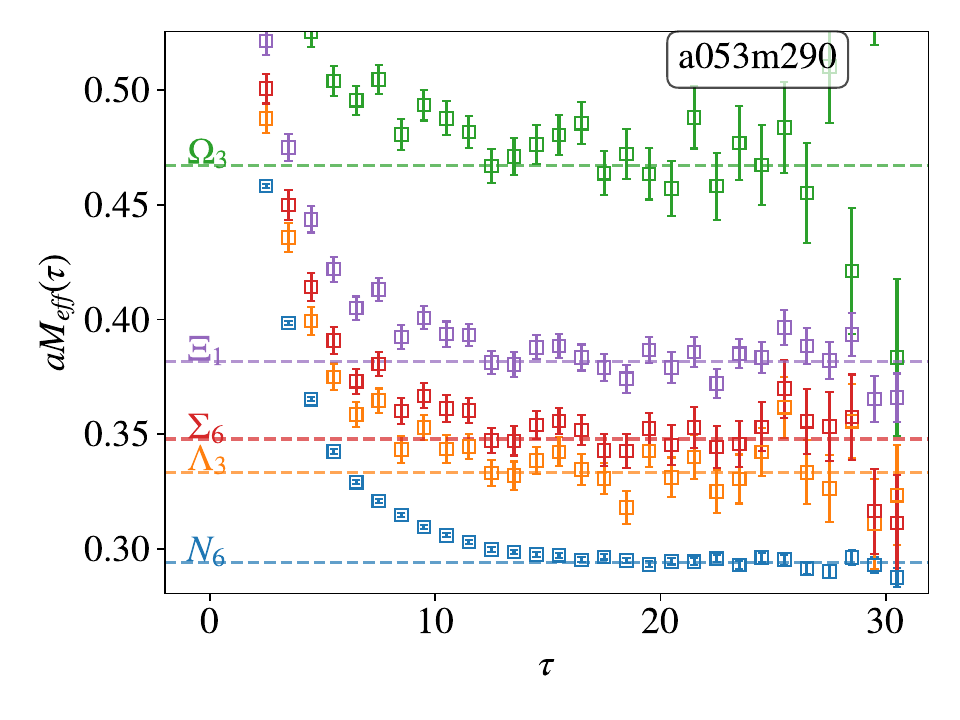} \hspace{0.5in}%
    \includegraphics[width=0.32\linewidth]{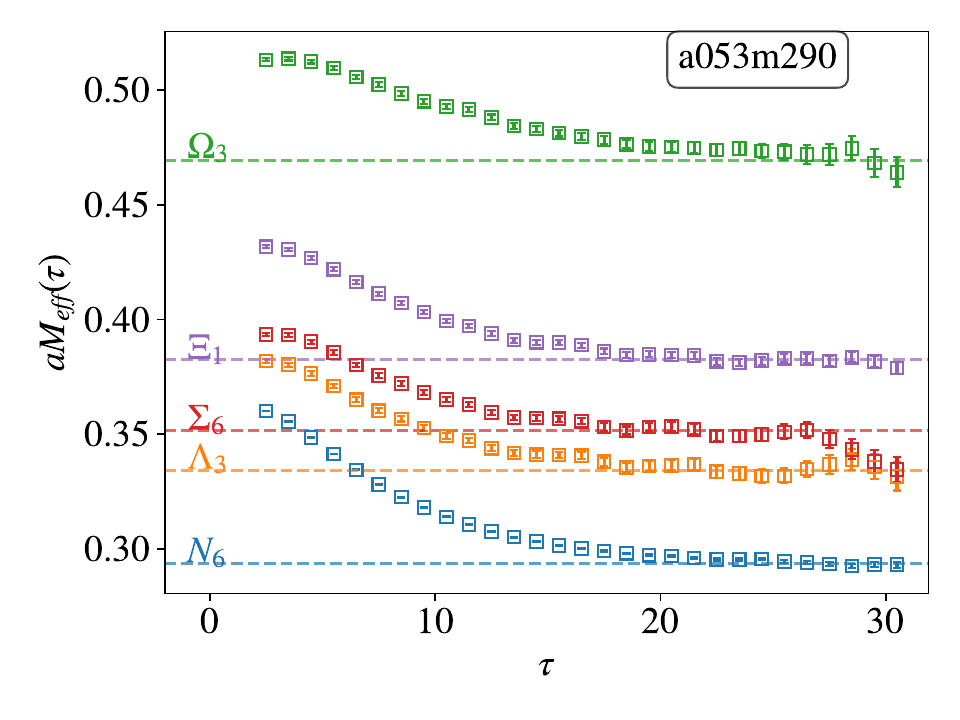}%
    \\
%
\vspace{-0.15in}
    \includegraphics[width=0.32\linewidth]{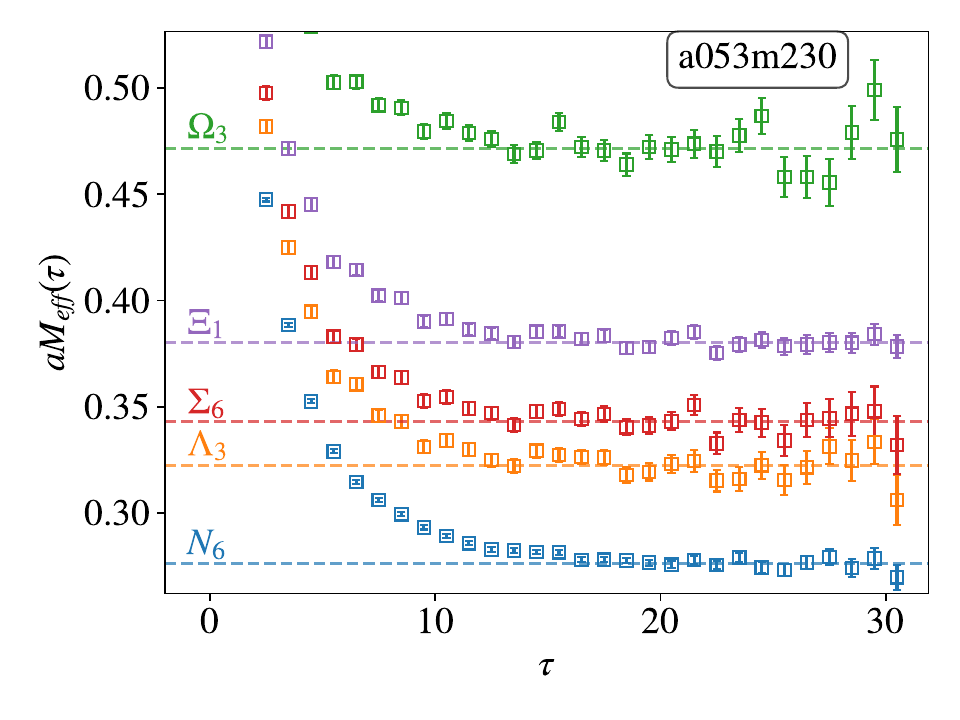} \hspace{0.5in}%
    \includegraphics[width=0.32\linewidth]{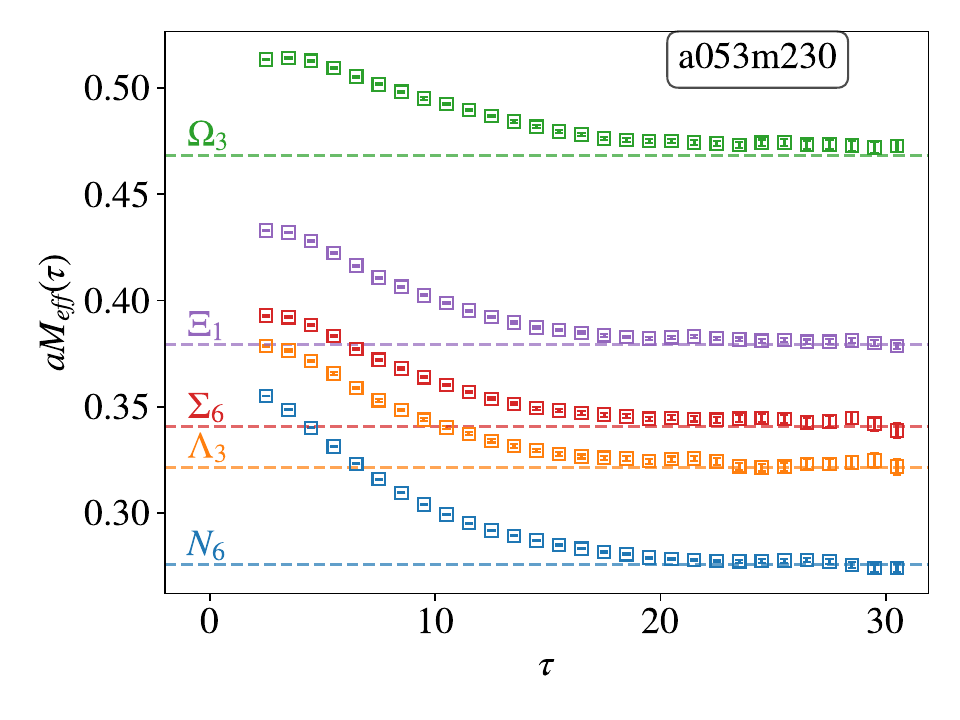}
%
\vspace{-0.2in}
    \caption{Effective mass plots for the five baryon states, $\{\Omega_3,\ \Xi_1, \ \Sigma_6,\ \Lambda_3, \ N_6 \}$, using the SS (left) and SP (right) correlators.  The labels specify the ensemble and the state, e.g., $\Omega_3$, specifies the interpolation operator used as defined in Table~\ref{tab:OP_list}. The horizontal dashed lines give the ground-state mass obtained 
    from 2-state fits.}
    \label{fig:meff_baryons_2}
\end{figure*}



\clearpage
\section{Meson and Baryon Interpolating Operators Used}
\label{sec:Operators}

The interpolating operators used to study the mesons and baryons are written as
\begin{equation}
    \chi_f^M = (\bar{q}_1 S q_2 ), \qquad\qquad  \chi_f^B = (\bar{q}_1 S q_2 ) q_3 \,.
\label{eq:opST}
\end{equation}
In terms of these, the two-point correlators are 
\begin{equation}
 C_{2pt}^M = \langle \chi_f \bar{\chi}_f \rangle  , \qquad\qquad  \hfill C_{2pt}^B = T \langle \chi_f \bar{\chi}_f \rangle
 \label{eq:2ptST}
\end{equation}
with the matrices $S$ and $T$ for the various operators are defined in Table~\ref{tab:OP_list}.
\end{widetext}

\newcommand{\gfour}{$\gamma_4$}
\newcommand{\Cgfive}{$ C \gamma_5$}
\newcommand{\CC}{$ C$}
\newcommand{\gminus}{$ \gamma_- $}

\begin{table}[]
    \centering
    \begin{tabular}{|l|c|c|}
    \hline 
State           & S                                   & T           \\
\hline
    $\pi_1$    &   $ \gamma_5 $                      &                                     \\
    $\pi_2$    &   $ \gamma_4\gamma_5 $              &                                     \\
    $\rho_1$    &   $\gamma_i$                        &                                     \\
    $\rho_2$    &   $  \gamma_4\gamma_i$              &                                     \\
    $N_1$       &   $ C\gamma_5$                      & $(1+\Sigma_3)\frac{1+\gamma_4}{2}$  \\
    $N_2$       &   $ C\gamma_5\gamma_4 $             &  $(1+\Sigma_3)\frac{1+\gamma_4}{2}$ \\
    $N_3$       &   $ C\gamma_5\frac{1+\gamma_4}{2}$  &  $(1+\Sigma_3)\frac{1+\gamma_4}{2}$ \\
    $N_4$       &   $ C\gamma_5$                      & $\frac{1+\gamma_4}{2}$              \\
    $N_5$       &   $ C\gamma_5\gamma_4 $             & $\frac{1+\gamma_4}{2}$              \\
    $N_6$       &   $ C\gamma_5\frac{1+\gamma_4}{2}$  &  $\frac{1+\gamma_4}{2}$             \\
        $\Delta_1$  &   $ C\gamma_-$                      &  $(1+\Sigma_3)\frac{1+\gamma_4}{2}$ \\
    $\Delta_2$  &   $ C\gamma_4\gamma_-$              &  $(1+\Sigma_3)\frac{1+\gamma_4}{2}$ \\
    $\Delta_3$  &   $ C\frac{1+\gamma_4}{2}\gamma_-$  &  $(1+\Sigma_3)\frac{1+\gamma_4}{2}$ \\
    $\Omega_1$  &   $ C\gamma_-$                      &  $(1+\Sigma_3)\frac{1+\gamma_4}{2}$ \\
    $\Omega_2$  &   $ C\gamma_4\gamma_-$              &  $(1+\Sigma_3)\frac{1+\gamma_4}{2}$ \\
    $\Omega_3$  &   $ C\frac{1+\gamma_4}{2}\gamma_-$  &  $(1+\Sigma_3)\frac{1+\gamma_4}{2}$ \\
\hline
$\Sigma_1$        & \Cgfive                     & ($1+\Sigma_3$)(1+\gfour)/2  \\
$\Sigma_2$        & \Cgfive  \gfour             & ($1+\Sigma_3$)(1+\gfour)/2  \\
$\Sigma_3$        & \Cgfive (1+\gfour)/2        & ($1+\Sigma_3$)*(1+\gfour)/2 \\
$\Sigma_4$        & \Cgfive                     & (1+\gfour)/2                \\
$\Sigma_5$        & \Cgfive  \gfour             & (1+\gfour)/2                \\
$\Sigma_6$        & \Cgfive (1+\gfour)/2        & (1+\gfour)/2                \\
$\Lambda_1$       & \Cgfive                     & ($1+\Sigma_3$)(1+\gfour)/2  \\
$\Lambda_2$       & \Cgfive  \gfour             & ($1+\Sigma_3$)(1+\gfour)/2 \\
$\Lambda_3$       & \Cgfive (1+\gfour)/2        & ($1+\Sigma_3$)(1+\gfour)/2 \\
$\Lambda_4$       & \Cgfive                     & (1+\gfour)/2                \\
$\Lambda_5$       & \Cgfive                     & ($1+\Sigma_3$)(1+\gfour)/2 \\
$\Xi_1$           & \Cgfive                     & (1+\gfour)/2                \\
$\Xi_2$           & \Cgfive                     & ($1+\Sigma_3$)(1+\gfour)/2 \\
$\Sigma^\ast_1$   & \CC \gminus                 & ($1+\Sigma_3$)(1+\gfour)/2. \\
$\Sigma^\ast_2$   & \Cgfive \gminus             & ($1+\Sigma_3$)(1+\gfour)/2 \\
$\Sigma^\ast_3$   & \Cgfive (1+\gfour)\gminus/2 & ($1+\Sigma_3$)(1+\gfour)/2 \\
$\Xi^\ast_1$      & \CC \gminus                 & ($1+\Sigma_3$)(1+\gfour)/2 \\
$\Xi^\ast_2$      & \CC \gfour \gminus          & ($1+\Sigma_3$)(1+\gfour)/2 \\
$\Xi^\ast_3$      & \CC (1+\gfour) \gminus/2    & ($1+\Sigma_3$)(1+\gfour)/2 \\
\hline
    \end{tabular}
    \caption{The  matrix $S$ in Eq.~\ref{eq:opST} used to define the meson and baryon interpolating 
    operators used in Eq.~\eqref{eq:2ptST}, and the projection matrix $T$ for the 2-point function. The pseudoscalar and vector mesons are 
    labeled, generically, by $\pi$ and $\rho$. Note that $C=\gamma_2 \gamma_4$, 
    $C \gamma_5 =  \gamma_1 \gamma_3$, $\Sigma_3 = - i \gamma_1 \gamma_2$ and $\gamma_- = \frac{1}{2}(\gamma_2 + i \gamma_1) $. These matrices are the same for the $\Delta$ and $\Omega$ states. }
    \label{tab:OP_list}
\end{table}

\clearpage
\bibliography{mybib}

\end{document}